%% file: main.tex
\documentclass[a4paper,11pt]{article}

\usepackage[utf8]{inputenc}
\usepackage[english]{babel}
\usepackage{amsmath}
\usepackage{auxi/jheppub}
\usepackage{amsfonts}
\usepackage{amssymb}
\usepackage{amsthm}
\usepackage{auxi/mhsetup}
\usepackage{auxi/mathtools}
\usepackage{graphicx}
\usepackage{pdfpages}
\usepackage{subfig}
\usepackage{multicol}
\usepackage{pdfpages}
\usepackage{float}
\usepackage[compat=1.1.0]{tikz-feynman} 
\usepackage{lscape}
\usepackage{bm}
\usepackage{bbm}

\usepackage{aligned-overset}
\usepackage{tikz}
\input{./auxi/commands.tex}

\newcommand{\be}{\begin{equation}}
\newcommand{\ee}{\end{equation}}
\newcommand{\beq}{\begin{eqnarray}}
\newcommand{\eeq}{\end{eqnarray}}
\newcommand{\bea}{\begin{eqnarray}}
\newcommand{\eea}{\end{eqnarray}}

\usepackage[T1]{fontenc}
\usepackage{slashed}

\newcommand{\CA}{\mathcal{A}}

\newcommand{\CD}{\mathcal{D}}
\newcommand{\CE}{\mathcal{E}}
\newcommand{\CF}{\mathcal{F}}

\newcommand{\CI}{\mathcal{I}}
\newcommand{\CJ}{\mathcal{J}}
\newcommand{\CK}{\mathcal{K}}
\newcommand{\CL}{\mathcal{L}}
\newcommand{\CM}{\mathcal{M}}
\newcommand{\CN}{\mathcal{N}}
\newcommand{\CO}{\mathcal{O}}
\newcommand{\CP}{\mathcal{P}}

\newcommand{\CX}{\mathcal{X}}

\newcommand{\g}{\gamma}

\DeclareMathOperator{\sech}{sech}
\DeclareMathOperator{\csch}{csch}

\newcommand\eq[1]{eq.~(\ref{eq:#1})}

\newcommand{\iu}{\text{i}}

\newcommand{\p}{\partial}

\renewcommand{\d}{\text{d}}

\title{Boundaries and defects in conformal field theories at finite temperature}

\input{./Tatreez/tatreez.tex}

\author[\motifTreeRed, \motifOliveGreen]{Daniele Artico,}
\author[\motifStarrRed, \motifBird]{Adam Chalabi}

\affiliation[\motifTreeRed]{Dipartimento SMFI, Universit\`a di Parma, Viale G.P. Usberti 7/A, 43100, Parma, Italy}
\affiliation[\motifOliveGreen]{INFN Gruppo Collegato di Parma}

\affiliation[\motifStarrRed]{Dipartimento di Fisica, Universit\`a di Torino, Via Pietro Giuria 1, 10125 Turin, Italy}
\affiliation[\motifBird]{INFN, Sezione di Torino}

\emailAdd{daniele.artico@unipr.it}
\emailAdd{adam.chalabi@unito.it}

\abstract{We consider conformal boundaries and defects in Euclidean conformal field theory (CFT) at finite temperature and infinite volume.
We distinguish defects that wrap the thermal circle and defects that are localised at a point on it. 
In both set-ups, and for any dimension and co-dimension, we classify bulk one-point functions of scalar, vector and spin-2 primaries.
We relate the low-temperature expansion of bulk scalar one-point functions to zero-temperature defect CFT data and new finite-temperature data of defect primaries.
For localised defects, we derive one-point bootstrap sum rules using periodicity and the defect operator expansion. 
We consider several examples in free bulk CFTs:
unitary and non-unitary boundary CFTs, a wrapping monodromy defect, and a localised magnetic line. 
We determine thermal defect CFT data in all examples.
For defects wrapping the thermal circle, we compute thermodynamic observables.
For localised defects, we observe a Hurwitz zeta function decomposition of one-point functions suggestive of an underlying dispersion relation.}

\begin{document}
\maketitle
\noindent

\section{Introduction}
\label{sec:intro}

Boundaries, impurities and finite temperature are ubiquitous features of real-world physical systems.  
Their interplay raises a basic question: 
\emph{how do spatial inhomogeneities modify a system's thermal properties}?
The answer to this question typically depends on microscopic physics.
At distances much larger than microscopic scales, many systems admit an effective description in terms of continuum quantum field theory (QFT) with microscopic details encoded in the couplings.
For systems at criticality, the continuum physics is described by a conformal field theory (CFT).
The leading large-distance behaviour is universal and constrained by conformal symmetry.
This provides a handle on the problem through symmetry, even when the underlying dynamics are strongly coupled.

At zero temperature, conformal symmetry provides an organising principle for the local observables of a CFT. 
Local operators form multiplets labelled by the scaling dimension and spin of their primaries~\cite{Mack:1975je}.
Their spectrum and three-point function coefficients -- the CFT data -- determine vacuum correlation functions through the convergent operator product expansion (OPE)~\cite{Mack:1976pa,Pappadopulo:2012jk}.
Consistency between different OPE channels imposes crossing equations on these data, forming the basis of the conformal bootstrap~\cite{Ferrara:1973yt,Polyakov:1974gs,Mack:1975jr}.
Together with unitarity, these equations impose non-perturbative constraints on the CFT data, which can be analysed numerically using positivity of the squared OPE coefficients~\cite{Rattazzi:2008pe,Poland:2011ey,Simmons-Duffin:2015qma} and analytically~\cite{Fitzpatrick:2012yx,Komargodski:2012ek,Caron-Huot:2017vep,Carmi:2019cub,Caron-Huot:2020adz,Mazac:2019shk}.
These methods constrain the space of consistent CFTs and, in cases such as the 3d Ising model, determine CFT data with remarkable precision~\cite{El-Showk:2012cjh,Kos:2016ysd}.

Bulk criticality alone does not determine the behaviour of a system near a boundary or impurity.
Boundary or impurity couplings flow under the renormalisation group (RG), even when the bulk remains at criticality.
The fixed points of these flows define distinct boundary or defect universality classes~\cite{Cardy:1984bb,Diehl:1996kd,Affleck:1995ge}.
When these fixed points are conformal, the resulting boundary or defect conformal field theory (B/DCFT) preserves a subgroup of the bulk conformal transformations.
This residual symmetry constrains correlation functions involving both bulk operators and local operators supported on the defect~\cite{McAvity:1993ue,McAvity:1995zd,Billo:2016cpy,Lauria:2018klo}.
Defect local operators organise into conformal multiplets and possess their own spectrum and OPE coefficients.
Their coupling to bulk operators is encoded in the convergent boundary or defect operator expansion (B/DOE), which expresses bulk operators in terms of defect primaries and their descendants~\cite{Cardy:1991tv,McAvity:1995zd,Billo:2016cpy,Lauria:2017wav}.
The bulk CFT data, defect spectrum, defect OPE coefficients and B/DOE coefficients constitute the B/DCFT data.
Together they determine local vacuum correlation functions in the presence of the boundary or defect.
Consistency between different operator expansions imposes bootstrap equations on these data~\cite{Liendo:2012hy,Billo:2016cpy}.
Unlike in ordinary CFT, however, the DCFT data entering some OPE channels need not be positive.
This limits positivity-based numerical methods, though some numerical studies have been carried out~\cite{Liendo:2012hy,Gaiotto:2013nva,Gliozzi:2015qsa}.
Analytic tools such as inversion formulas and dispersion relations are available~\cite{Lemos:2017vnx,Bissi:2018mcq,Mazac:2018biw,Kaviraj:2018tfd,Liendo:2019jpu,Bianchi:2022ppi,Barrat:2022psm}.
These methods constrain the spectra and couplings that distinguish boundary and impurity universality classes in a given bulk CFT.

A finite temperature introduces a length scale $\beta=1/T$ and reduces the symmetry of the equilibrium state.
Consider first a CFT without boundaries or defects on $S^1_\beta\times\mathbb{R}^{d-1}$, with imaginary time compactified to a circle of circumference $\beta$.
The scaling dimensions and OPE coefficients of local operators remain unchanged~\cite{El-Showk:2011yvt} while the broken conformal symmetries constrain thermal observables through modified Ward identities~\cite{Marchetto:2023fcw}.
Due to the additional scale $\beta$, non-zero thermal one-point functions are now allowed and fixed by translational and spatial rotational invariance up to a dimensionless coefficient.
Together with the zero-temperature CFT data, these coefficients determine thermal correlators at short distances through the OPE~\cite{El-Showk:2011yvt,Katz:2014rla}.
Periodicity around the thermal circle is encoded in the Kubo-Martin-Schwinger (KMS) condition~\cite{Kubo:1957mj,Martin:1959jp}.
Compatibility of this condition with the OPE imposes consistency equations on the thermal one-point coefficients, forming the basis of the thermal bootstrap programme~\cite{El-Showk:2011yvt,Iliesiu:2018fao}.
Similarly to DCFT, the coefficients entering these bootstrap equations need not be positive.
Analytic methods, including inversion formulas and dispersion relations~\cite{Iliesiu:2018fao,Petkou:2018ynm,David:2023uya,Manenti:2019wxs,Alday:2020eua,Marchetto:2023xap,Barrat:2025nvu,Barrat:2025twb,Barrat:2026jfg,Guo:2026xyl}, and numerical approaches~\cite{Iliesiu:2018zlz,Barrat:2025wbi,Niarchos:2025cdg} nonetheless constrain the thermal CFT data.
Among these is the stress-tensor one-point coefficient, which determines the energy density and pressure. 
It thus connects the thermal bootstrap directly to macroscopic thermodynamics.

We can now sharpen our opening question for systems at criticality:
\emph{how do the data of a conformal defect enter thermal observables, and which constraints do conformal symmetry and the KMS condition impose on them?}
Early investigations considered thermal effects of quantum fields in the presence of boundaries~\cite{Dowker:1978md,Kennedy:1979ar}.
For conformal defects, however, the orientation relative to the thermal circle introduces a further distinction.
Two simple configurations are defects wrapping the circle, which describe static impurities, and localised defects supported on a fixed time slice, which define extended operator insertions.
E.g.\ wrapped line defects describe impurities that are point-like in space and have been studied in quantum critical systems~\cite{Sachdev:1999,Vojta:2000tld}.
In $\CN=4$ super-Yang-Mills (SYM) theory, line defects in both configurations have been studied at strong coupling via the AdS/CFT correspondence~\cite{Witten:1998zw,Brandhuber:1998bs,Rey:1998bq}. 
These works, however, are model-specific and do not fully exploit the constraints of conformal symmetry and the KMS condition.

For the first configuration, ref.~\cite{Barrat:2024aoa} established a thermal defect bootstrap framework for line defects.
In this setting, bulk one-point functions depend non-trivially on the transverse distance measured in units of $\beta$, while defect primaries can acquire constant defect thermal one-point functions.
These thermal coefficients complement the zero-temperature DCFT data and determine the spatial dependence of bulk one-point functions through the DOE.
Combining this expansion with the bulk OPE and the KMS condition yields sum rules which constrain the thermal DCFT data.
Through the stress tensor one-point function, ref.~\cite{Barrat:2024aoa} also relate these data to the thermal energy, free energy and entropy.

\newpage

In this work, we extend the analysis of ref.~\cite{Barrat:2024aoa} to higher-dimensional defects wrapping the thermal circle as well as defects localised on it.
We formulate a one-point bootstrap for localised defects, and study both configurations in explicit free field examples.

\paragraph{Higher-dimensional defects wrapping the thermal circle. }
We consider conformal defects of arbitrary dimension and co-dimension.
Using the preserved symmetries, we classify the constant thermal one-point functions of defect-local operators.
We identify the allowed representations and the coefficients that define the new thermal DCFT data.
We then classify bulk scalar, vector and spin-2 one-point functions, paying special attention to parity-odd structures.
For conserved currents and the stress tensor, we derive additional constraints from conservation.
We derive a Euclidean inversion formula in the defect channel that extract the thermal DCFT data and discuss how the KMS condition on bulk two-point functions constrains these data.

\paragraph{Localised defects and a one-point function bootstrap. }
We extend the analysis to defects localised on a fixed Euclidean time slice.
We classify the thermal one-point functions of defect primaries and identify the corresponding thermal DCFT data.
For bulk operators, we classify scalar, vector and spin-2 one-point functions and derive a Euclidean inversion formula in the defect channel.
In this configuration, bulk one-point functions depend on the Euclidean time separation from the defect, so the KMS condition imposes non-trivial constraints already at the level of a single bulk insertion.
Combining this condition with the DOE yields novel bootstrap sum rules relating bulk-to-defect couplings and thermal one-point coefficients.
Their derivation requires assumptions about DOE convergence that we test in an explicit example.

\paragraph{Explicit examples and thermodynamics. }
We illustrate these results in free-field examples of increasing complexity:
scalar, fermion, hypermultiplet and higher-derivative scalar BCFTs, a monodromy defect wrapping the thermal circle, and a magnetic line defect localised on the thermal circle.
These examples realise the tensor structures allowed by the symmetry analysis. We extract thermal DCFT data from these correlators and comment on their analytic structure.
For the co-dimension two monodromy defect we observe that one-point functions have branch cuts in the complex plane, even for a non-interacting theory.
The magnetic line defect provides an explicit test of the KMS sum rules.
Its bulk one-point functions admit Hurwitz zeta function decompositions analogous to those of thermal two-point functions without defects~\cite{Barrat:2025nvu}.
This suggests an underlying dispersion relation for one-point functions of localised defects.
For defects wrapping the thermal circle, the stress tensor one-point function gives access to thermodynamic observables in the presence of boundaries or impurities.
We compare these quantities at the endpoints of boundary and defect RG flows.
In all examples studied, the free energies are negative and obey $(-\mathsf{F}^{UV})>(-\mathsf{F}^{IR})$.
This is reminiscent of unitary $1+1$d BCFT where the boundary contribution to the thermal free energy is $\mathsf{F}_\partial=-\beta^{-1}\log g$ and obeys $g^{UV}>g^{IR}$ under a boundary RG flow~\cite{Affleck:1991tk,Friedan:2003yc}.
\\

Our paper is organised as follows.
In section~\ref{sec:review} we fix our conventions and briefly review zero-temperature DCFT and thermal CFT.
Section~\ref{sec:Kinematics} develops the kinematics of both defect configurations.
We classify the one-point functions of defect and bulk primaries, identify the thermal DCFT data and derive Euclidean inversion formulas that extract them.
In section~\ref{sec:Bootstr} we turn to the constraints imposed by the KMS condition. 
We derive novel bootstrap sum rules for bulk one-point functions in the presence of defects localised on the thermal circle.
The remaining sections are devoted to examples.
Section~\ref{sec:ex_bdy} studies free BCFTs at finite temperature: free scalars, free fermions, a 4d hypermultiplet and a non-unitary higher-derivative scalar.
For each of them, we extract thermal BCFT data and compute thermodynamic quantities. 
Whenever possible, we compare the endpoints of boundary RG flows.
In section~\ref{sec:mon} we consider a monodromy defect wrapping the thermal circle in the theory of a free massless complex scalar.
This example is richer than the boundary ones.
It depends on a continuous parameter, and its one-point functions exhibit branch cuts in the complex plane.
We again extract thermal DCFT data, study thermodynamic quantities and compare their behaviour under defect RG flows.
Section~\ref{sec:ex_line} discusses a magnetic line defect localised on the thermal circle in the theory of a free massless scalar. 
We test our KMS sum rules and find signatures of a dispersion relation.
Finally we conclude in section~\ref{sec:Disc} with a discussion of open questions and future directions.
Some technical derivations and details are collected in three appendices.
Appendix~\ref{app:boots} contains a derivation of the one-point bootstrap sum rules that uses milder assumptions than the ones in the main text. 
In appendix~\ref{app:mon} we collect lengthy expressions for thermal one-point functions with a monodromy defect. 
Appendix~\ref{app:Borel} contains the detailed derivation of the Hurwitz zeta function decomposition of a scalar one-point function with a localised magnetic line.
\\

\emph{Note added:} 
While this manuscript was being completed ref.~\cite{Li:2026udh} appeared. 
It has substantial overlap with section~\ref{subsec:kinematics_wrapping} on the kinematics of defects wrapping the thermal circle and sections~\ref{sec:free_scalar} and~\ref{sec:free_fermion} on the free scalar and free fermion BCFTs, respectively.
Where our results overlap, they agree.
\\

\section{Review}
\label{sec:review}

In this section, we briefly review some aspects of conformal defects at zero temperature and of finite-temperature CFTs without defects.
We set out our notation and conventions here.

\subsection{Conformal defects at zero temperature}
\label{sec:DCFT}

Consider a (Euclidean) CFT at zero temperature, which we place on Euclidean space $\Rds^d$.
We will use Euclidean coordinates $x^\mu$ with $\mu=0,\ldots, d-1$.
In this theory we introduce a non-local extended operator $\CD$ with support on $\Rds^p\subset\Rds^d$.
The $p$ coordinates parallel to the defect are collectively denoted by $x^a_\parallel$, while the $d-p$ coordinates transverse to the defect are $x^i_\perp$.
The indices $a$ and $i$ run over parallel and transverse directions, respectively.
The defect sits at $x^i_\perp=0$.
The extended operator $\CD$ is a $p$-dimensional conformal defect if it breaks the conformal group of the $d$-dimensional CFT to a subgroup
\begin{equation}
\label{eq:defect_sym}
SO(d+1,1)\to SO(p+1,1)\times SO(q)\,,
\end{equation}
where $SO(p+1,1)$ is the conformal group in $p$ dimensions. 
In the present work we assume that all transverse rotations around the defect are preserved such that $q\equiv d-p$ is the co-dimension.\footnote{
This need not always be the case.
See e.g.\ ref.~\cite{Gukov:2006jk} for an example of a surface defect that breaks transverse rotations. }
The $q=1$ case includes conformal boundaries and interfaces. 

The presence of a conformal defect gives rise to several new features which were studied systematically in refs.~\cite{McAvity:1993ue,McAvity:1995zd, Billo:2016cpy,Lauria:2018klo}.
For instance, while scale and translational invariance constrain one-point functions in ordinary CFT to vanish, this is no longer the case in the presence of a conformal defect.
Since transverse translations are broken, the transverse distance to a defect $x_\perp^i$ supplies a scale that permits non-zero one-point functions.
The residual conformal symmetry fixes the zero-temperature one-point function of a scalar primary $\CO_\Delta$ with scaling dimension $\Delta$ to be
\begin{equation}
\label{eq:T0_1-pt_def}
\langle\CO_\Delta(x)\rangle=\frac{a_\CO}{|x_\perp|^\Delta}\,,
\end{equation}
where $|x_\perp|$ is the Euclidean norm of $x^i_\perp$ and $a_\CO$ is an operator-dependent dimensionless constant.\footnote{
Note that for $q=1$ our conventions differ from the ones of ref.~\cite{McAvity:1995zd} by a factor of $2^\Delta$, such that $a_\CO^\mathrm{here}=a_\CO^\mathrm{there}/2^\Delta$.
}
We will refer to it as the coefficient of the zero-temperature one-point function.
In \eq{T0_1-pt_def} normalisation by the partition function with defect insertion is understood,
\begin{equation}
\langle\CO_\Delta(x)\rangle_\CD\equiv\frac{\langle\CO_\Delta(x)\CD\rangle}{\langle\CD\rangle}\,.
\end{equation}
We will typically drop the subscript $\CD$ on the angled brackets for brevity, as in \eq{T0_1-pt_def}.

Unlike QFT in a Lorentz invariant vacuum, spinning operators can also acquire non-zero one-point functions in the presence of a defect due to the broken rotations mixing parallel and transverse directions.
For instance, a spin-2 operator $S^{\mu\nu}_\Delta$ can have the following one-point function in the presence of a co-dimension $q>1$ defect~\cite{Kapustin:2005py,Billo:2016cpy},
\begin{equation}
\label{eq:T_def_1-pt_fn}
\langle S^{\mu\nu}_\Delta
(x)\rangle
=
-a_S\frac{(q-1)\delta^{\mu\nu}
-d(
N^{\mu\nu}
-\hat{x}_\perp^\mu \hat{x}_\perp^\nu)}{|x_\perp|^\Delta}\,,
\end{equation}
where $N^{\mu\nu}$ is a projector onto the directions normal to the defect, and $\hat{x}_\perp^\mu=N^\mu{}_\nu x^\nu/|x_\perp|$.\footnote{
When $p$ is even and $S^{\mu\nu}_d$ is the stress tensor $T^{\mu\nu}$, $a_T$ has definite sign by the averaged null energy condition and is proportional to a defect Weyl anomaly coefficient. 
This has been shown rigorously for $p=2,4$ in refs.~\cite{Lewkowycz:2014jia,Bianchi:2015liz,Jensen:2018rxu,Chalabi:2021jud}, and is expected to hold more generally.}
For $q=1$ the normal projector becomes $N^{\mu\nu}=n^\mu n^\nu$, where $n_\mu$ is the unit normal. 
Since also $\hat{x}^\mu_\perp=n^\mu$, the one-point function vanishes identically.
More generally, the one-point function of any spinning operator vanishes when $q=1$~\cite{McAvity:1995zd}.

One-point functions of vectors $V^\mu_\Delta$ are more constrained. 
In fact, only when $q=2$ does one find a non-trivial one-point function
\begin{equation}
\label{eq:0T_V_1pt}
\langle V^{\mu}_\Delta
(x)\rangle=a_V\frac{n^\mu{}_\nu\hat{x}^\nu_\perp}{|x_\perp|^\Delta}\,,
\end{equation}
where $n_{\mu\nu}$ is the Levi-Civita tensor in the two transverse directions.\footnote{In polar coordinates in the transverse directions, $\langle V_\Delta^\theta\rangle=-a_V/|x_\perp|^{\Delta+1}$, with all other components vanishing.}
A vector one-point function therefore must be odd under parity in the transverse directions if it is to be non-zero.

Conformal defects not only modify correlation functions of bulk operators but they generically enlarge the spectrum of local operators of the theory.
In addition to the bulk operators of the CFT, there are defect-localised operators $\hat{\CO}$ transforming under the subgroup of $SO(d+1,1)$ preserved by the defect.
Operators are labelled by their scaling dimensions $\hat{\Delta}$, parallel $SO(p)$ spin $j$, and transverse $SO(q)$ spin $s$.
The spectrum of defect local operators in (unitary) DCFTs organises into primaries and descendants under $SO(p+1,1)$.

A prominent example of a defect local operator is the displacement operator $D^i$.
Since a conformal defect (generically) breaks translational invariance in the transverse directions, the Ward identity for translations is modified. 
The bulk stress tensor is no longer conserved at the location of the defect.
Instead,
\begin{equation}
\label{eq:broken_WI_disp}
\p_\mu T^{\mu i}(x)=\delta^{(q)}(x_\perp)\,D^i(x)\,,
\end{equation}
where $i$ runs over the $q$ transverse directions.
It is a defect scalar with protected scaling dimension $\hat{\Delta}=p+1$ and $s=1$.
Similarly, if a defect breaks a continuous global symmetry then the Ward identity for the Noether current is modified as follows
\begin{equation}
\label{eq:broken_WI_tilt}
\p_\mu J_\mathrm{a}^\mu(x)=\delta^{(q)}(x_\perp)\,t_\mathrm{a}(x)\,,
\end{equation}
where $\mathrm{a}$ is an index that runs over the broken generators of the symmetry group.
The operator $t_\mathrm{a}$ is the tilt operator~\cite{Drukker:2022pxk}.
It is a scalar under both parallel and transverse rotations, and has protected scaling dimension $\hat{\Delta}=p$.

The two-point functions of defect primaries $\hat{\CO}$ can be orthogonalised. 
If there is no way to fix the normalisation unambiguously, one often chooses to normalise operators such that the coefficient $c_{\hat{\CO}\hat{\CO}}$ of the two-point function is set to 1.\footnote{
Notably, this is not the case for the two defect primaries introduced in eqs.~\eqref{eq:broken_WI_disp} and~\eqref{eq:broken_WI_tilt}.
The broken Ward identities for broken transverse translations and broken global symmetries fix the normalisation of $D^i$ and $t_\mathrm{a}$ in terms of $T^{\mu\nu}$ and $J^\mu$, respectively.
}
In some of the examples in the present work, however, it will be convenient to not always do so.
Just as in ordinary CFT, conformal symmetry fixes the form of three-point function of defect primaries up to a single coefficient $\hat{\lambda}_{123}$.
Additionally, a defect local operator and a bulk local operator can have a non-trivial two-point function. 
For instance for a bulk scalar primary $\CO_\Delta$ and a defect scalar primary~$\hat{\CO}_{\hat{\Delta}}$,
\begin{equation}
\label{eq:zeroT_bulk_defect_2pt}
\langle\CO_\Delta(x)\hat{\CO}_{\hat{\Delta}}(y)\rangle=\frac{\mu_{\CO\hat{\CO}}}{|x_\perp|^{\Delta-\hat{\Delta}}(x-y)^{2\hat{\Delta}}}\,,
\end{equation}
where $y^i_\perp=0$ and the inner product $(x-y)^2$ is computed with the $d$-dimensional Euclidean metric.\footnote{
For $q=1$ our conventions for the bulk-boundary two-point function differ from those of ref.~\cite{McAvity:1995zd} by a factor of $2^{\Delta-\hat{\Delta}}$.
In particular, $\mu_{\CO\hat{\CO}}^\mathrm{here}=\mu_{\CO\hat{\CO}}^\mathrm{there}/2^{\Delta-\hat{\Delta}}$.}
The coefficient $\mu_{\CO\hat{\CO}}$ is an operator-dependent dimensionless constant.
Note that the one-point function~\eqref{eq:T0_1-pt_def} is a special case of \eq{zeroT_bulk_defect_2pt} where $\hat{\CO}$ is the defect identity $\hat{\mathbbm{1}}$.

Not all defect correlators are fixed by symmetry.
Already starting with the bulk-bulk two-point function, correlation functions can depend on dimensionless cross-ratios built out of the coordinates of the operator insertions.
Higher-point functions can be decomposed using the OPE of bulk local operators and the OPE of defect local operators.
Furthermore, the DOE allows for decomposing a bulk local operator in terms defect local operators. 
Schematically, for a bulk scalar primary
\begin{equation}
\label{eq:DOE}
\CO_\Delta(x_\parallel,x_\perp)=\sum_{\hat{\CO}}\frac{\mu_{\CO}{}^{\hat{\CO}}}{|x_\perp|^{\Delta-\hat{\Delta}}}\,\hat{x}_\perp^{i_1}\cdots \hat{x}_\perp^{i_s}\,\hat{\CO}_{\hat{\Delta},s}^{i_1\ldots i_s}(x_\parallel)+\ldots\,,
\end{equation}
where the sum is taken over all defect local primaries with only transverse spin.
Here $\mu_{\CO}{}^{\hat{\CO}}$ are the DOE coefficients, which are related to the coefficients $\mu_{\CO\hat{\CO}}$ defined in eq.~\eqref{eq:zeroT_bulk_defect_2pt} via $\mu_{\CO\hat{\CO}}=c_{\hat{\CO}\hat{\CO}}\mu_{\CO}{}^{\hat{\CO}}$. 
The contributions of defect descendants are contained in the ellipsis.
They are captured by a differential operator whose precise form will not be important for us.
At zero temperature, the DOE as well as the bulk and defect OPEs converge to the nearest other operator insertion~\cite{Mack:1976pa,Pappadopulo:2012jk,Billo:2016cpy,Lauria:2017wav}.
This can be shown by a standard radial quantisation argument.
Together with the spectrum and the coefficients of the bulk and defect three-point functions, the coefficients $\mu_{\CO}{}^{\hat{\CO}}$ make up the \emph{DCFT data}.
These data, however, are not independent.
Whenever a correlator can be decomposed in more than one way using the bulk and defect OPEs and the DOE, the DCFT data must satisfy a consistency condition, or bootstrap equation~\cite{Liendo:2012hy,Billo:2016cpy}. 
The simplest example is the two-point function of two bulk operators, which is a function of two conformal cross-ratios when $q>1$, and one cross-ratio when $q=1$.
It can either be computed by first using the bulk OPE and summing the resulting one-point functions, or by using the DOE of both operators. 
These bootstrap equations impose non-perturbative constraints on the DCFT data.

\subsection{CFTs at finite temperature}
\label{sec:thermal_CFT}

In many ways Euclidean finite-temperature CFTs without defects are similar to zero-temperature DCFT.
To put a CFT at finite temperature, we place it on the thermal manifold $\Mm_\beta=S^1_\beta\times\Rds^{d-1}$.
The Euclidean time direction is compactified to a circle of length $\beta=1/T$, where $T$ is the temperature.
We again use Euclidean coordinates $x^\mu=(x^0,\vec{x})$, where $\vec{x}$ denotes the spatial coordinates and $x^0\equiv \tau$ is the coordinate along the thermal circle.
Periodicity then implies that $\tau$ is compact with $\tau\sim\tau+\beta$.

Already in the absence of a defect, the compactification of Euclidean time explicitly breaks the conformal symmetry of the bulk theory. 
The Ward identities for dilatations, boosts and special conformal transformations are modified by thermal effects~\cite{Marchetto:2023fcw}.
The isometries of $\Mm_\beta$ are translations along $S^1_\beta$ and the (special) Euclidean group in $d-1$ dimensions, $ISO^+(d-1)$.\footnote{
The Euclidean group $ISO(d-1)$ is the semi-direct product of translations and rotations, including reflections.
We will not necessarily assume parity invariance here, so we restrict to the special Euclidean group $ISO^+(d-1)$ which is orientation preserving.}
There is an additional discrete symmetry of the theory:
simultaneously taking $\tau\to-\tau$ and performing a spatial reflection gives an $SO(d)$ transformation on the covering space $\Rds^d$.
E.g.\ $(\tau,x^1)\to(-\tau,-x^1)$ corresponds to a rotation by $\pi$ in the $(\tau,x^1)$-plane.
This additional $\mathbb{Z}_2$ symmetry descends to a symmetry of any thermal CFT, even if parity and Euclidean time reversal separately are not~\cite{Iliesiu:2018fao}.
The $ISO^+(d-1)$ symmetry is therefore enhanced to $ISO(d-1)$, where elements of $O(d-1)\subset ISO(d-1)$ that are not in the component connected to the identity also send $\tau\to-\tau$.

Similarly to zero-temperature DCFT, the additional scale $\beta$ allows local operators with non-zero scaling dimensions to acquire a one-point function~\cite{El-Showk:2011yvt}.
It is straightforward to see that they can only be non-zero for primaries.
Thermal one-point functions of descendants vanish by translational invariance since the one-point function is constant.
Moreover, only operators in symmetric-traceless representations with even $SO(d)$ spin $J$ can acquire a one-point function~\cite{Iliesiu:2018fao}.
Since the only available tensor structure is the unit vector $e^\mu$ along the thermal circle, only operators in symmetric $SO(d)$ representations can acquire one-point functions.
Such representations can be decomposed into symmetric traceless representations labelled by spin $J$.
Under the discrete $\mathbb{Z}_2$ symmetry, $e^\mu\to-e^\mu$, and so $J$ must be an even non-negative integer. 
The thermal one-point function of an even spin-$J$ primary then reads
\begin{equation}
\label{eq:no_defect_1-pt}
\langle \CO^{\mu_1\ldots\mu_J}_{\Delta}(x)\rangle_\beta = \frac{b_\CO}{\beta^\Delta}(e^{\mu_1}\cdots e^{\mu_J}-\text{traces})\,,
\end{equation}
where $b_\CO$ is a dimensionless operator-dependent constant.\footnote{
In a parity-breaking theory, a totally anti-symmetric rank-$d$ tensor can acquire a one-point function proportional to the Levi-Civita tensor $\epsilon_{\mu_1\ldots\mu_d}$.
Such an operator is Poincar\'e dual to a pseudo-scalar.
Similarly, pseudo-tensors with even spin $J$ can acquire a one-point function.
Naively, a totally anti-symmetric rank $(d-1)$ tensor also has the tensor structure $\epsilon_{\mu_1\ldots\mu_{d-1}\nu}e^\nu$ available.
However, it is excluded by the discrete $\mathbb{Z}_2$ symmetry. 
} 

Similarly to zero-temperature DCFT, the two-point function of bulk primaries in thermal CFT is no longer fixed by symmetry alone.
In addition to translational and rotational invariance, periodicity of Euclidean time imposes a constraint on the thermal correlation functions.
For the two-point function of two local operators $\CO_1$ and $\CO_2$, cyclicity of the thermal trace implies
\begin{equation}
\label{eq:KMS_no_defect}
\langle \CO_1(\tau,\vec x)\CO_2(0)\rangle_\beta
=
(-1)^{F_1F_2}
\langle \CO_2(\beta-\tau,\vec 0)\CO_1(0,\vec x)\rangle_\beta\,,
\end{equation}
where $0<\tau<\beta$.
Here $F_1=0$ ($F_1=1$) if $\CO_1$ has bosonic (fermionic) statistics, and similarly for $F_2$.
This is known as the Kubo--Martin--Schwinger (KMS) condition~\cite{Kubo:1957mj,Martin:1959jp}.
Note in particular that when the operators are identical the correlation function must be invariant under $\tau\to\beta-\tau$.

If the operators are sufficiently close together, one can use the zero-temperature OPE to compute the two-point function via a sum over thermal one-point functions~\cite{El-Showk:2011yvt,Katz:2014rla,Iliesiu:2018fao}. 
Since the OPE encodes short-distance physics, it is reasonable to expect it to apply on scales smaller than $\beta$.
In fact, a radial quantisation argument determines the radius of convergence to be exactly $\beta$~\cite{Iliesiu:2018fao}.
The OPE converges whenever both operators can be placed inside a sphere whose interior is flat and contains no other insertions~\cite{Pappadopulo:2012jk,Iliesiu:2018fao}.
The largest such sphere embedded in $\Mm_\beta$ without self intersection has radius $\beta/2$.
The largest distance between any two points inside the sphere is then given by the diameter $\beta$.
Using the OPE on both sides of the KMS condition for $\CO_1=\CO_2$ then gives a consistency condition that the zero-temperature coefficients of the CFT three-point functions and the thermal one-point functions in \eq{KMS_no_defect} need to satisfy~\cite{El-Showk:2011yvt,Iliesiu:2018fao}.
Therefore the KMS condition plays the role of a bootstrap equation for the thermal CFT two-point function.
It is similar to the crossing equation for the bulk two-point function in zero-temperature DCFT in that it does not involve positive-definite coefficients.
Strictly speaking, the coefficients $b_\CO$ of thermal one-point functions can be computed from the zero-temperature CFT data via the flat-space limit of thermal one-point functions on $S^1_\beta\times S^{d-1}$~\cite{El-Showk:2011yvt,Iliesiu:2018fao,Gobeil:2018fzy}.
In practice, however, this is difficult as it involves computing an infinite sum of three-point functions.
Instead it is useful to think of them as new data that can be constrained by a bootstrap problem.
We will refer to the operator spectrum together with the coefficients of the  zero-temperature three-point functions $\lambda_{123}$ and thermal one-point functions $b_\CO$ as \emph{thermal CFT data}.

\section{Kinematics of conformal defects at finite temperature}
\label{sec:Kinematics}

In this section we consider the kinematics of conformal defects at finite temperature.
We will focus on simple correlation functions of local operators in the presence of a single defect.

It was noted in ref.~\cite{Barrat:2024aoa} that there are three kinematically inequivalent ways of inserting a defect in a thermal CFT.
Requiring the $p$-dimensional defect to lift to $\Rds^p\subset\Rds^d$ in the decompactification limit $\beta\to\infty$ of $\Mm_\beta=S^1_\beta\times \mathbb{R}^{d-1}$, one finds the following configurations at finite $\beta$:
\begin{enumerate}
    \item[(a)] The defect wraps the thermal circle.
    \item[(b)] The defect is localised at a point on the thermal circle and extends only in the non-compact directions.
    \item[(c)] The defect is placed on a non-trivially embedded submanifold that has a component along the thermal circle and at least one component along a non-compact spatial direction. 
    E.g.\ for a line this could be a helix winding along the thermal cylinder.
\end{enumerate}
In this work we will discuss defects of type (a) and (b).
We will refer to them as wrapping and localised defects, respectively.
In a quantisation of the theory on constant $\tau$ slices, a wrapping defect of type (a) describes a static impurity, or a boundary when $q=1$, which modifies the Hilbert space on each spatial slice. 
A localised defect of type (b) on the other hand corresponds to an extended operator insertion in the thermal trace over the ordinary CFT Hilbert space. 

The two types of defects preserve different subgroups of the isometries of $\CM_\beta$.
A wrapping defect (a) preserves translations along the thermal circle and $ISO^+(p-1)\times SO(q)$ rotations.
Similarly to thermal CFT without defects, there is an additional discrete $\mathbb{Z}_2$ symmetry.
Since sending $\tau\to-\tau$ accompanied by a spatial reflection along the defect amounts to an $SO(p)$ rotation, $ISO^+(p-1)$ is enhanced to $ISO(p-1)$, irrespective of whether the defect is time-reversal or parity invariant.
An element of $O(p-1)\subset ISO(d-1)$ that is not in the component connected to the identity also acts by reversing $\tau$.
A localised defect (b), on the other hand, preserves $ISO^+(p)\times O(q-1)$, where an orientation reversing element in $O(q-1)$ also acts by $\tau\to-\tau$. 
These descend from genuine $SO(q)$ rotations around the defect.
Translations along the thermal circle are broken by the presence of the defect.

The combination of the conformal defect and finite temperature modifies the CFT's Ward identities for the broken symmetries.
In the case of a line defect wrapping the thermal $S^1_\beta$, ref.~\cite{Barrat:2024aoa} found modified Ward identities for broken translations and dilatations.
Their derivation generalises straightforwardly to higher-dimensional defects.
In particular, the broken translation Ward identity~\eqref{eq:broken_WI_disp} does not receive finite-temperature contributions, while for the dilatation Ward identity the displacement contribution vanishes.
Generators that are broken simultaneously by the defect and by the thermal compactification, such as rotations mixing the thermal and transverse directions and transverse special conformal transformations, receive both displacement and thermal contributions.\footnote{
\emph{Note added:} Ref.~\cite{Li:2026udh} provides an elegant alternative derivation of the broken Ward identities for defects of arbitrary dimension $p$ at finite temperature.
Instead of relying on a Lagrangian formulation of the DCFT like ref.~\cite{Barrat:2024aoa}, the argument of ref.~\cite{Li:2026udh} is based on topological operators that implement conformal transformations. 
Their derivation is independent of the orientation of the defect.
}

In the remainder of this section we discuss the kinematics of simple correlation functions in the presence of a defect wrapping the thermal $S^1_\beta$ or a defect localised on a point on it.
We classify the structures appearing in the thermal one-point functions of bulk scalar, vector and spin-2 primaries for either configuration, and clarify which defect local operators can acquire thermal one-point functions.
Additionally we discuss the properties of simple correlators under KMS.
For wrapping defects of type (a), this extends the analysis of ref.~\cite{Barrat:2024aoa} to $p>1$.
For localised defects of type (b), this classification is new.
While for a wrapping defect the simplest correlation function that obeys a non-trivial KMS condition is the two-point function, we will show that for localised defects already the bulk one-point function is non-trivially constrained.

\subsection{Defects wrapping the thermal circle}
\label{subsec:kinematics_wrapping}

Consider a defect wrapping $S^1_\beta$ and extending along $\Rds^{p-1}\subset \Rds^{d-1}$.
We use Euclidean coordinates $x^\mu=(\bm{x}_\parallel,\vec{x}_\perp)$.
The coordinates along the defect are denoted by $\bm{x}_\parallel=(\tau,\vec{x}_\parallel)$, where we use boldface when including $\tau$ and a vector arrow to label the spatial coordinates only.
The spatial transverse directions are then denoted $\vec{x}_\perp$.

\subsubsection{Thermal one-point functions}
\label{sec:wrapping_1pt}

We now study thermal one-point functions of bulk operators with low spin in the presence of the defect.
The symmetries preserved by the defect alone are translations along the thermal circle and $ISO(p-1)\times SO(q)$.
With an extra operator $\CO$ in the bulk, the configuration breaks these symmetries to the stabiliser $O(p-1)\times SO(q-1)$.
Again, an orientation reversing rotation in $O(p-1)$ also acts by reversing the sign of $\tau$.
The stabiliser determines which tensor structures can appear in the one-point function of $\CO$.
Generically, they must be built out of $\hat{x}^\mu_\perp$, $e^\mu$, $g_{\mu\nu}$ and $N_{\mu\nu}$.
Here, $N_{\mu\nu}$ is the projector onto transverse spatial directions.\footnote{
One can also define a projector $h_{\mu\nu}$ onto parallel spatial directions.
However, since $g_{\mu\nu}=e_\mu e_\nu + h_{\mu\nu} + N_{\mu\nu}$ one can express any structure built using $h_{\mu\nu}$ in terms of the quantities listed in the main text.}
Since $-\mathbbm{1}\in O(p-1)$ sends $e^\mu\to-e^\mu$, any tensor structure for general $p$ can only contain an even number of $e^\mu$.
Care is required for special values of $p$ and $q$.
When $p=1$, the $O(p-1)$ factor is absent, and so tensor structures containing an odd number of $e^\mu$ are no longer excluded.
When $p=2$, there is an additional distinguished spatial direction parallel to the defect, and so correlation functions can depend on the unit normal $t_\mu$.
When $q=1$, the transverse space is parametrised by a single coordinate $x^\mu_\perp=|x_\perp| \,n^\mu$, where $n_\mu$ is the unit normal such that $N_{\mu\nu}=n_\mu n_\nu$.
This reduces the number of independent structures.
Conversely, for given $p$ and $q$ there can be additional dimension-dependent parity breaking structures.
These are built out of the bulk and transverse space Levi-Civita tensors $\epsilon_{\mu_1\ldots\mu_d}$ and $n_{\mu_1\ldots\mu_q}$, respectively.

We will now enumerate all admissible tensor structures for the thermal one-point function of a bulk operator with spin up to $J=2$.

\paragraph{Scalars. }
First consider a scalar primary $\CO_\Delta$ with scaling dimension $\Delta$.
Its zero-temperature one-point function is given by \eq{T0_1-pt_def}.
At finite temperature, $\beta$ provides an additional length scale, such that one can construct the dimensionless ``cross-ratio''
\begin{equation}
z\equiv\frac{|\vec{x}_\perp|}{\beta}\,.
\end{equation}
The one-point function of $\CO_\Delta$ can then depend on a function $F_\CO(z)$,
\begin{equation}
\label{eq:kin_O}
\langle \CO_\Delta(\bm{x}_\parallel,\vec{x}_\perp)\rangle_\beta=\frac{F_\CO(z)}{\beta^\Delta}\,.
\end{equation}
Note that the thermal one-point function of a bulk descendant need not vanish in the presence of a defect.
The function $F_\CO$ can be constrained using kinematic limits.
In the zero-temperature limit, $\beta\to\infty$, the one-point function should reduce to the zero-temperature one-point function~\eqref{eq:T0_1-pt_def}.
At large distances $x_\perp\to\infty$ the effect of the impurity becomes negligible. 
The physics should be dominated by thermal effects which supply the no-defect thermal one-point function~\eqref{eq:no_defect_1-pt}.
The kinematic regimes $z\ll1$ and $z\gg1$ therefore suppress thermal and defect effects, respectively. 
Matching to \eq{T0_1-pt_def} and \eq{no_defect_1-pt} in those regimes then requires
\begin{align}
\label{eq:FO_asymp}
F_\CO(z)&\overset{z\to0}{\sim}\frac{a_\CO}{z^\Delta}\,,& F_\CO(z)&\overset{z\to\infty}{\to}b_\CO\,.
\end{align}

The function $F_\CO$ admits a useful decomposition under the DOE of $\CO_\Delta$.
At zero-temperature the DOE of a scalar is given by \eq{DOE} and has infinite radius of convergence in the absence of other operator insertions.
At finite temperature, the DOE takes the same form as in \eq{DOE} (with $|x_\perp| \to |\vec{x}_\perp|$ since all transverse coordinates are spatial in this configuration).
Thermal effects, however, change its convergence properties. 
A radial quantisation argument similar to the one for the OPE presented below \eq{KMS_no_defect} determines the radius of convergence of the DOE to be $\beta/2$.
This is the radius of the largest sphere with flat interior that can be embedded in $\CM_\beta$.
If the sphere is centred on a point on the defect, then the farthest a bulk operator can be while remaining inside the sphere is $\beta/2$.
Provided $|\vec{x}_\perp|<\beta/2$, and there are no other operators closer to the defect, we can use the DOE inside \eq{kin_O}.

At finite temperature, a defect local operator may also acquire a one-point function.
By translational invariance along the defect, it must be a constant. 
It follows that only defect primaries can have a non-vanishing one-point function.
Due to rotational invariance in the transverse directions to the defect, there is no tensor structure that can be used to construct a one-point function for operators with transverse spin $s>0$.
Therefore any defect primary with $s>0$ must have vanishing defect thermal one-point function.
Similarly to the bulk case in \eq{no_defect_1-pt}, only defect primaries in symmetric traceless $SO(p)$ spin-$j$ representations can generically acquire a one-point function.
Since $-\mathbbm{1}\in O(p-1)$ sends $e^\mu\to-e^\mu$, $j$ must be an even non-negative integer.
The defect thermal one-point function of an $SO(p)$ spin-$j$ primary with $s=0$ then reads
\begin{equation}
\label{eq:defect_thermal_1pt}
\langle\hat{\CO}^{a_1\ldots a_j}_{\hat{\Delta}}(\bm{x}_\parallel)\rangle_\beta=\frac{\hat{b}_{\hat{\CO}}}{\beta^{\hat{\Delta}}}(e^{a_1}\cdots e^{a_j}-\text{traces})\,,
\end{equation}
where $\hat{b}_{\hat{\CO}}$ is a dimensionless operator-dependent constant.
Together with the zero-temperature DCFT data and the no-defect thermal one-point functions $b_\CO$, the coefficients $\hat{b}_{\hat{\CO}}$ are part of the \emph{thermal DCFT data}.\footnote{Similarly to the no-defect bulk coefficients $b_\CO$, the coefficient $\hat{b}_{\hat{\CO}}$ of the defect thermal one-point function can in principle be computed from the zero-temperature CFT data.
This, however, requires knowledge of an infinite amount of zero-temperature DCFT data, and it is more practical to think of the $\hat{b}_{\hat{\CO}}$ as new data~\cite{Iliesiu:2018fao,Barrat:2024aoa}.
In particular, the thermal one-point function of $\hat{\CO}$ considered here can be obtained by computing it on $S^1_\beta\times S_R^{d-1}$ and then taking the infinite (spatial) volume limit $R\to\infty$.
At finite $R$, the one-point function is a Boltzmann-weighted sum over the diagonal matrix elements of $\hat\CO$ in the impurity Hilbert space. 
Thus, $\hat{b}_{\hat{\CO}}$ can be computed with all zero-temperature defect three-point function coefficients of the form $\hat{\lambda}_{\hat{\CO}\hat{\CO}'\hat{\CO}'}$ as input.
}
When $p=2$ there is an additional permissible structure that includes the tangent vector $t^\mu$ along the single spatial direction parallel to the defect,
\begin{equation}
\label{eq:defect_thermal_1pt_p2}
\langle\hat{\CO}^{a_1\ldots a_j}_{\hat{\Delta}}(\bm{x}_\parallel)\rangle_\beta=\frac{\hat{b}^{e}_{\hat{\CO}}}{\beta^{\hat{\Delta}}}(e^{a_1}\cdots e^{a_j}-\text{traces})+\frac{\hat{b}^{t}_{\hat{\CO}}}{\beta^{\hat{\Delta}}}(e^{(a_1}\cdots e^{a_{j-1}} t^{a_j)}-\text{traces})\,.
\end{equation}
Note that tensor structures with more than one factor of $t^a$ are not linearly independent since $t^at^b=\delta^{ab}-e^ae^b$.
For a co-dimension $q$ defect, also defect primaries in the totally anti-symmetric transverse $SO(q)$ representation with any admissible $j$ can acquire a one-point function.
These necessarily come with a factor of the transverse Levi-Civita tensor $n_{\mu_1\ldots\mu_q}$ and are therefore odd under parity in the transverse directions.\footnote{
Similarly to the no-defect case, also totally anti-symmetric $SO(p)$ tensors can acquire a  one-point function proportional to $\epsilon_{\mu_1\ldots\mu_p}=\epsilon_{\mu_1\ldots\mu_d}n^{\mu_{p+1}\ldots\mu_d}$.
This structure breaks parity on the defect, and the corresponding operator is a pseudo-scalar. 
Pseudo-tensors with even parallel spin $j$ can similarly get a one-point function.}

The DOE~\eqref{eq:DOE} of $\CO_\Delta$ contains symmetric traceless spin-$s$ operators.
Using it inside \eq{kin_O}, only defect scalar primaries with $s=0$ contribute.
We can then organise the thermal one-point function into an expansion over zero-temperature defect conformal families, or ``thermal blocks''.
The function $F_\CO$ is then given by the following expression
\begin{equation}
\label{eq:FO_DOE}
F_\CO(z)=\sum_{\hat{\CO}}\mu_{\CO}{}^{\hat{\CO}}\hat{b}_{\hat{\CO}}z^{\hat{\Delta}-\Delta}\,,
\end{equation}
where the series converges for $z<1/2$ and the sum runs over defect scalar primaries with $s=0$ in the DOE of $\CO_\Delta$. 

In some cases $F_\CO$ can be computed but finding its expansion into thermal blocks may not be immediate.
To extract the coefficients of the thermal block expansion~\eqref{eq:FO_DOE}, one can make use of the following Euclidean inversion formula,
\begin{equation}
\label{eq:wrapping_inversion}
\mu_{\CO}{}^{\hat{\CO}}\hat{b}_{\hat{\CO}}=-\Res_{\bar{\Delta}\to\hat{\Delta}-\Delta}\int_0^{\varepsilon} \d z \, z^{-{\bar{\Delta}}-1}\,F_\CO(z)\,.
\end{equation}
Here, $\Res_{{\bar{\Delta}}\to\hat{\Delta}{-\Delta}}$ denotes the residue of the integral at ${\bar{\Delta}}=\hat{\Delta}{-\Delta}$, and $\varepsilon>0$.
The residue is in fact independent of $\varepsilon$, and it is convenient to choose $\varepsilon<1/2$ such that the integral is over a range in which the DOE converges.
This formula was originally found by ref.~\cite{Barrat:2024aoa} for line defects.
It applies to a defect of any dimension $p$ that wraps the thermal circle without modifications.

\paragraph{Vectors. }
Next consider a vector $V^\mu_\Delta$ with scaling dimension $\Delta$.
We find (at most) three admissible tensor structures, which we list in table~\ref{tab:vector_one_point_structures}.
The entry in the second row is only admissible when $p=1$, since $e^\mu\to-e^\mu$ under $-\mathbbm{1}\in O(p-1)$ when $p>1$.
The third entry is parity-odd as it changes sign under orientation reversal of the transverse directions.
The coefficient of each structure is a function of $z$.
When $\beta\to\infty$, such that $z\to0$, the full thermal one-point function must reduce to the zero-temperature one-point function~\eqref{eq:0T_V_1pt}.
When $|\vec{x}_\perp|\to\infty$ instead, such that $z\to\infty$, the whole one-point function must vanishes since in the absence of a defect $\langle V^\mu\rangle_\beta=0$.
These requirements give rise to the conditions on the coefficient functions summarised in the last two columns of table~\ref{tab:vector_one_point_structures}.

Naively, one could have expected an additional structure when $p=2$.
In that case there is a single spatial direction parallel to the defect, whose unit vector is $t^\mu$.
However, under $-\mathbbm{1}\in O(1)\cong \mathbb{Z}_2$ the unit vector $t^\mu$ changes sign together with $e^\mu$.
Thus a structure proportional to $t^\mu$ alone cannot appear.

\begin{table}[t]
\centering
\begin{tabular}{|c|c|c|c|}
\hline
tensor structure
&
exists for
&
$z\to0$
&
$z\to\infty$
\\
\hline
\rule{0pt}{5ex}

$\displaystyle
\frac{\hat{x}_\perp^\mu}{{\beta^{\Delta}}}
F^\perp_V(z)$
&
all $p,q$
&
$F^\perp_V{\sim z^{>-\Delta}}$
&
$F^\perp_V{\to0}$
\\[4mm]

$\displaystyle
\frac{e^\mu}{\beta^\Delta}F^\tau_V(z)$
&
$p=1$ only
&
$\displaystyle
F^\tau_V\sim z^{>-\Delta}$
&
$F^\tau_V\to 0$
\\[4mm]

$\displaystyle
\frac{n^\mu{}_\nu \hat{x}_\perp^\nu}
{{\beta^{\Delta}}}
F^{q=2}_V(z)$
&
$q=2$ only
&
$F^{q=2}_V{\sim a_V z^{-\Delta}}$
&
$F^{q=2}_V{\to0}$
\rule[-3ex]{0pt}{3ex}
\\
\hline
\end{tabular}
\caption{
Tensor structures appearing in the one-point function of a vector in the presence of a defect wrapping the thermal circle.
}
\label{tab:vector_one_point_structures}
\end{table}

If the vector is a conserved current, $V^\mu_{d-1}=J^\mu$, the conservation equation constrains the first entry in table~\ref{tab:vector_one_point_structures} only.
Away from the defect, conservation implies
\begin{equation}
    F^\perp_J(z)
    =
    c^\perp_J z^{{1-q}}\,,
\end{equation}
Comparison with the requisite large-distance $z\to\infty$ asymptotics in table~\ref{tab:vector_one_point_structures} sets $c^\perp_J=0$ for $q=1$.
If $J^\mu$ is a current for a global symmetry broken by the defect, one can use the distributional identity
\begin{equation}
\label{eq:distrib}
\partial_\mu \left(\frac{\hat{x}^\mu_\perp}{|\vec{x}_\perp|^{q-1}}\right)=\frac{2\pi^\frac{q}{2}}{\Gamma\left(\frac{q}{2}\right)}\,\delta^{(q)}(\vec{x}_\perp)
\end{equation}
to show that $c^\perp_J$ is related to the defect thermal one-point function of the tilt operator $t$.
In particular, using the broken Ward identity~\eqref{eq:broken_WI_tilt} together with the form of the defect thermal one-point function given by \eq{defect_thermal_1pt} one finds 
\begin{equation}
\hat{b}_t=\frac{2\pi^\frac{q}{2}}{\Gamma\left(\frac{q}{2}\right)}\,c^\perp_J\,.
\end{equation}
If the symmetry is preserved, then $c^\perp_J=0$ also when $q>1$.
The remaining two structures in table~\ref{tab:vector_one_point_structures} are separately conserved.
The full thermal one-point function of a conserved current is then
\begin{equation}
    \label{eq:kin_conserved_J}
    \langle J^\mu(\bm{x}_\parallel,\vec{x}_\perp)\rangle_\beta
    =
   c^\perp_J
    \frac{x_\perp^\mu}
    {|\vec{x}_\perp|^q\beta^p} + \delta_{p,1}\,
    \frac{e^\mu}{\beta^{d-1}}F^\tau_J(z)
    +
    \delta_{q,2}\,
    \frac{n^\mu{}_\nu \hat{x}_\perp^\nu}
    {{\beta^{d-1}}}
    F^{q=2}_J(z)\,,
\end{equation}
where $c^\perp_J=0$ when the defect preserves the global symmetry. 
When $q=1$ the thermal one-point function vanishes identically.\footnote{
\emph{A priori} one would expect the structure proportional to $e^\mu$ to persist for a boundary in a 2d CFT on the cylinder.
A defect wrapping the thermal circle, however, can be mapped by a conformal transformation to the plane with a straight line defect.
This sets one-point functions of any spinning (Virasoro) primary to zero.}

\paragraph{Spin-2. }
Finally consider a spin-two operator $S^{\mu\nu}_\Delta$.
Its thermal one-point function is built out of seven different tensor structures, which cannot all appear simultaneously.
We list all manifestly symmetric traceless structures in table~\ref{tab:spin_two_one_point_structures}.
Note that when $p=1$, the metric decomposes as $g_{\mu\nu}=e_\mu e_\nu+N_{\mu\nu}$ such that the first two structures, with $F_S^{\tau\tau}$ and $F^{\perp\perp}_{S,1}$, become linearly dependent.
Similarly when $q=1$, $\hat{x}^\mu_\perp$ reduces to the unit normal $n^\mu$ along the single transverse direction. 
The normal projector $N^{\mu\nu}$ meanwhile becomes $n^\mu n^\nu$, and so the second and third structures, with $F^{\perp\perp}_{S,1}$ and $F^{\perp\perp}_{S,2}$, become identical.
When $p=2$, one may have expected an additional structure involving $t^\mu t^\nu$.
However, this is nothing but the parallel spatial projector $h_{\mu\nu}=g_{\mu\nu}-e_\mu e_\nu-N_{\mu\nu}$, and so it would not give a new linearly independent structure.
Consistency with the zero-temperature and large-distance behaviour imposes constraints on these functions.
In the $\beta\to\infty$ limit, such that $z\to0$, one should recover the zero-temperature one-point function of a spin-two operator given by \eq{T_def_1-pt_fn}.
Conversely, for $|\vec{x}_\perp|\to\infty$, such that $z\to\infty$, the thermal one-point function should reduce to the no-defect result \eq{no_defect_1-pt}.
This fixes the behaviour of the functions as listed in the last two columns of table~\ref{tab:spin_two_one_point_structures}.

\begin{table}[t]
\centering
\small
\setlength{\tabcolsep}{6pt}
\renewcommand{\arraystretch}{1.3}
\begin{tabular}{|c|c|c|c|}
\hline
tensor structure
&
exists for
&
$z\to0$
&
$z\to\infty$
\\
\hline
\rule{0pt}{5ex}

$\displaystyle
\frac{e^\mu e^\nu-\frac1d g^{\mu\nu}}
{\beta^\Delta}
F^{\tau\tau}_S(z)$
&
all $p,q$
&
$F^{\tau\tau}_S\sim z^{>-\Delta}$
&
$F^{\tau\tau}_S\to b_S$
\\[4mm]

$\displaystyle
\frac{N^{\mu\nu}-\frac qd g^{\mu\nu}}
{{\beta^\Delta}}
F^{\perp\perp}_{S,1}(z)$
&
all $p,q$
&
$F^{\perp\perp}_{S,1}{\,\sim  d\,a_S\,z^{-\Delta}}$
&
$F^{\perp\perp}_{S,1}{\,\to 0}$
\\[4mm]

$\displaystyle
\frac{
\hat{x}_\perp^\mu \hat{x}_\perp^\nu
-\frac1d g^{\mu\nu}}
{{\beta^\Delta}}
F^{\perp\perp}_{S,2}(z)$
&
all $p,q$
&
$F^{\perp\perp}_{S,2}{\,\sim  -d\,a_S\,z^{-\Delta}}$
&
$F^{\perp\perp}_{S,2}{\,\to 0}$
\\[4mm]

$\displaystyle
2\frac{e^{(\mu}\hat{x}_\perp^{\nu)}}
{{\beta^\Delta}}
F^{\tau\perp}_S(z)$
&
$p=1$ only
&
$F^{\tau\perp}_S{\,\sim z^{>-\Delta}}$
&
$F^{\tau\perp}_S{\,\to 0}$
\\[4mm]

$\displaystyle
2\frac{e^{(\mu}t^{\nu)}}{{\beta^\Delta}}
F^{\tau\parallel}_S(z)$
&
$p=2$ only
&
$F^{\tau\parallel}_S{\,\sim z^{>-\Delta}}$
&
$F^{\tau\parallel}_S{\,\to 0}$
\\[4mm]

$\displaystyle
2\frac{\hat{x}_\perp^{(\mu}n^{\nu)}{}_\rho \hat{x}_\perp^\rho}
{{\beta^\Delta}}
F^{q=2}_{S,1}(z)$
&
$q=2$ only
&
$F^{q=2}_{S,1}{\,\sim z^{>-\Delta}}$
&
$F^{q=2}_{S,1}{\,\to 0}$
\\[4mm]

$\displaystyle
2\frac{e^{(\mu}n^{\nu)}{}_\rho \hat{x}_\perp^\rho}
{{\beta^\Delta}}
F^{q=2}_{S,2}(z)$
&
$p=1$, $q=2$ only
&
$F^{q=2}_{S,2}{\,\sim z^{>-\Delta}}$
&
$F^{q=2}_{S,2}{\,\to 0}$
\rule[-3ex]{0pt}{3ex}
\\
\hline
\end{tabular}
\caption{
Tensor structures appearing in the one-point function of a spin-2 primary in the presence of a defect wrapping the thermal circle.
}
\label{tab:spin_two_one_point_structures}
\end{table}

If the spin-2 operator is the stress tensor, $S^{\mu\nu}_d=T^{\mu\nu}$, the $d$ conservation equations lead to non-trivial constraints on several of the functions in table~\ref{tab:spin_two_one_point_structures}.
The first conservation equation $\p_\mu T^{\mu\tau}=0$ is trivially satisfied for $p>1$. 
When $p=1$, the conservation equation away from the defect implies $F^{\tau\perp}_T(z)=c^{\tau\perp}_T\, z^{2{-d}}$ for some integration constant $c^{\tau\perp}_T$.
This is compatible with the requisite asymptotic behaviour when $d>2$.
The distributional identity~\eqref{eq:distrib} then implies that $\partial_\mu T^{\mu\tau}\propto c^{\tau\perp}_T/\beta^2\times\delta^{(d-1)}(\vec{x}_\perp)$.
However, the defect preserves translations along the thermal circle and the corresponding Ward identity is not modified.
Therefore, $c^{\tau\perp}_T=0$ and the structure with $F^{\tau\perp}_T$ is absent for the stress tensor.
Conservation along the spatial directions of the defect is trivial for all $p$.
Conservation along the transverse directions, however, leads to non-trival constraints. For the parity-even sector, it becomes the following first-order ordinary differential equation (ODE)
\begin{equation}
\label{eq:ODE}
\frac{\d}{\d z}F^{\tau\tau}_T
=
(d-q)\frac{\d}{\d z}
F^{\perp\perp}_{T,1}
+
d(q-1)z^{-1}F^{\perp\perp}_{T,2}
+
(d-1)\frac{\d}{\d z}
F^{\perp\perp}_{T,2}\,.
\end{equation}
This equation generalises the ODE found by ref.~\cite{Barrat:2024aoa} for $p=1$.\footnote{
In the case of a line defect, the functions $f_1(z)$ and $f_2(z)$ in ref.~\cite{Barrat:2024aoa} are related to ours as follows:
\begin{align*}
f_1
&=
\frac{{z^d}}{d}
\left(
F^{\perp\perp}_{T,1}
-
F^{\perp\perp}_{T,2}
-
F^{\tau\tau}_T
\right),
&
f_2&=-\frac{{z^d}}{2}F^{\perp\perp}_{T,2}\,.
\end{align*}
}
In the parity-odd sector, the conservation equation away from the defect implies $F^{q=2}_{T,1}=c_T^{q=2}/z^{2}$.
Distributionally, however, the divergence $\partial_\mu T^{\mu\nu}|_{\mathrm{odd}}\propto c_T^{q=2}/\beta^{d-2} \times n^{\nu\rho}\partial_\rho\delta^{(2)}(\vec{x}_\perp)$.\footnote{
This can be seen, for instance, by noting that $n^{\mu\nu}\hat{x}_\nu/|\vec{x}_\perp|=n^{\mu\nu}\partial_\nu \log|\vec{x}_\perp|$ such that $N^{\mu\nu}\partial_\mu\partial_\nu\log|\vec{x}_\perp|\propto \delta^{(2)}(x_\perp)$.}
If the broken translation Ward identity is to take the form given in \eq{broken_WI_disp} also at finite temperature, then $c_T^{q=2}=0$.
The remaining structures with $F^{\tau\parallel}_T$ and $F^{q=2}_{T,2}$ are identically conserved.

Putting everything together, the thermal one-point function of the stress tensor for general $p$ and $q$ reads
\begin{align}
\label{eq:T_def}
\langle T^{\mu\nu}(\bm{x}_\parallel,\vec{x}_\perp)\rangle_\beta
={}&
\frac{e^\mu e^\nu-\frac{1}{d}g^{\mu\nu}}{\beta^{d}}
F^{\tau\tau}_T(z)
+
\frac{N^{\mu\nu}-\frac{q}{d}g^{\mu\nu}}
{{\beta^d}}
F^{\perp\perp}_{T,1}(z)
+
\frac{\hat{x}_\perp^\mu \hat{x}_\perp^\nu
-\frac{1}{d}g^{\mu\nu}}
{{\beta^d}}
F^{\perp\perp}_{T,2}(z)\nonumber\\
&+2\delta_{p,2}\,\frac{e^{(\mu}t^{\nu)}}{{\beta^d}}
F^{\tau\parallel}_T(z)
+2\delta_{p,1}\delta_{q,2}\,\frac{e^{(\mu}n^{\nu)}{}_\rho \hat{x}_\perp^\rho}
{{\beta^d}}
F^{q=2}_{T,2}(z)\,,
\end{align}
where the functions have the asymptotic properties listed in table~\ref{tab:spin_two_one_point_structures} and obey \eq{ODE}.
Note that when $p=1$, we can eliminate one of $F^{\tau\tau}_{T}$ and $F^{\perp\perp}_{T,1}$ since they become linearly dependent.

We will now briefly comment on the special case $q=1$ corresponding to a boundary or interface.
When $q=1$ the tensor structures involving $F^{\perp\perp}_{T,1}$ and $F^{\perp\perp}_{T,2}$ become identical.
The ODE~\eqref{eq:ODE} can then be solved immediately to find
\begin{equation}
F^{\tau\tau}_T(z)=(d-1)F^{\perp\perp}_{T}(z)+c^{q=1}_T\,,
\end{equation}
where $F^{\perp\perp}_{T}=F^{\perp\perp}_{T,1}+F^{\perp\perp}_{T,2}$.
Compatibility with the asymptotic behaviour summarised in table~\ref{tab:spin_two_one_point_structures} fixes $c^{q=1}_T=b_T$, such that for $q=1$
\begin{equation}
\label{eq:kin_T}
\begin{split}
\langle T^{\mu\nu}
(\bm{x}_\parallel,x_\perp)\rangle_\beta
={}&
\frac{e^\mu e^\nu-\frac{1}{d}g^{\mu\nu}}
{\beta^d}\,b_T
+
\frac{(d-1)e^\mu e^\nu+n^\mu n^\nu-g^{\mu\nu}}
{{\beta^d}}
F^{\perp\perp}_{T}(z)
\\
&+
\delta_{p,2}\,
2\frac{e^{(\mu}t^{\nu)}}{\beta^d}
F^{\tau\parallel}_T(z)\,.
\end{split}
\end{equation}
Here $b_T$ is the thermal stress tensor one-point coefficient of the bulk CFT on the side containing the insertion, evaluated in the absence of the interface.
Concretely, $b_T=b_T^L$ for $x_\perp<0$ corresponding to the one-point function of $T^{\mu\nu}_L$ in the theory on the left of the interface, and $b_T=b_T^R$ for $x_\perp>0$ corresponding to the one-point function of $T^{\mu\nu}_R$ in the theory on the right.
Note that $\langle T^{\perp\perp}(x)\rangle_\beta$ is independent of $z$ and proportional to $b_T$ only.
The displacement operator is given by the difference of the two stress tensors on either side, $D(\bm{x}_\parallel)=T_R^{\perp\perp}(\bm{x}_\parallel,0)-T_L^{\perp\perp}(\bm{x}_\parallel,0)$.
Therefore, the defect thermal one-point function of the displacement operator is
\begin{equation}
\hat b_D=
-\frac{b^{R}_T-b^{L}_T}{d} \,,
\end{equation}
The boundary case with the BCFT on $x_\perp>0$ is obtained by setting $b^L_T=0$ in the formula above.
Note that for higher co-dimension $q>1$, the displacement carries transverse spin $s=1$, and so its defect thermal one-point function vanishes.

\subsubsection{Thermal two-point functions and KMS}

As discussed in section~\ref{sec:thermal_CFT}, the KMS condition imposes constraints on thermal correlation functions.
This is also the case in the presence of an impurity. 
A defect wrapping the thermal circle modifies the Hilbert space on spatial slices without inserting any additional operators inside the thermal trace.
Cyclicity of the thermal trace then gives rise to constraints on correlation functions in the presence of an impurity.

For thermal one-point functions these constraints turn out to be trivial.
The simplest correlator that is non-trivial under KMS is the two-point function of scalar operators.
If the scalars have bosonic statistics, the general KMS condition for two scalars reads
\begin{equation}
\label{eq:KMS_wrapping_2pt_general}
\left\langle
\CO_1\!\left(\frac{\beta}{2}+\tau,\vec{x}_{1\parallel},
\vec{x}_{1\perp}\right)
\CO_2\!\left(\bm{0}_\parallel,\vec{x}_{2\perp}\right)
\right\rangle_\beta
=
\left\langle
\CO_2\!\left(\frac{\beta}{2}-\tau,\vec{0}_\parallel,\vec{x}_{2\perp}\right)
\CO_1\!\left(0,\vec{x}_{1\parallel},
\vec{x}_{1\perp}\right)
\right\rangle_\beta .
\end{equation}
Here we used translational invariance along the thermal circle and the $p-1$ non-compact Eq.~\eqref{eq:KMS_wrapping_2pt_general} holds irrespective of whether $\CO_{1,2}$ are bulk or defect operators.
Now take both operators to be identical $\CO_1=\CO_2$ at equal transverse positions $\vec{x}_{1\perp}=\vec{x}_{2\perp}=\vec{x}_{\perp}$.
Since by translational invariance the correlator only depends on parallel spatial position through $|\vec{x}_{1\parallel}|$, the KMS condition equates the correlators at separations $\beta/2+\tau$ and $\beta/2-\tau$ on the thermal circle.\footnote{
Here we implicitly assume $p\neq2$. 
When $p=2$, there is a single spatial direction along the defect and so the correlator can depend on the sign of $x_{1\parallel}$.
For parity-breaking defects we must require $x_{1\parallel}=0$ to equate the correlators at $\beta/2+\tau$ and $\beta/2-\tau$.}
This imposes a non-trivial constraint on the correlation function.

For a line defect the configuration with $|\vec{x}_\perp|>0$ and the resulting constraints were studied in ref.~\cite{Barrat:2024aoa}.
In that case all spatial directions are transverse to the defect and so the correlator is a function of two variables, $\tau$ and $|\vec{x}_{\perp}|$.
When $p>1$, the correlator can also depend on $|\vec{x}_{1\parallel}|$, and so it is a function of three variables.

\subsection{Defects localised on the thermal circle}
\label{subsec:kinematics_localised}

Next consider a $p$-dimensional defect $\mathcal D$ that only extends along the non-compact directions of the thermal manifold $\CM_\beta$.
Such a defect is localised at a point on the thermal circle.
Without loss of generality we place it at $\tau=0$ and $\vec{x}_\perp=0$, where $\vec{x}_\perp$ labels the $q-1$ spatial transverse directions.
We use $\bm{x}_\perp=(\tau,\vec{x}_\perp)$ to collectively denote all transverse directions including the thermal circle.
The $p$ directions along the defect are denoted $\vec{x}_\parallel$.

When quantising on spatial slices, a defect $\CD$ at fixed $\tau$ corresponds to an extended operator acting on the CFT Hilbert space.
Such a defect should be interpreted as a non-local operator rather than an impurity that deforms the Hilbert space.
To emphasise this, we will keep the dependence on $\CD$ explicit and write
\begin{equation}
\label{eq:localised_XD}
\langle\mathcal{X}\,\mathcal{D}\rangle_\beta\,,
\end{equation}
for some collection of local operators $\mathcal{X}$.
Normalisation by $\langle\CD\rangle_\beta$ is implicit.
In the following subsection we will take $\CX$ to be a single local operator.
Since the correlator depends only on one bulk point, we will refer to the correlator in \eq{localised_XD} as a thermal one-point function.

\subsubsection{Thermal one-point functions and KMS}
\label{sec:localised_1pt_KMS}

To determine the tensor structures that can appear in the one-point function, we need to consider the stabiliser of this configuration.
At finite temperature, the defect preserves the (special) Euclidean group of $p$ dimensions as well as rotations in the $q-1$ transverse spatial directions, $ISO^+(p)\times O(q-1)$.
Orientation reversing elements of $O(q-1)$ also act by sending $\tau\to-\tau$.
Translations along the thermal circle are instead broken by the defect insertion.
For a bulk point at generic $\tau$ and $\vec{x}_\perp\neq0$, the stabiliser is $SO(p)\times SO(q-2)$.
Note that an orientation reversing $O(q-1)$ transformation also reflects the bulk point around $\tau=0$, and so the transverse rotation stabiliser is $SO(q-2)$ rather than $O(q-2)$.
It is enhanced to $O(q-2)$ if the point has $\tau=0$ or $\tau=\beta/2$, which are fixed points under $\tau\to-\tau$.
Moreover, if the bulk point is separated from the defect only along the thermal circle, i.e.\ $\vec{x}_\perp=0$, then the stabiliser is enhanced to $SO(p)\times SO(q-1)$.
The $SO(q-1)$ factor is further enhanced to $O(q-1)$ when $\tau=\beta/2$, or if the point is on the defect $\tau=0$.

The tensor structures in thermal one-point functions can then be built out of $\hat{x}^\mu_\perp$, $e^\mu$, $g_{\mu\nu}$, and $\tilde{N}_{\mu\nu}$, where $\tilde{N}_{\mu\nu}$ is a projector onto the spatial transverse directions not including $\tau$ and $\hat{x}_\perp^\mu=\tilde{N}^\mu{}_\nu x^\nu/|\vec{x}_\perp|$.
The norm $|\vec{x}_\perp|$ only includes the transverse spatial coordinates.
For given $p$ and $q$, one also has the bulk and transverse Levi-Civita tensors $\epsilon_{\mu_1\ldots\mu_d}$ and $\tilde{n}_{\mu_1\ldots\mu_{q-1}}=e^{\mu_q}n_{\mu_1\ldots\mu_q}$.
Note that for a generic bulk point the configuration is no longer fixed by a transverse reflection accompanied by $\tau\to-\tau$.
Thus the one-point function need not be invariant under $e^\mu\to-e^\mu$ unless $\tau=0,\beta/2$.

For special values of the co-dimension, the transverse part of the stabiliser simplifies. 
When $q=1$, the defect extends over the whole spatial $\Rds^{d-1}$.\footnote{
Typically we will consider a co-dimension one defect $\CD$ rather than a boundary here.
Imposing boundary conditions at $\CD$ would lead to a CFT on a strip of length $\beta$.
Though the kinematics is the same, the physics is different.
Instead of capturing thermal effects, this set-up describes the physics of the Casimir effect.}
Since there are no non-compact directions transverse to it, the vector $\hat{x}_\perp^\mu$ and spatial normal projector $\tilde{N}_{\mu\nu}$ are unavailable.
The tensor structures in the thermal one-point function should then become analogous to the no-defect case.
When $q=2$, there is a distinguished normal direction with unit vector $n_\mu$.
This is analogous to the co-dimension one case discussed in section~\ref{sec:wrapping_1pt}.
More generally, the  structures appearing for a co-dimension $q$ localised defect on the thermal circle are similar to the co-dimension $q-1$ wrapping defect.
However due to the different discrete symmetries the details will be different.
Note also that for $p=1$, the direction parallel to the defect is distinguished, and so one-point functions can depend on the unit vector $t^\mu$ along it.

\paragraph{Scalars. }
The simplest thermal one-point function is that of a scalar primary $\CO_\Delta$.
By invariance under the preserved translations and rotations, it must be a function of the following \emph{two} ``cross-ratios''
\begin{align}
\label{eq:localised_defect_cross_ratios}
w&=\frac{\tau}{\beta}\,,
&
z&=\frac{|\vec{x}_\perp|}{\beta}\,.
\end{align}
While the cross-ratio for the non-compact transverse distance takes the range $z\geq0$, the one for the distance along the thermal circle has periodicity equal to 1. 
We will restrict its range to $0<w<1$.
The thermal one-point function of a scalar then becomes
\begin{equation}
\langle\CO_\Delta(x)\CD\rangle_\beta=\frac{F_\CO(w,z)}{\beta^\Delta}\,.
\end{equation}
For $q=1$, the cross-ratio $z$ is absent and $F_\CO$ only depends on $w$.

The $O(q-1)$ symmetry of the defect for $q\geq2$ imposes constraints on the function $F_\CO$.
An orientation reversing $O(q-1)$ transformation flips the sign of $\tau$ and a spatial transverse coordinate.
Since the correlator only depends on the transverse spatial coordinates through $z$, the spatial part is left invariant. 
The reflection of $\tau$, however, reverses the order of the operators in the correlation function since they are placed under Euclidean time ordering,
\begin{equation}
\label{eq:localised_discrete_sym}
\left\langle
\CO_\Delta(\tau,\vec{x}_\parallel,\vec{x}_\perp)\mathcal D(0)
\right\rangle_\beta
=
\left\langle
\mathcal D(0)\CO_\Delta(-\tau,\vec{x}_\parallel,\vec{x}_\perp)
\right\rangle_\beta\,.
\end{equation}
Here we have used that $\CO$ is a scalar and transforms trivially under an $SO(d)$ rotation.
Since \eq{localised_discrete_sym} is a thermal correlator with two operator insertions a version of the KMS condition~\eqref{eq:KMS_no_defect} applies.
In particular we can use KMS to bring the transformed correlator~\eqref{eq:localised_discrete_sym} back to its original ordering,
\begin{equation}
\label{eq:KMS_localised_ordered}
\left\langle
\CO_\Delta(\tau,\vec{x}_\parallel,\vec{x}_\perp)\mathcal D(0)
\right\rangle_\beta
=
\left\langle
\CO_\Delta(\beta-\tau,\vec{x}_\parallel,\vec{x}_\perp)\mathcal D(0)
\right\rangle_\beta\,,
\end{equation}
where $0<\tau<\beta$.
This is the KMS condition for the thermal one-point function of a scalar with a defect localised on the thermal circle.
It follows directly from cyclicity of the thermal trace.
Thus, KMS implies the following reflection property
\begin{equation}
\label{eq:KMS_localised_reflection}
F_\CO(w,z)
=
F_\CO(1-w,z)\,.
\end{equation}
Note that we did not assume any additional symmetries.
In particular, \eq{KMS_localised_reflection} holds also for theories that are not time-reversal invariant.

In the above we implicitly assumed that $q>2$.
For $q=2$, the defect separates a spatial slice in two halves such that we can take $z=x_\perp/\beta$ rather than its absolute value.
The one-point function can then depend on the sign of $z$, indicating the side on which the local operator is inserted.
The element $-\mathbbm{1}\in O(1)\cong\mathbb{Z}_2$ then flips the sign of both $w$ and $z$ such that the KMS condition~\eqref{eq:KMS_localised_reflection} becomes $F_\CO(w,z)=F_\CO(1-w,-z)$.
When $q=1$, there is no $O(q-1)$ symmetry, and there is no analogue of \eq{KMS_localised_reflection} in general.
$F_\CO(w)=F_\CO(1-w)$ only holds for theories invariant under time-reversal.

The asymptotics of the function $F_\CO$ are fixed by comparing with the zero-temperature one-point function~\eqref{eq:T0_1-pt_def} in the $\beta\to\infty$ limit and the no-defect thermal one-point function~\eqref{eq:no_defect_1-pt} when $|\vec{x}_\perp|\to\infty$.\footnote{
The second condition assumes that the effect of the defect becomes negligible at large spatial separation.
In explicit examples in free CFTs we do not always find this to be the case. 
In particular, in section~\ref{sec:ex_line} we discuss a conformal line defect in the theory of a free massless scalar in $d=4$.
In this theory, the one-point function of certain bulk operators grows logarithmically at large distances.
However, with interactions, or in theories where a mass scale is dynamically generated, the large-distance behaviour is fixed by the no-defect thermal one-point function as described in the main text.}

It is useful to introduce radial and angular variables in the $q$ directions transverse to the defect,
\begin{align}
\label{eq:localised_defect_cross_ratios_polar}
\varrho&=\frac{r}{\beta}=\sqrt{w^2+z^2}\,,
&
\eta&=\frac{\tau}{r}=\frac{w}{\varrho}\,.
\end{align}
Then the zero-temperature limit $\beta\to\infty$ sends $\varrho\to0$ with fixed $\eta$.
Conversely, the no-defect $|\vec{x}_\perp|\to\infty$ limit sends $z\to\infty$ with $w$ fixed.
Therefore,
\begin{align}
\label{eq:localised_scalar_asymp}
F_\CO(w,z)
    &\overset{\varrho\to0}{\sim}\frac{a_\CO}{\varrho^\Delta}\,,
&
F_\CO(w,z)
    \overset{z\to\infty}{\longrightarrow}
   b_\CO\,.
\end{align}
The case $q=1$ is special.
Since the only transverse direction is compact there is no large-distance limit at fixed temperature.
Moreover, since the the defect intersects the thermal circle, the defect can be approached from two sides.
Thus there are two ways in which the $\beta\to\infty$ limit can be taken.
For a point $\tau=\epsilon$, where $\epsilon$ is fixed, the $\beta\to\infty$ limit sends $w\to0^+$.
However, for a point $\tau=\beta-\epsilon$ close to the other side of the defect, $w\to1^-$.
Unlike for higher co-dimension defects these two limits need not give the same answer.\footnote{
A simple example is a defect that cuts open the thermal circle.
This is equivalent to studying a CFT on a strip of width $\beta$.
For a free massless scalar, one could imagine imposing D BCs at $w=0$ and N BCs at $w=1$.
The two $\beta\to\infty$ limits then send one end or the other to infinity. 
}
In particular one can have
\begin{align}
\label{eq:localised_scalar_q1_limits}
    F_\CO(w)
    &\overset{w\to0^+}{\sim}
    \frac{a_\CO^+}{w^\Delta}\,,
&
    F_\CO(w)
    &\overset{w\to1^-}{\sim}
    \frac{a_\CO^-}{(1-w)^\Delta}\,,
\end{align}
where $a_\CO^+$ and $a_\CO^-$ are the zero-temperature one-point-function coefficients on the two sides of the defect.
If the theory is invariant under time reversal $\tau\to-\tau$, then $a_\CO^+=a_\CO^-$ since the $w\to1^-$ limit is equivalent to $w\to0^-$.

The function $F_\CO$ can be decomposed using the DOE.
It takes the same form as in \eq{DOE} with $x_\perp$ replaced by $\bm{x}_\perp=(\tau,\vec{x}_\perp)$ since now $\tau$ is one of the transverse directions.
The operators $\hat{\CO}^{i_1\ldots i_s}_{\hat{\Delta},s}$ appearing in the DOE~\eqref{eq:DOE} are defect primaries with scaling dimension $\hat{\Delta}$ and spin $s$ under the zero-temperature transverse rotation group $SO(q)$, as before.
The $i$ indices run over all $q$ transverse directions, including $\tau$.
At zero temperature the radius of convergence of the DOE is infinite in the absence of other operator insertions.
At finite temperature, however, a radial quantisation argument similar to the one presented in section~\ref{sec:wrapping_1pt} gives a radius of convergence of $\beta/2$.
Note that $\beta/2$ is in some sense a lower bound.
As we will see in a simple example in section~\ref{sec:ex_line}, the DOE may converge beyond $\beta/2$ to the nearest thermal image of the defect at distance $\beta$.

Before using the DOE to decompose $F_\CO$, let us briefly discuss which defect local operators can acquire one-point functions for a localised defect.
Given that the defect preserves $SO(p)$, there is no available tensor structure for an operator with parallel spin $j>0$ to acquire a defect thermal one-point function.\footnote{
As for the wrapping defects considered in section~\ref{sec:wrapping_1pt}, totally antisymmetric rank-$p$ representations are an exception.
These are again pseudo-scalars on the defect, and their one-point functions are proportional to $\epsilon_{\mu_1\ldots\mu_p}=\epsilon_{\mu_1\ldots\mu_d}n^{\mu_{p+1}\ldots\mu_d}$.}
For transverse spin, on the other hand, there are two building blocks, $e^\mu$ and $n_{\mu_1\ldots\mu_q}$.
This gives rise to three types of tensor structures:
The first one is built out of $e^\mu$ alone, corresponding to symmetric traceless primaries.
Since $-\mathbbm{1}\in O(q-1)$ flips a transverse direction together with $\tau$, the transverse spin $s$ needs to be even.
The defect thermal one-point then takes the following form
\begin{equation}
\label{eq:localised_defect_primary_one_point}
    \left\langle
    \hat\CO_{\hat\Delta}^{\,i_1\ldots i_s}(x)\CD
    \right\rangle_{\beta}
    =
    \frac{
    \hat b_{\hat\CO}
    }{
    \beta^{\hat\Delta}
    }
    \Pi_s^{\,i_1\ldots i_s}(e)\,,
\end{equation}
where the coefficient $\hat{b}_{\hat{\CO}}$ is an operator dependent constant.
The tensor structure $\Pi_s^{\,i_1\ldots i_s}(e)=\CN\,(e^{\mu_1}\cdots e^{\mu_s}-$traces$)$ for some normalisation $\CN$ that we will fix shortly.
Note that a non-zero thermal one-point function in the presence of a localised defect requires $j=0$, whereas the wrapped configuration of section~\ref{sec:wrapping_1pt} requires $s=0$ for symmetric traceless representations.
Operators that are scalars under both $SO(p)$ and $SO(q)$ can acquire thermal one-point functions in either configuration.
We use the same notation $\hat{b}_{\hat{\CO}}$ for their coefficients.
However, since they characterise different thermal configurations they need not coincide.\footnote{
As for wrapping defects, $\hat b_{\hat\CO}$ can in principle be computed from zero-temperature data, but the relevant data are different. 
Place the theory on $S^1_\beta\times S^{d-1}_R$, with $\CD$ at $\tau=0$ wrapping a great sphere $S^p\subset S^{d-1}_R$, and take $R\to\infty$ at the end. 
Since $\CD$ does not wrap the thermal circle, the trace now runs over the ordinary CFT Hilbert space. 
The defect with $\hat\CO$ inserted acts on this space as an operator at $\tau=0$. 
At finite $R$ the one-point function is therefore a Boltzmann-weighted sum over the diagonal matrix elements $\langle\CO'|\,\hat\CO\CD\,|\CO'\rangle$ in bulk states. 
By the state-operator map, each matrix element is a zero-temperature bulk-bulk-defect three-point function $\langle\CO'\,\hat\CO\,\CO'\rangle_{\CD}$. 
Thus $\hat b_{\hat\CO}$ can be obtained from these correlators if they are known for all bulk operators $\CO'$.
The correlators are fixed by bulk-to-defect couplings $\mu_{\CO'}{}^{\hat{\CO}'}$ and defect three-point coefficients $\hat{\lambda}_{\hat{\CO}\hat{\CO}'\hat{\CO}''}$, where both $\hat{\CO}'$ and $\hat{\CO}''$ are in the DOE of $\CO'$.}

The remaining two admissible tensor structures are odd under parity in the transverse space.
One of them involves a factor of $\tilde{n}_{\mu_1\ldots\mu_{q-1}}=n_{\mu_1\ldots\mu_q}e^{\mu_q}$ multiplied by an odd number of uncontracted $e^\mu$ direction vectors, such that the total number of $e^\mu$'s is even.
This corresponds to an operator with $q-1$ anti-symmetric indices and an odd number of symmetric traceless indices.
The other one involves a factor of $n_{\mu_1\ldots\mu_q}$ together with an even number of uncontracted $e^\mu$ direction vectors.
This corresponds to an operator with $q$ anti-symmetric indices and an even number of symmetric traceless indices.
All other transverse $SO(q)$ representations must have a vanishing one-point function.

Let us now return to the question of decomposing $F_\CO$ using the DOE.
Suppose for the moment that $q>2$.
The operators appearing in the DOE of a scalar are in symmetric traceless spin-$s$ representations, and their one-point functions are given by \eq{localised_defect_primary_one_point} for even $s$.
Let $\hat{\bm{x}}^{i}_\perp=\bm{x}_\perp^i/r$ such that $\hat{\bm{x}}_\perp\cdot e=\eta$.
Then it is convenient to choose $\CN$ such that $\hat{\bm{x}}^{i_1}_\perp\cdots\hat{\bm{x}}^{i_s}_\perp \,\Pi_s^{\,i_1\ldots i_s}(e)=\mathcal P_s^{(q)}(\eta)$, where 
\begin{equation}
\label{eq:normalized_Gegenbauer_defect}
    \mathcal P_s^{(q)}(\eta)
    \equiv
    \frac{
    C_s^{\left(\frac{q-2}{2}\right)}(\eta)
    }{
    C_s^{\left(\frac{q-2}{2}\right)}(1)
    }\,,
\end{equation}
and $C_s^{(n)}(\eta)$ are the Gegenbauer polynomials.
In particular, $\mathcal P_s^{(q)}(1)=1$.
Inserting the DOE~\eqref{eq:DOE} into the thermal one-point function of $\CO$, one finds the following thermal block expansion
\begin{equation}
\label{eq:thermal_DOE_point_defect}
    F_\CO(w,z)
    =
    \sum_{\substack{\hat\CO\\ s\ {\rm even}}}
    \mu_{\CO}{}^{\hat\CO}\,
    \hat b_{\widehat\CO}\,
    \varrho^{\hat\Delta-\Delta}
    \mathcal P_s^{(q)}(\eta)\,,
\end{equation}
Note that defect descendants do not contribute because the defect thermal one-point functions of primaries are constants.
The defect identity contribution reproduces the leading zero-temperature behaviour in \eq{localised_scalar_asymp} since $\CP_0^{(q)}=1$ and $\hat{b}_{\hat{\mathbbm{1}}}=1$.

When $q=2$, there is a distinguished non-compact direction transverse to the defect.
The tensor structure $n_{\mu\nu}e^\nu$ that we used to construct parity-odd one-point functions now just carries a single index, like $e^\mu$, and points along the unique transverse spatial direction.
Both $e$ and $v$ pick up a sign under the orientation reversing element of $O(q-1)\cong\mathbb{Z}_2$.
Thus the one-point function of symmetric traceless representations with even transverse spin $s$ can involve two independent structures,
\begin{equation}
\label{eq:localised_defect_primary_one_point_q=2}
    \left\langle
    \hat\CO_{\hat\Delta}^{\,i_1\ldots i_s}(x)\CD
    \right\rangle_{\beta}
    =
    \frac{
    \hat b^e_{\hat\CO}
    }{
    \beta^{\hat\Delta}
    }
    \Pi_s^{\,i_1\ldots i_s}(e)
    +    \frac{
    \hat b^v_{\hat\CO}
    }{
    \beta^{\hat\Delta}
    }
    \Pi_s^{\,i_1\ldots i_s}(e,v)\,.
\end{equation}
Here $\Pi_s^{\,i_1\ldots i_s}(e,v)$ is the symmetric traceless structure in the transverse directions involving $s-1$ factors of $e$ and one factor of $v$.
It is proportional to $e^{(i_1}\cdots e^{i_{s-1}}v^{i_s)}$ with all traces removed.
We choose its normalisation such that $\hat{\bm{x}}^{i_1}_\perp\cdots\hat{\bm{x}}^{i_s}_\perp \,\Pi_s^{i_1\ldots i_s}(e,v)= x_\perp/r \times U_{s-1}^{(q)}(\eta)$, where $U_{s-1}(\eta)$ is the Chebyshev polynomial of the second kind.
Here, $x_\perp$ is the distance along the spatial transverse direction and can have either sign.
Meanwhile, $\CP_s^{(2)}$ in \eq{normalized_Gegenbauer_defect} naively becomes singular.
It should instead be understood as the $q\to2$ limit of the right-hand side of \eq{normalized_Gegenbauer_defect} which reduces to the Chebyshev polynomial of the first kind $T_s(\eta)$.
The DOE of $\CO$ then contains generically two types of terms, one involving $\hat{b}_{\hat{\CO}}^e$ with $T_s(\eta)$ and the other involving $\hat{b}_{\hat{\CO}}^v$ with $x_\perp/r \times U_{s-1}(\eta)$.
If the theory is time reversal invariant, and $\hat{\CO}$ transforms trivially under that symmetry, then the terms involving $U_{s-1}(\eta)$ are absent.

For $q=1$, there is no continuous transverse-spin label and no angular polynomial.
There are instead two local DOEs, one for each of the two sides of the defect.
They take the form
\begin{align}
\label{eq:thermal_DOE_point_defect_q1}
F_\CO(w)
&=
\sum_{\hat\CO}
\mu^+_{\CO\hat\CO}\,
\hat b_{\hat\CO}\,
w^{\hat\Delta-\Delta}\,,
&
F_\CO(w)
&=
\sum_{\hat\CO}
\mu^-_{\CO\hat\CO}\,
\hat b_{\hat\CO}\,
(1-w)^{\hat\Delta-\Delta}\,,
\end{align}
where the expressions on the left and right hold for $0<w<1/2$ and $1/2<w<1$, respectively.
For a generic co-dimension one interface, the DOE coefficients $\mu^+_{\CO\hat\CO}$ and $\mu^-_{\CO\hat\CO}$ on the two sides need not coincide.

Given the thermal block expansion of the scalar one-point function in \eq{thermal_DOE_point_defect}, one can straightforwardly derive a Euclidean inversion formula analogous to \eq{wrapping_inversion} for a wrapping defect.
To do so, we first project the one-point function onto a fixed transverse spin.
After the projection we then simply apply the inversion formula~\eqref{eq:wrapping_inversion} to the resulting radial function.
The polynomials $\CP_s^{(q)}$ obey $\int_{-1}^{+1}\d\eta (1-\eta^2)^{(q-3)/2}\,\CP_s^{(q)}(\eta)\CP_{s'}^{(q)}(\eta)=\delta_{s,s'}\,\mathsf n_{s,q}$, where $\mathsf n_{s,q}$ is the square norm.
Then the spin-$s$ radial function can be taken to be
\begin{equation}
\label{eq:localised_spin_projection}
\varphi_{\CO,s}(\varrho)
=
\frac{1}{\mathsf n_{s,q}}
\int_{-1}^{1}\d\eta\,
(1-\eta^2)^{\frac{q-3}{2}}
\mathcal P_s^{(q)}(\eta)\,
F_\CO(\varrho,\eta)\,,
\end{equation}
and the inversion formula becomes
\begin{equation}
\label{eq:localised_radial_inversion}
\mu_{\CO}{}^{\hat\CO}\,
\hat b_{\hat\CO}
=
-\Res_{\bar\Delta\to\hat\Delta-\Delta}
\int_0^\varepsilon\d\varrho\,
\varrho^{-\bar\Delta-1}
\varphi_{\CO,s}(\varrho)\,.
\end{equation}
The residue of the integral is independent of $\varepsilon>0$ and we take $\varepsilon<1/2$ in order to avoid any potential singularities when reaching the boundary of the DOE's domain of convergence.
The formula picks out the coefficients corresponding to an operator $\hat{\CO}$ with transverse spin $s$ and scaling dimension $\hat{\Delta}$.
If several defect primaries have the same scaling dimension and transverse spin, the left-hand side is understood as the sum over contributions of all these operators.

Here we again assumed $q>2$.
For $q=2$, one can use the orthogonality properties of Chebyshev polynomials to define a radial function.\footnote{
Recall that $F_\CO$ depends on the sign of the single transverse spatial coordinate $x_\perp$.
One can then isolate the coefficients $\mu_{\CO}{}^{\hat{\CO}}\hat{b}^e_{\hat{\CO}}$ of the Chebyshev polynomials of the first kind $T_s(\eta)$ by taking the sum $F(w,z)+F(w,-z)$, where $z=x_\perp/\beta$.
Similarly, one can isolate the coefficients $\mu_{\CO}{}^{\hat{\CO}}\hat{b}^v_{\hat{\CO}}$ of $x_\perp/r\times U_{s-1}(\eta)$ by taking the difference $F(w,z)-F(w,-z)$.}
The inversion formula \eq{localised_radial_inversion} then applies unchanged.
For $q=1$, there is no transverse-spin.
Defining $\varphi_{\CO,0}^{+}(w)=F_\CO(w)$ and $\varphi_{\CO,0}^{-}(w)= F_\CO(1-w)$, the inversion formula can be applied to the two functions separately to extract $\mu^+_{\CO\hat\CO}\hat b_{\hat\CO}$ and $\mu^-_{\CO\hat\CO}\hat b_{\hat\CO}$, respectively.
To avoid picking up residues from both sides one should lower the upper bound of integration to some value $0<a<1$.

\paragraph{Vector operators. }
The simplest bulk operator with spin is a vector.
We find four different tensor structures which we list in table~\ref{tab:localised_vector_structures}.
The entry in the second row is built out of the spatial transverse coordinate vector and therefore only admissible when $q\geq2$.
The structure in the third row is built out of the two-index Levi-Civita tensor in the two transverse spatial directions $\tilde{n}_{\mu\nu}=n_{\mu\nu\rho}e^\rho$, which only exists when $q=3$.
When $p=1$, one can include a structure proportional to the unit tangent vector to the defect $t^\mu$.
It is only allowed if the line defect is not invariant under parity on the line.

\begin{table}[t]
\centering
\small
\setlength{\tabcolsep}{6pt}
\renewcommand{\arraystretch}{1.3}
\begin{tabular}{|c|c|c|c|c|}
\hline
structure
& exists for
& KMS
& $\varrho\to0$
& $z\to\infty$
\\
\hline
\rule{0pt}{4ex}
$\displaystyle \frac{e^\mu}{\beta^\Delta} F^\tau_V$
& all $p$, $q$
& odd
& $F^\tau_V\sim \varrho^{>-\Delta}$
& $F^\tau_V\to 0$
\\
$\displaystyle \frac{\hat{x}^\mu_\perp}{\beta^\Delta} F^\perp_V$
& $q\geq2$
& even
& $F^\perp_V\sim \varrho^{>-\Delta}$
& $F^\perp_V\to 0$
\\
$\displaystyle \frac{\tilde{n}^\mu{}_\nu \hat{x}^\nu_\perp}{\beta^\Delta} F^{q=3}_V$
& $q=3$
& odd
& $F^{q=3}_V\sim \varrho^{>-\Delta}$
&$ F^{q=3}_V\to 0$
\\
$\displaystyle \frac{t^\mu}{\beta^\Delta} F^\parallel_V$
& $p=1$
& even
& $F^\parallel_V\sim \varrho^{>-\Delta}$
& $F^\parallel_V\to 0$
\rule[-3ex]{0pt}{3ex}
\\
\hline
\end{tabular}
\caption{
Tensor structures appearing in the one-point function of a vector in the presence of a defect localised on the thermal circle.
}
\label{tab:localised_vector_structures}
\end{table}

KMS together with the $O(q-1)$ symmetry that also acts on $\tau$ imposes constraints on the coefficient functions of the tensor structures in table~\ref{tab:localised_vector_structures}.
Since the $O(q-1)$ symmetry descends from an ordinary $SO(q)\subset SO(d)$ rotation, a vector transforms covariantly.
For instance, for the $\tau$ component of $V$, covariance implies $\langle \CD(0) V^\tau (-\tau,\vec{x})\rangle_\beta=-\langle V^\tau(\tau,\vec{x})\CD(0)\rangle_\beta$ under the coordinate transformation $\tau\to-\tau$ and $\vec{x}_\perp\to R\vec{x}_\perp$, where $R\in O(q-1)$.
Here we used the fact that the correlation function depends on $\vec{x}_\perp$ through its norm only, such that it takes the same value at points $R\vec{x}_\perp$ and $\vec{x}_\perp$.
Using periodicity of the thermal trace one thus finds
$F_V^\tau(1-w,z)=-F_V^\tau(w,z)$.
We will refer to functions with this behaviour as being odd under KMS.
Note that such functions must vanish at $w=1/2$ for any $z$.
For $q=3$, $F_V^{q=3}$ is also odd since the tensor structure involves the $SO(3)$ invariant Levi-Civita tensor $n_{\mu\nu\rho}$ and the thermal direction vector $e^\mu$. 
The latter changes sign under an $O(2)\subset SO(3)$ transformation.
If the theory is separately time-reversal invariant, $F_V^{q=3}$ in fact has to vanish.
Under time reversal alone both $e^\mu$ and $n_{\mu\nu\rho}$ pick up a minus sign such that $F_V^{q=3}$ is also even. 
A function that is both even and odd must be zero. 
The remaining structures in table~\ref{tab:localised_vector_structures} are even under KMS, meaning that they obey $F_V(1-w,z)=F_V(w,z)$.
When $q=2$, $-\mathbbm{1}\in O(1)\cong\mathbb{Z}_2$ also reverses the sign of the single spatial transverse coordinate.
Thus, $F^\tau_V(1-w,-z)=-F^\tau_V(w,z)$ and similarly for $F^\parallel_V$ and $F^\perp_V$.

The asymptotic behaviour of the coefficient functions can be determined by matching with the zero-temperature one-point function~\eqref{eq:0T_V_1pt} for $\beta\to\infty$ and the no-defect thermal one-point function~\eqref{eq:no_defect_1-pt} for $|\vec{x}_\perp|\to\infty$.
Both vanish in general such that all $F_V$'s have the behaviour listed in the last two columns of table~\ref{tab:localised_vector_structures}.
When $q=1$ the $|\vec{x}_\perp|\to\infty$ limit does not exist and the constraints in the last column do not apply.
Meanwhile the entries in the fourth column refer to either of the two limits $w\to0^+$ or $w\to1^-$.
When $q=2$, a vector can acquire a one-point function at zero temperature.
This must be supplied by a linear combination of the structures in the first two rows. 
One therefore must have $F_V^\tau\sim a_V z/\varrho^{\Delta+1}$ and $F_V^\perp \sim -a_V w/\varrho^{\Delta+1}$ as $\varrho\to0$, where $a_V$ is the coefficient of the zero-temperature one-point function in \eq{0T_V_1pt}.

When the bulk vector is a conserved current for a continuous symmetry, $V^\mu_{d-1}=J^\mu$, then its thermal one-point takes the following form
\begin{equation}
\label{eq:localised_vector_one_point}
\begin{split}
    \left\langle
   J^\mu(\tau,\vec{x}_\parallel,\vec{x}_\perp)\CD
    \right\rangle_{\beta}
    =\frac{1}{\beta^{d-1}}
    \bigg[
    &e^\mu F^\tau_J(w,z)
    +\hat{x}_\perp^\mu F^\perp_J(w,z)
    \\
    &+\delta_{p,1}\,t^\mu F^\parallel_J(w,z)
    +\delta_{q,3}\,\tilde{n}^\mu{}_\nu \hat{x}^\nu_\perp F^{q=3}_J(w,z)
    \bigg]\,.
\end{split}
\end{equation}
Away from the defect, the conservation equation $\partial_\mu J^\mu =0 $ becomes the following partial differential equation (PDE)
\begin{equation}
\partial_w F^\tau_J(w,z)+\partial_z F^\perp_J(w,z)+\frac{q-2}{z} F_J^\perp(w,z)=0\,,
\end{equation}
since the structures involving $F_J^{q=3}$ and $F_J^\parallel$ are identically conserved.
This PDE has the general solution
\begin{align}
F_J^\tau(w,z)&=z^{2-q}\partial_z f_J(w,z)\,, & F_J^\perp=-z^{2-q}\partial_w f_J(w,z)\,,
\end{align}
where $f_J$ is an arbitrary function such that $F_J^\tau$ and $F_J^\perp$ have the periodicity properties listed in table~\ref{tab:localised_vector_structures}.
When $q=1$, conservation becomes the ODE $\partial_w F_J^\tau(w)=0$.
Its solution is $F_J^\tau(w)=c^{q=1}_J$, where $c_J^{q=1}$ is a constant.
The Ward identity~\eqref{eq:broken_WI_tilt} for a continuous symmetry relates the tilt operator to the jump across the defect in the $\tau$-component of the current, $J^\tau(\tau=0^+,\vec{x}_\parallel)-J^\tau(\tau=0^-,\vec{x}_\parallel)=t(\vec{x}_\parallel)$.
Since the conservation equation sets $\langle J^{\tau}\CD\rangle_\beta$ to a constant, the thermal one-point function of the tilt operator vanishes, $\hat{b}_t=0$.

\paragraph{Spin-2. }
Finally consider a symmetric traceless spin-2 operator $S^{\mu\nu}_\Delta$.
We find at most 9 permissible structures, which we list in table~\ref{tab:localised_spin_two_structures}.
When $q=2$ the structures in the third and fourth row with coefficient functions $F^{\perp\perp}_{S,1}$ and $F^{\perp\perp}_{S,2}$, respectively, become identical. 
We can thus eliminate one of the two.
When $q=1$ there are no non-compact transverse directions.
For general $p$ that removes all structures expect the first one with coefficient $F^{\tau\tau}_S$.
As for the vector, special structures appear when $q=3$ or $p=1$.
These are built out of the Levi-Civita tensor in the two transverse spatial directions $\tilde{n}_{\mu\nu}=n_{\mu\nu\rho}e^\rho$ and the unit vector $t^\mu$ along the direction parallel to the defect, respectively.
A term proportional to $t^\mu t^\nu$ would not be linearly independent as it can be expressed in terms of $g^{\mu\nu}$, $e^\mu e^\nu$, and the projector onto the spatial transverse directions $\tilde{N}^{\mu\nu}$.

\begin{table}[t]
\centering
\small
\setlength{\tabcolsep}{6pt}
\renewcommand{\arraystretch}{1.3}
\begin{tabular}{|c|c|c|c|c|}
\hline
structure
& exists for
& KMS
& $\varrho\to0$
& $z\to\infty$
\\
\hline
\rule{0pt}{4ex}
$\displaystyle \frac{e^\mu e^\nu -\frac{1}{d}g^{\mu\nu}}{\beta^\Delta}F^{\tau\tau}_S$
& all $p$, $q$
& even
& $F^{\tau\tau}_S\sim d\, a_S z^2\varrho^{-\Delta-2}$
& $F^{\tau\tau}_S\to b_S$ 
\\
$\displaystyle 2\frac{e^{(\mu}\hat{x}_\perp^{\nu)}}{\beta^\Delta}F^{\tau\perp}_S$
& $q\geq2$
& odd
& $F^{\tau\perp}_S\sim -d\, a_S wz \varrho^{-\Delta-2}$
& $F^{\tau\perp}_S\to 0$
\\
$\displaystyle \frac{\tilde{N}^{\mu\nu} -\frac{q-1}{d}g^{\mu\nu}}{\beta^\Delta}F^{\perp\perp}_{S,1}$
& $q\geq2$
& even
& $F^{\perp\perp}_{S,1}\sim d\, a_S\varrho^{-\Delta}$
& $F^{\perp\perp}_{S,1}\to 0$
\\
$\displaystyle \frac{\hat{x}_\perp^\mu \hat{x}_\perp^\nu -\frac{1}{d}g^{\mu\nu}}{\beta^\Delta}F^{\perp\perp}_{S,2}$
& $q\geq2$
& even
& $F^{\perp\perp}_{S,2}\sim -d\, a_S z^2\varrho^{-\Delta-2}$
& $F^{\perp\perp}_{S,2}\to 0$
\\
$\displaystyle 2\frac{e^{(\mu}t^{\nu)}}{\beta^\Delta}F^{\tau\parallel}_S$
& $p=1$
& odd
& $F^{\tau\parallel}_S\sim \varrho^{>-\Delta}$
& $F^{\tau\parallel}_S\to 0$
\\
$\displaystyle 2\frac{\hat{x}_\perp^{(\mu}t^{\nu)}}{\beta^\Delta}F^{\perp\parallel}_S$
& $p=1$, $q\geq2$
& even
& $F^{\perp\parallel}_S\sim \varrho^{>-\Delta}$
& $F^{\perp\parallel}_S\to 0$
\\
$\displaystyle 2\frac{e^{(\mu}\tilde{n}^{\nu)}{}_\rho \hat{x}^\rho_\perp}{\beta^\Delta}F^{e,q=3}_{S}$
& $q=3$
& even
& $F^{e,q=3}_{S}\sim \varrho^{>-\Delta}$
& $F^{e,q=3}_{S}\to 0$
\\
$\displaystyle 2\frac{\hat{x}_\perp^{(\mu}\tilde{n}^{\nu)}{}_\rho \hat{x}^\rho_\perp}{\beta^\Delta}F^{\perp,q=3}_{S}$
& $q=3$
& odd
& $F^{\perp,q=3}_{S}\sim \varrho^{>-\Delta}$
& $F^{\perp,q=3}_{S}\to 0$
\\
$\displaystyle 2\frac{t^{(\mu}\tilde{n}^{\nu)}{}_\rho \hat{x}^\rho_\perp}{\beta^\Delta}F^{\parallel,q=3}_{S}$
& $p=1$, $q=3$
& odd
& $F^{\parallel,q=3}_{S}\sim \varrho^{>-\Delta}$
& $F^{\parallel,q=3}_{S}\to 0$
\rule[-3ex]{0pt}{3ex}
\\
\hline
\end{tabular}
\caption{
Tensor structures appearing in the one-point function of a spin-2 primary in the presence of a defect localised on the thermal circle.
}
\label{tab:localised_spin_two_structures}
\end{table}

By covariance, an $O(q-1)$ transformation and KMS together imply some reflection properties for the coefficient functions $F_S$ around $w=1/2$. 
In the third column of table~\ref{tab:localised_spin_two_structures} we summarise which functions are even, $F_S(1-w,z)=F_S(w,z)$, and which are odd, $F_S(1-w,z)=-F_S(w,z)$.
Here we implicitly assume that $q>2$.
When $q=1$, the only transverse direction is the thermal circle, and so there is no dependence on $z$.
For $q=2$, there is only a single transverse spatial direction and the coefficient functions $F_S$ can depend on the sign of that coordinate.
Since $-\mathbbm{1}\in O(1)\cong\mathbb{Z}_2$ flips the sign of that coordinate as well as $\tau$, the reflection properties are modified by replacing $z\to-z$ on the left-hand side only.
Note that if the theory is additionally invariant under time-reversal, some structures are forced to vanish. 
In particular, $n_{\mu\nu\rho}$ picks up an additional sign and so functions that are odd by KMS must now also be even and \emph{vice versa}.
This removes the last three structures in table~\ref{tab:localised_spin_two_structures}.
Similarly, if a line defect is parity invariant, then all structures linear in $t^\mu$ must vanish.

The asymptotic behaviour of the coefficient functions as $\varrho\to0$ and $z\to\infty$ can be found by considering the $\beta\to\infty$ and $|\vec{x}_\perp|\to\infty$ limits, respectively.
In the $\beta\to\infty$ limit, the thermal one-point function should reduce to the zero-temperature one-point function~\eqref{eq:T_def_1-pt_fn}.
In that equation, $\hat{x}^\mu_\perp$ should be interpreted as $\hat{\bm{x}}^\mu_\perp=\frac{w}{\varrho} e^\mu+\frac{z}{\varrho} \hat{x}_\perp^\mu$ since the thermal circle is now part of the transverse directions.
Meanwhile the projector onto all $q$ transverse directions is $N^{\mu\nu}=e^\mu e^\nu+\tilde{N}^{\mu\nu}$.
The requisite behaviour in the fourth column of table~\ref{tab:localised_spin_two_structures} then follows immediately.
Here we implicitly assumed $q>1$. 
For $q=1$, no spinning operator can acquire a one-point function and so all coefficient functions $F_S\sim w^{>-\Delta}$ as $w\to0$.
In the opposite limit, $|\vec{x}_\perp|\to\infty$, the thermal one-point function should reduce to the no-defect result given by \eq{no_defect_1-pt}.
This determines the behaviour of the functions summarised in the last column.
When $q=1$, there is no such limit and the constraints in the last column do not apply.

Finally consider the case where the spin-2 operator is the stress tensor, $S_d^{\mu\nu}=T^{\mu\nu}$.
Its thermal one-point function consists of all the structures appearing in table~\ref{tab:localised_spin_two_structures} whenever they are admissible,
\begin{equation}
\label{eq:localised_T_one_point}
\begin{split}
\langle T^{\mu\nu}(x)\CD\rangle_\beta={}&
\frac{e^\mu e^\nu -\frac{1}{d}g^{\mu\nu}}{\beta^d}F^{\tau\tau}_T
+2\frac{e^{(\mu}\hat{x}_\perp^{\nu)}}{\beta^d}F^{\tau\perp}_T
+\frac{\tilde{N}^{\mu\nu} -\frac{q-1}{d}g^{\mu\nu}}{\beta^d}F^{\perp\perp}_{T,1}
+\frac{\hat{x}_\perp^\mu \hat{x}_\perp^\nu -\frac{1}{d}g^{\mu\nu}}{\beta^d}F^{\perp\perp}_{T,2}\\
&+2\delta_{p,1}\,\frac{e^{(\mu}t^{\nu)}}{\beta^d}F^{\tau\parallel}_T
+2\delta_{p,1}\,\frac{\hat{x}_\perp^{(\mu}t^{\nu)}}{\beta^d}F^{\perp\parallel}_T\\
&+\delta_{q,3}\,2\frac{e^{(\mu}\tilde{n}^{\nu)}{}_\rho \hat{x}^\rho_\perp}{\beta^d}F^{e,q=3}_{T}
+2\delta_{q,3}\,\frac{\hat{x}_\perp^{(\mu}\tilde{n}^{\nu)}{}_\rho \hat{x}^\rho_\perp}{\beta^d}F^{\perp,q=3}_{T}\\
&+2\delta_{p,1}\delta_{q,3}\,\frac{t^{(\mu}\tilde{n}^{\nu)}{}_\rho \hat{x}^\rho_\perp}{\beta^d}F^{\parallel,q=3}_{T}\,,
\end{split}
\raisetag{68pt}
\end{equation}
where all $F_T$'s are functions of $w$ and $z$.
Away from the defect the stress tensor is conserved, $\partial_\mu T^{\mu\nu}=0$.
Similarly to the case of the current, the conservation equations are PDEs when $q>1$.
We find two PDEs involving the functions appearing in the first line only, one PDE involving the functions in the second line, and another PDE involving the functions in the third line.
The function in the last term is independently conserved.
We have not found the PDEs particularly illuminating so we do not reproduce them here.
However, the conservation equations reduce the number of free functions from 9 to at most 5. 
When $q=1$, the PDEs reduce to the two ODEs, $\partial_w F^{\tau\tau}_T(w)=0$ and $\partial_w F^{\tau\parallel}_T(w)=0$, where the second one only exists when $p=1$.
Thus, both functions are constants, $F^{\tau\tau}_T(w)=c_T^{\tau, q=1}$ and $F^{\tau\parallel}_T(w)=c_T^{\parallel,q=1}$.
By the broken translation Ward identity~\eqref{eq:broken_WI_disp} for $q=1$, $T^{\tau\tau}(\tau=0^{+},\vec{x}_\parallel)-T^{\tau\tau}(\tau=0^{-},\vec{x}_\parallel)=D(\vec{x}_\parallel)$, where $D$ is the displacement operator.
Since $\langle T^{\tau\tau}\CD\rangle_\beta$ is a constant by the conservation equations, the thermal one-point function of the displacement vanishes, $\hat{b}_D=0$.

\section{Bootstrap sum rules}
\label{sec:Bootstr}

The KMS condition can be used to formulate bootstrap constraints for both classes of thermal defects introduced in section~\ref{sec:Kinematics}.
For defects wrapping the thermal circle, one-point functions are trivially invariant under thermal translations, and the first non-trivial KMS constraint arises for two-point functions.
The corresponding bootstrap problem for two bulk scalar operators in the presence of a wrapping line defect, $p=1$, was introduced in ref.~\cite{Barrat:2024aoa}.
For correlators whose bulk descendant contributions vanish in the collinear configuration, the construction extends directly to arbitrary $p$ and the resulting sum rules take the same form as in the line defect case.\footnote{
The assumption that bulk descendants can be neglected in the collinear configuration is non-trivial.
Translational invariance along the thermal circle eliminates descendants involving $\tau $-derivatives.
However, it does not in general eliminate all descendant contributions. 
In particular, a descendant $\partial^2\mathcal O$ will typically have a non-vanishing one-point function since the one-point function of the primary $\CO$ is a non-trivial function of the transverse coordinates.
The sum rules therefore apply only to correlators for which these additional descendant contributions vanish.
In the explicit generalised free-field example of ref.~\cite{Barrat:2024aoa} thermal corrections to the correlator do not depend on the transverse distance, and so the two-point function satisfies the sum rules.
We thank the authors of ref.~\cite{Barrat:2024aoa} for correspondence on this point.}

For a localised defect at a point on the thermal circle, the one-point function of scalar operators is the simplest quantity that satisfies a non-trivial KMS condition.
In the remainder of this section, we will show that we can set up a thermal bootstrap problem by treating the KMS condition~\eqref{eq:KMS_localised_ordered} analogously to a crossing equation.
We will derive a set of sum rules for the coefficients entering the thermal block expansion of the one-point function.
The derivation requires additional assumptions on the radius of convergence of the DOE as we shall explain.

Consider the dimensionless one-point function of a scalar bulk operator $\mathcal O$ of scaling dimension $\Delta_{\mathcal O}$,
\begin{equation}
 F_{\mathcal O}(w,z)
=
\beta^{\Delta_{\mathcal O}}
\vev{\mathcal O(\tau,y)\mathcal D}_\beta\,,
\label{eq:bootstrap_dimensionless_correlator}
\end{equation}
where $w=\tau/\beta$ and $z=|x_\perp|/\beta$.
Using the thermal block expansion~\eqref{eq:thermal_DOE_point_defect}, one has
\begin{equation}
 F_{\mathcal O}(w,z)
=
\sum_{\hat{\mathcal O}}
\mu_{\mathcal O}{}^{\hat{\mathcal O}}\,
\hat b_{\hat{\mathcal O}}\,
\varrho^{\hat\Delta-\Delta_{\mathcal O}}
\CP^{(q)}_s(\eta)\,,
\label{eq:generic_DOE_full_space}
\end{equation}
where $\varrho^2=w^2+z^2$ and $\eta=w/\varrho$. 
For a bosonic bulk scalar, the KMS condition acts at fixed spatial
separation and implies
\begin{equation}
 F_{\mathcal O}(w,z)
=
 F_{\mathcal O}(1-w,z)\,.
\label{eq:KMS_full_space}
\end{equation}
We would then like to plug in \eq{generic_DOE_full_space} into either side of the KMS condition.
However here we encounter a subtlety.
Let
\begin{align}
\bar\varrho
&=
\sqrt{(1-w)^2+z^2}\,,
&
\bar\eta
&=
\frac{1-w}{\bar\varrho}
\label{eq:bootstrap_image_variables}
\end{align}
denote the radial and angular variables associated with the KMS image, respectively.
Recall that the DOE is only guaranteed to converge up to $\varrho<1/2$. 
If $\varrho <1/2$ for some $0<w<1$ and $z>0$, then $\bar{\varrho}>1/2$. 
Thus, either $\varrho>1/2$ or $\bar{\varrho}>1/2$, or both, whereas to use KMS we must be in a configuration where the series on both sides converge.

To make progress we must make additional assumptions that may improve the radius of convergence of the thermal block expansion.
First consider the series with $z=0$ such that $\eta=1$ and $\varrho=w$. 
In that case the transverse-spin dependent structures in \eq{generic_DOE_full_space} become irrelevant since $\CP^{(q)}_s(1)=1$.
A sufficient assumption then is the following.
Suppose that the coefficients 
\begin{equation}
\label{eq:sum_rule_assumption}
c_{\hat{\CO}}=\mu_{\CO}{}^{\hat{\CO}}\hat{b}_{\hat{\CO}}>0
\end{equation}
at large $\hat{\Delta}$.
Landau's theorem on general Dirichlet series then states that the radius of convergence of the series is set by the singularities of $F_\CO$ on the positive real $\varrho$-axis.\footnote{
This is an extension of the Vivanti-Pringsheim theorem for ordinary power series to general Dirichlet series, see e.g.\ ref.~\cite{HardyRiesz1915}.}
In the absence of other operator insertions, that singularity is generically at the thermal image of $\CD$ at $\varrho=w=1$.
Therefore, the series converges absolutely for $0\leq \varrho<1$ and the thermal block expansions with $\varrho$ and $\bar{\varrho}$ both converge in an overlapping region.
Turning on small $z>0$ does not change this conclusion.
Regard \eq{generic_DOE_full_space} at $z>0$ as a general Dirichlet series in $\varrho$ labelled by a parameter $0\leq\eta\leq1$.
The coefficients of that series obey $|c_{\hat{\CO}}\CP_s^{(q)}(\eta)|\leq c_{\hat{\CO}}$ since $|\CP_s^{(q)}(\eta)|\leq1$ for $0\leq\eta\leq1$.
Since the series with $\eta=1$ converges absolutely for $\varrho<1$, then by the comparison test so does the series with $0\leq\eta<1$.
Thus, the radius of convergence of the thermal block expansion extends from $1/2$ to $\varrho=1$ if assumption~\eqref{eq:sum_rule_assumption} holds.
In the explicit example of a magnetic line defect in the theory of a free scalar presented in section~\ref{sec:ex_line}, we indeed find assumption~\eqref{eq:sum_rule_assumption} to be true and the thermal block expansions of simple operators to have radius of convergence $\varrho=1$.\footnote{
It would be interesting to check under which conditions \eq{sum_rule_assumption} can be proven.
Positivity of the thermal block expansion coefficients could then allow for estimating the spectrum of heavy operators via Tauberian theorems. 
See ref.~\cite{Marchetto:2023xap} for Tauberian theorems in thermal CFTs without defects.}

Substituting the thermal block expansion~\eqref{eq:generic_DOE_full_space} into both sides of the KMS condition~\eqref{eq:KMS_full_space}, one obtains the bootstrap equation
\begin{align}
0
=
\sum_{\hat{\mathcal O}}
\mu_{\CO}{}^{\hat{\CO}}\,
\hat b_{\hat{\mathcal O}}
\bigg[
\varrho^{\hat\Delta-\Delta_{\mathcal O}}
\CP^{(q)}_s(\eta)
-
\bar\varrho^{\hat\Delta-\Delta_{\mathcal O}}
\CP^{(q)}_s(\bar\eta)
\bigg]\,.
\label{eq:DOE_KMS_bootstrap_full}
\end{align}
The right-hand side is odd under $w\to1-w$ by construction.
Differentiating the bootstrap equation an odd number of times and evaluating it at the KMS symmetric point $w=1/2$ one obtains the infinite family of non-trivial sum rules
\begin{equation}
\sum_{\hat{\mathcal O}}
\mu_{\CO}{}^{\hat{\CO}}\,
\hat b_{\hat{\mathcal O}}
\left.
\partial_{w}^{\,2m+1}
\left[
\varrho^{\hat\Delta-\Delta_{\mathcal O}}
\CP^{(q)}_s(\eta)
\right]
\right|_{w=\frac12}
=0\,,
\label{eq:DOE_KMS_sum_rules_full}
\end{equation}
where $m$ is a non-negative integer.
Expanding each thermal block around $z=0$, one has
\begin{equation}
\varrho^{\hat\Delta-\Delta_{\mathcal O}}
\CP^{(q)}_s(\eta)
=
\sum_{n=0}^{\infty}
A^{(q)}_n\left(\hat\Delta-\Delta_{\mathcal O},s\right)
z^{2n}
w^{\hat\Delta-\Delta_{\mathcal O}-2n}\,,
\label{eq:thermal_block_transverse_expansion}
\end{equation}
where
\begin{equation}
\begin{split}
A^{(q)}_n(\alpha,s)
&=
\frac{1}{n!}
\left.
\frac{\d^n}{\d t^n}
\left[
(1+t)^{\alpha/2}
\CP^{(q)}_s\left(\frac{1}{\sqrt{1+t}}\right)
\right]
\right|_{t=0}\\
&=\frac{\Gamma (q-2) \Gamma (s+1) }{\Gamma (n+1) \Gamma \left(\frac{q-2}{2}\right) \Gamma (q+s-2)} \\
&\phantom{=}\times\sum _{k=0}^{\left\lfloor s/2\right\rfloor }\frac{(-1)^k 2^{s-2 k} \Gamma \left(\frac{q-2}{2}-k+s\right) \Gamma \left(\frac{1}{2} (2 k-s+\alpha )+1\right)}{\Gamma (k+1) \Gamma (-2 k+s+1) \Gamma \left(\frac{1}{2} (2 k-s+\alpha )-n+1\right)}\,.
\label{eq:angular_expansion_coefficients}
\end{split}
\end{equation}
The first two coefficients are
\begin{align}
A_0^{(q)}(\alpha,s)&=1\,,
&
A_1^{(q)}(\alpha,s)
&=
\frac{1}{2} \left(\alpha -\frac{s (q+s-2)}{q-1}\right)\,,
\label{eq:first_angular_expansion_coefficients}
\end{align}
and so dependence on transverse spin $s$ already appears at order $z^2$ in the small-$z$ expansion.
Substituting eq.~\eqref{eq:thermal_block_transverse_expansion} into eq.~\eqref{eq:DOE_KMS_sum_rules_full} and equating separately each power of $z$, we obtain the two-parameter family of sum rules
\begin{equation}
\sum_{\hat{\mathcal O}}
\mu_{\CO}{}^{\hat{\CO}}\,
\hat b_{\hat{\mathcal O}}\,
2^{2n-\hat\Delta+\Delta_{\mathcal O}}
A^{(q)}_n\left(\hat\Delta-\Delta_{\mathcal O},s\right)
\frac{
\Gamma\left(\hat\Delta-\Delta_{\mathcal O}-2n+1\right)
}{
\Gamma\left(\hat\Delta-\Delta_{\mathcal O}-2n-2m\right)
}
=0\,,
\label{eq:DOE_KMS_sum_rules_transverse}
\end{equation}
where $m,n$ are non-negative integers.
The ratios of Gamma functions in eq.~\eqref{eq:DOE_KMS_sum_rules_transverse} are understood by analytic continuation whenever necessary.
Note that the sum rules for $n=0$ constrain the thermal DCFT data of defect operators without being sensitive to their transverse spin. 
This corresponds to taking the transverse spatial separation $z$ to zero.
The higher-$n$ constraints in eq.~\eqref{eq:DOE_KMS_sum_rules_transverse} instead retain information about transverse spin and provide additional sum rules that are invisible to the thermal one-point function at zero spatial separation.

The derivation presented here relies on the assumption of positivity described around \eq{sum_rule_assumption}.
This is a sufficient condition that allowed us to extend the radius of convergence of the DOE beyond $\varrho=1/2$.
This condition, however, need not be necessary.
In appendix~\ref{app:boots} we present an alternative derivation of the sum rules~\eqref{eq:DOE_KMS_sum_rules_transverse} that makes weaker assumptions about convergence.
In particular, the derivation there requires only the coefficients of the $z^2$ expansion~\eqref{eq:thermal_block_transverse_expansion}, differentiated term-wise with respect to $w$, to admit a finite limit as $w\to1/2$.

\section{Boundaries in CFTs at finite temperature}
\label{sec:ex_bdy}

In this section we consider examples of simple CFTs with conformal boundaries at finite temperature.
We will compute thermal one-point functions of simple bulk operators from which we extract some thermal DCFT data. 
One of the operators that we consider is the stress tensor, through which we can compute certain thermodynamic quantities.
We will discuss theories of a free scalar, a free Dirac fermion, a free 4d hypermultiplet and a free higher-derivative scalar.

\subsection{Free scalar field}
\label{sec:free_scalar}

Consider a free massless scalar field $\phi$.
To set our conventions we first place the theory on $\mathbb{R}^d$ with $d>2$.
Let $G(x,y)\equiv\langle\phi(x)\phi(y)\rangle$ denote the two-point function, where we use Cartesian coordinates $x^\mu$ and $\mu=0,\ldots,d-1$ as in section~\ref{sec:Kinematics}.
We introduce a boundary by choosing $x_\perp\equiv x^{d-1}$ and restrict to the half-space $\mathbb{R}^d_{>0}$ with $x_\perp>0$.
Imposing Neumann (N) or Dirichlet (D) boundary conditions (BCs), $\p_\perp\phi|_{x_\perp=0}=0$ or $\phi|_{x_\perp=0}=0$, respectively, the propagator then takes the following form
\begin{equation}
\label{eq:flat_G}
G_\varsigma(x,y)=\kappa\left(\frac{1}{|x-y|^{d-2}}+\varsigma\,\frac{1}{|x-\tilde{y}|^{d-2}}\right),
\end{equation}
where $\varsigma=+1$ for N BCs and $\varsigma=-1$ for D BCs.
Here $x^\mu=(\tau_1,\vec{x}_\parallel, x_\perp)=(\bm{x}_\parallel, x_\perp)$, and similarly for $y^\mu$, while $\tilde{y}^\mu=(\bm{y}_\parallel,-y_\perp)$.
The constant
\begin{equation}
\label{eq:kappa}
\kappa=\frac{1}{4\pi^{d/2}}\,\Gamma\left(\frac{d-2}{2}\right)
\end{equation}
ensures that the Green's function is appropriately normalised. 

We now put the theory on the thermal manifold $S^1_\beta\times\mathbb{R}^{d-1}_{>0}$ obtained by compactifying the $\tau$-direction of $\mathbb{R}^d_{>0}$.
The propagator at finite temperature can be obtained through the method of thermal images.
In practice, one takes $\Delta\tau\equiv\tau_2-\tau_1 \to \Delta\tau + m\beta$ in the zero-temperature Green's function \eq{flat_G} and sums over all integer $m$,
\begin{equation}
\label{eq:thermal_G}
G^\beta_\varsigma(x,y)=\sum_{m=-\infty}^\infty G_\varsigma(x,y+m\beta \,e)\,,
\end{equation}
where $e^\mu=(1,0,\ldots,0)$ is the unit vector pointing along $\tau$.
To our knowledge these propagators were first written down by refs.~\cite{Dowker:1978md, Kennedy:1979ar}.

Equipped with these propagators, we now compute thermal one-point functions and extract thermal BCFT data.
In particular, we will consider the thermal one-point functions of $\phi^2$ and the stress tensor.
The latter will allow us to study some of the system's thermodynamic properties.

\subsubsection{Thermal one-point function of $\phi^2$}
\label{sec:scalar_phi2}

We begin with the thermal one-point function $\langle\phi^2\rangle_\beta$.
For a scalar operator, the thermal one-point function takes the form of \eq{kin_O} and admits an expansion as in \eq{FO_DOE}.
The thermal one-point function of $\phi^2$ can be computed by considering the coincident limit of the propagator in \eq{thermal_G}.
This limit is divergent and requires regularisation.
We will use point splitting regularisation,
\begin{equation}
\langle\phi^2(x)\rangle_\beta=\lim_{\epsilon^\mu\to0}G_\varsigma^\beta(x,x+\epsilon)|_\mathrm{finite}\,,
\end{equation}
where any divergence as $\epsilon^\mu\to0$ is subtracted.
There are several choices for the direction of the vector $\epsilon^\mu$, however, they all give a $1/\epsilon^{d-2}$ divergence and the same finite piece.
The divergence comes from the $m=0$ Matsubara mode.
It is the same divergence coming from the first term of \eq{flat_G} at zero temperature when $x^\mu\to y^\mu$.
We proceed by subtracting this divergence by hand following the prescription of ref.~\cite{Kennedy:1979ar}.
For concreteness, we choose to point split along the thermal circle, i.e.\ $\epsilon^\mu=(\epsilon, 0,\ldots,0)$.
The N and D propagators then become
\begin{equation}
G_\varsigma^\beta(x,x+\epsilon)=\kappa\sum_{m=-\infty}^\infty\left(\frac{1}{|\epsilon+m\beta|^{d-2}}+\varsigma\, \frac{1}{((\epsilon+m\beta)^2+4x_\perp^2)^\frac{d-2}{2}}\right).
\end{equation}
Removing the $1/\epsilon^{d-2}$ divergence from the first term of the $m=0$ mode and taking the $\epsilon\to0$ limit, we find
\begin{equation}
\label{eq:phi2}
\langle\phi^2(x)\rangle_\beta=\varsigma\,\frac{\kappa}{(2x_\perp)^{d-2}}+2\kappa\sum_{m=1}^\infty\left(\frac{1}{(m^2\beta^2)^\frac{d-2}{2}}+ \varsigma\, \frac{1}{(m^2\beta^2+4x_\perp^2)^\frac{d-2}{2}}\right).
\end{equation}
The sum over $m$ converges provided that $d>3$ for N BCs.
For D BCs convergence only requires $d>1$ due to the cancellation of leading large-$m$ terms.
However, the form of the zero-temperature propagator in \eq{flat_G} is only valid for $d>2$.
The term proportional to $\varsigma$ is the one-dimensional inhomogeneous Epstein\footnote{Not that one.} zeta function.
For even integer $d\geq4$, the sum can be performed to give elementary hyperbolic functions.
E.g.\ when $d=4$ one finds for the function $F_{\phi^2}^\varsigma(z)$ of the cross-ratio $z\equiv x_\perp/\beta$ defined in \eq{kin_O}
\begin{equation}
\label{eq:F_phi2}
F^\varsigma_{\phi^2}(z)= \frac{2 \pi ^2   z+ 3 \pi  \varsigma  \coth (2 \pi  z)}{6z }\kappa\,,
\end{equation}
where $\varsigma=1$ for N BCs and $\varsigma=-1$ for D BCs.
In even dimensions $d>4$, the sum over thermal images is similarly immediate to perform.
However, expressions for $F_{\phi^2}$ quickly become cumbersome. 
One can write them compactly using the following identity
\begin{equation}
\sum_{m=1}^\infty \frac{1}{(m^2+a^2)^{n+1}}=\frac{(-1)^{n}}{n!}\left[\frac{\p^{n}}{\p A^{n}}\frac{\pi  \sqrt{a^2+A} \coth \left(\pi  \sqrt{a^2+A}\right)-1}{2 \left(a^2+A\right)}\right]_{\!A=0}\,,
\end{equation}
where $A$ is a real number that is set to zero after differentiating.
One then has
\begin{equation}
\label{eq:F_phi2_d}
\begin{split}
F^\varsigma_{\phi^2}(z)&=\kappa\,\Bigg(2\zeta(2n+2)\\
&\quad+\varsigma\frac{(-1)^n}{n!}
\frac{\partial^n}{\partial A^n}\Bigg[\frac{\pi  \sqrt{A+4 z^2} \coth \big(\pi  \sqrt{A+4 z^2}\big)}{ \left(A+4 z^2\right)}\Bigg]_{\!A=0}\Bigg),
\end{split}
\end{equation}
where $n=(d-4)/2$ and $\zeta(s)$ is the Riemann zeta function.

Note that while eq.~\eqref{eq:F_phi2} and its generalisation~\eqref{eq:F_phi2_d} to even $d>4$ are well-behaved for $z>0$, they have singularities in the complex $z$-plane. 
In the present example, and in fact for all analogous functions found in this section, they are poles on the imaginary axis with half-integer spacing.
At $z=0$ \eq{F_phi2_d} has a pole of order $d-2$, while at $z=\iu k/2$, with $k$ a non-zero integer, the function has a pole of order $(d-2)/2$. 

Given these closed-form expressions for the one-point functions, it is straightforward to check that they have the asymptotic properties expected from \eqref{eq:FO_asymp}.
Taking the large distance limit $z\gg1$ of \eq{F_phi2_d}, we find that the leading behaviour is given by $F^\varsigma_{\phi^2}(z) \overset{z\to\infty}{\sim} 2\kappa\zeta(d-2)$.
This value is nothing but the no-boundary thermal one-point function of $\phi^2$, $b_{\phi^2}$ as defined in \eq{no_defect_1-pt}.
Computing $\langle \phi^2\rangle_\beta$ in the absence of a boundary amounts to dropping all terms in \eq{phi2} that are proportional to $\varsigma$.
The remaining sum over $m$ then gives $2\kappa \zeta(d-2)/\beta^{d-2}$ such that $b_{\phi^2}=2\kappa\zeta(d-2)$. 
Note that $b_{\phi^2}$ diverges for $d\to3$, reflecting the logarithmic divergence of the thermal image sum.
In the opposite limit $z\to0$, we find from \eq{F_phi2_d} that $F^\varsigma_{\phi^2}(z)\overset{z\to0}{\sim} {\frac{\varsigma \kappa}{(2z)^{d-2}}}$ for any even integer $d$.
The coefficient agrees precisely with the zero-temperature one-point functions $a_{\phi^2}^\varsigma$ for N and D BC scalars computed by ref.~\cite{McAvity:1995zd}.\footnote{
Our conventions differ from the ones of ref.~\cite{McAvity:1995zd} by a factor of $2^\Delta$, such that $a_\CO^\mathrm{here}=a_\CO^\mathrm{there}/2^\Delta$, see section~\ref{sec:review}.}
Indeed, it is immediate to see from the first term of \eq{phi2} that the zero-temperature ($\beta\to\infty$) one-point function is $a_{\phi^2}^\varsigma= {\frac{\varsigma\kappa}{2^{d-2}}}$.

Away from $z=0$, we can expand for small $z$ to extract thermal BCFT data.
E.g.\ when $d=4$ we can use the Laurent expansion of $\coth (2\pi z)$ inside \eq{F_phi2}.
Comparing with \eq{FO_DOE}, we see that only boundary local operators with even scaling dimension ${\hat{\Delta}}$ in the BOE of $\phi^2$ can acquire a boundary thermal one-point function.
Here we assume that there are no degeneracies and consequently there can be no cancellations between boundary primaries of the same scaling dimension.
We will justify this assumption below \eq{thermal_BCFT_data}.
For the coefficients we find
\begin{equation}
\mu_{\phi^2}{}^{\hat{\CO}_{{\hat{\Delta}}}}\hat{b}_{\hat{\CO}_{{\hat{\Delta}}}}=\varsigma\,\frac{(4\pi)^{{\hat{\Delta}} } B_{{\hat{\Delta}} } }{4{\hat{\Delta}} !}\kappa
\end{equation}
for ${\hat{\Delta}}=0$ and for even integer ${\hat{\Delta}}\geq4$.
The ${\hat{\Delta}}=0$ case $\mu_{\phi^2}{}^{\hat{\CO}_{{\hat{\Delta}}}}\hat{b}_{\hat{\CO}_{{\hat{\Delta}}}}={\frac{\varsigma\kappa}{4}}$ corresponds to the zero-temperature one-point function $a^\varsigma_{\phi^2}$.
For N BCs there is an additional operator of dimension ${\hat{\Delta}}=2$ with $\mu_{\phi^2}{}^{\hat{\CO}_{{\hat{\Delta}}}}\hat{b}_{\hat{\CO}_{{\hat{\Delta}}}}=\frac{2\pi^2}{3}\kappa$.
Note that the Laurent expansion of $\coth(2\pi z)$ converges provided that $|z|<1/2$.
This is precisely the radius of convergence of the BOE at finite temperature, and coincides with the distance to the nearest pole in the complex $z$-plane at $z=\pm\iu/2$.

We could similarly expand $F^\varsigma_{\phi^2}(z)$ as given by \eq{F_phi2_d} for any even integer dimension $d>4$ to obtain thermal BCFT data.
However, we can do better and extract thermal boundary one-point functions $\hat{b}_{\hat{\CO}}$ for \emph{any} $d>3$.
Even though the summation over thermal images in \eq{phi2} is challenging, finding a series expansion of $F^\varsigma_{\phi^2}(z)$ is straightforward for general $d$.
Using the generalised binomial theorem
\begin{equation}
\label{eq:gen_bin}
(1+y)^{-s}=\frac{1}{\Gamma(s)}\sum_{n=0}^\infty(-1)^n\frac{\Gamma(s+n)}{\Gamma(n+1)}y^n
\end{equation}
inside \eq{phi2}, we obtain
\begin{equation}
\langle\phi^2(x)\rangle_\beta=\frac{\varsigma\,\kappa}{(2x_\perp)^{d-2}}+\frac{2\kappa}{\beta^{d-2}}\sum_{m=1}^\infty\left(\frac{1}{m^{d-2}}+\varsigma \sum_{n=0}^\infty(-1)^n \frac{2^{2n}\Gamma(n+\frac{d-2}{2})}{\Gamma(\frac{d-2}{2})\Gamma(n+1)}\frac{z^{2n}}{m^{d+2n-2}}\right).
\end{equation}
Exchanging the order of summation, and performing the sum over $m$ first, we find that
\begin{equation}
\label{eq:Fphi2}
\begin{split}
F^\varsigma_{\phi^2}(z)&=\kappa {\frac{1}{z^{d-2}}}\Bigg({\frac{\varsigma}{2^{d-2}}}+2(1+\varsigma)\zeta(d-2)z^{d-2}\\
&\quad+ \varsigma\sum_{n=1}^\infty(-1)^n \frac{\Gamma(n+\frac{d-2}{2})}{\Gamma(\frac{d-2}{2})\Gamma(n+1)}{2^{2n+1}}\zeta(d+2n-2)z^{d+2n-2}\Bigg)\,.
\end{split}
\end{equation}
The sum over $n$ converges for any $d$ provided that $|z|<1/2$, which is the radius of convergence of the BOE.
Comparing with \eq{FO_DOE} it is then straightforward to read off the thermal boundary one-point functions for non-trivial boundary local operators
\begin{equation}
\label{eq:thermal_BCFT_data}
\mu^{\varsigma,\hat{\CO}}_{\phi^2}\hat{b}^\varsigma_{\hat{\CO}}=
\begin{cases}
2(1+\varsigma)\zeta({\hat{\Delta}})\,\kappa & \text{ if } {\hat{\Delta}}=d-2\,,\\
-{2^{2n+3}}\varsigma\cos\left(\pi\frac{d-{\hat{\Delta}}}{2}\right)\dfrac{ \Gamma \left(\frac{{\hat{\Delta}} }{2}\right) \zeta ({\hat{\Delta}} )}{\Gamma \left(\frac{d-2}{2}\right) \Gamma \left(\frac{{\hat{\Delta}}-d}{2} +2\right)}\,\kappa  & \text{ if }{\hat{\Delta}}=d+2n\,,
\end{cases}
\end{equation}
where $n$ is a non-negative integer, and zero otherwise.
Note that all ${\hat{\Delta}}$'s have the same parity modulo 2 as $d$, and so the cosine factor is just a sign.
In principle the left-hand side of \eq{thermal_BCFT_data} could involve a sum over all primaries $\hat{\CO}$ with scaling dimension ${\hat{\Delta}}$.
For a free massless scalar field with N or D BCs, however, it is simple to show that there is a unique boundary primary $\hat{\CO}$ for each ${\hat{\Delta}}$.
This follows from the BOE of $\phi$, which contains only one boundary primary.
For N BCs this unique boundary primary is just the boundary value of $\phi|_{x_\perp=0}\equiv \hat{\phi}^{(0)}$, while for D BCs it is $\p_\perp\phi|_{x_\perp=0}\equiv\phi^{(1)}$.
Operators $\hat{\CO}$ in the BOE of $\phi^2$ are then ``double-trace'' operators quadratic in $\phi^{(0)}$ or $\phi^{(1)}$.
Schematically, $:\hat{\phi}^{(0)}\p^{2n+2}_\parallel\hat{\phi}^{(0)}:$ for N BCs and $:\hat{\phi}^{(1)} \p^{2n}_\parallel\hat{\phi}^{(1)}:$ for D BCs, where $n$ is a non-negative integer and colons denote normal ordering.
For N BCs we also allow for $n=-1$ corresponding to an operator of scaling dimension ${\hat{\Delta}}=d-2$.
These boundary primaries can be constructed explicitly by demanding that they are annihilated by the special conformal generators preserved by the boundary.
Given a seed scalar primary in a free CFT, refs.~\cite{Heemskerk:2009pn,Penedones:2010ue,Fitzpatrick:2011dm} constructed all primary ``double-trace'' operators.
Since only scalar primaries appear in the BOE of a scalar in BCFT, this selects a subset of ``double-trace'' primaries.
These primaries all have non-degenerate scaling dimensions which precisely match the ones contributing non-trivially to the BOE channel of the thermal one-point function of $\phi^2$ in \eq{thermal_BCFT_data}.
This implies that there can be no sum over boundary primaries on the left-hand side of \eq{thermal_BCFT_data}.
Moreover, our calculation shows that all operators in the BOE of $\phi^2$ also acquire boundary thermal one-point functions.

To extract the finite-temperature boundary one-point functions $\hat{b}_{\hat{\CO}}$, we need to divide out by the zero-temperature bulk-boundary two-point functions $\mu_{\phi^2}{}^{\hat{\CO}}$.
These were computed by ref.~\cite{Liendo:2012hy} from the zero-temperature two-point function of $\phi^2$ in the BOE channel.
The non-vanishing bulk-boundary two-point functions are $(\mu_{\phi^2}{}^{\hat{\CO}})^2=4(1+\varsigma)\kappa^2$ for ${\hat{\Delta}}=d-2$ and
\begin{equation}
\label{eq:lambda2}
(\mu_{\phi^2}{}^{\hat{\CO}})^2=\frac{\sqrt{\pi } 2^{6-d }  \Gamma ({\hat{\Delta}} -1) \Gamma ({\hat{\Delta}} )}{\Gamma \left(\frac{d}{2}-1\right) \Gamma (d-2) \Gamma (-d+{\hat{\Delta}} +3) \Gamma \left(-\frac{d}{2}+{\hat{\Delta}} +\frac{1}{2}\right)}\kappa^2
\end{equation}
for ${\hat{\Delta}}=d+2n$ with non-negative integer $n$.\footnote{
Here we assumed that the propagator is normalised as in \eq{flat_G}.
Note that our normalisation differs from that of ref.~\cite{Liendo:2012hy} by a factor of $\kappa^2$.
Additionally, ref.~\cite{Liendo:2012hy} implicitly normalises the operator $\phi^2$ with an extra factor of $1/\sqrt{2}$.
Consequently, our \eq{lambda2} is ${2^{2({\hat{\Delta}}-d)+5}}\kappa^2$ their result.}\textsuperscript{,}\footnote{
More general two-point functions of the form $\langle \phi^k (x) \phi^k(y)\rangle$ for generalised free fields $\phi$ with N or D BCs as well as for the critical $O(N)$ BCFT were recently studied by ref.~\cite{Sun:2026mib}.}
Without loss of generality we can choose the sign of $\mu_{\phi^2}{}^{\hat{\CO}}$ to be positive.
Choosing the opposite sign just amounts to flipping the sign of $\hat{\CO}$.
This does not affect orthonormality of the boundary-boundary two-point function but changes the sign of the thermal boundary one-point function $\hat{b}_{\CO}$.
The $\hat{b}_{\CO}$'s can then be isolated by dividing the right-hand side of \eq{thermal_BCFT_data} by the square root of \eq{lambda2}.
We thus obtain an infinite amount of thermal BCFT data.

\subsubsection{Thermal one-point function of $T_{\mu\nu}$}
\label{sec:scalar_T}

Next we consider the thermal one-point function of the stress tensor of a free massless scalar field in $d$ dimensions.
Classically, the stress tensor is given by
\begin{equation}
\label{eq:scalar_T}
T_{\mu\nu}=\partial_\mu\phi\partial_\nu\phi-\frac{d-2}{4(d-1)}\left(\partial_\mu\partial_\nu+\frac{\delta_{\mu\nu}}{d-2}\partial^2\right)\phi^2\,.
\end{equation}
It is symmetric-traceless and conserved on-shell.
Its thermal one-point function can be computed as for the scalar case.
For the first term $\langle\partial_\mu\phi\partial_\nu\phi\rangle_\beta$ we use the propagator in \eq{thermal_G} in the coincident limit with UV divergences subtracted.
For the second term $\propto \left(\partial_\mu\partial_\nu+\frac{\delta_{\mu\nu}}{d-2}\partial^2\right)\langle\phi^2\rangle_\beta$ we can use the thermal one-point function of $\phi^2$ computed in \eq{F_phi2_d}.
Putting everything together, we find that the result takes the form of \eq{kin_T}.
In general even dimensions $d\geq4$, we can express it as
\begin{subequations}
\label{eq:T_S_d}
\begin{align}
b_T&=-4n (2n+2) \,\zeta (2n+2)\,\kappa\,,\\
\begin{split}
F^{\perp\perp}_T&=-\varsigma\,\kappa\,\frac{\pi 2^{2n+2-d} n(2n+2) \,}{2n+1}\, \frac{(-1)^n}{(n+1)!}\\
&\quad\times\left[\frac{\p^n}{\p A^n} \left(\frac{\coth \left(\pi  \sqrt{4z^2+A}\right)}{\sqrt{4z^2+A}}-\pi  \csch^2\left(\pi  \sqrt{4z^2+A}\right)\right)\right]_{\!A=0}\,,
\end{split}
\end{align}
\end{subequations}
where $n=(d-2)/2$ and $\varsigma=+1$ ($\varsigma=-1$) is again for N (D) BCs.
Note that eq.~\eqref{eq:T_S_d} is regular for $z=0$, and has poles of order $(d+2)/2$ for $z=\iu k/2$ with $k\in \mathbb{Z}\setminus\{0\}$.
The absence of a pole at $z=0$ is due to a cancellation of the poles of the $\coth$ and $\csch$ terms, which both have poles at the same location in the complex plane.
Physically this comes from the fact that the zero-temperature one-point function vanishes for a spinning primary in the presence of a boundary.

For $d=4$, $\langle T_{\mu\nu}\rangle_\beta$ was computed by ref.~\cite{Kennedy:1979ar}.
Substituting $d=4$ in \eq{T_S_d}, we find 
\begin{align}
\label{eq:T_S_4d}
b_T&=-\frac{2\pi^2}{45}\,,&
F^{\perp\perp}_T&={\frac{\varsigma}{16 z^{3}}} \,\frac{  2 \pi  z (4 \pi  z \coth (2 \pi  z)-1) \csch^2(2 \pi  z)-\coth (2 \pi  z)}{6 \pi }\,.
\end{align}

Let us briefly discuss the asymptotic properties of $F^{\perp\perp}_T$ for general even integer $d$ and check that it matches the expected behaviour from table \ref{tab:spin_two_one_point_structures}.
For $z\gg1$, we have $F^{\perp\perp}_T(z)\overset{z\to\infty}{\sim}z^{{1-d}}$ for any $d$.
This ensures that $\langle T_{\mu\nu}\rangle_\beta$ reduces to the no-defect one-point function when $x_\perp\to\infty$.
For $z\ll1$, it is easy to see that $F^{\perp\perp}_T(z)$ approaches a constant as $ z\to0$ such that the zero-temperature one-point function of $T_{\mu\nu}$ vanishes in the $\beta\to\infty$ limit.
In principle one could expand $F^{\perp\perp}_T(z)$ as a series around $z=0$ in order to extract thermal BCFT data of spinning primaries in the BOE of $T_{\mu\nu}$.
However, we will not pursue this here and leave the thermal bootstrap of spinning operators for future work.

\begin{figure}[h]
\centering
\includegraphics[width=12cm]{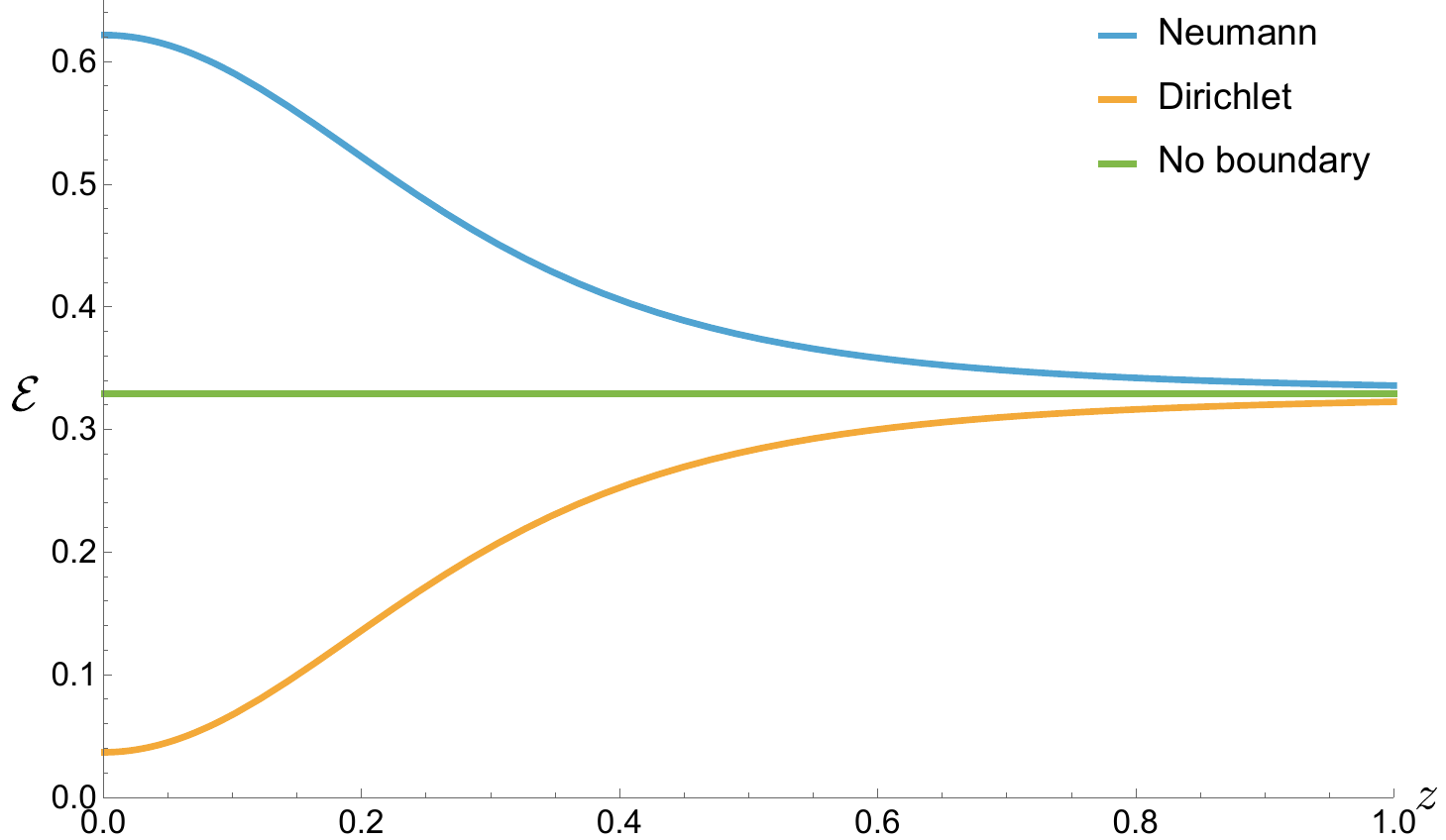}
\caption{
Plot of the dimensionless energy density $\CE(z)$ for a free scalar field with N (blue) and D (orange) BCs as a function of the cross-ratio $z=x_\perp/\beta$, and the value of $\CE(z)$ for a free scalar field without boundary (green).
}
\label{fig:E_4d}
\end{figure}

Instead we will study thermodynamic properties of the scalar with N or D BCs.
Given a thermal one-point function of the stress tensor $T^{\mu\nu}$ we can define the energy density
\begin{equation}
\label{eq:E}
E=\frac{\CE(z)}{\beta^d}=-\langle T^{\tau\tau}\rangle_\beta\,,
\end{equation}
where the minus sign ensures that after Wick rotation to Lorentzian signature the energy density is exactly equal to the stress tensor one-point function.\footnote{
If there are decoupled degrees of freedom on the boundary, the right-hand side of \eq{E} should be understood as the total stress tensor, including a separately conserved boundary stress tensor.}
It is convenient to introduce the dimensionless function $\CE(z)$, which we will refer to as the dimensionless energy density.\footnote{
Note that we use different conventions compared to ref.~\cite{Barrat:2024aoa}.
While ref.~\cite{Barrat:2024aoa} makes densities dimensionless by multiplying with an appropriate factor of the transverse distance, we find it more natural to use factors of $\beta$.
Consequently, our $\CE^\mathrm{here}=z^{-d}\CE^\mathrm{there}$.}
For concreteness we take $d=4$.
The qualitative features we shall describe below persist in higher even integer dimensions as well.
Using \eq{kin_T} with \eq{T_S_4d}, we have 
\begin{equation}
\label{eq:Ttt_S}
\CE(z) =\frac{ \pi ^2}{30}+\frac{\varsigma}{48\pi z^3}  \left(\coth (2 \pi  z)+2 \pi  z (1-4 \pi  z \coth (2 \pi  z)) \csch^2(2 \pi  z)\right).
\end{equation}
In figure~\ref{fig:E_4d} we plot $\CE$ as a function of $z$ for a free massless scalar with N or D BCs together with the constant no-boundary value $\CE=-3b_T/4$.
We observe that $\CE(z)$ for N (D) BCs is a monotonically decreasing (increasing) function of $z$. 
Far from the boundary $z\gg1$, boundary effects become negligible and $\CE(z)$ for both N and D BCs asymptotes to the no-boundary value $\frac{\pi ^2}{30}$.
The deviation of the one-point functions from the no-boundary value becomes $\CO(1)$ in units of $\kappa$ when $2\pi z\sim1$.
As $z\to0$, the one-point function approaches a finite non-zero value, $\CE(0) =\frac{1}{270} \pi ^2 (9+ 8\varsigma)$.
This is a thermal effect which is entirely due to the presence of the boundary and signals the presence of ${\hat{\Delta}}=4$ operators in the BOE of $T_{\tau\tau}$ which acquire a non-zero defect thermal one-point function.
Even though $\CE(0)\neq0$, the full one-point function $\langle T_{\tau\tau}\rangle_\beta\to0$ as $\beta\to\infty$, which is consistent with the fact that spinning operators in BCFT cannot acquire one-point functions at zero temperature.
Note that the energy densities remain positive for all $z$, and $\CE_+(z)>\CE_-(z)$ for all $z$, where $\CE_{+ (-)}$ is the energy density for the scalar with N (D) BCs.
Since D BCs can be reached from Neumann BCs via a boundary RG flow, we observe that $\CE(z)$ is greater in the UV than in the IR.

Given the energy density $E$, we can define the thermal free energy density $F$ via the thermodynamic relation
\begin{equation}
\label{eq:F_E-TS}
F=\frac{f(z)}{\beta^d}=E-TS\,,
\end{equation}
where $T=1/\beta$ and $S$ is the entropy density
\begin{equation}
\label{eq:S}
S=\frac{s(z)}{\beta^{d-1}}=\beta^2\frac{\d F}{\d \beta}\,.
\end{equation}
The functions $f$ and $s$ are dimensionless functions of the cross-ratio $z$, and the derivative in \eq{S} is taken at fixed $x_\perp$.\footnote{
Similarly to the dimensionless energy density $\CE(z)$ we define $s(z)$ and $f(z)$ by multiplying with an appropriate factor of $\beta$ rather than a factor of the transverse distance $|x_\perp|$.
Our quantities therefore differ from those defined by ref.~\cite{Barrat:2024aoa} as $s^\mathrm{here}=z^{1-d}s^\mathrm{there}$ and $f^\mathrm{here}=z^{-d}f^\mathrm{there}$.}
The defining equation of the free energy can be recast as 
\begin{equation}
zf'(z)=(1-d)f(z)-\CE(z)\,.
\end{equation}
Its formal solution is
\begin{equation}
\label{eq:f_int}
f(z)=z^{1-d}\left( c-\int_1^z\d y\,y^{d-2}\CE(y)\right),
\end{equation}
such that
\begin{equation}
\label{eq:s_int}
s(z)=\CE(z)-z^{1-d}\left(c-\int_1^z \d y\,y^{d-2}\CE(y)\right),
\end{equation}
where $c$ is an integration constant that amounts to a temperature-independent shift of the entropy density $S$. 
To fix $c$, ref.~\cite{Barrat:2024aoa} argued that one should compare to the entropy density $S$ at zero-temperature.
Physically, as $\beta\to\infty$ the system settles into its ground state. 
The entropy then measures the logarithm of the ground state degeneracy. 
If the ground state is unique, which is the case for the free scalar BCFT, then the entropy must vanish as $\beta\to\infty$.

We will now compute $s(z)$ and $f(z)$ for the free scalar BCFT when $d=4$.
Since $\CE(z)$ is finite as $z\to0$, the limit $S\to0$ as $\beta\to\infty$ requires
\begin{equation}
\label{eq:c_S}
c=-\int_0^1\d y\, y^{2} \CE(y)\,.
\end{equation}
The indefinite version of this integral is the same as the one that appears in eqs.~\eqref{eq:s_int} and \eqref{eq:f_int} for $s(z)$ and $f(z)$, respectively.
We have not been able to perform this integral analytically.
Consequently, we will compute $s(z)$ and $f(z)$ numerically.

\begin{figure}[t]
\centering
  \begin{minipage}{0.49\textwidth}
  \centering
    \includegraphics[width=\linewidth]{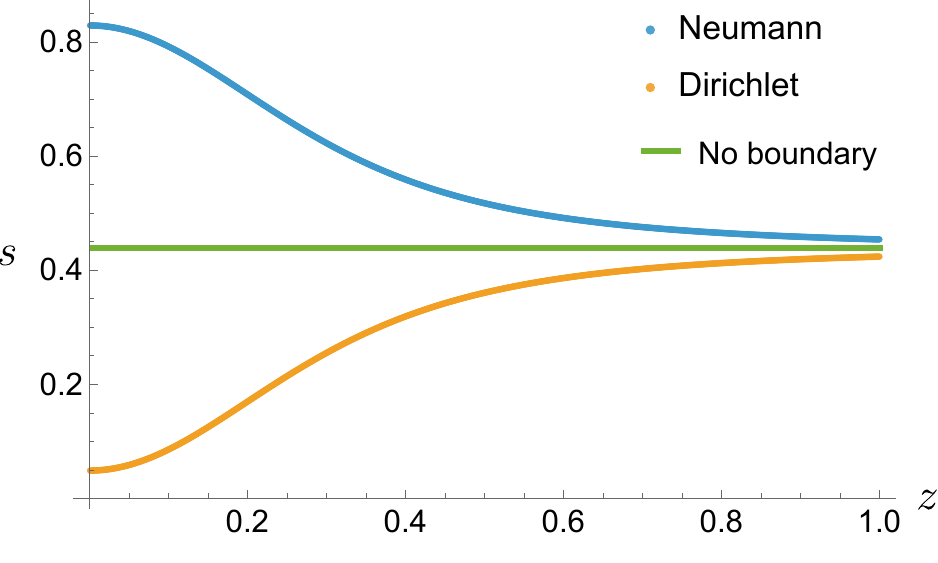}
  \end{minipage}\hfill
  \begin{minipage}{0.49\textwidth}
  \centering
    \includegraphics[width=\linewidth]{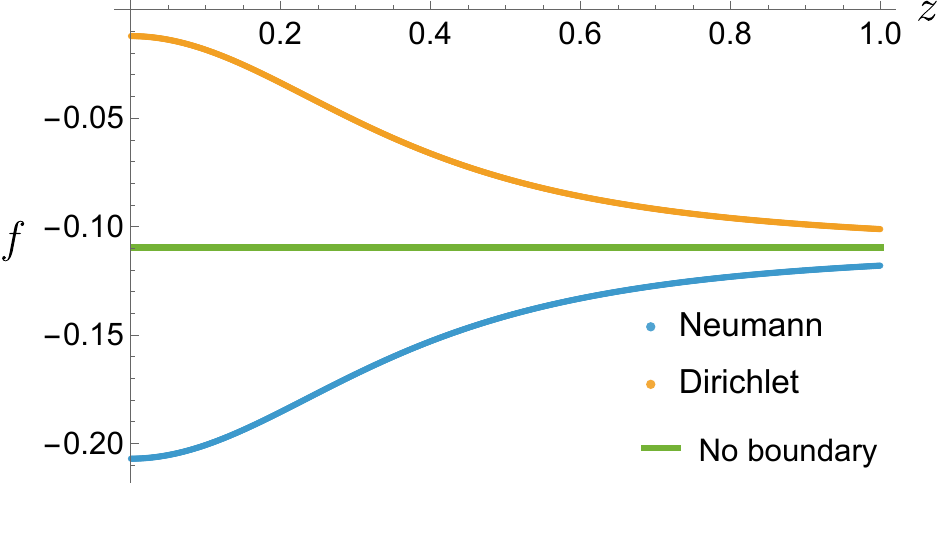}
  \end{minipage}
\caption{\emph{Left:} Plot of the dimensionless entropy density $s(z)$ for a free scalar field with N (blue) and D (orange) BCs as a function of the cross-ratio $z=x_\perp/\beta$ and the value of $s(z)$ for a free scalar field without boundary (green). \\
\emph{Right:} Plot of the dimensionless free energy $f(z)$ for a free scalar field with N (blue) and D (orange) BCs as a function of $z$ and the value of $f(z)$ for a free scalar field without boundary (green).}
\label{fig:s_f_4d}
\end{figure}

We plot the dimensionless entropy density $s(z)$ for N and D BCs as a function of $z$ in figure~\ref{fig:s_f_4d} on the left, together with the no-boundary value.
The latter can be computed analytically, and we find $s(z)=\frac{2 \pi ^2}{45}$.
We observe that N BCs have greater entropy than D BCs.
Similarly, we can compute the dimensionless free energy density $f(z)$.
We plot the numerical results in figure~\ref{fig:s_f_4d} on the right, together with the analytic no-boundary result $f(z)=-\frac{ \pi ^2}{90}$.
The dimensionless free energy density is negative for all values of $z$.
Moreover, we see that $f(z)$ is larger in magnitude for N than D BCs.
Consequently, $f(z)$ is larger in magnitude in the UV than in the IR at every $z$.
Note that the functions of $\CE(z)$, $s(z)$ and $f(z)$ are qualitatively similar, with the roles of N and D BCs reversed for $f(z)$ compared with $\CE(z)$ and $s(z)$.
However, the functions are not identical, and we have not been able to find analytic expressions for $f(z)$ and $s(z)$.

In addition to the local densities $\CE$, $s$ and $f$ we can also compute the the full energy, entropy and free energy, $\mathsf{E}$, $\mathsf{S}$ and $\mathsf{F}$, respectively. 
E.g.\ to obtain the energy $\mathsf{E}$ we can simply integrate the energy density $E$ over space
\begin{equation}
\mathsf{E}=\int \d^{d-1} \vec{x} \,E(x) = \frac{A}{\beta^{d-1}}\int_0^\infty \d z \, \CE(z)\,,
\end{equation}
where $A$ is the $(d-2)$-dimensional area of the non-compact boundary directions.
From the general form of the stress tensor one-point function \eq{kin_T} it is clear that $\mathsf{E}$ diverges due to the constant contribution proportional to $b_T$.
To obtain a finite quantity, we subtract the no-boundary thermal one-point function $\langle T_{\tau\tau}\rangle_\beta^{(0)}$.
We therefore define the boundary contribution to the energy as
\begin{equation}
\label{eq:int_thermo_E_bdy}
\mathsf{E}_\p \equiv \frac{A}{\beta^{d-1}}\int_0^\infty \d z \, \left(\CE(z)+\frac{d-1}{d}b_T\right),
\end{equation}
which is manifestly finite.
Instead of directly integrating the stress tensor thermal one-point function computed for even $d$, we can obtain $\mathsf{E}$ analytically for any $d$ by first integrating over $z$ before performing the sum over thermal images.
The one-point function with the no-boundary piece subtracted is
\begin{equation}
\langle T_{\tau\tau}\rangle_\beta - \langle T_{\tau\tau}\rangle_\beta^{(0)} = - {\kappa}\varsigma \frac{d (d-2)^2}{d-1}\,{\sum_{m=-\infty}^{+\infty}}\frac{(m\beta)^2}{\left(m^2\beta ^2 +4 x_\perp^2\right)^{\frac{d}{2}+1}}\,.
\end{equation}
Integrating $x_\perp$ first over the half-line and then performing the sum over Matsubara modes gives
\begin{equation}
\label{eq:total_E_S}
\mathsf{E}_\p=\varsigma\frac{A}{\beta^{d-1}}\frac{(d-2)\Gamma \left(\frac{d-1}{2}\right)}{4 \pi ^{\frac{d-1}{2}}}\zeta (d-1)
\end{equation} 
for $d>2$.

The full thermodynamic free energy is proportional to the logarithm of the partition function, $\mathsf{F}=-\beta^{-1}\log Z$.
It is related to the energy $\mathsf{E}$ via a version of eqs.~\eqref{eq:F_E-TS} and~\eqref{eq:S} involving the full thermodynamic quantities $\mathsf{E}$, $\mathsf{S}$ and $\mathsf{F}$.
In particular, the boundary contribution to the free energy $\mathsf{F}_\p$ is related to $\mathsf{E}_\p$ as $\mathsf{E}_\partial=\partial_\beta(\beta\mathsf{F}_\partial)$.
If $\mathsf{F}_\p$ is extensive, i.e.\ it is proportional to the area of the boundary, then scale invariance fixes $\mathsf{F}_\p$ up to a dimensionless constant.\footnote{
Note that $\mathsf{F}_\p$ could include an additional term $\mathsf{c}/\beta$ for some area-independent integration constant $\mathsf{c}$.
In the infinite area limit this term can be neglected compared to the leading extensive contribution provided that $d>2$.
For $d=2$, $\mathsf{E}_\p=0$ since $\langle T_{\tau\tau}\rangle_\beta = \langle T_{\tau\tau}\rangle_\beta^{(0)}$.
Integrating $\mathsf{E}_\partial=\partial_\beta(\beta\mathsf{F}_\partial)$ then gives only the contribution proportional to $1/\beta$. 
Comparing with the boundary contribution to the partition function, the coefficient of the $1/\beta$ term is identified with minus the Affleck-Ludwig boundary entropy $\log g$~\cite{Affleck:1991tk}.}
Therefore, $\mathsf{E}_\p=-(d-2)\mathsf{F}_\p$.
Similarly, the boundary contribution to the entropy $\mathsf{S}_\p$ is obtained from $\mathsf{F}_\p$ via \eq{S}, with $S$ and $F$ replaced by $\mathsf{S}_\partial$ and $\mathsf{F}_\p$, respectively.
If $\mathsf{S}_\partial$ is extensive, then $\mathsf{S}_\p=-(d-1)\beta\mathsf{F}_\p$.
Thus, \eq{total_E_S} for the boundary contribution to the energy also determines $\mathsf{S}_\p$ and $\mathsf{F}_\p$ analytically.
It is simple to check numerically that the area under the curves of $s$ and $f$ in figure~\ref{fig:s_f_4d}, with no-boundary contribution subtracted, reproduce the analytic expressions when $d=4$.
When $\beta\to\infty$, the boundary contribution to the entropy $\mathsf{S}_\partial\to0$ as a consequence of fixing the integration constant $c$ as in \eq{c_S}.
Moreover, since $(-f^{UV})>(-f^{IR})$ holds point-wise for every $z$, the same relation holds for the integrated quantities.
Consequently, the free energy  is negative and greater in magnitude in the UV than in the IR.

We emphasise the behaviour of the free energy in the free scalar BCFT under boundary RG flows for the following reason.
For RG flows in ordinary QFT without defects, $\beta^d F$ is a dimensionless constant at the CFT fixed points.
Ref.~\cite{Appelquist:1999hr} argued that $-\beta^d F$ is  greater in the UV than in the IR, provided that the UV is asymptotically free.
More recently, ref.~\cite{Cheung:2026dng} showed rigorously in effective field theory that corrections from interactions to minus the free energy in the IR must be positive in a low temperature expansion.
Their result holds provided that the theory is free of long range forces and the UV fixed point is perturbative.
For BCFT, $\beta^d F$ is no longer a dimensionless constant but rather the function $f(z)$.
We will not attempt to generalise the argument of ref.~\cite{Appelquist:1999hr} to the case with boundaries here.
Instead, we just note that the two naive generalisations to the boundary case, i.e.\ $(-f^{UV})>(-f^{IR})$ point-wise at every $z$ and the integrated version $(-\mathsf{F}^{UV})>(-\mathsf{F}^{IR})$ both hold for the boundary RG flow from N to D BCs.\footnote{
When the UV fixed point is strongly coupled there are known counter-examples to the proposal of ref.~\cite{Appelquist:1999hr}, see e.g.\ refs.~\cite{Chubukov:1993aau,Sachdev:1993pr}.
A more robust quantity for monotonicity theorems in QFT was later shown to be the universal part of the CFT free energy on $S^d$, $\CF\propto \log Z[S^d]$, rather than the thermodynamic free energy on $S^1\times \mathbb{R}^{d-1}$.
It involves the A-type trace anomaly coefficients featuring in the 2d $c$-theorem~\cite{Zamolodchikov:1986gt} and 4d $a$-theorem~\cite{Cardy:1988cwa, Komargodski:2011vj} and the quantity entering the 3d $\CF$-theorem~\cite{Jafferis:2011zi,Klebanov:2011gs}.
See also refs.~\cite{Casini:2004bw, Casini:2012ei, Casini:2016udt, Casini:2017vbe} for quantum information theoretic proofs for integer $2\leq d \leq4$.
A generalised $\CF$-theorem for any continuous $d$ was conjectured in ref.~\cite{Giombi:2014xxa}.
Similarly, the known monotonicity theorems for boundary or defect RG flows generically do \emph{not} involve the thermodynamic free energy.
The sole exception occurs for boundaries of 2d CFTs, where the boundary contribution to the thermal free energy obeys the so-called $g$-theorem of refs.~\cite{Affleck:1991tk,Friedan:2003yc}.
For higher-dimensional boundaries or defects, the robust quantity again is the sphere free energy.
See ref.~\cite{Cuomo:2021rkm} for lines, refs.~\cite{Jensen:2015swa,Shachar:2022fqk} for surfaces, refs.~\cite{Nozaki:2012qd,Gaiotto:2014gha} for 3d boundaries in 4d CFTs, and ref.~\cite{Wang:2021mdq} for 4d defects.
See also refs.~\cite{Casini:2016fgb,Casini:2018nym,Casini:2022bsu,Casini:2023kyj} for quantum information theoretic proofs for $1\leq p\leq 4$ and any co-dimension.
A generalised defect $\CF$-theorem was conjectured by ref.~\cite{Kobayashi:2018lil}.
}

\subsection{Free Dirac fermion}
\label{sec:free_fermion}

In this subsection we consider a free massless Dirac fermion in $d$ dimensions.
We will impose mixed BCs,
\begin{equation}
\label{eq:mixed_BCs}
(\mathbbm{1}-\g_\perp)\psi|_{x_\perp=0}=0\,,
\end{equation}
where $\g_\perp$ is one of the $d$ Dirac matrices.\footnote{
We use a stationary frame such that we can identify indices of the Dirac matrices with spacetime indices.}
The Dirac matrices obey the Clifford algebra $\{\gamma_\mu,\gamma_\nu\}=2\delta_{\mu\nu}\mathbbm{1}_{\mathfrak{d}}$.
Here $\mathfrak{d}=2^{\lfloor d/2\rfloor}$ denotes the dimension of the Clifford algebra representation.
The zero-temperature propagator respecting these BCs is 
\begin{equation}
\label{eq:F_prop}
G(x,y)=\langle\psi(x)\bar{\psi}(y)\rangle=\kappa_f\left(\frac{\g\cdot(x-y)}{|x-y|^d}+\frac{\g_\perp\g\cdot(\tilde{x}-y)}{|\tilde{x}-y|^d}\right),
\end{equation}
where $\kappa_f=(d-2)\kappa$ with $\kappa$ defined in \eq{kappa}.
To obtain the finite-temperature Green's function $G^\beta(x,y)$ we replace $\Delta\tau\to\Delta\tau+m\beta$.
Unlike the scalar case, we sum over $m$ with alternating sign $(-1)^m$,
\begin{equation}
\label{eq:thermal_G_F}
G^\beta(x,y)=\sum_{m=-\infty}^\infty (-1)^m\,G(x,y+m\beta \,e)\,,
\end{equation}
where $e^\mu=(1,0,\ldots,0)$.
This ensures that the fermion has anti-periodic BCs along the thermal circle.
The resulting correlation functions are genuine finite-temperature observables.

In the following subsections, we will compute the one-point functions of simple operators that are quadratic in fields. 
In particular, we will consider the scalar fermion bilinear $\bar{\psi}\psi$ and the stress tensor. 
The $U(1)$ vector current as well as the pseudoscalar fermion bilinear and the axial current for even $d$ are also quadratic in fields. 
However, it is straightforward to show that their thermal one-point functions vanish.

\subsubsection{Thermal one-point function of $\bar{\psi}\psi$}

The simplest observable to evaluate is the thermal one-point function of the fermion bilinear $\bar{\psi}\psi(x)$. 
Due to standard Dirac matrix trace identities, the thermal one-point function is finite and does not need to be regularised.
We find
\begin{equation}
\label{eq:F_bilin}
\langle \bar{\psi}\psi(x)\rangle= - \Tr G^\beta(x,x)=2\mathfrak{d}\,\kappa_f\,\sum_{m=-\infty}^\infty (-1)^m \frac{x_\perp}{|m^2\beta^2+4x_\perp^2|^{d/2}}\,,
\end{equation}
where the factor of $(-1)^m$ under the sum ensures anti-periodic BCs.
As for the scalar, this sum can be performed in closed form only when $d\geq2$ is an even integer.
We find the following way to compactly write the thermal one-point function for general even integer $d$,
\begin{equation}
\label{eq:F_F_d}
F_{\bar{\psi}\psi}(z)=\mathfrak{d}\,\kappa_f{2z} \frac{(-1)^n}{n!}\!\left[\frac{\p^{n}}{\p A^{n}}\, \frac{\pi  \sqrt{4z^2+A} \csch\left(\pi  \sqrt{4z^2+A}\right)}{4z^2+A}\right]_{\!A=0},
\end{equation}
where $n=(d-2)/2$.
Similarly to the scalar BCFT of section~\ref{sec:free_scalar}, this function has poles on the imaginary axis when $z=\iu k/2$ for integer $k$.

It is straightforward to check that \eq{F_F_d} has the requisite asymptotic properties given by \eq{FO_asymp}.
For $z\gg1$, $F_{\bar{\psi}\psi}(z)\overset{z\to\infty}{\sim}0$ non-perturbatively fast.
This is consistent with the fact that the no-boundary thermal one-point function $b_{\bar{\psi}\psi}=0$.
This can be seen from the fermion propagator, which in the absence of a boundary just contains the first term in \eq{F_prop} with a single Dirac matrix.
The no-boundary one-point function then vanishes as a consequence of a Dirac matrix trace identity, also at finite temperature.
In the opposite regime $z\to0$, we see that $F_{\bar{\psi}\psi}(z)\overset{z\to0}{\sim}{\frac{\mathfrak{d}\,\kappa_f}{(2z)^{d-1}}}$ for any even integer $d$.
The coefficient is nothing but the zero-temperature one-point function $a_{\bar{\psi}\psi}= {\frac{\mathfrak{d}\,\kappa_f}{2^{d-1}}}$.

Beyond the strict $z\to0$ limit, we can find a small $z$ expansion of $F_{\bar{\psi}\psi}(z)$ starting from \eq{F_bilin} and using the generalised binomial theorem in \eq{gen_bin}.
This allows us to extract thermal BCFT data for any (not necessarily even integer) $d$. 
We find the following series expansion in terms of thermal blocks
\begin{equation}
F_{\bar{\psi}\psi}(z)= {\frac{\mathfrak{d}\,\kappa_f}{(2z)^{d-1}}} \left(1-2\sum_{n=0}^\infty(-1)^n\frac{ \left(2^{d+2 n}-2\right)   \Gamma \left(\frac{d}{2}+n\right)\zeta (d+2 n)\,z^{d+2 n}}{\Gamma\left(\frac{d}{2}\right)\Gamma (n+1)}\right),
\end{equation}
where we used $\sum_{m=1}^\infty \frac{(-1)^m}{m^s}=\left(2^{1-s}-1\right) \zeta (s)$.
The first term corresponds to the zero-temperature one-point function $a_{\bar{\psi}\psi}= {\frac{\mathfrak{d}\,\kappa_f}{2^{d-1}}}$.
Note that the radius of convergence of the series is $|z|<1/2$, which is the radius of convergence of the BOE.
From the sum over $n$ we can read off the thermal defect one-point functions of $\hat{\CO}$ in the BOE of $\bar{\psi}\psi$,
\begin{equation}
\label{eq:thermal_BCFT_data_F}
\mu_{\bar{\psi}\psi}{}^{\hat{\CO}}\hat{b}_{\hat{\CO}}=- {2^{{\hat{\Delta}}-d+1}}\cos \left(\pi  \frac{d-{\hat{\Delta}} }{2}\right)\,\mathfrak{d}\, \frac{ \left(2-2^{2-{\hat{\Delta}} }\right)  \Gamma \left(\frac{{\hat{\Delta}} }{2}\right)\zeta({\hat{\Delta}})}{\Gamma\left(\frac{d}{2}\right)\Gamma \left(\frac{1}{2} ({\hat{\Delta}}-d +2)\right)}\,\kappa_f\,,
\end{equation}
where ${\hat{\Delta}}=d+2n$ for non-negative integer $n$, and zero otherwise.
Note that since ${\hat{\Delta}}$ has the same parity as $d$ modulo 2, the cosine factor is just an alternating sign.
A priori one would expect a sum over degenerate boundary primaries on the left-hand side of \eq{thermal_BCFT_data_F}.
However, we shall now argue that this is not the case.
Ref.~\cite{Herzog:2022jlx} showed that there is only one boundary primary $\rho$ in the BOE of $\psi$.
Similarly, there is only one primary $\bar{\rho}$ in the BOE of $\bar{\psi}$.
Therefore one would expect an argument similar to the scalar case to show that there is a unique boundary fermion bilinear that is a scalar boundary primary.
The two boundary primaries $\rho$ and $\bar{\rho}$ live in orthogonal subspaces under the decomposition of the bulk Dirac representation into boundary spinor representations.
Any bilinear involving them must include an odd number of parallel Dirac matrices since each parallel Dirac matrix maps one subspace into the other.
Taking the product of the BOEs of $\psi$ and $\bar{\psi}$ determined by ref.~\cite{Herzog:2022jlx}, one finds that only operators of the schematic form $\bar{\rho}\slashed{\p}_\parallel \p_\parallel^{2n}\rho$ survive, where $\slashed{\p}_\parallel=\gamma^a\partial_a$.
Quotienting by descendants leaves only one boundary primary at each derivative level as can be verified explicitly.
Since $\rho$ and $\bar{\rho}$ have scaling dimensions $(d-1)/2$, the resulting bilinear has ${\hat{\Delta}}=d+2n$.
These precisely match the scaling dimensions of operators that acquire a non-zero boundary thermal one-point function in the BOE channel of the thermal one-point function of $\bar{\psi}\psi$ determined in \eq{thermal_BCFT_data_F}.

\subsubsection{Thermal one-point function of $T_{\mu\nu}$}
\label{sec:F_T}

We now compute the thermal one-point function of the stress tensor $T_{\mu\nu}$.
For a stationary frame the spin connection is trivial such that the stress tensor of a free Dirac fermion is just
\begin{equation}
T_{\mu\nu}=\frac{1}{2}\left(\bar{\psi}\g_{(\mu}\p_{\nu)}\psi-\p_{(\mu}\bar{\psi}\,\g_{\nu)}\psi\right).
\end{equation}
We then compute its thermal one-point function using Wick's theorem.
Due to the trace identities of the Dirac matrices one immediately notices that the contributions coming from the second term in \eq{F_prop} vanish in general $d>3$ such that the thermal one-point function in the presence of a boundary reduces to the no-boundary result.
Concretely,
\begin{equation}
\langle T_{\mu\nu}(x)\rangle_\beta|_{\mathrm{reg.}} =d\,\mathfrak{d}\,\kappa_f\sum_{m=-\infty}^\infty (-1)^m\frac{e_\mu e_\nu-\frac{g_{\mu\nu}}{d}}{|m\beta-\epsilon|^d}\,,
\end{equation}
where we have employed point splitting along the $\tau$ direction.
Removing the divergence in the $m=0$ mode and summing over Matsubara modes one finds
\begin{align}
\label{eq:T_F}
b_T&=2(2^{1-d}-1)d\,\mathfrak{d}\,\kappa_f \,\zeta(d) \,,& F_T^{\perp\perp}&=0
\end{align}
in the notation of \eq{kin_T}.
This trivially satisfies the requisite asymptotic behaviour given in table \ref{tab:spin_two_one_point_structures}.
The fact that $F_T^{\perp\perp}=0$ need not be surprising.
The mixed BCs in \eq{mixed_BCs} impose D BCs on half of the spinor's components. 
The Dirac equation then dynamically enforces N BCs on the remaining components.
Since the boundary-dependent contributions to $T_{\mu\nu}$ of a free massless scalar with N or D BCs come with opposite signs, see \eq{T_S_d}, it seems reasonable that they should vanish for a Dirac fermion.
Of course, a Dirac fermion is not merely a direct sum of free scalars, and indeed the boundary-independent contributions to $\langle T_{\mu\nu}\rangle_\beta$ are not the sum of the respective ones for a free scalar.

When $d=3$, the trace identities of the $\gamma$ matrices allow for an $x_\perp$-dependent contribution.
Using $\Tr \gamma^\mu\gamma^\nu\gamma^\rho=2\iu\epsilon^{\mu\nu\rho}$, we find that the $\tau x_\parallel$-component of the stress tensor can acquire a non-trivial one-point function.
Here $x_\parallel$ is the single spatial direction along the boundary with unit tangent vector $t^\mu$.
For 2d boundaries or defects, the stress tensor one-point function admits an extra tensor structure that is odd under parity along $x_\parallel$.
It is given by the term in the second line of \eq{kin_T}. 
Computing $\langle T_{\tau\parallel}(x)\rangle_\beta$ explicitly, we find that this contribution has a purely imaginary coefficient and can be represented by the following series
\begin{equation}
F^{\tau\parallel}_T=\frac{3\iu}{2\pi}\sum^\infty_{m=1} (-1)^m\frac{m^2}{(m^2+4z^2)^{5/2}}\,,
\end{equation}
where we used $\kappa_f=1/4\pi$ when $d=3$.
This series converges but we have not been able to express it in closed form.

Finally, we can compute some thermodynamic quantities using the stress tensor one-point function \eqref{eq:T_F}. 
Defining dimensionless energy, entropy and free energy densities as in eqs.~\eqref{eq:E}, \eqref{eq:s_int} and \eqref{eq:f_int}, respectively, we find for any $d$
\begin{subequations}
\begin{align}
\CE&=2\left(1-2^{1-d}\right)  (d-1) \mathfrak{d}\, \kappa_f\,  \zeta (d)\,, \\
s&=2 \left(1-2^{1-d}\right) d  \,\mathfrak{d} \,\kappa_f\,  \zeta (d)\,, \\
f&=-2 \left(1-2^{1-d}\right)  \mathfrak{d}\, \kappa_f\,  \zeta (d)\,.
\end{align}
\end{subequations}
Here we fixed the integration constant $c$ by demanding that $S\to0$ as $\beta\to\infty$ motivated by the fact that the free Dirac fermion BCFT has a unique ground state.
All thermodynamic quantities therefore coincide with the no-boundary results.
Note that $\CE$ and $s$ are always positive for $d>1$ whereas $f$ is negative.

\subsection{Free hypermultiplet}

In this subsection we consider the stress tensor one-point function for a simple supersymmetric example: a free massless $\CN=2$ hypermultiplet in 4d.
Such a hypermultiplet consists of one Dirac fermion and two complex scalars.
At zero-temperature we can impose supersymmetric BCs, which amount to giving mixed BCs to the Dirac fermion, and D and N BCs to each of the complex scalars.
When putting the theory on a thermal manifold, one has a choice of BCs along the thermal circle for the Dirac fermion. 
We will impose anti-periodic BCs for the fermions, which correspond to putting the theory at finite temperature.
These, however, break supersymmetry since the scalars must necessarily be periodic around the thermal circle.

Since the theory is free, the stress tensor is just the sum of the individual contributions.
Adding \eq{T_S_4d} twice with each sign and \eq{T_F} with $d=4$, we find
\begin{align}
b_T&=-\frac{\pi ^2}{3} \,,& F_T^{\perp\perp}&=0\,.
\end{align}
Therefore the stress tensor thermal one-point function is a non-zero constant and equal to the no-boundary result.\footnote{
If we impose periodic BCs along the thermal circle for the fermion, supersymmetry is preserved.
However, the correlators no longer describe the ordinary thermal ensemble.
Correlation functions computed with periodic BCs involve an additional insertion of the fermion parity operator.
Nonetheless we can repeat the computation of subsection~\ref{sec:F_T} for a fermion with periodic BCs.
Computing the one-point function of $T_{\mu\nu}$ similarly as before, one finds
\begin{align*}
\label{eq:T_F_PBC}
b_T&=2d\,\mathfrak{d}\,\kappa_f \,\zeta(d) \,,& F_T^{\perp\perp}&=0\,.
\end{align*}
Setting $d=4$ and adding to this \eq{T_S_4d} twice with each sign, we find that all contributions exactly cancel. 
Thus, $\langle T_{\mu\nu}\rangle_\beta=0$ with supersymmetric BCs.}
It is then straightforward to compute the dimensionless energy, entropy and free energy densities. 
We find
\begin{align}
\CE=\frac{\pi^2}{4}\,, && s=\frac{\pi^2}{3}\,, && f=-\frac{\pi^2}{12}\,.
\end{align}

\subsection{Free higher-derivative scalar}

In this subsection we consider the simplest free higher-derivative scalar theory at finite temperature with a conformal boundary.
Free higher-derivative scalar CFTs are non-unitary theories with an action of the schematic form
\begin{equation}
\label{eq:boxk}
\int\d^dx\, \phi\Box^k\phi\,,
\end{equation}
where $\Box$ denotes the Laplacian and $k\geq2$ is a positive integer.
The scaling dimension of $\phi$ is $\Delta_\phi=(d-2k)/2$, which violates the unitarity bound for $k>1$.
Even though these theories are non-unitary, they have wide-ranging applications, famously including the theory of elasticity, and have received renewed attention~\cite{Osborn:2016bev,Brust:2016gjy,Stergiou:2022qqj}.\footnote{
See also refs.~\cite{Herzog:2024zxm,Guo:2025edk,Guo:2026vmq} for interacting generalisations.}
In the present work we will restrict to the free four-derivative theory with $k=2$.
It naturally lives in dimensions $d>2k=4$ such that scaling dimensions remain positive.\footnote{The case $d=2k$ is analogous to an ordinary free scalar in $d=2$ and requires special consideration, see e.g.\ refs.~\cite{Gaikwad:2023gef,Paci:2025pxo}.}

The free conformal BCs of free higher-derivative scalar theories were classified in ref.~\cite{Chalabi:2022qit}.
The $k=2$ theory has four conformal BCs, which are described by the absence of certain boundary primaries in the BOE of $\phi$. 
Varying the action in \eq{boxk} for $k=2$, one finds that the boundary terms arrange themselves into two pairs of boundary primaries, $(\hat{\phi}^{(0)},\hat{\phi}^{(3)})$ and $(\hat{\phi}^{(1)},\hat{\phi}^{(2)})$, where $\hat{\phi}^{(n)}$ is the boundary primary of the schematic form $\p_\perp^n\phi|_{x_\perp=0}$.
To ensure a well-posed variational problem, one must set one of each pair to zero.
This gives rise to the following four BCs
\begin{subequations}
\label{eq:BCs_HD_k2}
\begin{align}
\text{NN : } \quad\hat{\phi}^{(3)}=\hat{\phi}^{(2)}=0\,,\\
\text{ND : } \quad\hat{\phi}^{(3)}=\hat{\phi}^{(1)}=0\,,\\
\text{DN : } \quad\hat{\phi}^{(0)}=\hat{\phi}^{(2)}=0\,,\\
\text{DD : } \quad\hat{\phi}^{(0)}=\hat{\phi}^{(1)}=0\,.
\end{align}
\end{subequations}
If the $i\textsuperscript{th}$ letter labelling the BC is N (D) we set the second (first) element of the $i\textsuperscript{th}$ pair of boundary primaries to zero. 
The zero-temperature propagators for these theories were found in ref.~\cite{Chalabi:2022qit} to be
\begin{equation}
\label{eq:HD_prop}
G_{\varrho,\varsigma}(x,y)=\kappa_2 \left(\frac{1}{|x-y|^{d-4}}+\varrho\,\frac{1}{|x-\tilde{y}|^{d-4}}+(\varrho+\varsigma)(d-4)\,\frac{x_\perp y_\perp}{|x-\tilde{y}|^{d-2}}\right),
\end{equation}
where $\varrho,\varsigma\in\{+1,-1\}$ and $+1$ $(-1)$ is identified with N (D).
E.g.\ the ND propagator is then given by setting $(\varrho,\varsigma)=(+1,-1)$.
Here, $\kappa_2=\Gamma(\frac{d}{2}-2)/16\pi^{d/2}$ and ensures that the propagator is canonically normalised.

At finite temperature, we obtain the propagators by replacing $\Delta\tau\equiv\tau_2-\tau_1 \to \Delta\tau + m\beta$ and introducing a sum over images labelled by integer $m$, just as in \eq{thermal_G} for the ordinary free massless scalar.
We denote the resulting propagators $G_{\varrho,\varsigma}^\beta(x,y)$.
The sum over thermal images for NN and ND BCs converges provided that $d>5$.
For DN and DD BCs the leading term at large $|m|$ cancels such that the image sum converges for $d>3$. 
However, since \eq{HD_prop} assumes $\Delta_\phi=(d-4)/2>0$ we will always take $d>4$.
Equipped with these propagators, we will now compute simple observables.
Again, we will consider the one-point functions of $\phi^2$ and the stress tensor, and discuss some thermodynamic properties.

\subsubsection{Thermal one-point function of $\phi^2$}

First consider the thermal one-point function of $\phi^2$.
It can be straightforwardly obtained from the coincident limit of the propagator $G_{\varrho,\varsigma}^\beta(x,y)$ using point splitting regularisation.
For even $d$, the sums over $m$ can be performed in closed form.
From a computation analogous to the one for ordinary scalar BCFT that led to \eq{F_phi2_d}, we find for general even integer $d$
\begin{equation}
\begin{split}
F^{\varrho,\varsigma}_{\phi^2}(z)&=\kappa_2\, \left[2\zeta (2n+2)+\frac{(-1)^n}{n!} \left(\varrho\CA^{(n)}(0)-2 z^2 (\varrho +\varsigma ) \CA^{(n+1)}(0)\right)\right],
\end{split}
\end{equation}
where $n=(d-6)/2$ and 
\begin{equation}
\CA(A)=\frac{\pi  \sqrt{A+4 z^2} \coth \left(\pi  \sqrt{A+4 z^2}\right)}{A+4 z^2}\,.
\end{equation}
We use the shorthand $\CA^{(n)}$ for $\d^n\CA/\d A^n$. 

It is simple to check that the function $F_{\phi^2}^{\varrho, \varsigma}(z)$ has the expected  asymptotic behaviour of \eq{FO_asymp}.
At large distances $z\gg1$ the function approaches $F_{\phi^2}^{\varrho, \varsigma}(z)\overset{z\to\infty}{\sim}b_{\phi^2} $, where $b_{\phi^2}=2\kappa_2\zeta(d-4)$ is the no-boundary thermal one-point function.
As $d\to5$, $b_{\phi^2}$ diverges. 
This reflects the logarithmic divergence of the thermal image sum.
This can for instance be seen by setting $\varrho=\varsigma=0$ on the right-hand side of \eq{HD_prop}, introducing a sum over thermal images as in \eq{thermal_G} and taking the coincident limit while removing the divergence from the $m=0$ mode.
In the $z\to0$ limit, on the other hand, we instead have $F_{\phi^2}^{\varrho, \varsigma}(z)\overset{z\to0}{\sim}\left(\varrho+\frac{d-4}{4}(\varrho+\varsigma)\right){\frac{\kappa_2}{(2z)^{d-4}}}$ whose coefficient we identify with the zero-temperature one-point function $a_{\phi^2}^{\varrho,\varsigma}$.
Away from $z=0$, we can use the generalised binomial theorem \eq{gen_bin} to expand the finite-temperature propagators $G_{\varrho,\varsigma}^\beta$ for small $z$ before taking the coincident limit.
We find the following expansion into thermal blocks for the thermal one-point function of $\phi^2$,
\begin{equation}
\begin{split}
F_{\phi^2}^{\varrho,\varsigma}={\frac{1}{(2z)^{d-4}}}\kappa_2\Bigg[&\varrho+\frac{d-4}{4}(\varrho+\varsigma)+2\zeta(d-4)(2z)^{d-4}\\
&+\frac{2\varrho}{\Gamma\left(\frac{d-4}{2}\right)}\sum_{n=0}^\infty(-1)^n \frac{\Gamma\left(\frac{d-4}{2}+n\right)}{\Gamma(n+1)}\zeta(d+2n-4)(2z)^{d+2n-4}\\
&+\frac{\varrho+\varsigma}{\Gamma\left(\frac{d-4}{2}\right)}\sum_{n=0}^\infty(-1)^n \frac{\Gamma\left(\frac{d-2}{2}+n\right)}{\Gamma(n+1)}\zeta(d+2n-2)(2z)^{d+2n-2}\Bigg]\,.
\end{split}
\end{equation}
The radius of convergence of both series is $|z|<1/2$, which coincides with the radius of convergence of the BOE.
Using these expressions it is straightforward to extract thermal BCFT data by comparing with \eq{FO_DOE}.
We find
\begin{equation}
\label{eq:thermal_BCFT_data_HD}
\mu^{\varrho,\varsigma;\hat{\CO}}_{\phi^2}\hat{b}^{\varrho,\varsigma}_{\hat{\CO}}=
\begin{cases}
2(1+\varrho)\zeta({\hat{\Delta}})\,\kappa_2 & \text{ if } {\hat{\Delta}}=d-4\,,\\
{2^{2n+1}}\cos \left(\pi\frac{d-{\hat{\Delta}} }{2} \right)\dfrac{((d-{\hat{\Delta}} ) (\varrho +\varsigma )-4 \varsigma )\Gamma \left(\frac{{\hat{\Delta}} }{2}\right) \zeta ({\hat{\Delta}} )  }{\Gamma \left(\frac{d}{2}-2\right) \Gamma \left(\frac{1}{2} ({\hat{\Delta}}-d +6)\right)}\,\kappa_2 & \text{ if } {\hat{\Delta}}=d+2n-2\,,
\end{cases}
\end{equation}
where $n$ is a non-negative integer, and zero otherwise.
Note that the left-hand side contains an implicit sum over boundary scalar primaries $\hat{\CO}$ in the BOE of $\phi^2$ with the same scaling dimension ${\hat{\Delta}}$.
Since $\phi$ contains two primaries in its BOE for any of the four BCs in \eq{BCs_HD_k2}, there can now be more than one scalar ``double-trace'' boundary primary $\hat{\CO}$ for any given ${\hat{\Delta}}$.
Schematically, these are operators of the form $\hat{\phi}^{(i)}\p_\parallel^{2n}\hat{\phi}^{(j)}$, where $i$ and $j$ run over the two boundary primaries present in the BOE of $\phi$ that specify the BC.
However, for $i\neq j$ the boundary limit of $G^\beta_{\varrho,\varsigma}$ vanishes and therefore the boundary thermal one-point functions of all such operators must be zero.
Thus, only operators with $i=j$ can contribute in the BOE channel of the thermal one-point function of $\phi^2$.

\subsubsection{Thermal one-point function of $T_{\mu\nu}$}
\label{sec:HD_T}

Next we move on to a discussion of the stress tensor thermal one-point function and associated thermodynamic quantities.
Given the action of a QFT, one can find the stress tensor by computing the Noether currents associated with translational invariance.
These $d$ currents, however, are generally not traceless, nor do they form a spin-2 primary if the QFT is conformal. 
For free higher-derivative scalar CFTs with action \eqref{eq:boxk}, refs.~\cite{Osborn:2016bev,Stergiou:2022qqj} argued that improvement terms can be added in a unique way to produce a stress tensor that is traceless and a spin-2 primary.\footnote{
The improved stress tensor coincides with the metric variation of a Weyl invariant action of a scalar field with $2k$ derivatives on a general curved background, provided that $d>2k$.
This action is obtained by promoting $\Box^k$ to the GJMS operator~\cite{Graham:1992}, which transforms covariantly under Weyl transformations.}
For the $k=2$ theory in $d$-dimensional flat space, the improved stress tensor reads
\begin{equation}
\begin{split}
T_{\mu\nu}&=\frac{d(d+2)}{2(d-2)(d-1)}\p_\mu\p_\nu\phi\p^2\phi-\frac{d+2}{d-1}\p_{(\mu}\phi\p_{\nu)}\p^2\phi+\frac{d-4}{2(d-1)}\phi\p_\mu\p_\nu\p^2\phi\\
&+\frac{2}{d-1}\p_\mu\p_\nu\p_\rho\phi\p^\rho\phi-\frac{2d}{(d-2)(d-1)}\p_\mu\p_\rho\phi\p_\nu\p^\rho\phi-\frac{d+2}{2(d-2)(d-1)}g_{\mu\nu}(\p^2\phi)^2\\
&+\frac{1}{d-1}g_{\mu\nu}\p_\rho\phi\p^\rho\p^2\phi-\frac{d-4}{2(d-1)}g_{\mu\nu}\phi\p^4\phi+\frac{2}{(d-2)(d-1)}g_{\mu\nu}\p_\rho\p_\sigma\phi\p^\rho\p^\sigma\phi\,.
\end{split}
\end{equation}
Its thermal one-point function can be computed using the thermal propagators $G_{\varrho,\varsigma}^\beta(x,y)$ with point splitting regularisation.
One can straightforwardly obtain closed-form expressions when $d$ is an even integer.
However, they are rather cumbersome and we will only report results for $d=6$. 
This is the lowest even integer spacetime dimension in which $\phi$ has positive scaling dimension.
We then find
\begin{subequations}
\label{eq:T_HD_6}
\begin{align}
b_T&=-\frac{8 \pi ^3}{315}\,,\\
\begin{split}
F_{T,\varrho\varsigma}^{\perp\perp}&={\frac{1}{640\pi^2z^5}} \left(-3 \varsigma  \coth (2 \pi  z)
+3 \pi  z \csch^2(2 \pi  z) (\pi  z (5 \varrho +\varsigma ) \coth (2 \pi  z)-2 \varsigma )\right.\\
&\quad+2 \pi ^3 z^3 \csch^4(2 \pi  z) ((15 \varrho +11 \varsigma ) (\cosh (4 \pi  z)+2)+88 \pi  \varsigma  z \coth (2 \pi  z))\\
&\quad+16 \pi ^4 \varsigma  z^4 \cosh (6 \pi  z) \csch^5(2 \pi  z)\\
&\quad\left.-8 \pi ^5 z^5 (\varrho +\varsigma ) (26 \cosh (4 \pi  z)+\cosh (8 \pi  z)+33) \csch^6(2 \pi  z)\right)
\end{split}
\end{align}
\end{subequations}
in the notation of \eq{kin_T}.
Note that $F^{\perp\perp}_{T,++}=-F^{\perp\perp}_{T,--}$ and $F^{\perp\perp}_{T,+-}=-F^{\perp\perp}_{T,-+}$.

\begin{figure}[h]
\centering
\includegraphics[width=12cm]{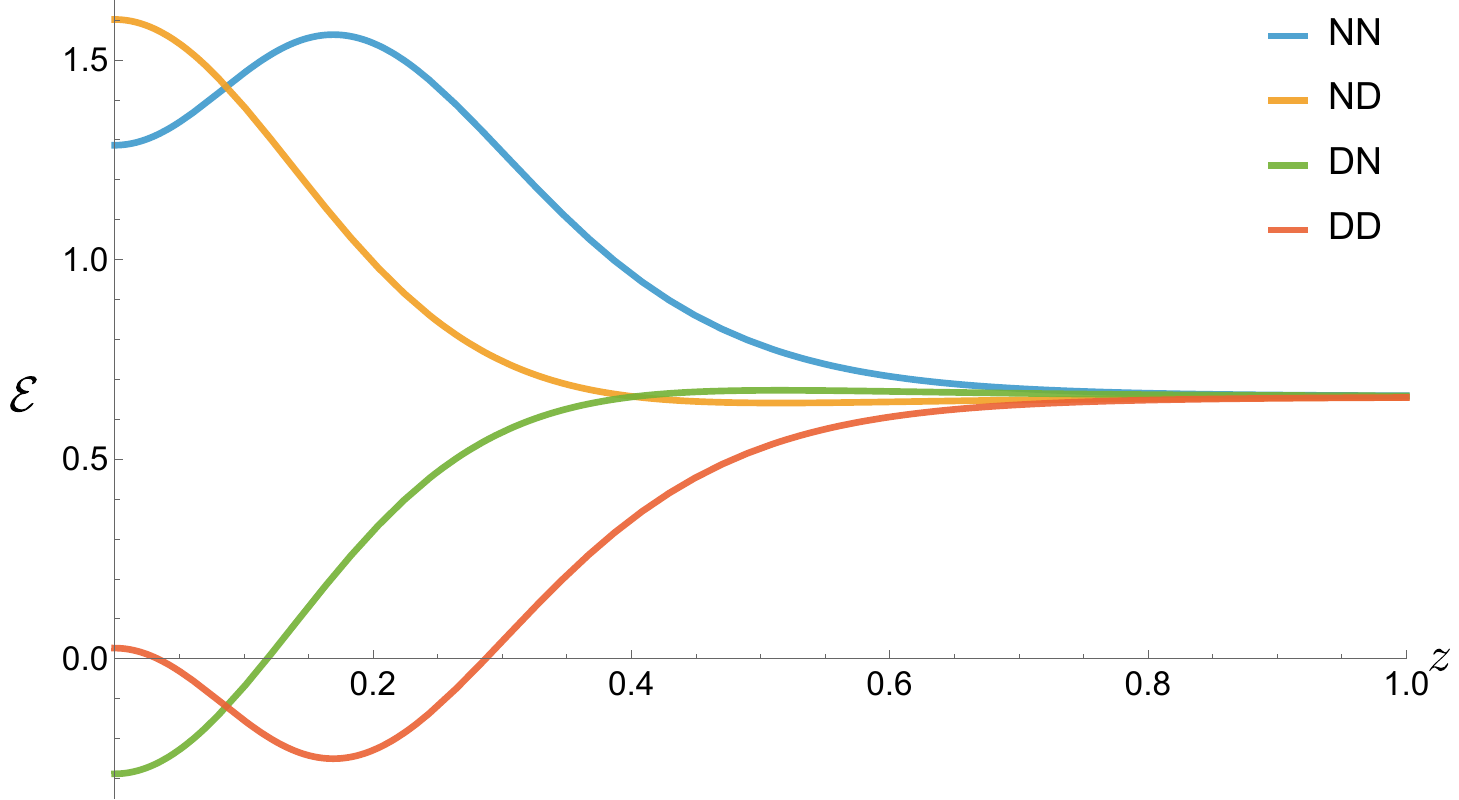}
\caption{
Plot of the dimensionless energy density $\CE(z)$ for the free four-derivative scalar BCFTs with BCs given by \eq{BCs_HD_k2} as a function of the thermal cross-ratio $z=x_\perp/\beta$ when $d=6$.
}
\label{fig:E_HD_6d}
\end{figure}

Using these expressions we can compute the dimensionless energy, entropy and free energy densities when $d=6$.
In figure~\ref{fig:E_HD_6d} we plot the dimensionless energy density $\CE(z)$ defined as in \eq{E} for the four BCs listed in \eq{BCs_HD_k2}.
Unlike the case of an ordinary free massless scalar, $\CE(z)$ is not a monotonic function for higher-derivative BCFTs.
$\CE(z)$ has a maximum for both NN and DN BCs, whereas there is a minimum for ND and DD BCs. 
While $\CE(z)$ is always positive for NN and ND BCs, the function is negative for a range of $z$ for DN and DD BCs.
It is tempting to attribute the negative energies to non-unitarity, however, the function remains bounded from below.
As $z\to0$, $\CE(z)$ approaches a non-zero constant value given by $4 \pi ^3 (30 \rho -6 \sigma +25)/4725$.
This signals the presence of an operator with ${\hat{\Delta}}=6$ in the BOE of $T_{\tau\tau}$ that acquires a non-zero boundary thermal one-point function.
At large distances $z\gg1$, on the other hand, boundary effects become negligible and $\CE(z)$ approaches the no-boundary value $-5b_T/6=4 \pi ^3/189$.

It was argued in ref.~\cite{Chalabi:2022qit} that the four BCs in \eq{BCs_HD_k2} are connected by boundary RG flows, similarly to the N and D BCs of an ordinary free massless scalar.
NN BCs flow to ND or DN BCs, depending on the relevant quadratic deformation on the boundary, while ND and DN BCs flow to DD BCs.
From figure~\ref{fig:E_HD_6d} we see that the dimensionless energy density need not always decrease at a given $z$ under a boundary RG flow.
E.g.\ comparing the values of $\CE(z)$ at the endpoints of the flow from NN to ND, one has $\CE_{UV}>\CE_{IR}$ for $z\gtrsim0.086$ but $\CE_{UV}<\CE_{IR}$ for $0<z\lesssim 0.086$.

Next we would like to compute the dimensionless entropy density $s$ and dimensionless free energy density $f$ defined in eqs.~\eqref{eq:s_int} and \eqref{eq:f_int}, respectively.
Recall that for the ordinary scalar with N and D BCs discussed in section~\ref{sec:free_scalar}, we fixed the integration constant $c$ in those equations by demanding that $S\to0$ as $\beta\to\infty$.
Intuitively, this is because the low-temperature limit selects the ground state of the Hamiltonian, which is unique.
For higher-derivative theories, the classical Hamiltonian is unbounded from below, reflecting the Ostrogradski instability.
When quantising the theory, one can nonetheless obtain a positive semi-definite energy spectrum by allowing for negative norm states~\cite{vanTonder:2006ye,Brust:2016gjy}.
In this non-unitary quantisation the vacuum is unique.
Motivated by this we can employ the same prescription for the entropy and free energy densities as for ordinary scalar BCFT.
In particular, to fix the integration constant $c$ in eqs.~\eqref{eq:s_int} and \eqref{eq:f_int}, we will choose
\begin{equation}
\label{eq:c_HD_BCFT}
c=-\int_0^1\d y\, y^{4} \CE(y)\,,
\end{equation}
such that the entropy density $S$ vanishes in the zero-temperature limit, as before.

\begin{figure}[ht]
\centering
  \begin{minipage}{0.49\textwidth}
  \centering
    \includegraphics[width=\linewidth]{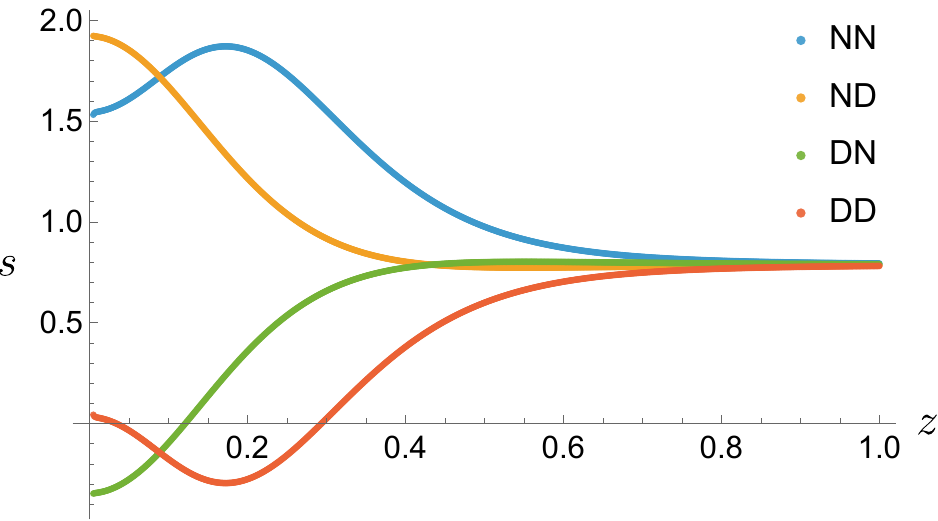}
  \end{minipage}\hfill
  \begin{minipage}{0.49\textwidth}
  \centering
    \includegraphics[width=\linewidth]{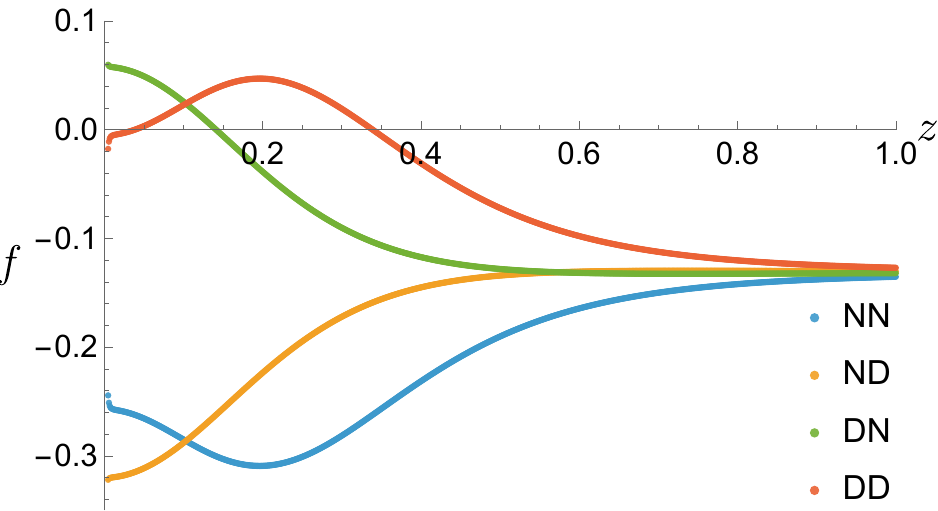}
  \end{minipage}
\caption{\emph{Left:} Plot of the dimensionless entropy density $s(z)$ for the free four-derivative scalar BCFTs with BCs given by \eq{BCs_HD_k2} as a function of $z=x_\perp/\beta$ when $d=6$. \\
\emph{Right:} Plot of the dimensionless free energy density $f(z)$ for free four-derivative scalar BCFTs when $d=6$.
} 
\label{fig:s_f_HD_6d}
\end{figure}

We have not been able to analytically perform the integral in \eq{c_HD_BCFT} or the $y$-integral appearing in eqs.~\eqref{eq:s_int} and \eqref{eq:f_int}.
Instead we will compute them numerically.
We plot the dimensionless entropy density $s(z)$ for the four BCs in figure~\ref{fig:s_f_HD_6d} on the left.
We observe that it has qualitatively the same features as $\CE(z)$.
At large $z$, boundary effects become negligible and $s(z)$ approaches the no-boundary value $8\pi^3/315$, which can be found analytically.
We plot the dimensionless free energy density $f$ in figure~\ref{fig:s_f_HD_6d} on the right.
For $z\gg1$, the boundary is negligible and the function approaches the no-boundary value $-4\pi^3/945$.
For $2\pi z\sim\CO(1)$, $f(z)$ behaves qualitatively like $-\CE(z)$.
For instance, $f(z)$ is positive for DD and DN BCs for a range of $z$, and we observe that $(-f^{UV})\ngtr (-f^{IR})$ for all $z$ for some boundary RG flows.
In particular, for the flows from NN to ND and from DN to DD we observe that close to the boundary, i.e.\ for small $z$, minus the dimensionless free energy density is in fact greater in the UV than in the IR.
This is in contrast to the ordinary free massless scalar discussed in section~\ref{sec:scalar_T}, for which $(-f^{UV}) > (-f^{IR})$ under a boundary RG flow from N to D BCs.

We can similarly compute the integrated energy $\mathsf{E}_\p$ defined in \eq{int_thermo_E_bdy}.
When $d=6$, $\mathsf{E}_\p=-4\mathsf{F}_\p$ as argued in the paragraph below \eq{total_E_S}.
We then find
\begin{equation}
\mathsf{F}_\p=-3 A(3\varrho+\varsigma)\frac{\zeta(5)}{16\pi^2\beta^5}\,,
\end{equation}
where $A$ is the 4d area of the boundary.
Here we assumed that the boundary contribution to the free energy is extensive.
The integrated $\mathsf{F}_\p$ agrees with integral over all of space of $f(z)/\beta^6$ with no-boundary value subtracted, where the integration constant $c$ is fixed as in \eq{c_HD_BCFT}.
In particular, $\mathsf{F}_\p$ reproduces the area under the curves on the right-hand side of figure~\ref{fig:s_f_HD_6d} with no-boundary value subtracted.
Unlike the dimensionless free energy density $f(z)$, the integrated free energy including both bulk and boundary contributions is negative and does obey $(-\mathsf{F}^{UV})>(-\mathsf{F}^{IR})$ for all admissible flows, i.e.\ NN to ND, NN to DN, ND to DD, and DN to DD.
The boundary generalisation of the proposal of ref.~\cite{Appelquist:1999hr} with integrated thermodynamic quantities appears to hold in this non-unitary example when $d=6$.
This is in contrast to established monotonicity theorems.
For instance, when $d=5$, ref.~\cite{Chalabi:2022qit} showed that this BCFT generally violates the boundary $a$-theorem~\cite{Wang:2021mdq} as a consequence of bulk non-unitarity.\footnote{
Recently, ref.~\cite{Diatlyk:2026oxm} studied the RG flow of an interacting non-unitary $PT$-symmetric QFT at finite temperature in dimensions $2\geq d\geq6$.
While the generalised $\CF$-theorem~\cite{Giombi:2014xxa} is violated due to the lack of unitarity, the authors find that the thermodynamic free energy obeys $(-\mathsf{F}^{UV})>(-\mathsf{F}^{IR})$ in this non-unitary model.}

\section{Monodromy defect in free scalar CFT at finite temperature}
\label{sec:mon}

In this section we consider a monodromy defect at finite temperature.
Monodromy defects are co-dimension two defects that generalise the twist defects of the 3d Ising CFT~\cite{Billo:2013jda,Gaiotto:2013nva}.
They exist in any QFT with a global symmetry as the $(d-2)$-dimensional boundaries of the topological operators implementing symmetry transformations~\cite{Gaiotto:2014kfa}.\footnote{
See ref.~\cite{Copetti:2026ncv} for a recent discussion of monodromy defects for anomalous symmetries.}
The boundaries of these operators are typically dynamical.
Here we will consider a monodromy defect in the theory of a free complex scalar field.
Although the theory in our example is free, the monodromy defect at finite temperature turns out to be considerably richer than the conformal boundaries of section~\ref{sec:ex_bdy}.
We will compute thermal one-point functions of simple bulk operators, extract thermal DCFT data, and study thermodynamic quantities.

We begin with a brief review of the zero-temperature case where we place the theory on $\mathbb{R}^d$.
The action of a conformally coupled free complex scalar field  $\Phi$ in $d$ dimensions is
\begin{equation}
\label{eq:action_mon}
S=\int\d^d x\sqrt{g}\left(\nabla^\mu \Phi\nabla_\mu\Phi^\dagger+\frac{d-2}{4(d-1)}R|\Phi|^2\right),
\end{equation}
where $g=\det g_{\mu\nu}$ for a background metric $g_{\mu\nu}$, $\nabla$ is the Levi-Civita connection and $R$ is the Ricci scalar of the corresponding curvature.
We then insert a monodromy defect for the global $U(1)$ symmetry in this theory.
Take the defect to extend along $d-2$ directions for which we use Cartesian coordinates $\tau$ and $\vec{x}_\parallel$.
At this point we are arbitrarily singling out one of the $d-2$ directions to call it $\tau$.
Later, we will compactify the $\tau$-direction when studying the defect at finite temperature.
In the two transverse directions to the defect we introduce cylindrical coordinates $x^\mu=(\tau, \vec{x}_\parallel, \rho, \theta)$.
The defect sits at the origin of the 2d plane parametrised by the radial coordinate $\rho$ and the angular coordinate $\theta\in[0,2\pi)$.
Under a $2\pi$-rotation around the defect, the scalar field comes back to itself up to a global $U(1)$ transformation,
\begin{equation}
\label{eq:monodromy_BCs}
\Phi(\tau, \vec{x}_\parallel, \rho, \theta+2\pi)=e^{-2\pi\iu\alpha}\,\Phi(\tau, \vec{x}_\parallel, \rho, \theta)\,,
\end{equation}
where $\alpha\in[0,1]$ is a parameter that specifies the defect.
We will refer to it as the \emph{monodromy parameter}.

The quantisation of this defect was performed in refs.~\cite{Giombi:2021uae, Bianchi:2021snj}.
To this end it is convenient to perform a singular background gauge transformation $\Phi\to \tilde{\Phi} =e^{\iu\alpha\theta}\Phi$.
The transformed field is then $2\pi$ periodic, which comes at the cost of introducing a non-trivial background $U(1)$ gauge field $A=\alpha\,\d\theta$.
Covariant derivatives $\nabla_\mu$ are thus promoted to gauge covariant derivatives $D_\mu=\nabla_\mu-\iu A_\mu$.
One can then canonically quantise $\tilde{\Phi}$ by solving the equation of motion (EOM) and then promoting the coefficients to creation and annihilation operators.
For an appropriate ansatz with factorised coordinate dependence, the EOM reduces to a second-order ODE for the factor depending on the radial coordinate $\rho$.
The two independent solutions are Bessel functions of the first kind $J_{\pm(n- \alpha)}(k_\rho \rho)$, where $n\in\mathbb{Z}$ and $k_\rho$ is a constant that is integrated over in the mode expansion of $\tilde{\Phi}$.
Note that for every $n$ one of the solutions is regular at the origin $\rho=0$ while the other one is divergent. 
Naively one would regard the singular solutions as unphysical and discard them.
However, as argued in refs.~\cite{Bianchi:2019sxz, Bianchi:2021snj}, it is consistent with unitarity to include an admixture of regular and singular modes for the $n=0$ and $n=1$ modes only.
The relative coefficients between regular and singular modes were denoted $\xi,\tilde{\xi}\in[0,1]$ in ref.~\cite{Bianchi:2021snj} for the $n=0$ and $n=1$ modes, respectively, and we will adopt this notation here.
For instance when $\xi=0$, the regular solution for the $n=0$ mode with $J_\alpha(k_\rho\rho)$ is included in the mode expansion whereas the singular solution $J_{-\alpha}(k_\rho \rho)$ is not.
When $\xi=1$, the singular solution $J_{-\alpha}(k_\rho\rho)$ is included whereas the regular one $J_{\alpha}(k_\rho \rho)$ is not.
For values in-between, an appropriate linear combination appears in the mode expansion, see ref.~\cite{Bianchi:2021snj} for details.

Canonically quantising the mode expansion, it is then straightforward to compute the propagator $G(x,x')=\langle\Phi(x)\Phi(x')^\dagger\rangle$.
After undoing the singular gauge transformation to consider the original field $\Phi$ with twisted boundary conditions~\eqref{eq:monodromy_BCs}, ref.~\cite{Bianchi:2021snj} found the following integral representation for the zero-temperature propagator
\begin{subequations}
\label{eq:prop_0T}
\begin{equation}
\label{eq:G_exp}
\begin{split}
G_{\alpha,\xi,\tilde \xi} (x,x')  
= \;&\sum_{n =1}^{+\infty} e^{\iu (n-\alpha) (\theta-\theta')} G^{(n-\alpha)} (x,x')
+ \sum_{n =0}^{+\infty} e^{-\iu (n+\alpha) (\theta-\theta')} G^{(n+\alpha)} (x,x') \\
&+ \xi \, e^{-\iu \alpha (\theta-\theta')}\left[G^{(-\alpha)} (x,x') - G^{(\alpha)} (x,x')\right] \\
&+ \tilde \xi \, e^{\iu (1-\alpha)(\theta-\theta')}\left[G^{(\alpha-1)} (x,x') - G^{(1-\alpha)} (x,x')\right] ,
\end{split}
\end{equation} 
where
\begin{equation}
\label{eq:heat_kernel_prop_no_A}
\begin{split}
G^{(\nu)} (x,x') 
&= \frac{1}{(4\pi)^{d/2}}  \int_0^\infty \d s \, s^{d/2-2} e^{-s(\rho^2+\rho'^2+(\bm{x}_\parallel-\bm{x}_\parallel')^2)/4 }     I_{\nu} \left(\frac{s\,\rho\,\rho'}{2}\right).
\end{split}
\end{equation}
\end{subequations}
Here, $\bm{x}_\parallel=(\tau,\vec{x}_\parallel)$ and $I_\nu$ is the modified Bessel function of the first kind.
The integral in \eq{heat_kernel_prop_no_A} can be performed analytically in closed form \cite{Bianchi:2021snj}.
Together with the exponential factors in \eq{G_exp}, each $G^{(\nu)}$ gives rise to a defect channel conformal block.
The propagator thus decomposes in a conformal block expansion where the exchanged defect local operators are scalars under rotations along the defect with scaling dimensions $\hat{\Delta}^{(\nu)}=\frac{d}{2}-1+\nu$ and carry transverse $SO(2)$ spin.
The transverse spin $s$ can be read off from the exponentials.
It appears as $e^{\iu s(\theta-\theta')}$.
One thus finds two sets of defect operators in the DOE of $\Phi$, denoted $\hat{\CO}^\pm_s$.
The operators $\hat{\CO}^+_s$ have scaling dimensions $\hat{\Delta}^+_s=\frac{d}{2}-1+|s|$ with transverse spin $s\in\mathbb{Z}-\alpha$ while $\hat{\CO}^-_s$ have $\hat{\Delta}^-_s=\frac{d}{2}-1-|s|$ with $s=-\alpha, 1-\alpha$.
Note that $\hat{\Delta}^-_s$ satisfies the unitarity bound $\hat{\Delta}\geq\frac{d-4}{2}$ for these values of $s$ for the full range of $\alpha\in[0,1)$ provided that $d>4$.
For $d\leq4$ the unitarity bound $\hat{\Delta}\geq0$ requires that $|s|<\frac{d-2}{2}$.
The defect local operators in the DOE of $\Phi^\dagger$ have the same scaling dimensions $\hat{\Delta}^\pm_s$ but transverse spin $-s$.

We now place the theory on the thermal manifold $\CM_\beta = S^1_\beta\times \mathbb{R}^{d-1}$.
This amounts to compactifying the $\tau$-direction to a circle of circumference $\beta$. 
The monodromy defect extends along the $\bm{x}_\parallel=(\tau,\vec{x}_\parallel)$ directions, as before, with $\tau$ now compact.
The finite-temperature propagator can then be obtained from \eq{prop_0T} as usual through the method of images: one replaces $\tau-\tau'\to\tau-\tau'+m\beta$ and sums over thermal images $m\in\mathbb{Z}$, just like we did for free scalar BCFT in \eq{thermal_G}.
We denote the resulting propagator $G^\beta_{\alpha,\xi,\tilde \xi} (x,x')$.
This prescription agrees with the computation of the propagator directly on $\CM_\beta$, where one chooses one of the non-compact directions among $\vec{x}_\parallel$ as time and canonically quantises on spatial slices with cylinder geometry $S^1_\beta\times\mathbb{R}^{d-2}$.
Equipped with this thermal propagator, we can now compute correlation functions at finite temperature using Wick's theorem.
In particular, we will consider the thermal one-point functions of $|\Phi|^2$, the $U(1)$ current and the stress tensor.
The latter will further allow us to study some of the system's thermodynamic properties.

\subsection{Thermal one-point function of \texorpdfstring{$|\Phi|^2$}{|Φ|²}} 

The simplest observable to compute is the thermal one-point function $\langle|\Phi|^2\rangle_\beta$, which is simply given by the renormalised coincident limit of the thermal propagator.
For the moment we set $\xi=\tilde{\xi}=0$, i.e.\ we consider regular modes only.
Replacing $\tau\to\tau+m\beta$ in \eq{prop_0T}, taking the coincident limit and summing over thermal images, one finds
\begin{equation}
\label{eq:Phi2_int_s}
\langle|\Phi|^2(x)\rangle^{\xi=\tilde{\xi}=0}_\beta=\sum_{m=-\infty}^\infty\frac{1}{(4\pi)^{d/2}}
\int_{0}^\infty\d s\,e^{-\frac{2\rho^2+(m\beta)^2}{4s}}s^{-d/2}\CI_{\alpha}\left(\frac{\rho^2}{2s}\right)\,,
\end{equation}
where $\mathcal{I}_\alpha (\zeta) \equiv\sum_{n} I_{|n -\alpha | }(\zeta)$.
The $m=0$ term is the same one that appears in the zero-temperature computation performed in ref.~\cite{Bianchi:2021snj}.
The $m=0$ integral is UV divergent at $s=0$,
and one can regularise it by computing the integral with a short-distance cut-off $\varepsilon^2$.
Since the divergence as $\varepsilon^2\to0$ is independent of the monodromy parameter $\alpha$, it can be subtracted unambiguously.
The integrals appearing for the images with $m\neq0$, on the other hand, are convergent due to the exponential suppression from the $e^{-(m\beta)^2/4s}$ factor in the integrand.

Computing the integrals for $m\neq0$ is challenging and we have not been able to find a closed form expression that captures the full correlator at arbitrary transverse distance $\rho$ valid for any $d$.
When $d$ is even, however, we are able to find a rapidly converging integral representation that captures the full thermal one-point function.
Within the convergence domain of the DOE, we are also able to compute the one-point function as a series in the thermal cross-ratio $z=\rho/\beta$ without any restrictions on the dimension.
This allows us to extract the thermal DCFT data of operators exchanged in the DOE of $|\Phi|^2$ for any $d$.
We will discuss both computations in turn.
\paragraph{Integral representation. }
First consider the computation of the full thermal one-point function.
For now we keep $d$ arbitrary before specialising to even $d$.
Using the following integral representation of the modified Bessel function of the first kind
\begin{equation}
\label{eq:BesselI_int}
I_\nu(\zeta)=\frac{1}{\pi}\int_0^\pi \d \theta\,e^{\zeta \cos \theta}\cos(\nu \theta)-\frac{\sin (\nu\pi)}{\pi}\int_0^\infty \d t\, e^{-\zeta \cosh t-\nu t}\,,
\end{equation}
it is straightforward to show that\footnote{
The $e^\zeta$ term in \eq{sumI} arises from the $\theta$-integral in \eq{BesselI_int} after employing the distributional identity
\begin{equation*}
\int_0^\pi\d \theta\, f(\theta) \sum_{n=-\infty}^\infty e^{\iu n \theta}=\pi f(0)
\end{equation*}
for some test function $f$.
This follows from the Dirac comb identity
$\sum_{n} e^{\iu n x}=2\pi \sum_{k} \delta(x-2\pi k)$.
The $t$-integrand can be massaged into a geometric series, which gives rise to the ratio of hyperbolic cosines in \eq{sumI}.
}
\begin{equation}
\label{eq:sumI}
\mathcal{I}_\alpha (\zeta) \equiv\sum_{n} I_{|n -\alpha | }(\zeta) = e^\zeta-\frac{\sin\pi\alpha}{\pi}\int_0^\infty\d t\, \frac{\cosh((1/2-\alpha)t)}{\cosh(t/2)} e^{-\zeta \cosh t}\,.
\end{equation}
Then, performing the $s$-integral first, we find
\begin{equation}
\label{eq:Phi2_step}
\begin{split}
\langle|\Phi|^2(x)\rangle^{\xi=\tilde{\xi}=0}_\beta
&=\langle|\Phi|^2(x)\rangle^{\xi=\tilde{\xi}=0}_{\beta\to\infty}
+\frac{1}{2\pi ^{\frac{d}{2}}}  \Gamma \left(\frac{d-2}{2}\right) \frac{\zeta (d-2)}{\beta^{d-2}}\\
&-\frac{\sin \pi\alpha}{2 \pi^{d/2+1}}\Gamma \left(\frac{d}{2}-1\right)
\sum_{m=1}^\infty\int_0^\infty\d t\, \frac{\cosh ((1/2-\alpha)t)\sech(t/2)}{\left(\beta ^2 m^2+4 \rho ^2\cosh^2 (t/2)\right)^{\frac{d}{2}-1}}\,,
\end{split}
\end{equation}
where the \(s\)-integral converges if \(d>2\).
The first term is the renormalised zero-temperature defect contribution.
This is obtained from the \(m=0\) image of the $t$-integral term in \(\mathcal I_\alpha(\rho^2/2s)\) after subtracting the defect-independent vacuum divergence.
It was found by refs.~\cite{Bianchi:2021snj,Giombi:2021uae} to be
\begin{equation}
\label{eq:Phi2_0T}
\langle|\Phi|^2(x)\rangle^{\xi=\tilde{\xi}=0}_{\beta\to\infty}=
-\frac{\Gamma\left(\frac{d}{2}-\alpha\right)\Gamma\left(\frac{d}{2}+\alpha-1\right)\sin (\pi\alpha)}{2^{d-1}\pi^{\frac{d+1}{2}}(d-2)\Gamma\left(\frac{d-1}{2}\right)}\frac{1}{\rho^{d-2}}\,.
\end{equation}
The second term in \eq{Phi2_step} is exactly the thermal one-point function of \(|\Phi|^2\) in the absence of a defect, \(2\kappa\zeta(d-2)/\beta^{d-2}\).
It originates from the \(e^{\rho^2/2s}\) term in \(\mathcal I_\alpha\): its \(m=0\) image is the vacuum short-distance divergence removed by renormalisation, while the $m\neq0$ thermal images give the no-defect thermal one-point function.
The last term in \eq{Phi2_step} contains the defect-dependent contribution from the non-zero thermal images.
Whenever \(d\) is even, the remaining \(t\)-integrand is a rational function of \(m^2\), and the sum over thermal images can be performed explicitly.
E.g.\ when $d=4$, the last term in \eq{Phi2_step} becomes
\begin{equation}
-\frac{\sin (\pi  \alpha )}{8 \pi ^2}\frac{z}{\rho^2}\int_0^\infty\d t\,\sech^2\left(\frac{t}{2}\right) \cosh \left(\!\left(\frac{1}{2}-\alpha \right) t\!\right)   \coth \left(2 \pi  z  \cosh \left(\!\frac{t}{2}\right)\!\right)
-\frac{(\alpha -1) \alpha }{8 \pi ^2 \rho ^2}\,,
\end{equation}
where we introduced the cross-ratio $z=\rho/\beta$.
The last term exactly cancels against the zero-temperature result in \eq{Phi2_0T} for $d=4$.
The full one-point function when $d=4$ then is
\begin{equation}
\label{eq:Phi2_4d}
\begin{split}
&\langle|\Phi|^2(x)\rangle^{\xi=\tilde{\xi}=0}_\beta=\frac{1}{12\beta^2}\\
&\qquad-\frac{\sin (\pi  \alpha )}{8 \pi ^2}\frac{z}{\rho^2}
\int_0^\infty\d t\,\sech^2\left(\frac{t}{2}\right) \cosh \left(\left(\frac{1}{2}-\alpha \right) t\right) \coth \left(2 \pi z \cosh \left(\frac{t}{2}\right)\right),
\end{split}
\end{equation}
where the first term is the no-defect thermal one-point function of $|\Phi|^2$ and all effects due to the defect are contained in the integral.
Note that both zero and finite-temperature one-point functions are invariant under $\alpha\to1-\alpha$.
This follows from the form of the propagator in \eq{prop_0T}, which for $\xi=\tilde{\xi}=0$ is invariant under this transformation in the coincident limit.

\begin{figure}[ht]
\centering
  \begin{minipage}{0.49\textwidth}
  \centering
    \includegraphics[width=\linewidth]{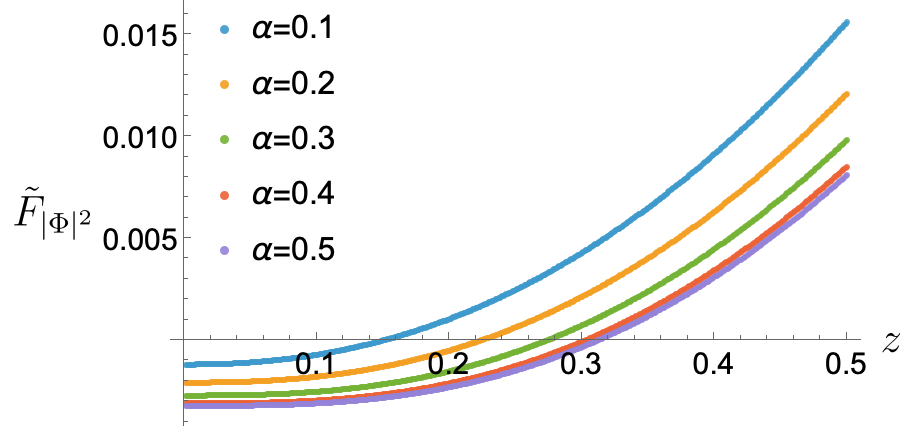}
  \end{minipage}\hfill
  \begin{minipage}{0.49\textwidth}
  \centering
    \includegraphics[width=\linewidth]{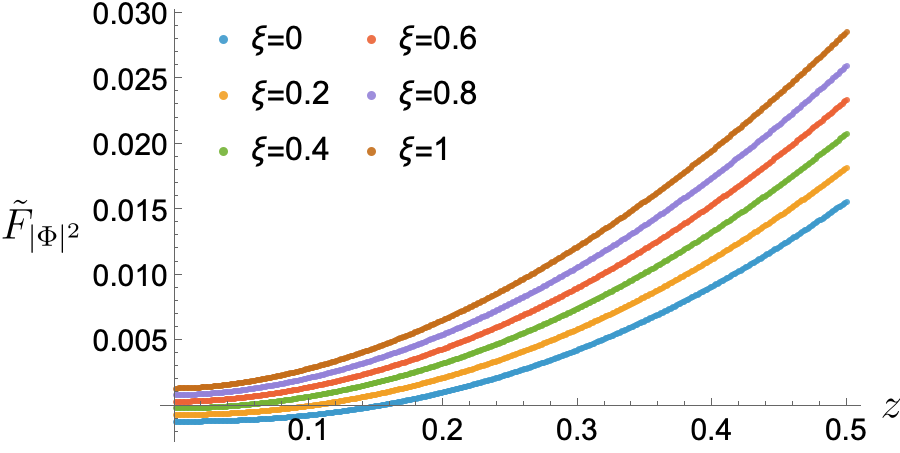}
  \end{minipage}
\caption{\emph{Left:} Plot of $\tilde{F}_{|\Phi|^2}(z)$ when $d=4$ and $\xi=\tilde{\xi}=0$ for different values of $\alpha$. 
At large $z$ the function grows quadratically with $z$, where the slope is determined by the no-defect thermal one-point function of $|\Phi|^2$.
Since $\tilde{F}_{|\Phi|^2}(z)$ is invariant under $\alpha\to1-\alpha$, we do not plot values of $\alpha>1/2$.\\
\emph{Right:} Plot of $\tilde{F}_{|\Phi|^2}(z)$ when $d=4$ and $\tilde{\xi}=0$ at $\alpha=0.1$ for different values of $\xi$.
The blue curves in both panels represent the same function, $\tilde{F}_{|\Phi|^2}(z)$ with $\alpha=0.1$ and $\xi=\tilde{\xi}=0$.
}
\label{fig:Phi2_mon_4d}
\end{figure}

Even though we are not able to perform the integral analytically for general $z$, we can extract its asymptotic behaviour to check that it has the properties~\eqref{eq:FO_asymp} required of a thermal one-point function in the presence of a defect.
For $z\ll1$, we can approximate $\coth x \approx 1/x$ where $x=2\pi z\cosh(t/2)$.
The integral can then be performed analytically and gives precisely the zero-temperature one-point function \eqref{eq:Phi2_0T} with $d=4$.
For $z\gg1$, we can approximate $\coth x \approx 1$.
The resulting integral is then independent of $z$, and together with the prefactor gives a contribution $\sim z/\rho^2=1/\beta\rho$.
This vanishes in the large distance limit $\rho\to\infty$.
Thus the one-point function approaches the no-defect thermal one-point function at large distances, as expected.
For finite $z$ we observe that the integrand is smooth everywhere and decays exponentially at large $t$.
We can therefore cut off the $t$-integral at some large but finite value and only make an exponentially small error.\footnote{
For even integer dimension $d>4$, convergence improves with increasing $d$.
It is straightforward to argue from \eq{Phi2_step} that the integrand for general even integer $d$ decays as $ e^{-\left(\frac{d-2}{2}-\left| \frac{1}{2}-\alpha \right| \right)t}$ at large $t$.} 
We plot $\tilde{F}_{|\Phi|^2}(z)=\rho^2 \langle|\Phi|^2(x)\rangle^{\xi=\tilde{\xi}=0}_\beta$ when $d=4$ in figure~\ref{fig:Phi2_mon_4d} on the left.
When plotting $\tilde{F}_{|\Phi|^2}$, rather than $F_{|\Phi|^2}(z)=\beta^2\langle|\Phi|^2(x)\rangle_\beta^{\xi=\tilde{\xi}=0}$, the vertical axis intercept becomes the zero-temperature one-point function.

It is straightforward to include singular modes in the computation of $\langle|\Phi|^2\rangle_\beta$.
For instance for the $\xi$ mode, the contribution takes the form
\begin{equation}
\label{eq:Phi2_int_s_xi}
\langle|\Phi|^2(x)\rangle^{\xi}_\beta=
\xi\sum_{m=-\infty}^\infty\frac{1}{(4\pi)^{d/2}}
\int_{0}^\infty\d s\,e^{-\frac{2\rho^2+(m\beta)^2}{4s}}s^{-d/2}\left(I_{-\alpha} \left(\frac{\rho^2}{2s}\right)-I_{\alpha} \left(\frac{\rho^2}{2s}\right)\right).
\end{equation}
Unlike the regular contribution, this expression has no UV divergence.
Refs.~\cite{Bianchi:2021snj,Giombi:2021uae} found the zero-temperature result
\begin{equation}
\label{eq:Phi2_xi_0T}
\langle|\Phi|^2(x)\rangle^{\xi}_{\beta\to\infty}=
\xi\,\frac{\Gamma\left(\frac{d}{2}-1-\alpha\right)\Gamma\left(\frac{d}{2}-1+\alpha\right)\sin(\pi\alpha)}{2^{d-1}\pi^{\frac{d+1}{2}}\Gamma\left(\frac{d-1}{2}\right)}\,\frac{1}{\rho^{d-2}}\,.
\end{equation}
The finite temperature contributions come from the $m\neq0$ terms. 
We have not been able to find a compact closed-form expression for the sum over thermal images.
However, when $d$ is even we can find a convenient integral representation for the full one-point function by employing the integral representation \eqref{eq:BesselI_int}.
Performing the $s$-integral one finds
\begin{equation}
\label{eq:Phi2_xi}
\langle|\Phi|^2(x)\rangle^{\xi}_\beta=\xi\,\frac{\Gamma \left(\frac{d}{2}-1\right)}{2\pi ^{\frac{d+2}{2}}}
\sum_{m=-\infty}^\infty\int_0^\infty\d t\, \frac{ \cosh (\alpha  t) \sin (\pi  \alpha )}{\left(\beta ^2 m^2+4 \rho ^2 \cosh ^2\left(\frac{t}{2}\right)\right)^{\frac{d}{2}-1}}\,.
\end{equation}
When $d$ is even the sum over thermal images can be performed in closed form.
E.g.\ when $d=4$ the result is 
\begin{equation}
\label{eq:Phi2_4d_xi}
\langle|\Phi|^2(x)\rangle^\xi_\beta=
\xi \,\frac{z}{4 \pi ^2 \rho ^2}\int_0^\infty\d t\,\frac{\sin (\pi  \alpha )  \cosh (\alpha  t)  \coth \left(2 \pi  z \cosh \left(\frac{t}{2}\right)\right)}{\cosh\left(\frac{t}{2}\right)}\,.
\end{equation}
An analogous expression for $\langle|\Phi|^2(x)\rangle^{\tilde{\xi}}_\beta$ is obtained from \eq{Phi2_4d_xi} by $\alpha\to1-\alpha$.
The full one-point function then is the sum of eqs.~\eqref{eq:Phi2_4d}, \eqref{eq:Phi2_4d_xi} and its analogue for $\tilde{\xi}$.

We now briefly discuss the behaviour of \eq{Phi2_4d_xi} at short and large distances.
In the near-defect regime $z\ll1$, we can approximate $\coth x\approx 1/x$
with $x=2\pi z\cosh(t/2)$.
The resulting integral can be performed analytically and agrees with the zero-temperature contribution \eqref{eq:Phi2_xi_0T}, as required by \eq{FO_asymp}.
In the opposite regime $z\gg1$, we can approximate $\coth x\approx 1$.
The only $\rho$-dependence then comes from the pre-factor of the integral, $\sim z/\rho^2=1/\beta\rho$.
This vanishes in the $\rho\to\infty$ limit, as expected.
All purely thermal no-defect contributions are already contained in the $\xi=\tilde{\xi}=0$ result in \eq{Phi2_4d}.

A comment is in order about the convergence of the integral representation~\eqref{eq:Phi2_4d_xi}.
Inspecting the integrand, one finds that it behaves asymptotically $\sim e^{-(\frac{1}{2}-\alpha)t}$ at large $t$.
For $\alpha\geq1/2$, the $t$ integral diverges due to the exponentially growing tail.
This divergence comes from the summation over $m$ in \eq{Phi2_xi}.
Concretely, it is due to the fact that
\begin{equation}
\label{eq:mon_conv1}
\sum_m \frac{1}{m^2+4 \cosh ^2\left(\frac{t}{2}\right)}= \frac{1}{2} \pi  \sech\left(\frac{t}{2}\right) \coth \left(2 \pi  \cosh \left(\frac{t}{2}\right)\right)\sim e^{-t/2}
\end{equation}
at large $t$, ignoring multiplicative constants, whereas each term in the sum individually decays as $\sim e^{-t}$.
For general $d$ one can check for convergence by changing variables in the $s$-integral in \eq{Phi2_int_s_xi}.
In particular defining $u=(2\rho^2+(m\beta)^2)/4s=a_m^2/s$, one finds
\begin{equation}
\label{eq:mon_conv2}
\langle|\Phi|^2(x)\rangle^\xi_\beta\propto \sum_{m=-\infty}^\infty a_m^{2-d}\int_0^\infty\d u\, u^{d/2-2}e^{-u}\left(I_{-\alpha} \left(\frac{\rho^2}{2a_m^2}u\right)-I_{\alpha} \left(\frac{\rho^2}{2a_m^2}u\right)\right).
\end{equation}
For large $|m|$, one can use the asymptotic behaviour of the modified Bessel function, $I_\nu(\zeta)\sim \zeta^\nu$ for $\zeta\ll1$, where we ignore numerical factors.
Keeping only the leading large-$|m|$ behaviour coming from $I_{-\alpha}$, one finds that each term in the sum over $m$ goes like $|m|^{2-d+2\alpha}$
For the sum $\sum_{m=1}^\infty m^{2-d+2\alpha}$ to converge, we must restrict $\alpha<\frac{d-3}{2}$.
Therefore, $\langle|\Phi|^2\rangle_\beta$ diverges when $\alpha\geq\frac{d-3}{2}$.
Therefore, when  $d\leq5$ the range of $\alpha$ needs to be restricted beyond $[0,1)$ to ensure convergence.
Let us briefly comment on the physical origin of this divergence.
The bound on $\alpha<\frac{d-3}{2}$ ensures that all defect primaries in the DOE of $\Phi$ have $\hat{\Delta}>1/2$.
When $\alpha\geq\frac{d-3}{2}$ the operator $\hat{\CO}^-_{-\alpha}$ has scaling dimension $\hat{\Delta}^-_{-\alpha}\leq 1/2$.
Computing the two-point function $\langle\hat{\CO}^-_{-\alpha}(\bm{x}_1)\hat{\CO}^{\dagger-}_{\alpha}(\bm{x}_2)\rangle_\beta$ via a sum over thermal images of zero-temperature defect two-point function, one finds that the terms at large $|m|$ in the sum can be approximated by $\sim |m|^{-2\hat{\Delta}^-_{-\alpha}}$.
Here, $\hat{\CO}^{\dagger-}_{\alpha}$ is the operator in the DOE of $\Phi^\dagger$ with the same scaling dimension as $\hat{\CO}^-_{-\alpha}$ but opposite transverse spin.
As $\alpha$ increases beyond $\frac{d-3}{2}$, the power of $|m|$ becomes greater or equal to $-1$, and so the sum diverges.
This is morally the same IR divergence encountered in the two-point function of a free massless scalar field in $d\leq 3$ at finite temperature \cite{Iliesiu:2018fao,Laine:2016hma}.
The discussion so far applied to the $-\alpha$ mode with parameter $\xi$.
For the $-1+\alpha$ mode with parameter $\tilde{\xi}$, the regime for which the thermal one-point function is well-defined is $\alpha>\frac{5-d}{2}$.
Consequently when $d\leq4$ we cannot turn on both singular modes simultaneously at finite temperature without encountering IR divergences.

In the following we will restrict to ranges of $\alpha$ where these IR divergences are absent.
In figure~\ref{fig:Phi2_mon_4d} on the right we plot $\tilde{F}_{|\Phi|^2}(z)=\rho^2 \langle|\Phi|^2(x)\rangle_\beta$ when $d=4$ at fixed $\alpha<1/2$ as we vary $\xi$.
Turning on the singular modes primarily changes the near defect behaviour by increasing the value of $\tilde{F}_{|\Phi|^2}(z)$.
At large distances $z\gg1$, however, the effects due to the singular modes are overwhelmed by the no-defect thermal one-point functions.
Thermal no-defect effects are quadratic in $z$ whereas the term proportional to $\xi$ is linear in $z$.

Before moving on let us make a short remark about the analytic structure of the integrals in the cross-ratio $z$ encountered here. 
Consider e.g.\ the integral in \eq{Phi2_4d}.
While for physical separations $z>0$ the integral is perfectly well-behaved, it develops singularities in the complex plane.
This is most clearly seen after a change of variables $u=\cosh(t/2)$.
The integral then becomes
\begin{equation}
2\int_1^\infty \d u\,\frac{\cosh\!\big((1-2\alpha)\operatorname{arccosh}u\big)}{u^2\sqrt{u^2-1}}\,\coth(2\pi z u)\,,
\end{equation}
with the $(u^2-1)^{-1/2}$ factor in the integrand coming from the Jacobian.
The $\coth(2\pi z u)$ factor becomes singular whenever its argument is equal to an integer multiple of $\iu\pi$.
This occurs for $u_k=\iu k/2z$ for some non-zero integer $k$.
For most values of $z$ these poles are away from the integration contour in the complex $u$-plane.
When $z$ becomes purely imaginary, a given $u_k$ pole can approach the integration contour.
Most poles of the integrand can be avoided by deforming the contour in the complex plane and picking up residues as a pole $u_k$ crosses the contour.
However, when a pole approaches the endpoint of the integral at $u=1$ the singularity becomes unavoidable, leading to a genuine divergence of the integral.
For the pole $u_k$ this happens when $z\to \iu k/2$ for any non-zero integer $k$. 
Since the integrand contains a factor of $(u^2-1)^{-1/2}$ as $u\to1$, the singularities on the imaginary $z$-axis at half-integer spacing are in fact square-root branch points.
Since the integral jumps whenever the pole $u_k$ crosses the integration contour, the branch cut in the $z$-plane emanating from $z=\iu k/2$ must end at the origin $z=0$ where the $u_k\to+\infty$.
A similar discussion applies to the integral in \eq{Phi2_4d_xi} and in fact all analogous integrals that we will encounter in this section.
They all share the same analytic structure.

\paragraph{DOE limit. }
Instead of trying to compute the full thermal one-point function, we now study the near-defect expansion.
This will allow us to extract the thermal DCFT data of operators exchanged in the DOE of $|\Phi|^2$ for arbitrary dimension $d$.
Starting from the expression for the one-point function in \eq{Phi2_int_s} with $\xi=\tilde{\xi}=0$, we can use a different integral representation for the sum over Bessel functions
\begin{equation}
\begin{split}
\mathcal{I}_\alpha (\zeta) \equiv\sum_{n} I_{|n -\alpha | }(\zeta) =\;&  \frac{1}{2 \alpha}  \left[ e^{\zeta}  \int_0^\zeta e^{-x} I_{-\alpha }(x)\, dx - \zeta I_{-\alpha }(\zeta) - \zeta I_{1-\alpha}(\zeta)  \right] + \\
& +  \frac{1}{2(1-\alpha)}  \left[ e^{\zeta}  \int_0^\zeta e^{-x} I_{\alpha-1}(x)\, dx - \zeta I_{\alpha-1}(\zeta) - \zeta I_{\alpha}(\zeta)  \right] .
\end{split}
\end{equation}
In order to perform the resulting integral over $s$ in \eq{Phi2_int_s}, we use the following trick.
We Taylor expand the factor $e^{-\frac{(m\beta)^2}{4s}}$ in the integrand and exchange the summation and the $s$-integral, i.e. we replace the $s$-integral of the series with the series of the $s$-integrals.
While keeping a UV regulator to keep track of spurious divergences generated in this non-uniform expansion, we perform these integrals in closed form for fixed $m$ and finally re-sum the resulting series.

Consider a term with $m\neq0$.
Expanding $e^{-\frac{(m\beta)^2}{4s}}=\sum_{n=0}^\infty (-1)^n(m\beta)^{2n}/n!(4s)^{n}$
we find that the $s$-integral of each term in the sum has the following finite piece
\begin{equation}
-\frac{\left(\csc \left(\frac{1}{2} \pi  (d-2 \alpha )\right)-\csc \left(\pi  \alpha +\frac{\pi  d}{2}\right)\right)  \Gamma \left(-\frac{d}{2}-n+\frac{3}{2}\right) (m\beta)^{2n}}{n! (d+2 n-2) \pi ^{\frac{d}{2}-\frac{1}{2}} 2^{d+2 n} \Gamma \left(-\frac{d}{2}-n-\alpha +2\right) \Gamma \left(-\frac{d}{2}-n+\alpha +1\right)\rho^{d+2n-2}}\,.
\end{equation}
Re-summing over $n$ to obtain the $s$-integral of the exponential that we had Taylor expanded, we find
\begin{equation}
\label{eq:Phi2_step2}
-\frac{ \sin (\pi  \alpha )  \Gamma \left(\frac{d}{2}-\alpha \right) \Gamma \left(\frac{d}{2}+\alpha -1\right) \, _3F_2\left(\frac{d-2}{2},\frac{d}{2}-\alpha ,\frac{d-2}{2}+\alpha;\frac{d-1}{2},\frac{d}{2};-\frac{m^2}{4z^2}\right)}{\pi(4\pi)^{(d-1)/2}(d-2) \Gamma \left(\frac{d-1}{2}\right)}\frac{1}{\rho^{d-2}}\,,
\end{equation}
where $_3F_2$ is the generalised hypergeometric function, and we have introduced the thermal cross-ratio $z\equiv\rho/\beta$.

Let us now discuss the divergences.
The first type of divergences comes from the zero-temperature contribution $m=0$, whose $s$-integral has a UV divergence at $s=0$.
As explained in ref.~\cite{Bianchi:2021snj}, it can be regulated by introducing a short-distance cut-off $\varepsilon^2$.
Since the divergence as $\varepsilon^2\to0$ is independent of $\alpha$, we can subtract it unambiguously and remove the regulator.
The second type of divergences comes from the $m\neq0$ integrals.
Unlike the $m=0$ integral, the integrals for $m\neq0$ in \eq{Phi2_int_s} are finite prior to Taylor expanding due to the exponential suppression near $s=0$.
By expanding $e^{-\frac{(m\beta)^2}{4s}}$ inside the integral and commuting the sum with the integral, however, all the resulting integrals develop a power-law divergence at $s=0$ and require regularisation.
In terms of the UV regulator $\varepsilon^2$, we have
\begin{equation}
\label{eq:divergences_m}
 \frac{ (-1)^n (m\beta)^{2n}}{n! (d+2 n-2)2^{d+2n-1} \pi ^{\frac{d}{2}} }\,\frac{1}{\varepsilon^{d+2n-2}}\,.
\end{equation}
Re-summing over $n$, we find
\begin{equation}
\label{eq:resum_div}
\frac{1}{4\pi ^{\frac{d}{2}}}  (m\beta)^{2-d}\, \gamma \left(\frac{d}{2}-1,\frac{m^2 \beta ^2}{4 \varepsilon^2}\right),
\end{equation}
where $\gamma$ is the lower incomplete gamma function,
\begin{equation}
\gamma(s,x)=\int_0^x\d t \,t^{s-1}e^{-t}\,.
\end{equation}
Since $\lim_{x\to\infty}\gamma(s,x)=\Gamma(s)$, we can safely remove the regulator and take $\varepsilon^2\to0$ for all terms with $m\neq 0$.
Once re-summed, the power law divergences in \eq{divergences_m} give a finite result and encode no-defect thermal effects.
This is consistent with the fact that we started with an integral that was convergent before Taylor expanding and exchanging the integral and the series.
For $\alpha=0$, we can evaluate the integral in \eq{Phi2_int_s} in closed form using $\CI_0(\zeta)=e^\zeta$.
The result precisely agrees with \eq{resum_div} after taking $\varepsilon^2\to0$ and performing the sum over thermal images with $m\neq0$.
In both cases one finds
\begin{equation}
\label{eq:no_def_1pt}
2\kappa \frac{ \zeta (d-2)}{\beta ^{d-2}}\,.
\end{equation}
Thus, re-summing the spurious divergences for the $m\neq0$ terms produces the no-defect thermal one-point function of $|\Phi|^2$.

The full thermal one-point function of $|\Phi|^2$ in the presence of the defect with $\xi=\tilde{\xi}=0$ is thus given by the sum over thermal images of \eq{Phi2_step2} plus the no-defect thermal effect \eq{no_def_1pt}.
The sum over thermal images converges for all $\alpha$ and all $z$ when $d>3$ but we have not been able to find a closed form for it.
Nonetheless we can check its asymptotic properties.
When $z\ll1$, the $_3F_2$'s go to zero whenever $m\neq0$.
The leading contribution comes from the $m=0$ mode.
It is easy to check that for $m=0$ \eq{Phi2_step2} reduces to the zero-temperature one-point function in \eq{Phi2_0T}. 
When $z\gg1$, we observe numerically that the sum grows $\sim z/\rho^{d-2}$.
Therefore, the no-defect piece, which goes like $z^{d-2}/\rho^{d-2}$, dominates.

Though we cannot find a closed-form expression for the sum over $m$ of \eq{Phi2_step2}, we can expand the $_3F_2$'s for small $z$ and perform the sum over $m$ term-by-term in the series in $z$.
In doing so we find three types of contributions.
The first one precisely cancels the contribution of the no-defect thermal one-point function in \eq{no_def_1pt}.
The remaining two types of terms involve in general non-integer powers of $z$.
After performing the sum over thermal images, we find
\begin{equation}
\label{eq:F_Phi2}
\begin{split}
&F_{|\Phi|^2}^{\xi=\tilde{\xi}=0}(z)=\frac{a^{\xi=\tilde{\xi}=0}_{|\Phi|^2}}{{z^{d-2}}}\\
&\!+\sum_{n=0}^\infty (-1)^n\frac{  \Gamma \left(n+\alpha +\frac{1}{2}\right) \Gamma \left(\frac{d-2}{2}+n+\alpha\right)  \zeta (d+2n-2+2\alpha)}{2^{2 (1-\alpha -n)}\pi ^{\frac{d+1}{2}}(\alpha +n) \Gamma (n+1) \Gamma (n+2 \alpha )}z^{2n+2\alpha }\\
&\!+\sum_{n=0}^\infty \,(\alpha\to 1-\alpha)\,,
\end{split}
\end{equation}
where the summand in the second line is given by the summand in the first line with $\alpha \to 1-\alpha$, and $a^{\xi=\tilde{\xi}=0}_{|\Phi|^2}$ is the coefficient of $\rho^{2-d}$ in \eq{Phi2_0T}. 
Both series have a finite radius of convergence $|z|<1/2$, which is precisely the radius of convergence of the DOE.
Within its radius of convergence and when $d=4$, \eq{F_Phi2} matches the integral representation~\eqref{eq:Phi2_4d} to excellent numerical accuracy.

The sums in the second and third lines of \eq{F_Phi2} correspond to operators in the DOE of $|\Phi|^2$ that acquire defect thermal one-point functions.
Comparing with the general form of the expansion in \eq{FO_DOE}, one finds that the exchanged operators form two sets with scaling dimensions $\hat{\Delta}=d+2n-2\alpha$ and $\hat{\Delta}=d+2n-2+2\alpha$, where $n$ is a non-negative integer, such that $\hat{\Delta}\geq d-2$ for both sets. 
In free theory one can think of these operators as composite operators consisting of the operators in the DOEs of $\Phi$ and $\Phi^\dagger$.
As discussed below \eq{prop_0T}, the DOE of $\Phi$ contains primary operators $\hat{\CO}^+_s$ with transverse spin $s\in\mathbb{Z}-\alpha$ and scaling dimension $\hat{\Delta}^+_s=\frac{d}{2}-1+|s|$. 
In addition to the primaries, the DOE also includes their descendants, which schematically look like $\p_\parallel^{2p}\hat{\CO}^+_s$ for positive integer $p$.
The DOE of $\Phi^\dagger$ has primary operators $\hat{\CO}^{\dagger+}_{-s}$ with spin $-s$ and scaling dimension $\hat{\Delta}^+_{s}=\frac{d}{2}-1+|s|$ together with their descendants.
The DOE of $|\Phi|^2$ then contains scalar operators with zero total transverse spin of the schematic form $:\p_\parallel^{2p}\hat{\CO}^+_s\p_\parallel^{2q}\hat{\CO}^{\dagger+}_{-s}:$, where $\p^2_\parallel$ denotes the Laplacian along the $d-2$ parallel directions and $p,q$ are non-negative integers.
For given $p+q$, there is a unique scalar primary since different distributions of derivatives are equivalent modulo descendants~\cite{Heemskerk:2009pn,Penedones:2010ue,Fitzpatrick:2011dm}.
These operators have scaling dimension $\hat{\Delta}=2\hat{\Delta}^+_s+2(p+q)$.
There can, however, be several operators with a given $\hat{\Delta}$.
E.g.\ for $\hat{\Delta}=d+2\alpha$ and generic $\alpha$ there are two operators with that scaling dimension: 
(1) the operator with $s=-1-\alpha$ and $p=q=0$, and (2) the scalar primary with $s=-\alpha$ and $p+q=1$. 
Each term in \eq{F_Phi2} then receives contributions from all primaries of the schematic form $:\p_\parallel^{2p}\hat{\CO}^+_s\p_\parallel^{2q}\hat{\CO}^{\dagger+}_{-s}:$ with a given $\hat{\Delta}$.

In order to distinguish the contributions of composite operators built out of $\hat{\CO}^+_s$ (and $\hat{\CO}^{+\dagger}_{-s}$) with different $s$, one needs to compute the one-point function mode-by-mode.
This amounts to repeating the computation in the DOE limit starting with \eq{Phi2_int_s} but with $\CI_\alpha$ now replaced by $I_{\nu}$, where $\nu=|s|\in|\mathbb{Z}-\alpha|$.
This computation can be done but requires a careful treatment of the divergences that appear when performing the $s$-integral for each term of the Taylor expansion of $e^{-\frac{(m\beta)^2}{4s}}$.
Instead of Taylor expanding, one can perform the integration using the Bessel function identity
\begin{equation}
e^{-u} I_\nu (u)=\frac{(u/2)^\nu}{\Gamma(\nu+1)}\,{}_1F_1(\nu+1/2,2\nu+1;-2u)\,,
\end{equation}
where $_1F_1$ is the confluent hypergeometric function.
Since the zero-temperature contributions are well-understood we will take $m\neq0$.
Using the power series definition of the confluent hypergeometric, $_1F_1(a,b;z)=\sum_{n=0}^\infty (a)_n z^n/(b)_n n!$, performing the $s$-integral term-by-term in the series and then re-summing gives
\begin{equation}
(-1)^n \frac{ \Gamma \left(n+\nu +\frac{1}{2}\right) \Gamma \left(\frac{d}{2}+n+\nu -1\right) |m\beta|^{2 -d{-2\nu} -2n} \rho ^{2 (\nu +n)}}{2^{1-2 \nu -2 n}\pi ^{\frac{d+1}{2}}  \,\Gamma (n+1) \Gamma (n+2 \nu +1)}\,.
\end{equation}
The integrals converge provided that $\nu>1-d/2-n$, which holds in all cases of interest.
Finally performing the sum over thermal images (omitting $m=0$) and multiplying by $\beta^{d-2}$ to facilitate comparison with \eq{F_Phi2}, one finds
\begin{equation}
\label{eq:F_Phi2_mode}
\sum_{n=0}^\infty(-1)^n \frac{ \Gamma \left(n+\nu +\frac{1}{2}\right) \Gamma \left(\frac{d-2}{2}+n+\nu\right)  \zeta (d+2 n+2 \nu -2)}{2^{1-2 \nu -2 n}\pi ^{\frac{d+1}{2}} \Gamma (n+1) \Gamma (n+2 \nu +1)}z^{2 \nu +2 n}\,.
\end{equation}
Note that the terms in the sum have the right powers of $z$ to account for primaries of the schematic type $:\p_\parallel^{2p}\hat{\CO}^+_{\nu}\p_\parallel^{2q}\hat{\CO}^{+\dagger}_{-\nu}:$ if $\nu\in\mathbb{Z}-\alpha\geq0$ or $:\p_\parallel^{2p}\hat{\CO}^+_{-\nu}\p_\parallel^{2q}\hat{\CO}^{+\dagger}_{\nu}:$ if $\nu\in\mathbb{Z}+\alpha\geq0$, where in both cases $n=p+q$.\footnote{
As a consistency check, one can compare the result of the all-mode computation in \eq{F_Phi2} to the computation for a single mode in \eq{F_Phi2_mode}.
Consider the $z^{2+2\alpha}$ term in \eq{F_Phi2}: it is a sum over the contributions of scaling dimension $\hat{\Delta}=d+2\alpha$ from the two operators $:\hat{\CO}^+_{-1-\alpha}\hat{\CO}^{+\dagger}_{1+\alpha}:$ and the primary involving $:\p_\parallel^2\hat{\CO}^+_{-\alpha}\hat{\CO}^{+\dagger}_{\alpha}:$ and $:\hat{\CO}^+_{-\alpha}\p_\parallel^2\hat{\CO}^{+\dagger}_{\alpha}:$.
The first operator corresponds to the $n=0$ term in \eq{F_Phi2_mode} with $\nu=|-1-\alpha|=1+\alpha$, while the latter is captured by the $n=1$ term in \eq{F_Phi2_mode} with $\nu=|-\alpha|=\alpha$.
Summing these two terms, we precisely reproduce the $z^{2+2\alpha}$ term in \eq{F_Phi2}.
For any $\hat{\Delta}$, the coefficient of $z^{\hat{\Delta}{-d+2}}$ in \eq{F_Phi2} can be reproduced by summing over all coefficients in \eq{F_Phi2_mode} with $\nu$ and $n$ such that $d+2\nu+2n-2=\hat{\Delta}$.}
We can then read off the thermal DCFT data of operators in the DOE of $|\Phi|^2$ from \eq{F_Phi2_mode}.
We find
\begin{equation}
\label{eq:DT1pt_mon}
\mu_{|\Phi|^2}{}^{\hat{\CO}}\hat{b}_{\hat{\CO}}=
(-1)^n\frac{  \Gamma \left(n+\nu +\frac{1}{2}\right) \Gamma \left(\frac{d-2}{2}+n+\nu\right)  \zeta (d+2 n+2 \nu -2)}{2^{1-2 \nu -2 n}\pi ^{\frac{d+1}{2}} \Gamma (n+1) \Gamma (n+2 \nu +1)} \,,
\end{equation}
where $\hat{\CO}$ on the left-hand side refers to the unique scalar primary of the schematic form $:\p_\parallel^{2p}\hat{\CO}^+_{\nu}\p_\parallel^{2q}\hat{\CO}^{+\dagger}_{-\nu}:$ if $\nu\in\mathbb{Z}-\alpha\geq0$ or $:\p_\parallel^{2p}\hat{\CO}^+_{-\nu}\p_\parallel^{2q}\hat{\CO}^{+\dagger}_{\nu}:$ if $\nu\in\mathbb{Z}+\alpha\geq0$ with $n=p+q$ in either case.
The scaling dimensions of these operators are $\hat{\Delta}=d+2\nu+2n-2$.

We can repeat the calculation above in the presence of singular modes that correspond to scalars with dimensions $\hat{\Delta}^-_s=\frac{d}{2}-1-|s|$ in the DOE of $\Phi$ with $s=-\alpha$ or $s=1-\alpha$.
For the $\hat{\Delta}^-_{-\alpha}$ contributions with coefficient $\xi\neq0$, we start with \eq{Phi2_int_s_xi} and apply the procedure described above. 
We then find for the contribution proportional to $\xi$
\begin{equation}
\label{eq:F_Phi2_xi}
\begin{split}
&F^\xi_{|\Phi|^2}(z)=\frac{a^\xi_{|\Phi|^2}}{{z^{d-2}}}\\
&+\xi\sum_{n=0}^\infty(-1)^n\frac{  \Gamma \left(n-\alpha +\frac{1}{2}\right) \Gamma \left(\frac{d-2}{2}+n-\alpha \right)  \zeta (d+2 n-2 \alpha -2)}{2^{1+2 \alpha -2 n} \pi ^{\frac{d+1}{2}}\Gamma (n+1) \Gamma (n-2 \alpha +1)}z^{2 n-2 \alpha}\\
&-\xi\sum_{n=0}^\infty(-1)^n\,(\alpha\to-\alpha)\,,
\end{split}
\end{equation}
where $a^\xi_{|\Phi|^2}$ is the coefficient of $\rho^{2-d}$ in \eq{Phi2_xi_0T}
and the summand in the third line is obtained from the one in the second line by replacing $\alpha$ with $-\alpha$.
The first term here captures the unique primary for given $n$ of the schematic form $:\p_\parallel^{2p}\hat{\CO}^-_{-\alpha}\p_\parallel^{2q}\hat{\CO}^{-\dagger}_{\alpha}:$ whereas the second one captures $:\p_\parallel^{2p}\hat{\CO}^+_{-\alpha}\p_\parallel^{2q}\hat{\CO}^{+\dagger}_{\alpha}:$, with $p+q=n$ in both cases.
From this we can read off the defect thermal one-point functions
\begin{equation}
\label{eq:DT1pt_mon_xi}
\mu_{|\Phi|^2}{}^{\hat{\CO}}\hat{b}_{\hat{\CO}}=\xi
(-1)^n\frac{  \Gamma \left(n-\alpha +\frac{1}{2}\right) \Gamma \left(\frac{d-2}{2}+n-\alpha \right)  \zeta (d+2 n-2 \alpha -2)}{2^{1+2 \alpha -2 n} \pi ^{\frac{d+1}{2}}\Gamma (n+1) \Gamma (n-2 \alpha +1)} \,,
\end{equation}
for the operators $:\p_\parallel^{2p}\hat{\CO}^-_{-\alpha}\p_\parallel^{2q}\hat{\CO}^{-\dagger}_{\alpha}:$.
The defect thermal one-point functions of operators $:\p_\parallel^{2p}\hat{\CO}^+_{-\alpha}\p_\parallel^{2q}\hat{\CO}^{+\dagger}_{\alpha}:$, which are in \eq{DT1pt_mon} for $\nu=|-\alpha|=\alpha$, meanwhile receive an additional contribution such that
\begin{equation}
\label{eq:DT1pt_mon_xi2}
\mu_{|\Phi|^2}{}^{\hat{\CO}}\hat{b}_{\hat{\CO}}=(1-\xi)
(-1)^n\frac{  \Gamma \left(n+\alpha +\frac{1}{2}\right) \Gamma \left(\frac{d-2}{2}+n+\alpha\right)  \zeta (d+2 n+2 \alpha -2)}{2^{1-2 \alpha -2 n}\pi ^{\frac{d+1}{2}} \Gamma (n+1) \Gamma (n+2 \alpha +1)} \,.
\end{equation}
The discussion for the $\tilde{\xi}\neq0$ modes proceeds analogously.
We find defect thermal one-point functions for the primaries schematically represented by $:\p_\parallel^{2p}\hat{\CO}^-_{1-\alpha}\p_\parallel^{2q}\hat{\CO}^{-\dagger}_{-1+\alpha}:$ which are of the form in \eq{DT1pt_mon_xi} with $\xi\to\tilde{\xi}$ and $\alpha\to1-\alpha$.
Similarly, the defect thermal one-point functions of $:\p_\parallel^{2p}\hat{\CO}^+_{1-\alpha}\p_\parallel^{2q}\hat{\CO}^{+\dagger}_{-1+\alpha}:$ are corrected to take the form of \eq{DT1pt_mon_xi2} together with the same replacement.

\subsection{Thermal one-point function of \texorpdfstring{$J_\mu$}{J μ}}

Next we turn to the $U(1)$ current 
\begin{equation}
J_\mu=-\iu\left(\Phi\p_\mu\Phi^\dagger-\p_\mu\Phi\Phi^\dagger\right).
\end{equation}
Its thermal one-point function $\langle J_\mu\rangle_\beta$ can be straightforwardly computed using Wick's theorem. 
Taking a single derivative of the propagator in \eq{prop_0T} followed by the coincident limit, ref.~\cite{Bianchi:2021snj} showed that at zero temperature only the $J_\theta$ component may acquire a one-point function.
Here $\theta$ labels the angular direction transverse to the defect.
This matches precisely the form of the one-point of a vector primary for co-dimension two defects in \eq{0T_V_1pt} after changing to Euclidean coordinates.
At finite temperature we replace $\tau'-\tau\to\tau'-\tau+m\beta$ in \eq{prop_0T} and sum over thermal images $m$.
We find that also at finite temperature only the $\theta$-component can acquire a one-point function.
Comparing with the allowed structures in \eq{kin_conserved_J} for a thermal one-point function of a conserved current, it is clear that only the last term can be non-zero. 
The first term must be absent because the monodromy preserves the global $U(1)$ symmetry.
The second term is absent because the defect is invariant under $\tau\to-\tau$ as can be seen from the propagator in \eq{prop_0T}.\footnote{
This transformation is a combination of time reversal and charge conjugation.
Since time reversal is implemented by an anti-unitary operator $T$, time reversal alone also sends $\alpha\to-\alpha$. 
Charge conjugation $C$ reverses the sign of $\alpha$ again such that under $CT$ only the $\tau$ coordinate changes sign.} 
Setting $\xi=\tilde{\xi}=0$ for now, we find from the coincident limit of the propagator
\begin{equation}
\label{eq:Jth_int_s}
\langle J_\theta(x)\rangle^{\xi=\tilde{\xi}=0}_\beta=-\sum_{m=-\infty}^\infty\frac{2}{(4\pi)^{d/2}}
\int_{0}^\infty\d s\,e^{-\frac{2\rho^2+(m\beta)^2}{4s}}s^{-d/2}\CJ_{\alpha}\left(\frac{\rho^2}{2s}\right),
\end{equation}
where $\CJ_\alpha(\zeta)=\sum_{n=-\infty}^\infty(n-\alpha)I_{|n-\alpha|}(\zeta)$.
To perform the integration it is useful to note the following Bessel function identity
\begin{equation}
\label{eq:Bessel_id_J}
\CJ_\alpha(\zeta)=\frac{\zeta}{2}\left[I_{1-\alpha}(\zeta)+I_{-\alpha}(\zeta)-I_{1+\alpha}(\zeta)-I_\alpha(\zeta)\right]-\alpha I_\alpha(\zeta)\,.
\end{equation}
For the $m=0$ mode, ref.~\cite{Bianchi:2021snj} used this identity to compute the zero-temperature one-point function.
The integral is convergent and gives
\begin{equation}
\label{eq:Jth_0T}
\langle J_\theta(x)\rangle_{\beta\to\infty}^{\xi=\tilde{\xi}=0}=\frac{(1-2\alpha)\Gamma\left(\frac{d}{2}-\alpha\right)\Gamma\left(\frac{d}{2}+\alpha-1\right)\sin(\pi\alpha)}{2^d \pi^{\frac{d+1}{2}}\Gamma\left(\frac{d+1}{2}\right)}\,\frac{1}{\rho^{d-2}}\,.
\end{equation}

At finite temperature, the integrals for $m\neq0$ are also convergent.
They can be performed, however, we are not able to find a closed form for the sum over thermal images.
As for $\langle|\Phi|^2\rangle_\beta$ we will instead find an integral representation for the sum that is rapidly convergent.
Using the integral representation~\eqref{eq:BesselI_int} of the modified Bessel function of the first kind inside the Bessel identity \eq{Bessel_id_J}, we find
\begin{equation}
\begin{split}
&\langle J_\theta(x)\rangle_{\beta}^{\xi=\tilde{\xi}=0}=\sum_{m=-\infty}^\infty\frac{\Gamma\left(\frac{d}{2}-1\right)\sin(\pi\alpha)}{2\pi^{\frac{d+2}{2}}}\Bigg[ \frac{1}{\left(\beta ^2 m^2+4 \rho ^2\right)^{\frac{d-2}{2}}}\\
&- \int_0^\infty\d t\,\Bigg(
\frac{(d-2)\rho^2(1-e^{-t})\cosh(t\alpha)}{\left(m^2\beta^2+4\rho^2\cosh^2\left(\frac{t}{2}\right)\right)^{d/2}}+  \frac{\alpha e^{-t\alpha}}
      {\left(m^2\beta^2+4\rho^2\cosh^2\left(\frac{t}{2}\right)\right)^{\frac{d-2}{2}}}\Bigg)\Bigg]\,,
\end{split}
\end{equation}
where the first term in square brackets comes from performing the $\theta$-integral in \eq{BesselI_int}.
The sum over thermal images can be performed when $d$ is even.
This allows us to write the thermal one-point function as an integral that is amenable to rapidly converging numerical integration.
E.g.\ when $d=4$, we find a one-point function of the form \eqref{eq:kin_conserved_J} with 
\begin{equation}
\label{eq:FJ_mon}
\begin{split}
&F^{q=2}_J (z)= \frac{\sin(\pi\alpha)}{4\pi^2{z^2}}
\Bigg[-\coth\left(2\pi z\right)\\
&+ \int_0^\infty \d t\,
  \frac{\left(1-e^{-t}\right)\cosh (\alpha  t)}{4}     \left( \frac{2 \pi  z\csch^2\left(2 \pi  z \cosh \left(\frac{t}{2}\right)\right)}{\cosh^2\left(\frac{t}{2}\right)}+\frac{\coth \left(2 \pi  z \cosh \left(\frac{t}{2}\right)\right)}{\cosh^3\left(\frac{t}{2}\right)}\right)\\
&+\alpha\int_0^\infty \d t\,  e^{-\alpha t} \frac{ \coth \left(2 \pi  z \cosh \left(\frac{t}{2}\right)\right)}{\cosh\left(\frac{t}{2}\right)}
\Bigg]\,.
\end{split}
\end{equation}
Here we used that $\langle J_\theta(x)\rangle_\beta=-zF^{q=2}_J(z)/\beta^2$ after changing to polar coordinates in the transverse plane and lowering the index with the metric.
Let us briefly check some asymptotic properties of this expression.
When $z\ll1$ we can approximate $\coth x \approx \csch x \approx 1/x$ with $x=2\pi z\cosh(t/2)$.
In that limit the $t$-integrals can be performed analytically, and the one-point function reduced to zero-temperature result in \eq{Jth_0T}.
In the opposite limit $z\gg1$, we approximate $\coth x\approx1$ and $\csch x\approx 0$.
It is then easy to see that $F_J^{q=2}(z)\sim z^{{-2}}$ such that $\langle J_\theta\rangle_\beta\sim 1/\beta\rho$, which vanishes in the $\rho\to\infty$ limit.
This is consistent with the fact that the thermal one-point function of a current in the absence of a defect vanishes.

\begin{figure}[ht]
\centering
  \begin{minipage}{0.49\textwidth}
  \centering
    \includegraphics[width=\linewidth]{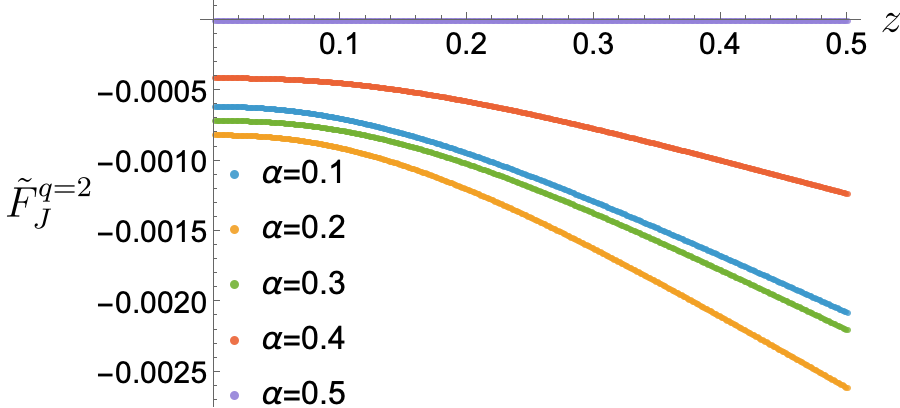}
  \end{minipage}\hfill
  \begin{minipage}{0.49\textwidth}
  \centering
    \includegraphics[width=\linewidth]{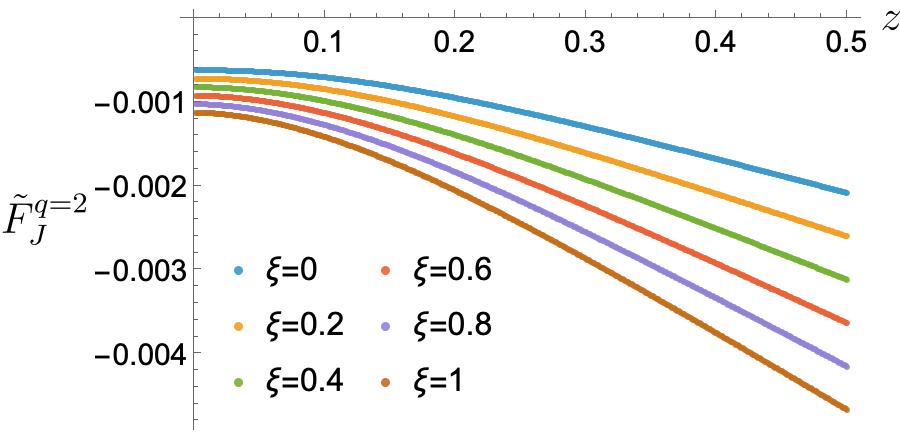}
  \end{minipage}
\caption{\emph{Left.} Plot of $\tilde{F}^{q=2}_{J}(z)$ when $d=4$ and $\xi=\tilde{\xi}=0$ for different values of $\alpha$. 
At large $z$ the function grows linearly in $z$. 
It is slower than quadratic because the no-defect thermal one-point function of $J_\theta$ vanishes.
Since $\tilde{F}^{q=2}_{J}(z)\to-\tilde{F}^{q=2}_{J}(z)$ under $\alpha\to1-\alpha$, we do not plot values of $\alpha>1/2$.\\
\emph{Right.} Plot of $\tilde{F}^{q=2}_{J}(z)$ when $d=4$ and $\tilde{\xi}=0$ at $\alpha=0.1$ for different values of $\xi$.
The blue curves in both panels represent the same function, $\tilde{F}^{q=2}_{J}(z)$ with $\alpha=0.1$ and $\xi=\tilde{\xi}=0$.
} 
\label{fig:FJ_mon_4d}
\end{figure}

In figure~\ref{fig:FJ_mon_4d} on the left we plot $\tilde{F}_J^{q=2}(z)=z^3 F^{q=2}_J(z)$ for various values of $\alpha$.
With this re-scaling of \eq{FJ_mon} by $z^{d-1}$, the vertical axis intercept is given by the zero-temperature one-point function~\eqref{eq:Jth_0T}.
Note that $\tilde{F}_J^{q=2}=F_J^{q=2}=0$ for $\alpha=1/2$, which follows from $\CJ_{\alpha=1/2}(\zeta)=0$.
Physically, this is a consequence of the invariance of the action \eqref{eq:action_mon} under $\theta\to-\theta$ and $\alpha\to-\alpha$ once spacetime covariant derivatives have been promoted to gauge covariant derivatives, as explained below \eq{monodromy_BCs}. 
In the absence of singular modes, a monodromy of $-\alpha$ is identified with $1-\alpha$.
If $\alpha=1/2$, then $\theta\to-\theta$ becomes a symmetry of the system with $\xi=\tilde{\xi}=0$.
Since $J_\theta\to-J_\theta$ under this action, we must have $\langle J_\theta\rangle_\beta^{\xi=\tilde{\xi}=0}=0$ when $\alpha=1/2$.

We can repeat the calculation when singular modes are turned on in the DOE of $\Phi$.
The contribution proportional to $\xi$ is
\begin{equation}
\langle J_\theta(x)\rangle^{\xi}_\beta=\alpha\,\xi\sum_{m=-\infty}^\infty\frac{2}{(4\pi)^{d/2}}
\int_{0}^\infty\d s\,e^{-\frac{2\rho^2+(m\beta)^2}{4s}}s^{-d/2}\left(I_{-\alpha}\left(\frac{\rho^2}{2s}\right)-I_{\alpha}\left(\frac{\rho^2}{2s}\right)\right).
\end{equation}
For the $m=0$ mode ref.~\cite{Bianchi:2021snj} found
\begin{equation}
\label{eq:Jth_0T_xi}
\langle J_\theta(x)\rangle^\xi_{\beta\to\infty}=\xi\frac{\alpha \Gamma\left(\frac{d-2}{2}-\alpha\right)\Gamma\left(\frac{d-2}{2}+\alpha\right)\sin(\pi\alpha)}{2^{d-2}\pi^{\frac{d+1}{2}}\Gamma\left(\frac{d-1}{2}\right)}\,\frac{1}{\rho^{d-2}}\,.
\end{equation}
For the $m\neq0$ terms, we use \eq{BesselI_int} to find
\begin{equation}
\langle J_\theta(x)\rangle^{\xi}_\beta=\alpha\, \xi\, \Gamma \left(\frac{d}{2}-1\right) \sin (\pi  \alpha )\sum_{m=-\infty}^\infty\int_0^\infty\d t\, \frac{  \cosh (\alpha  t) }{\pi ^{\frac{d+2}{2}}\left(\beta ^2 m^2+4 \rho ^2 \cosh ^2\left(\frac{t}{2}\right)\right)^{\frac{d-2}{2}}}\,.
\end{equation}
When $d$ is an even integer we can perform the sum over thermal images in closed form.
E.g. for $d=4$ we find
\begin{equation}
\label{eq:FJ_mon_xi}
F^{q=2, \xi}_J(z)=-\alpha\,\xi\int_0^\infty\d t\,\frac{  \cosh (\alpha  t) \coth \left(2 \pi  z \cosh \left(\frac{t}{2}\right)\right)\sin (\pi  \alpha )}{2 \pi ^2 \cosh\left(\frac{t}{2}\right){z^2}}\,.
\end{equation}
Note that this integral is rapidly converging at large $t$ provided that $\alpha<1/2$.
For $\alpha\geq 1/2$ the integral diverges.
This is the same IR divergence encountered for $\langle|\Phi|^2\rangle_\beta^\xi$ discussed around eqs.~\eqref{eq:mon_conv1} and \eqref{eq:mon_conv2}.
Assuming $\alpha<1/2$, it is then straightforward to check the asymptotic properties of \eq{FJ_mon_xi} using the same approximations used below \eq{FJ_mon}.
In the $z\ll1$ limit it reduces to the zero-temperature result \eq{Jth_0T_xi}.
Meanwhile for $z\gg1$, $F^{q=2, \xi}_J(z)\sim z^{{-2}}$ such that $\langle J_\theta(x)\rangle^\xi_{\beta}\sim 1/\beta\rho$.
This vanishes in the large distance $\rho\to\infty$ limit.

In figure~\ref{fig:FJ_mon_4d} on the right we plot $\tilde{F}^{q=2}_{J}(z)=z^3(F^{q=2,0}_{J}(z)+F^{q=2,\xi}_{J}(z))$ at fixed $\alpha<1/2$ when $d=4$ as we vary $\xi$.
Here we added a superscript $0$ to the quantity defined in \eq{FJ_mon} to emphasise that $F^{q=2,0}(z)$ is the thermal one-point function with $\xi=\tilde{\xi}=0$.
The re-scaling by $z^{d-1}$ ensures that the vertical axis intercept is given by the zero-temperature one-point function, which is $(-\rho^2)$ times the sum of eqs.~\eqref{eq:Jth_0T} and~\eqref{eq:Jth_0T_xi} for $d=4$.
Turning on the singular modes changes the near defect behaviour by decreasing the value of $\tilde{F}^{q=2}_{J}(z)$.
Note that the term proportional to $\xi$ is linear in $z$ for $z\gg1$, just like the $\xi=\tilde{\xi}=0$ piece in \eq{FJ_mon}.
The full thermal one-point function of $J_\theta$, however, decays to zero as $1/\rho$ for any $\xi$ as $\rho\to\infty$.

The discussion for the contribution proportional to $\tilde{\xi}$ proceeds analogously, with the convergence condition now being $\alpha>1/2$.
It is obtained from \eq{FJ_mon_xi} by replacing $\alpha\to1-\alpha$ and multiplying by $(-1)$.
The extra sign here comes from the relative sign in the exponentials in the second and third lines of \eq{G_exp} after the replacement $\alpha\to1-\alpha$.

\subsection{Thermal one-point function of \texorpdfstring{$T_{\mu\nu}$}{T μv}}
\label{sec:mon_T}

Finally we consider the thermal one-point function of the stress tensor in the presence of the monodromy defect.
The symmetric traceless stress tensor for massless complex scalar field is
\begin{equation}
T_{\mu\nu}=\p_\mu\Phi\p_\nu\Phi^\dagger+\p_\mu\Phi^\dagger\p_\nu\Phi-\frac{d-2}{2(d-1)}\left[\nabla_\mu\nabla_\nu+\frac{g_{\mu\nu}}{d-2}\nabla^2\right]|\Phi|^2\,,
\end{equation}
where $\nabla$ is the Levi-Civita connection on $S^1\times\mathbb{R}^{d-1}$ using polar coordinates on transverse space.
Its one-point function can be computed using Wick's theorem and the scalar propagator \eqref{eq:prop_0T} in the coincident limit. 
In the absence of singular modes, ref.~\cite{Bianchi:2021snj} found that at zero temperature it takes the form of \eq{T_def_1-pt_fn} with $\Delta=d$, $q=2$ and
\begin{equation}
\label{eq:T_mon_0T}
a_T=\frac{\alpha(1-\alpha)\Gamma\left(\frac{d}{2}-\alpha\right)\Gamma\left(\frac{d}{2}+\alpha-1\right)\sin(\pi \alpha)}{2^{d-1}\pi^{\frac{d+1}{2}}\,d\,\Gamma\left(\frac{d+1}{2}\right)}\,.
\end{equation}
This is the finite piece after subtracting a short-distance divergence at $s=0$ in the integral.

At finite temperature we replace $\tau'-\tau\to\tau'-\tau+m\beta$ in \eq{prop_0T} and sum over integer $m$.
Unlike the zero-temperature case, the thermal one-point function is no longer determined by a single coefficient but rather by the \emph{a priori} five independent functions defined in \eq{T_def}.
Since the monodromy defect is manifestly invariant under $\tau\to-\tau$, only the three structures in the first line can contribute.
Using Wick's theorem, it is easy to see that only $\langle T_{\tau\tau}\rangle_\beta$, $\langle T_{xx}\rangle_\beta$, $\langle T_{\rho\rho}\rangle_\beta$, and $\langle T_{\theta\theta}\rangle_\beta$ can be non-zero.
Here, $T_{xx}$ is the component of the stress tensor along any of the non-compact directions parallel to the defect.
By rotational invariance these are all equal.
The thermal one-point function must obey the traceless condition $g^{\mu\nu}\langle T_{\mu\nu}\rangle_\beta=0$ as well as the conservation equation $\nabla^\mu\langle T_{\mu\nu}\rangle_\beta=0$ which takes the form of \eq{ODE}.

In the following we will only present the computation of the $\tau\tau$-component. 
The other components are discussed in appendix~\ref{app:mon}.
We begin with the parameters $\xi=\tilde{\xi}=0$ such that the field $\Phi$ only has regular modes.
Using the one-point function of $|\Phi|^2$ in \eq{Phi2_int_s}, we can write
\begin{equation}
\label{eq:Ttt_step}
\begin{split}
\langle T_{\tau\tau}\rangle_\beta=&\;\sum_{m=-\infty}^\infty\frac{1}{(4\pi)^{d/2}}\int_0^\infty\d s\, e^{-\frac{2\rho^2+(m\beta)^2}{4s}}s^{-\frac{d+2}{2}}\left(1-\frac{(m\beta)^2}{2s}\right)\CI_\alpha\left(\frac{\rho^2}{2s}\right)\\
&-\frac{1}{2(d-1)}\left(\p_\rho^2+\frac{1}{\rho}\p_\rho\right)\langle|\Phi|^2\rangle_\beta\,,
\end{split}
\end{equation}
where $\CI_\alpha(\zeta)=\sum_n I_{|n-\alpha|}(\zeta)$ and $z=\rho/\beta$.
As before, the $s$-integral of the $m=0$ term is divergent and requires regularisation.
Since the divergence is independent of $\alpha$, it can be subtracted unambiguously to give the zero-temperature result in \eq{T_mon_0T}.
The integrals for $m\neq0$ are convergent. 
Using the integral representation~\eqref{eq:sumI} of the sum over Bessel functions we find for the first integral in \eq{Ttt_step}
\begin{equation}
\!\!\!\!\frac{\Gamma\!\left(\frac{d}{2}\right)}{\pi^{d/2}}\!\sum_{m=-\infty}^\infty\!\left[-\frac{d-1}{|m\beta|^d}+\frac{\sin(\pi\alpha)}{\pi}\int_0^\infty\!\d t\,
\frac{\left[(d-1)(m\beta)^2 - 4\rho^2\cosh^2\!\frac{t}{2}\right]\cosh\left(\frac{t}{2}-t\alpha\right)}
     {\cosh\frac{t}{2}\,\left(m^2\beta^2+4\rho^2\cosh^2\frac{t}{2}\right)^{1+d/2}}\right]\!,
\end{equation}
The sum over $m$ can be performed in closed form when $d$ is an even integer. 
E.g.\ when $d=4$,
\begin{equation}
\label{eq:Ttt_mon_4d}
\begin{split}
\langle T_{\tau\tau}\rangle_\beta=-&\frac{\pi^2}{15\beta^4}-\frac{\sin (\pi  \alpha )}{2\beta^3\rho}\int_0^\infty\d t\,\frac{  \cosh \left(\left(\alpha -\frac{1}{2}\right) t\right) \coth \left(2 \pi  z  \cosh \left(\frac{t}{2}\right)\right) }{ \cosh^2\left(\frac{t}{2}\right)\sinh^2\left(2 \pi  z \cosh \left(\frac{t}{2}\right)\right)}\\
-&\frac{1}{6}\left(\p_\rho^2+\frac{1}{\rho}\p_\rho\right)\langle|\Phi|^2\rangle_\beta\,,
\end{split}
\end{equation}
where the one-point function of $|\Phi|^2$ was computed in \eq{Phi2_4d}.
It is understood that $\rho$-derivatives are taken before integrating over $t$.

Since we cannot perform the remaining integrals in closed form let us briefly discuss the thermal one-point function's asymptotic properties.
For $z\ll1$, we can approximate $\coth x\approx 1/x$ and $\sinh x\approx x$ in the integral in \eq{Ttt_mon_4d} and in \eq{Phi2_4d} after acting with $\p_\rho^2+\rho^{-1}\p_\rho$.
The resulting expression can be integrated exactly. 
We find $-a_T/\rho^4$ with $a_T$ given by \eq{T_mon_0T}.
This is precisely the zero-temperature one-point function of $T_{\mu\nu}$.
For $z\gg1$, it is straightforward to see that the explicit $t$-integral in \eq{Ttt_mon_4d} is exponentially suppressed, whereas the term involving $|\Phi|^2$ gives the power-law correction $\sim1/(\beta\rho^3)$.
In the $\rho\to\infty$ limit the leading contribution is given by the constant first term of \eq{Ttt_mon_4d}, which is just the no-defect thermal one-point function~\eqref{eq:no_defect_1-pt} with 
\begin{equation}
\label{eq:bT_mon_4d}
b_T=-\frac{4\pi^2}{45}\,.
\end{equation}

If singular modes are included, their contributions can be computed straightforwardly.
The contribution proportional to $\xi$ is
\begin{align}
\label{eq:Ttt_step_xi}
\langle T_{\tau\tau}\rangle^\xi_\beta=&\;\xi\sum_{m=-\infty}^\infty\frac{1}{(4\pi)^{d/2}}\int_0^\infty\d s\, e^{-\frac{2\rho^2+(m\beta)^2}{4s}}s^{-\frac{d+2}{2}}\left(1-\tfrac{(m\beta)^2}{2s}\right)\left(I_{-\alpha}\left(\tfrac{\rho^2}{2s}\right)-I_{\alpha}\left(\tfrac{\rho^2}{2s}\right)\right)\nonumber\\
&\quad-\frac{1}{2(d-1)}\left(\p_\rho^2+\frac{1}{\rho}\p_\rho\right)\langle|\Phi|^2\rangle^\xi_\beta\,,
\end{align}
where $\langle|\Phi|^2\rangle^\xi_\beta$ is given by \eq{Phi2_int_s_xi}.
For the $m=0$ term, ref.~\cite{Bianchi:2021snj} found the following finite piece
\begin{equation}
\label{eq:T_mon_0T_xi}
a_T^\xi=\xi\,\frac{\alpha^2\Gamma\left(\frac{d}{2}-\alpha-1\right)\Gamma\left(\frac{d}{2}+\alpha-1\right)\sin(\pi \alpha)}{2^{d-1}\pi^{\frac{d+1}{2}}\Gamma\left(\frac{d+1}{2}\right)}\,,
\end{equation}
where $a_T^\xi$ is the part of $a_T$ proportional to $\xi$.
For the $m\neq0$ terms, the $s$-integrals are convergent: using the integral representation of the modified Bessel function in \eq{BesselI_int} and performing the $s$-integral, one finds for the first line in \eq{Ttt_step_xi}
\begin{equation}
-4\xi\sum_{m=-\infty}^\infty \frac{\Gamma\left(1+\frac{d}{2}\right)\sin(\pi\alpha)}{d\,\pi^{1+\frac{d}{2}}}\int_0^\infty\d t \,\frac{\cosh(t\alpha)
\left[(d-1)m^{2}\beta^{2}-4\rho^{2}\cosh^{2}\left(\frac{t}{2}\right)\right]}
{\left[m^{2}\beta^{2}+4\rho^{2}\cosh^{2}\left(\frac{t}{2}\right)\right]^{1+\frac{d}{2}}}\,.
\end{equation}
When $d$ is an even integer, we can perform the sum over thermal images.
E.g.\ when $d=4$ we find
\begin{equation}
\label{eq:Ttt_mon_4d_xi}
\begin{split}
\langle T_{\tau\tau}(x)\rangle^\xi_\beta &= \xi\,\frac{z^3 \sin (\pi  \alpha ) }{\rho ^4}\int_0^\infty\d t\,\frac{ \cosh (\alpha  t) \coth \left(2 \pi  z \cosh \left(\frac{t}{2}\right)\right) }{\cosh\left(\frac{t}{2}\right)\sinh^2\left(2 \pi  z \cosh \left(\frac{t}{2}\right)\right)} \\
&\quad-\frac{1}{6}\left(\p_\rho^2+\frac{1}{\rho}\p_\rho\right)\langle|\Phi|^2\rangle^\xi_\beta\,,
\end{split}
\end{equation}
where $\langle|\Phi|^2\rangle^\xi_\beta$ was computed in \eq{Phi2_4d_xi}.
Similarly, the contributions proportional to $\tilde{\xi}$ can be obtained from \eq{Ttt_mon_4d_xi} by replacing $\xi\to\tilde{\xi}$ and $\alpha\to1-\alpha$.

Note that the $t$-integral in \eq{Ttt_mon_4d_xi} is rapidly convergent.
The $1/\sinh^2$ factor becomes doubly exponentially suppressed at large $t$ for any $\alpha$.
The part dependent on the one-point function of $|\Phi|^2$, however, diverges when $\alpha\geq1/2$.
This is the IR divergence discussed around eqs.~\eqref{eq:mon_conv1} and \eqref{eq:mon_conv2}.
Assuming $\alpha<1/2$, we can then check the asymptotic properties of \eq{Ttt_mon_4d_xi}. 
For $z\ll1$, we find that $\langle T_{\tau\tau}(x)\rangle^\xi_\beta$ reduces to $-a_T^\xi/\rho^4$ with $a_T^\xi$ given by \eq{T_mon_0T_xi}.
For $z\gg1$, $\langle T_{\tau\tau}\rangle_\beta^\xi\sim1/\beta\rho^3$, which vanishes in the $\rho\to \infty$ limit. 
Therefore the large-distance limit of the full stress tensor one-point function is controlled by the no-defect thermal part given by \eq{bT_mon_4d}.

The other components of $T_{\mu\nu}$ can be computed analogously.
In appendix~\ref{app:mon} we sketch the computations and give rapidly converging integral representations when $d=4$ for all of the non-zero components.
Summing all equations in~\eqref{eq:T_mon_4d} (with the correct $d$-dependent multiplicity for $T_{xx}$) it is straightforward to check numerically that the thermal one-point function of $T_{\mu\nu}$ is traceless.
This reduces the number of independent functions to three.
To check conservation, we can pick one way to package up $\langle T_{\tau\tau}\rangle_\beta$, $\langle T_{xx}\rangle_\beta$, $\langle T_{\rho\rho}\rangle_\beta$, and $\langle T_{\theta\theta}\rangle_\beta$ into the three functions in \eq{T_def},
\begin{subequations}
\label{eq:F_structs_for_T_mon}
\begin{align}
F^{\tau\tau}_{T} &= \beta^d\left(T_{\tau\tau}-T_{xx}\right), \\
F^{\perp\perp}_{T,1} &=-{\beta^d}\left(T_{xx}-\frac{1}{\rho^2}T_{\theta\theta}\right), \\
F^{\perp\perp}_{T,2} &={\beta^d}\left( T_{\rho\rho}-\frac{1}{\rho^2}T_{\theta\theta}\right).
\end{align}
\end{subequations}
Using eqs.~\eqref{eq:T_mon_4d} and \eqref{eq:T_mon_4d_xi} for $d=4$ one can check numerically that the conservation equation~\eqref{eq:ODE} is satisfied.

\paragraph{Defect thermodynamics. }
We now turn our attention to the computation of some thermodynamic quantities in the presence of a monodromy defect.
Given the one-point function of $\langle T_{\tau\tau}\rangle_\beta$, one can construct the energy density $E$, free energy density $F$ and entropy density $S$ defined in eqs.~\eqref{eq:E}, \eqref{eq:F_E-TS} and \eqref{eq:S}, respectively.
Here we will study their dimensionless versions $\CE(z)$, $f(z)$ and $s(z)$ which are defined by dividing by an appropriate power of $\beta$.

\begin{figure}[h]
\centering
\includegraphics[scale=0.5]{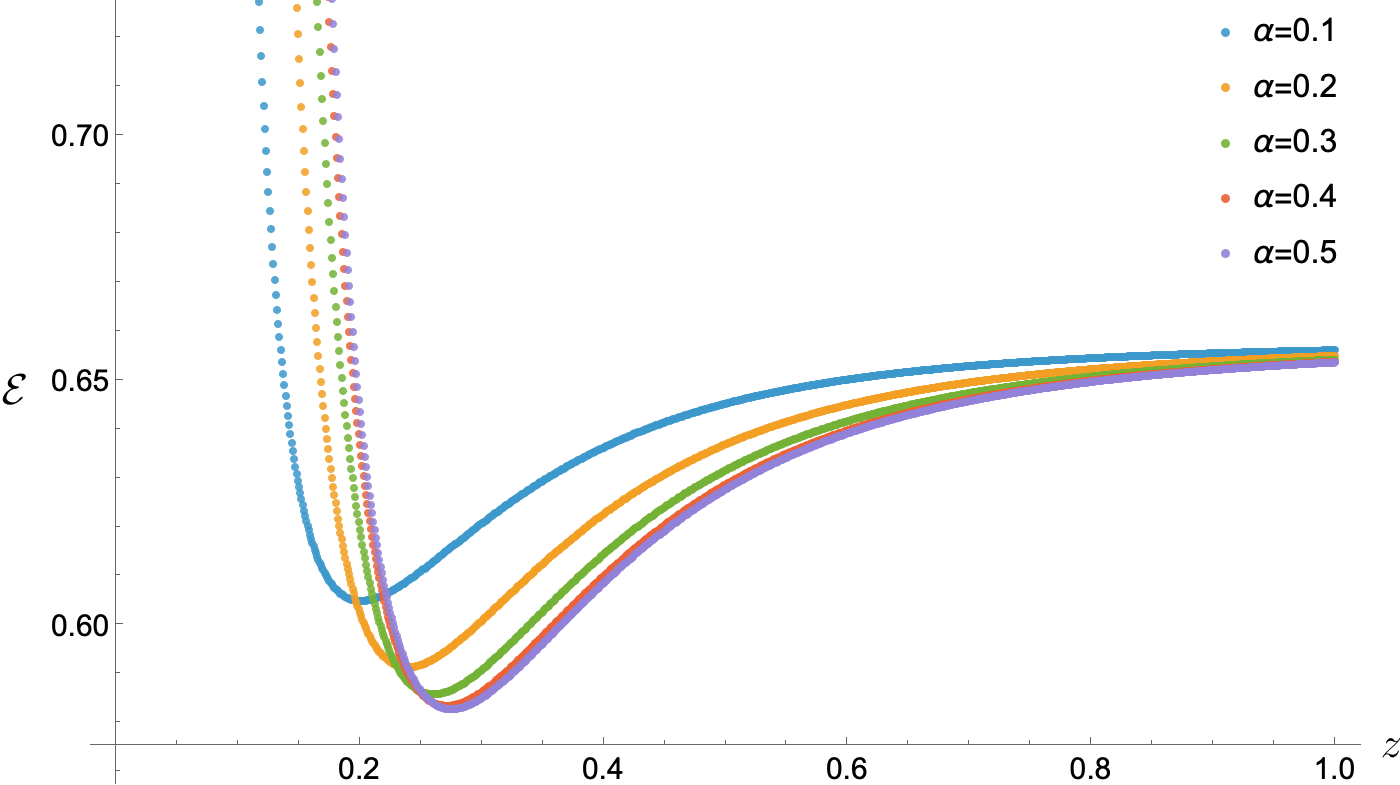}
\caption{
Plot of the dimensionless energy density $\CE(z)$ when $d=4$ and $\xi=\tilde{\xi}=0$ for various values of $\alpha$.
Since $\CE(z)$ is invariant under $\alpha\to1-\alpha$ we restrict to values $0<\alpha\leq1/2$.
}
\label{fig:E_mon_4d}
\end{figure}

We begin with the dimensionless energy density $\CE(z)=-\beta^d \langle T_{\tau\tau}\rangle_\beta$.
When $d$ is an even integer we can express it as a rapidly convergent integral that can be evaluated numerically. 
For $d=4$, and in the absence of singular modes, this integral representation can be obtained from \eq{Ttt_mon_4d} together with \eq{Phi2_4d}, such that
\begin{equation}
\label{eq:E_mon_4d}
\begin{split}
\CE(z)&=\frac{\pi^2}{15}-\frac{\sin(\pi\alpha)}{48 \pi ^2 z^3}\int_0^\infty\d t\,\frac{\cosh \left(\frac{1}{2} (t-2 \alpha  t)\right) }{\cosh\left(\frac{t}{2}\right) }\left[\frac{\coth \left(2 \pi  z \cosh \left(\frac{t}{2}\right)\right)}{\cosh\left(\frac{t}{2}\right) }\right.\\
&\left.\quad+2 \pi  z \csch^2\left(2 \pi  z \cosh \left(\frac{t}{2}\right)\!\right) \left(2 \pi  z \frac{(\cosh (t)-5)  \coth \left(2 \pi  z \cosh \left(\frac{t}{2}\right)\!\right)}{\cosh\left(\frac{t}{2}\right)}+1\right)\right]\!.
\end{split}
\end{equation}
We plot this function in figure~\ref{fig:E_mon_4d} for various values of $\alpha$.
We find that, as expected from the zero temperature limit, the dimensionless energy density diverges as $\CO(z^{-4})$ to positive infinity at $z=0$.
The coefficient is given by the zero-temperature one-point function of $T_{\mu\nu}$ defined in \eq{T_mon_0T}.
For $z\gg1$, the function asymptotes to a constant value.
This constant is $-3b_T/4$, where $b_T$ is the no-defect thermal one-point function of $T_{\mu\nu}$ given by \eq{bT_mon_4d} when $d=4$.
In-between these two regimes $\CE(z)$ has a minimum.
The existence of a minimum can be argued analytically from \eq{E_mon_4d}.
At large $z$, we can approximate $\coth x\approx 1$ and $\csch x\approx 0$ with $x=2\pi z\cosh(t/2)$.
The resulting $t$-integral can be performed in closed form. 
The leading correction to the constant is
\begin{equation}
\label{eq:E_largez}
-\frac{(1-2 \alpha ) \tan (\pi  \alpha )}{48 \pi  z^3}\,,
\end{equation}
which is negative for all values of $\alpha\in[0,1)$.
Since the divergence at $z=0$ is to $+\infty$, as dictated by the sign of $a_T$, there must exist a minimum for finite $z$.
As $\alpha\to0$, the minimum gets pushed towards $z=0$ while the difference in magnitude between the minimum and the large-$z$ asymptote goes to zero.
Physically, this minimum is a result of competing defect and thermal effects, although at present we lack an intuitive understanding.

\begin{figure}[ht]
\centering
  \begin{minipage}{0.49\textwidth}
  \centering
    \includegraphics[width=\linewidth]{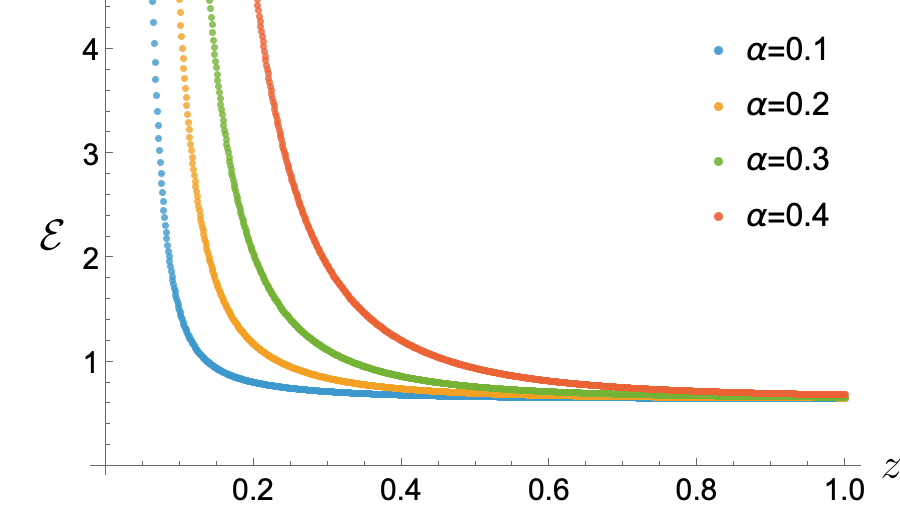}
  \end{minipage}\hfill
  \begin{minipage}{0.49\textwidth}
  \centering
    \includegraphics[width=\linewidth]{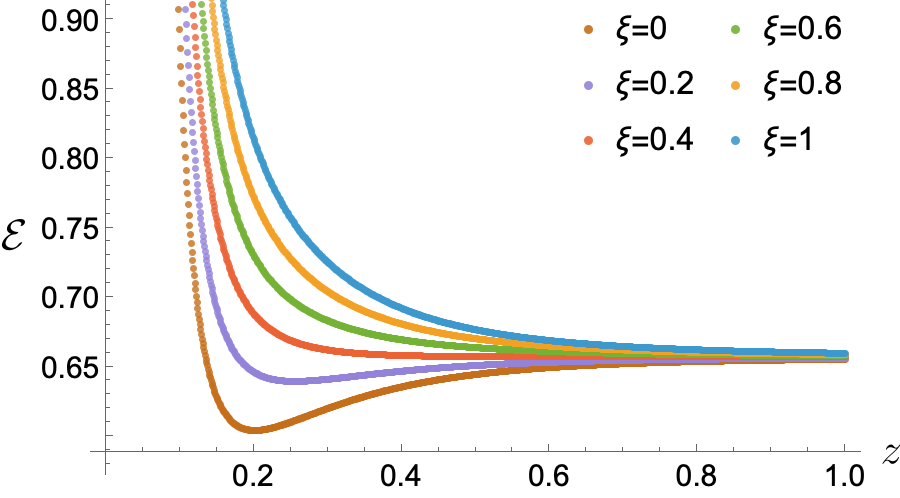}
  \end{minipage}
\caption{\emph{Left.} Plot of the dimensionless energy density $\CE(z)$ when $d=4$ and $\tilde{\xi}=0$ with fixed $\xi=1$ for various values of $\alpha$.\\
\emph{Right.} Plot of the dimensionless energy density $\CE(z)$ when $d=4$ and $\tilde{\xi}=0$ with fixed $\alpha=0.1$ for various values of $\xi$.
The blue curves in the left and right panels represent the same function.
}
\label{fig:E_mon_4d_xi}
\end{figure}

When singular modes are included, we can use \eq{Ttt_mon_4d_xi} together with \eq{Phi2_4d_xi} to deduce their effects.
In particular for the contribution to $\CE(z)$ proportional to $\xi$ we find
\begin{equation}
\label{eq:E_mon_4d_xi}
\begin{split}
\CE^\xi(z)&=\xi\,\frac{\sin(\pi\alpha)}{48\pi^{2}z^{3}}\int_0^\infty\d t\,\frac{\cosh(t\alpha)}
{\cosh\left(\frac{t}{2}\right)}\\
&\quad\left[\frac{4\pi z\cosh\left(\frac{t}{2}\right)
+\left(16\pi^{2}z^{2}\cosh^{2}\left(\frac{t}{2}\right)-48\pi^{2}z^{2}\right)
\coth\left(2\pi z\cosh\left(\frac{t}{2}\right)\right)}
{\sinh^{2}\left(2\pi z\cosh\left(\frac{t}{2}\right)\right)}\right.\\
&\quad+\left.2\coth\left(2\pi z\cosh\left(\frac{t}{2}\right)\right)\right].
\end{split}
\end{equation}
The contribution proportional to $\tilde{\xi}$ can be obtained by replacing $\xi\to\tilde{\xi}$ and $\alpha\to1-\alpha$.
As discussed below \eq{Ttt_mon_4d_xi}, the $t$-integral in \eq{E_mon_4d_xi} converges provided that $\alpha<1/2$. 
We will restrict to these values in the following discussion.
In figure~\ref{fig:E_mon_4d_xi}, we plot the effect on $\CE(z)$ of turning on $\xi>0$.
On the left we plot $\CE(z)$ at fixed $\xi=1$ for various values of $\alpha$.
The divergence is still $\CO(z^{-4})$ and fixed by the zero-temperature one-point function $a_T$, which is the sum of eqs.~\eqref{eq:T_mon_0T} and \eqref{eq:T_mon_0T_xi}.
At large distances $\CE(z)$ asymptotes to $-3b_T/4$ for any $\xi$.
Note that for $\xi=1$, $\CE(z)$ is a monotonically decreasing function in $z$.
On the right we plot $\CE(z)$ at fixed $\alpha$ for various values of $\xi$ between 0 and 1.
Increasing $\xi$ we see that the location of the minimum in $\CE(z)$ gets pushed towards $z\to\infty$ and disappears completely at some value $0<\xi_*<1$.
We can estimate this critical value by looking at the large-$z$ behaviour of \eq{E_mon_4d_xi}.
Approximating $\coth x\approx 1$ and $\csch x\approx 0$ with $x=2\pi z\cosh(t/2)$, we can perform the $t$-integral in closed form.
We find
\begin{equation}
\xi\,\frac{  \tan (\pi  \alpha )}{24 \pi  z^3}\,,
\end{equation}
such that the sub-leading large-$z$ behaviour is given by the sum of this expression and \eq{E_largez}. 
We thus see that for $\xi=\xi_*=(1-2\alpha)/2$ the sub-leading correction changes sign.
Numerically we observe that the minimum disappears approximately when we increase $\xi$ past $\xi_*$.

From \eq{E_mon_4d} for $\CE(z)$ we can compute the dimensionless entropy density $s(z)$ and dimensionless free energy density $f(z)$ via eqs.~\eqref{eq:s_int} and~\eqref{eq:f_int}, respectively.
Unlike $\CE(z)$, they depend on the integration constant $c$ that arises from solving eqs.~\eqref{eq:F_E-TS} and~\eqref{eq:S}.
It is fixed by demanding that the entropy density $S=\beta^{1-d}s(z)\to0$ in the zero-temperature limit $\beta\to\infty$.
In that limit the system should settle into its ground state which is unique. 
Since large $\beta$ corresponds to small $z$, we must ensure that $s(z)$ diverges with a power of $z$ strictly greater than $(1-d)$.
By considering \eq{s_int}, setting $\lim_{z\to0}z^{d-1}s(z)=0$ to zero and solving for $c$ we find after integrating by parts
\begin{equation}
\label{eq:c_mon}
c=\CE(1)-\int_0^1\d y\,y^{d-2}\left(y\CE'(y)+d\,\CE(y)\right).
\end{equation}
Note that the dimensionless energy density $\CE(z)$ diverges $\sim z^{-d}$ at $z=0$ since $T_{\mu\nu}$ has a non-trivial one-point function at zero temperature.
However, the combination of $\CE(z)$ with its first derivative in \eq{c_mon} removes the leading divergence. 
The expression for $c$ is sensitive to any sub-leading divergences at $z=0$.
In particular, the $y$-integral exists provided that the sub-leading divergence in $\CE(z)$ is $\CO(z^{>-d+1})$.

We will now estimate this sub-leading divergence in $\CE(z)$ for the monodromy defect when $d=4$. 
Since for small $z$ the one-point function is dominated by the DOE, the sub-leading divergence must correspond to the lightest non-trivial exchanged operator.
Such an operator must be a composite operator quadratic in $\hat{\CO}^\pm_s$.
If $\xi=\tilde{\xi}=0$, the lightest such operator (for $\alpha \leq 1/2$) is $:\hat{\CO}^+_{-\alpha}\hat{\CO}^{\dagger +}_{\alpha}:$ with scaling dimension $2\hat{\Delta}^+_{-\alpha}=2+2\alpha$.
If e.g.\ $\xi\neq0$ and $\tilde{\xi}=0$, the lightest operator is $:\hat{\CO}^-_{-\alpha}\hat{\CO}^{\dagger -}_{\alpha}:$ with scaling dimension $2\hat{\Delta}^-_{-\alpha}=2-2\alpha$.
A defect primary of dimension $\hat\Delta$ contributes to the dimensionless stress-tensor one-point function as $z^{\hat\Delta-d}$. 
Therefore, the sub-leading divergence in the absence of singular modes is $\CO(z^{-2+2\alpha})$ whereas it is $\CO(z^{-2-2\alpha})$ when $\xi\neq0$.
For the case of regular modes, the exponent lies between $-2$ and $-1$ for any $\alpha\in[0,1)$.
When $\xi\neq0$, the exponent is strictly greater than $-d+1=-3$ when $\alpha<1/2$, approaching $-3$ as $\alpha\to1/2$ from below.
Numerical analysis of eqs.~\eqref{eq:E_mon_4d} and~\eqref{eq:E_mon_4d_xi} confirms this behaviour.
An identical discussion applies when $\tilde{\xi}\neq0$ and $\xi=0$ instead, with $\alpha\to1-\alpha$.
We thus conclude that the integration constant $c$ in \eq{c_mon} is well-defined whenever there are no IR divergences.
We will compute it numerically for each $\alpha$.

\begin{figure}[h]
\centering
\includegraphics[scale=0.5]{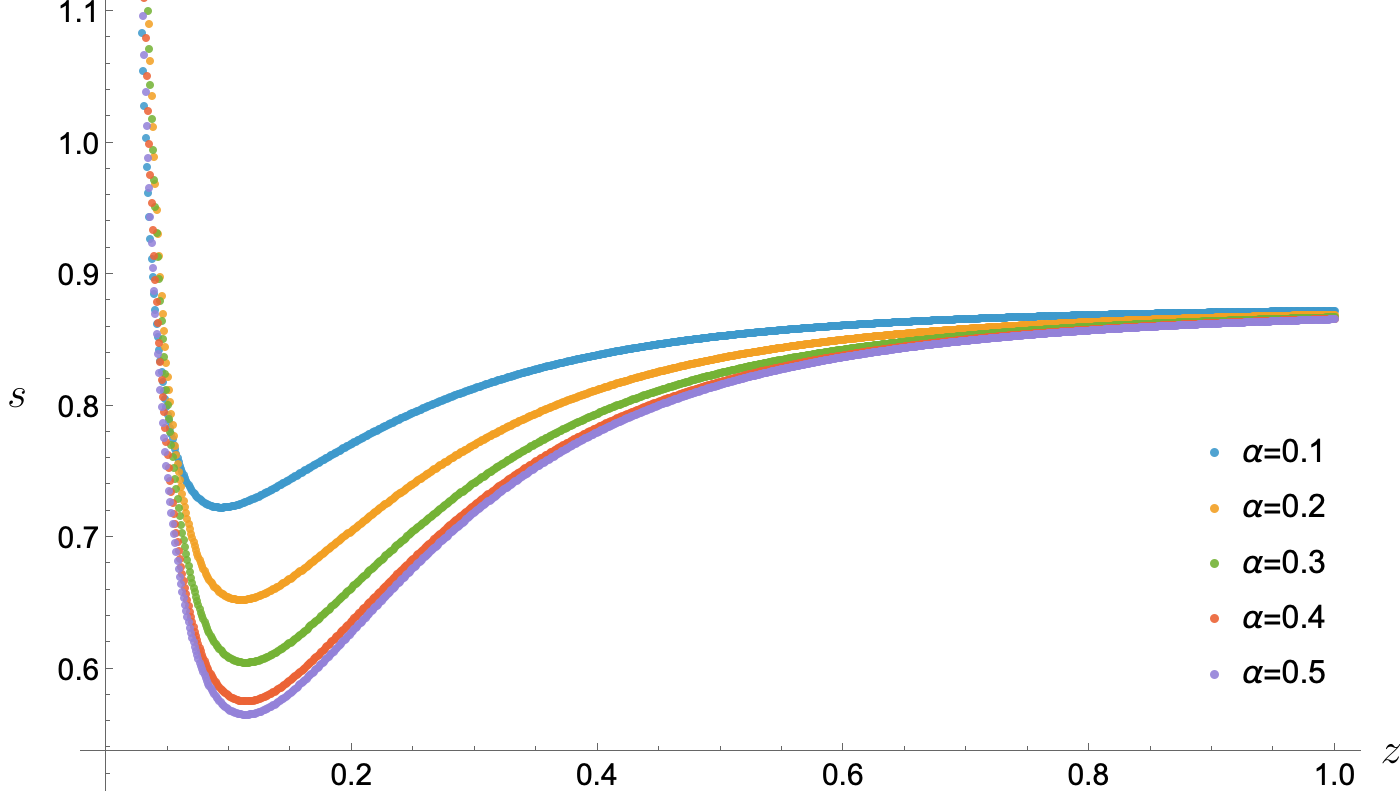}
\caption{
Plot of the dimensionless entropy density $s(z)$ when $d=4$ and $\xi=\tilde{\xi}=0$ for various values of $\alpha$.
}
\label{fig:s_mon_4d}
\end{figure}

Now that we have fixed $c$ we can compute the dimensionless entropy density $s(z)$ and dimensionless free energy density $f(z)$.
Let us start with $s(z)$ defined in \eq{s_int}.
Computing $s(z)$ involves \emph{two} integrations, one over a finite interval and the other over the semi-infinite line, which we will both perform numerically.
For $d=4$ we plot $s(z)$ for $\xi=\tilde{\xi}=0$ and a range of $\alpha$ in figure~\ref{fig:s_mon_4d}.
Near $z=0$ the function diverges.
By construction, the singularity is $\CO(z^{>-3})$ and is $\alpha$-dependent.
At large distances $s(z) \sim  -b_T$.
Similarly to $\CE(z)$, $s(z)$ also has a minimum.

\begin{figure}[ht]
\centering
  \begin{minipage}{0.49\textwidth}
  \centering
    \includegraphics[width=\linewidth]{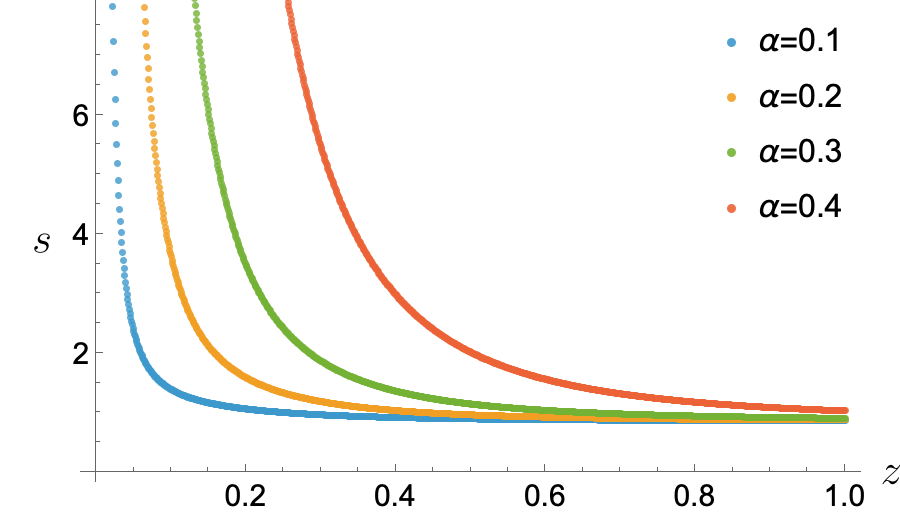}
  \end{minipage}\hfill
  \begin{minipage}{0.49\textwidth}
  \centering
    \includegraphics[width=\linewidth]{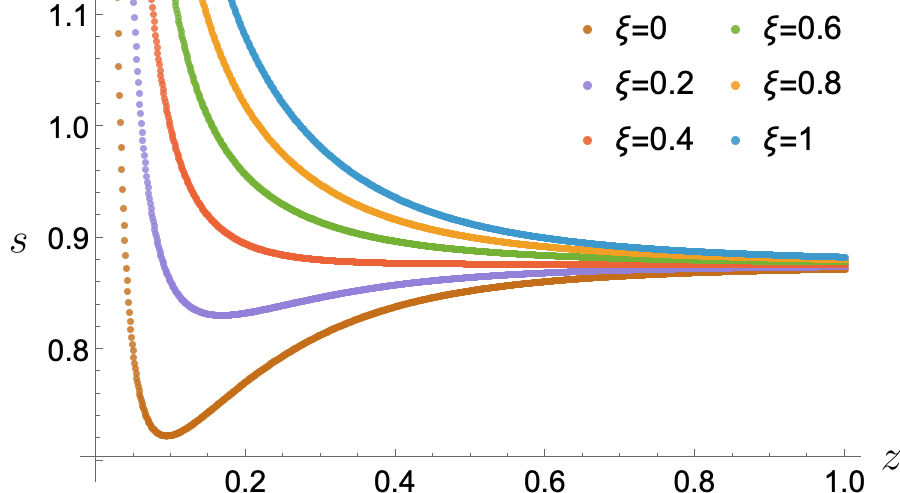}
  \end{minipage}
\caption{\emph{Left.} Plot of the dimensionless entropy density $s(z)$ when $d=4$ and $\tilde{\xi}=0$ with fixed $\xi=1$ for various values of $\alpha$.\\
\emph{Right.} Plot of the dimensionless entropy density $s(z)$ when $d=4$ and $\tilde{\xi}=0$ with fixed $\alpha=0.1$ for various values of $\xi$.
The blue curves in the left and right panels represent the same function.
}
\label{fig:s_mon_4d_xi}
\end{figure}

Turning on singular modes for $\Phi$, we observe a similar behaviour for $s(z)$ to that of $\CE(z)$.
In figure~\ref{fig:s_mon_4d_xi} on the left we plot $s(z)$ with $\xi=1$ and $\tilde{\xi}=0$ for various values of $0<\alpha<1/2$.
The singular mode increases the order of the divergence near the defect, however, it remains milder than $\CO(z^{-3})$ for all $\alpha$.
Meanwhile the large distance behaviour is unaffected.
Note that the minimum present in figure~\ref{fig:s_mon_4d} has disappeared.
In figure~\ref{fig:s_mon_4d_xi} on the right we plot $s(z)$ with $\tilde{\xi}=0$ and fixed $\alpha$ for various values of $0<\xi<1$.
Similarly to $\CE(z)$ we also observe that as we dial up $\xi$ there is a critical value for which the minimum disappears.
We have not found a good way of estimating this value analytically but numerically we observe that it differs from the critical value $\xi_*$ found for $\CE(z)$.

\begin{figure}[h]
\centering
\includegraphics[scale=0.5]{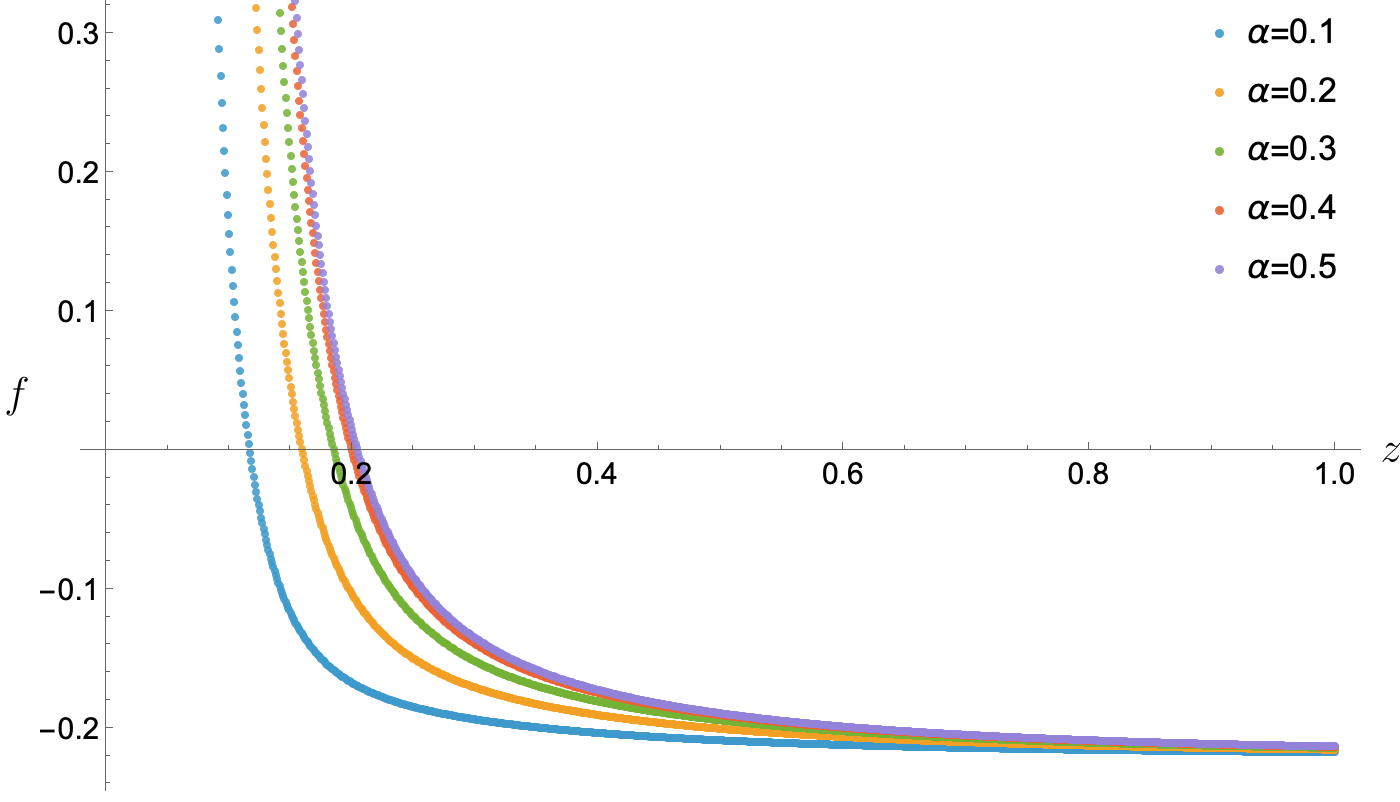}
\caption{
Plot of the dimensionless free energy density $f(z)$ when $d=4$ and $\xi=\tilde{\xi}=0$ for various values of $\alpha$.
}
\label{fig:f_mon_4d}
\end{figure}

The dimensionless free energy $f(z)$ is plotted in figure~\ref{fig:f_mon_4d}.
Near $z=0$ the function diverges as $\CO(z^{-4})$, just like $\CE(z)$.
At large distances it asymptotes to $f(z) \sim  b_T/4$.
Unlike $\CE(z)$ and $s(z)$, $f(z)$ is monotonically decreasing.

\begin{figure}[ht]
\centering
  \begin{minipage}{0.49\textwidth}
  \centering
    \includegraphics[width=\linewidth]{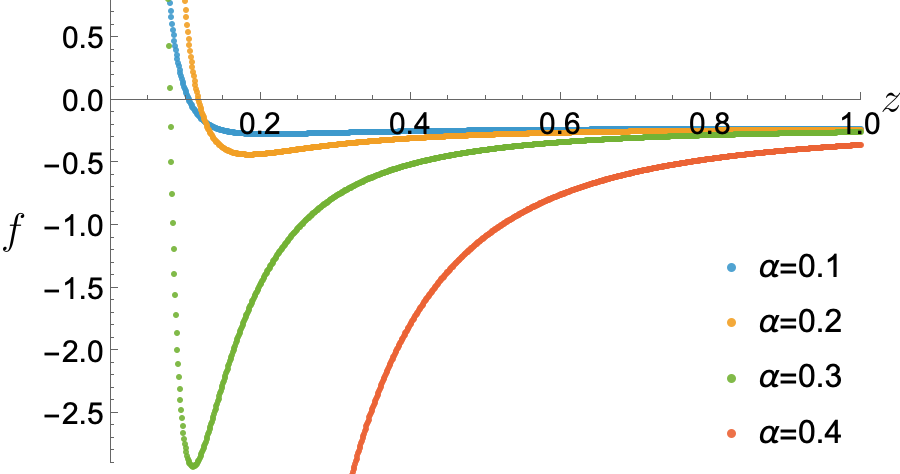}
  \end{minipage}\hfill
  \begin{minipage}{0.49\textwidth}
  \centering
    \includegraphics[width=\linewidth]{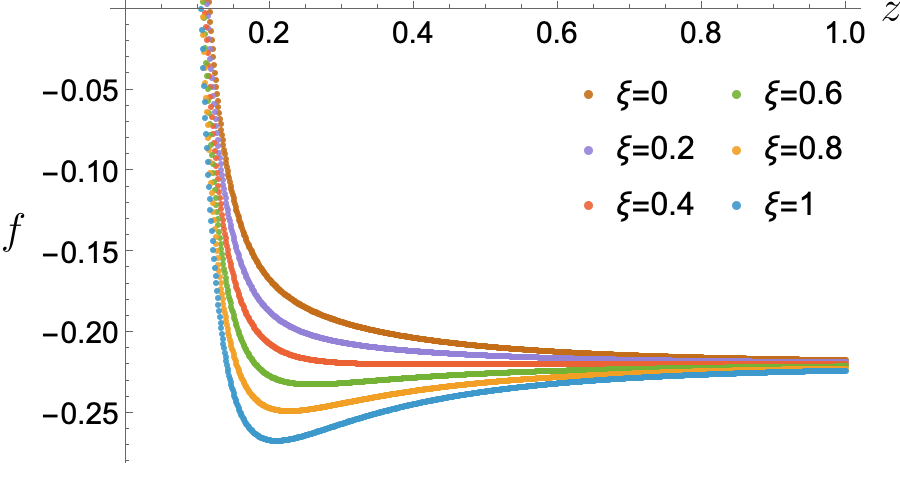}
  \end{minipage}
\caption{\emph{Left.} Plot of the dimensionless free energy density $f(z)$ when $d=4$ and $\tilde{\xi}=0$ with fixed $\xi=1$ for various values of $\alpha$.\\
\emph{Right.} Plot of the dimensionless free energy density $f(z)$ when $d=4$ and $\tilde{\xi}=0$ with fixed $\alpha=0.1$ for various values of $\xi$.
The blue curves in the left and right panels represent the same function.
}
\label{fig:f_mon_4d_xi}
\end{figure}

Including singular modes we observe the opposite behaviour compared to $\CE(z)$ and $s(z)$.
In figure~\ref{fig:f_mon_4d_xi} on the left we plot $f(z)$ with $\xi=1$ and $\tilde{\xi}=0$ for various values of $\alpha$.
The short and large distance power-law behaviour is unchanged compared to the $\xi=\tilde{\xi}=0$ case.
However, the behaviour in-between these two regimes changes.
$f(z)$ now has a minimum in $z$ whose numerical value becomes larger in magnitude and negative as $\alpha\to1/2$.
In figure~\ref{fig:f_mon_4d_xi} on the right we plot $f(z)$ with $\tilde{\xi}=0$ and fixed $\alpha=0.1$ as we vary $0<\xi<1$.
As we dial up $\xi$ a minimum develops at some critical value. 
Numerically we observe that this value does not coincide with the ones for the transitions in $\CE(z)$ and $s(z)$.

Given the thermodynamic densities $\CE(z)$, $s(z)$ and $f$, one may wonder about their full versions $\mathsf{E}$, $\mathsf{S}$ and $\mathsf{F}$.
Assuming that the defect contributions to the entropy and free energy are extensive, one has that $\mathsf{E}_\CD=-(p-1)\mathsf{F}_\CD$ and $\mathsf{S}_\CD=-p\beta\mathsf{F}_\CD$, where $p=d-2$ for a co-dimension two defect.
We can therefore straightforwardly obtain all three from the energy density.

From the plots in figures~\ref{fig:E_mon_4d} and~\ref{fig:E_mon_4d_xi}, however, it is clear that the integral of $\CE(z)$ over transverse space is ill-defined.
There are two sources of divergences:
The first one is the large-distance divergence that arises because the functions attain a non-zero constant plateau at large $z$.
It can be removed by subtracting the constant no-defect thermal one-point function of the stress tensor, just as in the boundary examples in section~\ref{sec:ex_bdy}.
The second divergence is short-distance and arises from integrating near $z=0$.
In this region, the functions diverge because the stress tensor has a zero-temperature one-point function that diverges like $\rho^{-4}$ for small $\rho$.
This leading divergence can be removed.
First consider $\CE(z)$ in the absence of singular modes given by \eq{E_mon_4d}. 
In the Taylor expansion of $\CE(z)$ the leading term $\CE^{\mathrm{lead}}(z)$ is $\CO(z^{-4})$ and gives rise to the zero-temperature one-point function upon integrating over $t$.
We subtract this term from $\CE(z)$ together with the constant in \eq{E_mon_4d}.
The remaining $z$ integral with measure $2\pi\int_0^\infty\d z\, z$ is convergent because the sub-leading divergence in $\CE(z)$ is $\CO(z^{>-2})$.
Remarkably, the integrals can be performed in closed form.
Integrating the function over $z$ and then $t$ one finds for the monodromy defect's contribution to the energy
\begin{equation}
\label{eq:integrated_E}
\mathsf{E}_\CD=\frac{2\pi}{\beta^2}\int_{-\infty}^{+\infty}\d x\int_0^\infty\d z\, z\left(\CE(z)-\CE^{\mathrm{lead}}(z)-\frac{\pi^2}{15}\right)=-{L}\frac{\pi \alpha (1-\alpha)}{6\beta^2} \,,
\end{equation}
where $L$ is an IR regulator for the integral over the single spatial direction along the defect denoted $x$.
Note that the energy is negative, meaning that the presence of the defect lowers the energy compared to the no-defect thermal state.

When singular modes are turned on, the sub-leading divergence in $\CE(z)$ is $\CO(z^{<-2})$, which is not integrable with the measure $\int_0^\infty\d z\, z$.
The energy therefore has an additional UV divergence coming from points near the defect.
For $\xi\neq0$ and $\tilde{\xi}=0$, consider $\CE^\xi(z)$ defined in \eq{E_mon_4d_xi} and subtract its leading singularity.
Assuming that $0<\alpha<1/2$, the $z$-integral with a UV cut-off $\varepsilon$ can then be performed in closed form.
We have not been able to perform the remaining integral over $t$ at finite $\varepsilon$.
However, removing the regulator $\varepsilon\to0$ at fixed $t$ before performing the $t$-integral gives
\begin{equation}
\label{eq:integrated_E_xi}
\mathsf{E}_\CD^\xi=\xi L \frac{\pi\alpha}{3\beta^2}\,.
\end{equation}
This procedure is a regularisation prescription.
The result is finite because taking the $\varepsilon\to0$ limit before integrating over $t$ removes contributions from the region of the integral where $t\sim2\log(1/\varepsilon)$.
These contributions scale like $\varepsilon^{-2\alpha}$, leading to a power-law divergence as $\varepsilon\to0$.
The right-hand side of \eq{integrated_E_xi} is the finite piece once this divergence is removed.
The easiest way to see the $\varepsilon^{-2\alpha}$ divergence is to compute $\mathsf{E}_\CD^\xi$ by integrating \eq{Ttt_step_xi} over $\rho$ with measure $2\pi\rho\,\d\rho$ first.
The terms in the first line of \eq{Ttt_step_xi} give the finite piece~\eqref{eq:integrated_E_xi}, whereas the divergence comes from the term in the second line involving $\langle|\Phi|^2\rangle^\xi_\beta$.
Including the extra factor of $\rho$ from the integration measure, that term becomes a total derivative $-\pi/3\times\p_\rho(\rho\p_\rho\langle|\Phi|^2\rangle^\xi_\beta)$.
The divergence then comes from the lower bound of integration since the leading finite temperature term in the DOE~\eqref{eq:F_Phi2_xi} diverges like $\varepsilon^{-2\alpha}$ for small $\rho$, where $\varepsilon$ is a short-distance cut-off.
The full energy for a monodromy defect with singular modes is then given by adding eqs.~\eqref{eq:integrated_E} and~\eqref{eq:integrated_E_xi}, provided that the divergence can be unambiguously subtracted.
Note that the finite contribution proportional to $\xi$ is positive.
The defect with singular modes then has a greater energy than the defect with regular modes only.

It was pointed out in ref.~\cite{Bianchi:2021snj} that the defect with $\xi>0$ or $\tilde{\xi}>0$ and the defect with regular modes only are connected by a defect RG flow.
Whenever $\xi>0$ the operator $:\hat{\CO}^-_{-\alpha}\hat{\CO}^{\dagger-}_{\alpha}:$ has scaling dimension $\hat{\Delta}=d-2-2\alpha<d-2$.
This operator therefore gives rise to a relevant quadratic deformation on the defect.
The endpoint of the defect RG flow triggered by this deformation was argued by ref.~\cite{Bianchi:2021snj} to coincide with monodromy defect with $\xi=0$.
Thus, $\xi>0$ corresponds to the UV whereas $\xi=0$ is the IR fixed point.
Comparing the integrated free energy in the UV and IR, we then find that the full free energy including bulk and defect contributions is negative and obeys $(-\mathsf{F}^{UV})>(-\mathsf{F}^{IR})$, since $\mathsf{F}_\CD=-\mathsf{E}_\CD$ when $d=4$.
The same relation, however, does not hold point-wise for the free energy density $f(z)$.
From the right panel of figure~\ref{fig:f_mon_4d_xi} it is evident that $(-f^{UV})>(-f^{IR})$ for large enough $z$.
However, for small $z$ we observe that the curves cross over and $(-f^{UV})<(-f^{IR})$.
For the dimensionless free energy densities at $\alpha=0.1$ plotted in figure~\ref{fig:f_mon_4d_xi}, this occurs for $z\approx0.09$.
Therefore, while $-\mathsf{F}_\CD$ decreases under a defect RG flow, $-f(z)$ at any given point need not.

\section{Line defects in scalar CFTs at finite temperature}
\label{sec:ex_line}

In this section we consider a pinning field, or magnetic, line defect in the theory of a free massless scalar field $\phi$ in four dimensions.
We begin by putting the theory on $S^1_\beta \times \mathbb{R}^3$, with Cartesian coordinates $x^\mu=(\tau,x_\parallel,\vec{x}_\perp)$.
We take the line defect to extend along the non-compact direction with coordinate $x_\parallel$, while $\vec{x}_\perp$ are the coordinates for the two non-compact directions transverse to the defect.
We define the non-local operator as the integral of the bulk scalar along a line,
\begin{equation}
\label{eq:Line_FFT}
\mathcal{L} = \exp\Bigg(\lambda \int_{-\infty}^{+\infty}\d x_\parallel \;\phi(0,x_\parallel,\vec{0}_\perp)\Bigg)\,,
\end{equation}
where $\lambda$ is a dimensionless coupling since $\phi$ has $\Delta_\phi=1$ when $d=4$.
If we quantise on hypersurfaces of constant $\tau$, $\CL$ is an operator on the CFT Hilbert space on equal-time slices.\footnote{
This is in contrast to the configurations considered in sections~\ref{sec:ex_bdy} and~\ref{sec:mon}, where the defect wraps the thermal circle and therefore deforms the Hilbert space on constant $\tau$ slices.
In a slight abuse of terminology we will use \emph{defect} and \emph{non-local operator} interchangeably to refer to the non-compact line defect considered in the present section.}
In the following we will consider simple finite-temperature correlation functions involving a single bulk local operator and $\CL$.
We will refer to these correlators as thermal \emph{one-point} functions as they contain a single local operator.
In particular, we will consider thermal one-point functions of $\phi$, $\phi^2$ and the stress tensor.

\subsection{Thermal one-point function of \texorpdfstring{$\phi$}{φ}}
\label{ssec:1ptPhiLine}

We begin with the thermal one-point function of $\phi(x)$.
Since $\CL$ is an operator inserted at $\tau=\vec{x}_\perp=0$, the kinematic structures will depend on the insertion point $x^\mu$ of $\phi$.

\paragraph{Zero space separation. }
Before considering a general insertion point $x^\mu$ of $\phi$, consider the special kinematic configuration $x^\mu=(\tau,0,0,0)$ with $\tau\neq0$.
The operators $\phi$ and $\CL$ are separated only along the thermal circle.
As discussed in section \ref{sec:localised_1pt_KMS}, the one-point function of a local operator in the presence of a defect at a point on $S^1_\beta$ obeys a non-trivial KMS condition, see eq.~\eqref{eq:KMS_localised_reflection}.
The present example is therefore arguably the simplest correlator that transforms non-trivially under KMS.

The scalar propagator at finite temperature is given by a sum over thermal images,
\begin{align}
G(x_1;x_2)_{\beta} &= \kappa \sum_{m = -\infty}^{+\infty} \frac{1}{(\tau_1-\tau_2+m\beta)^2+ \vec{x}_{12}^2}\,, & \kappa &= \frac{1}{4\pi^2}\,,
\label{eq:ThermalProp}
\end{align}
where $\vec{x}_{12}=\vec{x}_1-\vec{x}_2$ involves all three spatial directions.
The one-point function of $\phi$ can then be computed using Wick's theorem.
We will denote it as $\langle \phi\CL\rangle_\beta$, keeping the defect $\CL$ explicit.
The only contributing Feynman diagram gives\footnote{
There are other disconnected diagrams involving propagators on the defect.
These, however, exponentiate and cancel when dividing by the expectation value $\langle\CL\rangle_\beta$ of the line defect.
This choice of normalisation is implicit in our formulas.}
\begin{equation}
\begin{split}
\vev{\phi(\tau)\mathcal{L}}_\beta &=\ThermalLineOnePointScalar
=
\lambda \int_{-\infty}^{+\infty} \d y_\parallel\,G(\tau,0_\parallel,\vec{0}_\perp;0,y_\parallel,\vec{0}_\perp)_{\beta}
= \\ &=
\lambda \kappa \int_{-\infty}^{+\infty} \d y_\parallel \sum_{m = -\infty}^{+\infty}
\frac{1}{(\tau+m \beta)^2+ y_\parallel^2}
=
\lambda \kappa \pi\sum_{m=-\infty}^{+\infty}
\frac{1}{|\tau+m\beta|}\,.
\end{split}
\label{eq:OnePointIntegral}
\end{equation}
In the diagram the thermal manifold $\CM_\beta$ is represented as a cylinder, while the defect $\CL$ is the black horizontal line.
The propagator $G$ is the solid line joining the external operator $\phi$, represented by the white point, with the integrated field $\int\phi$ in the definition of $\CL$, represented by the orange dot. 
The sum over thermal images on the right-hand side of the last equality is logarithmically divergent, since the terms with image numbers $m$ and $-m$ behave as $1/|m|$ at large $m$.
To regularise the series, we cut off the sum at a finite integer.
We will choose a symmetric cut-off $\pm M$ and subtract the divergence as the regulator is removed, i.e.\ $M\to\infty$.\footnote{
This is a choice of renormalisation scheme.
We keep the sum symmetric and subtract only the divergent part as in the minimal subtraction scheme.}
This amounts to taking the Hadamard finite piece, which we denote by
\begin{equation}
\operatorname{FP}
\sum_{m\in\mathbb Z}\frac{1}{|m+w|}
\equiv
\lim_{M\rightarrow+\infty}
\left[
\sum_{m=-M}^{M}\frac{1}{|m+w|}
-2\log M
\right].
\label{eq:HadamardOnePoint}
\end{equation}
Here we introduced the dimensionless cross-ratio $w=\tau/\beta$.
Since $0<w<1$, the regulated sum can be written as
\begin{equation}
\sum_{m=-M}^{M}\frac{1}{|m+w|}
=
\psi(M+1+w)
+
\psi(M+1-w)
-
\psi(w)
-
\psi(1-w)\,,
\end{equation}
where $\psi$ is the digamma function, $\psi(z)\equiv \Gamma'(z)/\Gamma(z)$.
Using $\psi(M+a)=\log M+\CO(M^{-1})$ at large $M$, we obtain\footnote{
We remark that the divergent behaviour of the large-$m$ thermal images is captured by a continuous version of the sum, where the series in $m$ is replaced by an integral.
This corresponds to isolating the Fourier zero-mode on the compactified time direction. 
For a sufficiently well-behaved function $f(\tau)$, the Poisson formula reads
\begin{equation*}
\sum_{m \in \mathbb{Z}} f(\tau + m\beta)
=
\frac{1}{\beta}
\sum_{n \in \mathbb{Z}}
\hat{f}\left(\frac{2\pi n}{\beta}\right)
e^{2\pi \iu n \tau/\beta},
\end{equation*}
where $\hat{f}(k)$
is the Fourier transform of $f$.
The zero mode is given by
\begin{equation*}
\hat{f}(0)
=
\int_{-\infty}^{\infty}\d y\,f(y)\,.
\end{equation*}
For $f(\tau)=|\tau|^{-1}$, the Fourier zero-mode $\hat{f}(0)/\beta$ involves a continuous integral $\int\d m\, |\tau+m\beta|^{-1}$, where we have shifted the integration variable by $\tau/\beta$. 
This integral has the same logarithmic IR divergence as the discrete sum.
It also includes a short-distance UV divergence at $m=-\tau/\beta$.
The latter, however, is absent for general kinematics. 
It is therefore just an artefact of the special kinematic configuration where $\phi$ is inserted at $x^\mu=(\tau,\vec{0})$.
For general kinematics the finite part of the zero mode depends on $|\vec{x}_\perp|$ and leads to logarithmic growth at large transverse distance.
}
\begin{equation}
\vev{\phi(\tau)\mathcal{L}}_\beta
=
-\frac{\lambda\kappa\pi}{\beta}
\left(
\psi\left(1-w\right)
+
\psi\left(w\right)
\right).
\label{eq:OnePointFinite}
\end{equation}
It is immediate to see that this function satisfies the KMS condition \eqref{eq:KMS_localised_reflection}.
Expanding \eq{OnePointFinite} in the DOE limit gives
\begin{equation}
\label{eq:small_tau_exp_zero_space}
\vev{\phi(\tau)\mathcal{L}}_\beta
=\lambda \kappa\pi\left(\frac{1}{\tau}+\frac{2}{\beta}\sum_{k=0}^\infty\zeta(2k+1)w^{2k}\right).
\end{equation}
The first term is just the zero-temperature one-point function~\eqref{eq:T0_1-pt_def} with coefficient $a_\phi=\lambda\kappa\pi$.
The $\zeta(1)$ appearing in the sum should be understood as the finite part left over from subtracting the pole in $\zeta(s)$ as $s\to1$.
This finite part is FP$(\zeta(1))=\gamma$, where $\gamma$ is the Euler-Mascheroni constant.
The coefficients of the $w^{\hat{\Delta}-1}$ under the sum over $k$ contain the thermal DCFT data of operators in the DOE of $\phi$.
In this kinematics, however, contributions of operators of different transverse spin $s$ cannot be resolved.
To do so we will need to consider a non-zero transverse spatial separation.

\paragraph{Full spacetime dependence. }
We now consider a general kinematic configuration for the thermal one-point function of $\phi(x)$.
Using translational invariance along the defect and transverse rotations, we can set $x^\mu=(\tau,0,x_\perp,0)$ with $x_\perp>0$ without loss of generality.
We then find for the one-point function
\begin{align}
\vev{\phi\mathcal{L}}_\beta 
=  \int_{-\infty}^{+\infty}\d y_\parallel \sum_{m = -\infty}^{+\infty} \frac{\lambda \kappa}{(\tau+m \beta)^2+ x_\perp^2+y_\parallel^2}  = \lambda \kappa \sum_{m = -\infty}^{+\infty} \frac{\pi}{\sqrt{(\tau+m \beta)^2+ x_\perp^2}}\,,
\label{eq:OnePointIntegralFull}
\end{align}
after exchanging the sum with the integral.
The series is logarithmically divergent, just like in the zero space separation case, and can be regularised by introducing a cut-off.
In this representation each thermal image vanishes as $x_\perp\to\infty$, in agreement with the no-defect value $\langle\phi\rangle_\beta=0$. 
However, this limit does not commute with the removal of the regulator, and the renormalised one-point function grows logarithmically at large $x_\perp$ as we will see below.

The series in \eq{OnePointIntegralFull} can equivalently be written as an integral of a Jacobi theta function.
Using the identity
\begin{equation}
\frac{1}{\sqrt{a}} = \frac{1}{\sqrt{\pi }}\int_0^{\infty}\d s\, \frac{e^{-a s}}{ \sqrt{s}}\,,
\end{equation}
we obtain the integral form
\begin{equation}
\vev{\phi\mathcal{L}}_\beta = \lambda \kappa \pi \int_0^{\infty} \d s\, \frac{e^{-x_\perp^2 s}}{\sqrt{\pi } \sqrt{s}} \sum_{m = -\infty}^{\infty} e^{-(\tau+m\beta)^2 s} = \lambda \kappa \int_0^{\infty} \d s\, \frac{\pi  e^{-s x_\perp^2} \vartheta _3\Big(\frac{\pi  \tau}{\beta },e^{-\frac{\pi ^2}{s \beta ^2}}\Big)}{\beta  s}\,,
\label{eq:OnePointIntegralFullResult}
\end{equation}
where $\vartheta_3(u,v)\equiv\sum_{n=-\infty}^{+\infty} v^{n^2} e^{2 \iu n u}$.
Since $\vartheta_3(u,0)=1$, the logarithmic divergence of the sum over thermal images in \eq{OnePointIntegralFull} has now been recast as a logarithmic divergence of the integral at $s=0$.
We renormalise by again subtracting this divergence.
From the definition of $ \vartheta _3\left(u,v\right)$ it follows immediately that
\begin{equation}
\vartheta _3\left({\pi  w},e^{-\frac{\pi ^2}{s \beta ^2}}\right) =
\vartheta _3\left({\pi  (1-w)},e^{-\frac{\pi ^2}{s \beta ^2}}\right),
\end{equation}  
where again $w=\tau/\beta$.
Therefore, the one-point function satisfies the KMS relation~\eqref{eq:KMS_localised_reflection}.

The integral representation~\eqref{eq:OnePointIntegralFullResult} can be used to expand the correlator in different limits of the kinematic variables. 
E.g.\ when $x_\perp\ll\beta$, we can expand the exponential $e^{-sx_\perp^2}$ and then perform the $s$-integral term-by-term to find
\begin{equation}
\vev{\phi\mathcal{\CL}}_\beta
= \frac{\lambda \kappa \sqrt{\pi }}{\beta} \sum_{n=0}^\infty  \frac{(-1)^n \Gamma \left(n+\frac{1}{2}\right)}{n!} \,z^{2n}  \left(\zeta (2 n+1,1-w)+\zeta (2 n+1,w)\right) ,
\label{eq:OnePointIntegralSmallY}
\end{equation}
where $\zeta(s,a)$ is the Hurwitz zeta function and $z\equiv x_\perp/\beta$ is a dimensionless cross-ratio.
The series converges for $z<\min(w,1-w)$.
Note that the $n=0$ term is ill-defined because it naively contains $\zeta(1,a)$.
We regularise it by taking the Hadamard finite part, $\text{FP}(\zeta(1,a))=-\psi(a)$, which amounts to subtracting the simple pole at $s=1$.
This choice of scheme then agrees with the one used for the computation at zero space separation. 
In particular, the $n=0$ term in \eq{OnePointIntegralSmallY} reduces to the zero space separation result given by \eq{OnePointFinite}.

In principle, one can use \eq{OnePointIntegralSmallY} to extract thermal DCFT data by expanding it for small $w$ and $z$. 
We find it more convenient, however, to consider the last series in eq.~\eqref{eq:OnePointIntegralFull} and to expand it in Gegenbauer polynomials.
Regularising \eq{OnePointIntegralFull} as in \eq{HadamardOnePoint}, and separating the zero-temperature contribution from the thermal images, one finds
\begin{equation}
\label{eq:1-pt_scheme}
\vev{\phi\mathcal{L}}_\beta
= \lambda \kappa \pi
\left(
\frac{1}{r}
+
\lim_{M\to\infty}\underset{m\neq0}{\sum_{m=-M}^M}
\frac{1}{\sqrt{(\tau+m\beta)^2+x_\perp^2}}- \frac{ 2\log(M)}{\beta}
\right) \,.
\end{equation}
This choice of scheme agrees with taking $\text{FP}(\zeta(1,a))=-\psi(a)$ in \eq{OnePointIntegralSmallY}.
For $m>0$, the pair of images $m$ and $-m$ can be expanded using
\begin{equation}
\label{eq:pair_m}
\sum_{\sigma=\pm1}\frac{m\beta}{\sqrt{(\tau+\sigma\, m\beta)^2+x_\perp^2}}
=
\sum_{\sigma=\pm1}\frac{1}{\sqrt{1+\sigma\, 2\eta\frac{ r}{m\beta}+\big(\frac{r}{m\beta}\big)^2}}\,,
\end{equation}
where $r=\sqrt{\tau^2+x_\perp^2}$ and $\eta=\tau/r$.
The right-hand side of \eq{pair_m} can be recognised as the generating function of the Legendre polynomials,
\begin{equation}
\frac{1}{\sqrt{1-2q\eta+q^2}}
=
\sum_{\ell=0}^{\infty}q^\ell P_\ell(\eta)\,.
\end{equation}
Note that the odd powers of $q=r/m\beta$ cancel between the two images.
Summing over $m\neq0$, we thus find the thermal block expansion
\begin{equation}
\vev{\phi\mathcal{L}}_\beta
=
\lambda \kappa \pi
\left(
\frac{1}{r}
+
\frac{2}{\beta}
\sum_{k=0}^{\infty}
\zeta(2k+1)
\varrho^{2k}
P_{2k}\left(\eta\right)
\right),
\label{eq:OnePointGegenbauerExpansion}
\end{equation}
where $\varrho=r/\beta$.
This series converges provided that $r<\beta$.
Note that this is larger than the radius of convergence guaranteed by the radial quantisation argument discussed in section~\ref{sec:localised_1pt_KMS}, which gives $\beta/2$.
Regularisation of the $\zeta(1)$ factor in the $k=0$ term is understood.
In particular, taking $\text{FP}(\zeta(1))= \gamma$ selects the same scheme as in \eq{1-pt_scheme}.
In the limit where $\phi$ is separated only along $\tau$, $r\to\tau$ and $\eta\to1$ such that $P_{2k}(\eta)\to1$.
In this kinematic limit eq.~\eqref{eq:OnePointGegenbauerExpansion} reduces to \eq{small_tau_exp_zero_space}.

In order to extract the finite-temperature DCFT data, we compare eq.~\eqref{eq:OnePointGegenbauerExpansion} with the general DOE channel decomposition of the thermal one-point function in eq.~\eqref{eq:thermal_DOE_point_defect}.
In the present case $d=4$, $p=1$ and hence $q=3$, such that $\mathcal P_s^{(3)}(\eta)=P_s(\eta)$, while $\Delta_\phi=1$.
The first term in \eq{OnePointGegenbauerExpansion} corresponds to the defect identity. 
It is therefore equal to the zero-temperature one-point function $a_\phi=\lambda\kappa\pi$.
The remaining terms give
\begin{equation}
\mu_{\phi}{}^{\hat{\mathcal O}}
\hat b_{\hat{\mathcal O}}
=
2\lambda\kappa\pi\zeta(2k+1)
\label{eq:phi_thermal_data}
\end{equation}
for non-negative integer $k$, corresponding to defect primaries $\hat{\CO}$ with $\hat{\Delta}=2k+1$ and transverse spin $s=2k$.
It is understood that for $k=0$, $\zeta(1)$ is replaced by $\text{FP}(\zeta(1))=\gamma$ in our choice of scheme.
This IR scheme dependence affects the thermal defect one-point function of the defect primary $\hat{\phi}=\phi|_\CL$ and originates from the zero-mode divergence of the free massless scalar theory. 
In interacting theories, this divergence is expected to be regulated by the generation of a thermal mass~\cite{Dolan:1973qd,Parwani:1991gq,Laine:2016hma}. 
For our free example, after subtracting the divergent part, the remaining finite freedom parametrises a family of thermal DCFT correlators distinguished by the value of $\langle\hat\phi\mathcal L\rangle_\beta$. 

\emph{A priori} the left-hand side of eq.~\eqref{eq:phi_thermal_data} involves a sum over all primaries in the DOE of $\phi$ with given quantum numbers.
In the free theory, however, the primary for a given set of quantum numbers is unique.
It takes the form $
\hat{\CO}_s^{\,i_1\ldots i_s}
=
\p^{\langle i_1}\cdots\p^{i_s\rangle}\phi\big|_{\mathcal L}$, 
where brackets denote the transverse symmetric traceless projection. 
Its scaling dimension is $\hat{\Delta}=s+1$, where $s$ is an even non-negative integer and denotes the spin under the zero-temperature transverse spin group $SO(3)$.
In this normalisation, one can compute $\mu_{\phi}{}^{\hat{\CO}_s}$ and $\hat b_{\hat{\CO}_s}$ independently to find
\begin{align}
\mu_{\phi}{}^{\hat{\CO}_s}&=\frac{1}{s!}\,,
&
\hat b_{\hat{\CO}_s}
&=
2\lambda\kappa\pi\,s!\,\zeta(s+1)\,,
\end{align}
where $s=0,2,4,\ldots$ and $\zeta(1)$ is replaced by its scheme-dependent finite part. Their product reproduces eq.~\eqref{eq:phi_thermal_data}.
Note that only defect primaries with even transverse $SO(3)$ spin $s$ contribute.
This is consistent with the fact that only primaries in symmetric traceless representations with even $s$ can acquire a defect thermal one-point function as explained in section~\ref{sec:localised_1pt_KMS}.
Defect thermal one-point functions of odd transverse spin primaries therefore vanish, although such operators may still be present in the DOE.

\paragraph{Bootstrap sum rules. }
Given the OPE data in \eq{phi_thermal_data}, we can check that the sum rules of  section~\ref{sec:Bootstr} are satisfied.
In deriving these sum rules we assumed without justification that the radius of convergence of the thermal block expansion is $\varrho=1$, which is larger than the radius of convergence guaranteed by radial quantisation, $\varrho=1/2$.
Our explicit result \eq{OnePointGegenbauerExpansion} shows that the thermal block expansion indeed converges up to $\varrho=1$.
The radius is set by the singularity associated with the nearest thermal image.
Therefore the thermal DCFT data contained in the one-point function of $\phi$ should satisfy the sum rules given by \eq{DOE_KMS_sum_rules_transverse}.

We will now verify this explicitly for all the sum rules in \eq{DOE_KMS_sum_rules_transverse}.
As a warm-up consider the sum rule for $n=0$ and non-negative integer $m$.
From the thermal DCFT data in \eq{phi_thermal_data}, we see that exchanged operators $\hat{\CO}_s$ have $\hat{\Delta}=2k+1$ and $s=2k$ for non-negative integer $k$.
The sum rule~\eqref{eq:DOE_KMS_sum_rules_transverse} then becomes
\begin{equation}
\label{eq:sum_rule_phi_n0}
\frac{\Gamma(\hat{\Delta}_{\hat{\mathbbm{1}}})}{\Gamma(\hat{\Delta}_{\hat{\mathbbm{1}}}-2m-1)}+\sum_{k=0}^\infty\frac{\zeta(2k+1)}{4^k}\,\frac{\Gamma(2k+1)}{\Gamma(2k-2m)}=0\,,
\end{equation}
where $\zeta(1)$ is again replaced by its scheme-dependent finite part.
The choice of scheme does not matter here:
for any non-negative integer $m$, the factor $1/\Gamma(-2m)$ vanishes and so the scheme-dependent $k=0$ term is absent from the sum.
In the above expression $\hat{\Delta}_{\hat{\mathbbm{1}}}$ denotes the scaling dimension of the identity, which is taken to zero uniformly in both numerator and denominator to give $-\Gamma(2m+2)$.
Eq.~\eqref{eq:sum_rule_phi_n0} holds as a consequence of the following identity
\begin{equation}
\label{eq:identity_sum_rules}
S_R\equiv\sum_{k=\frac{R+1}{2}}^{\infty}
\frac{\zeta(2k+1)}{4^k}
\frac{\Gamma(2k+1)}
{\Gamma(2k-R+1)} = \Gamma(R+1)\,,
\end{equation}
where $R=2m+1$ is an odd integer and we have omitted all vanishing terms from the sum.\footnote{
This identity can be proved in a number of ways. 
The most immediate way is to recognise that the series is related to the generating function $A(x)=\sum_{k=1}^\infty\zeta(2k+1)x^{2k}=-\gamma-(\psi(1-x)+\psi(1+x))/2$.
Taking $R$ derivatives of the series and evaluating at $x=1/2$, one finds $S_R=2^{-R}A^{(R)}(1/2)$ for odd $R$.
The identity~\eqref{eq:identity_sum_rules} then follows from $\psi^{(R)}(z)=(-1)^{R+1}\Gamma(R+1)\zeta(R+1,z)$ with $\zeta(s,3/2)-\zeta(s,1/2)=-2^s$.}

Using this same identity we can show that the sum rule is satisfied for all non-negative integers $n$ and $m$.
From \eq{angular_expansion_coefficients}, one finds
\begin{align}
A^{(3)}_n(-1,0)&=\frac{(-1)^n \Gamma \left(n+\frac{1}{2}\right)}{\sqrt{\pi } \Gamma (n+1)}\,,& A^{(3)}_n(2k,2k)&=\frac{\Gamma\left(n+\frac{1}{2}-k\right)\Gamma(k+1)}{\Gamma\left(\frac12-k\right)\Gamma(k-n+1)\Gamma(n+1)^2}\,.
\end{align}
Substituting the thermal DCFT data in \eq{phi_thermal_data} into the sum rule~\eqref{eq:DOE_KMS_sum_rules_transverse} gives
\begin{equation}
\label{eq:sum_rule_phi_nm}
\Gamma(2n+1)\left(-\frac{\Gamma(2n+2m+2)}{\Gamma(2n+1)}\right)+\sum_{k=n+m+1}^\infty \frac{\zeta(2k+1)}{4^{k}} \frac{ \Gamma (2 k+1)}{ \Gamma (2 k-2 m-2 n)}=0\,.
\end{equation}
The expression in parentheses is the $\hat{\Delta}_{\hat{\mathbbm{1}}}\to0$ limit of $\Gamma(\hat{\Delta}_{\hat{\mathbbm{1}}}-2n)/\Gamma(\hat{\Delta}_{\hat{\mathbbm{1}}}-2n-2m-1)$, which is part of the contribution of the defect identity.
It is then immediate to see that the sum rule is satisfied as a consequence of \eq{identity_sum_rules} with $R=2n+2m+1$.
Again, the scheme-dependent term does not contribute to any of the sum rules.
The sum rules are satisfied by every member of the family of thermal DCFTs parametrised by $\hat{b}_{\hat{\phi}}=\vev{\hat{\phi}\CL}_\beta$.

\paragraph{High-temperature limit and dimensional reduction. }
Finally consider the high-temperature limit $\beta\to0$ of the integral representation~\eqref{eq:OnePointIntegralFullResult}. 
We take this limit at fixed $w=\tau/\beta$, with $0<w<1$, and fixed non-zero transverse separation $x_\perp$. 
In this regime it is convenient to use the Fourier series representation of the Jacobi theta
function,
\begin{equation}
    \vartheta_3
    \left(
        \pi w,
        e^{-\frac{\pi^2}{s\beta^2}}
    \right)
    =
    1
    +
    2\sum_{n=1}^{\infty}
    e^{-\frac{\pi^2n^2}{s\beta^2}}
    \cos(2\pi n w)\,.
    \label{eq:theta3_Fourier_expansion}
\end{equation}
Substituting this expression into eq.~\eqref{eq:OnePointIntegralFullResult} separates the Fourier
zero-mode from the non-zero ones:
\begin{align}
    \vev{\phi\mathcal L}_\beta
    =
    \frac{\lambda \kappa \pi}{\beta}
    \operatorname{FP}
    \int_0^{\infty}\frac{\d s}{s}\,e^{-s x_\perp^2}
    +
    \frac{2\lambda \kappa \pi}{\beta}
    \sum_{n=1}^{\infty}
    \cos(2\pi n w)
    \int_0^{\infty}\frac{\d s}{s}
    e^{
        -s x_\perp^2
        -\frac{\pi^2n^2}{s\beta^2}}\,.
    \label{eq:OnePointHighTemperatureIntermediate}
\end{align}
The logarithmic divergence of the $s$-integral discussed below \eq{OnePointIntegralFullResult} is now fully contained in the integral over the Fourier zero-mode.
We regularise it in the same scheme employed for the original image sum.\footnote{
This is not equivalent to a minimal subtraction performed directly on the $s$-integral. 
Restricting the range of integration to $s>\epsilon$ and subtracting $\log(1/\epsilon\beta^2)$ gives $-\gamma-2\log z$ for the zero mode instead of $2\log(2/z)$. The two schemes differ by a constant and amount to a shift in the value of $\hat b_{\hat\phi}$.}
For the non-zero modes, the exponential factor cures the logarithmic divergence from integrating $1/s$.
The integrals can be performed to give the following exact Poisson re-summed expression
\begin{equation}
    \vev{\phi\mathcal L}_\beta
    =
    \frac{2\lambda\kappa \pi}{\beta}
    \log\left(\frac{2}{z}\right)
    +
    \frac{4\lambda\kappa\pi}{\beta}
    \sum_{n=1}^{\infty}
    \cos(2\pi n w)
    K_0\left(2\pi n z\right),
    \label{eq:OnePointPoissonResummed}
\end{equation}
where $K_0$ is the modified Bessel function of the second kind. 
For fixed $x_\perp=z\beta>0$, the non-zero Fourier modes are exponentially suppressed as $\beta\rightarrow0$.
This high-temperature expansion has an immediate interpretation in terms of dimensional reduction.
Dimensionally reducing a massless scalar field in four dimensions produces a massless scalar in three dimensions and a tower of massive fields.
These arise from the Fourier zero-mode and non-zero modes, respectively.
Indeed, the 3d massless scalar has a propagator $\sim 1/|\Delta x|$.
After integrating the propagator along the line defect to compute the one-point function of $\phi$, the integrated Green's function gives rise to a logarithm.
This matches the leading logarithmic term $T\log(Tx_\perp)$ in \eq{OnePointPoissonResummed}, where $T=1/\beta$. 
Note that this limit does not go to zero for $x_\perp \to +\infty$ as one would expect.
Instead it grows logarithmically in any scheme. 
This does not present a contradiction: the mode contributing to the logarithmic term is the
same zero mode responsible for the IR divergence.
The coefficient of $\log x_\perp$ is scheme independent.
The IR divergence is generated by the fact that in a massless theory the zero Fourier mode still contributes at an arbitrary distance from the defect, regardless of how large this distance is.
Derivatives of the profile, e.g.\ $\partial_{\perp}\langle\phi\mathcal{L}\rangle_\beta\simeq-\lambda T/(2\pi x_\perp)$, do vanish at large distance, and so does the defect contribution to the
stress tensor, as we will see below.
The non-zero Fourier modes correspond to massive fields in the dimensionally reduced theory.
Their propagators decay exponentially fast in distance with the mass scale set by $T$. 
Their contributions give rise to the sum over Bessel functions.

\subsection{Thermal one-point function of \texorpdfstring{$\phi^2$}{φ²}}

Next consider the thermal one-point function of $\phi^2$.
We again start with a simple kinematic configuration, where the operators are only separated along Euclidean time, before considering general kinematics.

\paragraph{Zero space separation. }
Place the bulk operator $\phi^2(x)$ at $x^\mu= (\tau,0,0,0)$ in the presence of a straight line defect $\CL$ at $\tau=\vec{x}_\perp=0$. 
Although the diagrammatic computation is straightforward, this example presents a less immediate thermal block decomposition than that of $\vev{\phi\mathcal L}_\beta$. 
Using Wick’s theorem, the normalised one-point function receives contributions from two diagrams,
\begin{equation}
\label{eq:phi2_diagrams}
\vev{\phi^2\mathcal{L}}_\beta
 =
\ThermalLineOnePointPhiTwoConnected
\;+\;
\ThermalLineOnePointPhiTwoDisconnected=
(\vev{\phi\mathcal{L}}_{\beta})^2
+
\vev{\phi^2}_{\beta}\,.
\end{equation}
The first diagram has both external fields contracted with the line defect. 
Thus it factorises into the square of the thermal one-point function of $\phi$ given by \eq{OnePointFinite}.
The second diagram is disconnected and given by the no-defect thermal one-point function $\langle \phi^2\rangle_\beta$.
It can be obtained from the coincident limit of the bulk propagator in \eq{ThermalProp} after subtracting the zero-temperature divergence.
Therefore
\begin{equation}
 \vev{\phi^2\mathcal{L}}_\beta = \frac{\pi^2}{\beta^2}\left(\frac{\kappa}{3} + \lambda^2\kappa^2(\psi(w)+\psi(1-w))^2\right)\,,
 \label{eq:phi2_1pt_result1}
\end{equation}
where we used that $\langle\phi^2\rangle_\beta=2\kappa\zeta(d-2)/\beta^{d-2}$, which in $d=4$ reduces to $\langle\phi^2\rangle_\beta=\pi^2\kappa/(3\beta^2)$.
The one-point function can be straightforwardly expanded for small $w$ to find
\begin{equation}
\label{eq:phi2_1pt_w_exp}
\begin{split}
\vev{\phi^2\mathcal{L}}_\beta =\frac{1}{\beta^2}\Bigg(&\frac{1}{12}+\frac{\lambda^2}{16\pi^2w^2}+\frac{\lambda^2}{4\pi^2}\sum_{k=0}^\infty w^{2k-1}\zeta(2k+1)\\
&+\frac{\lambda^2}{4\pi^2}\sum_{k=0}^\infty\sum_{j=0}^k\zeta(2j+1)\zeta(2k-2j+1)w^{2k}\Bigg)\,,
\end{split}
\end{equation}
where it is understood that $\zeta(1)$ is replaced by FP$(\zeta(1))=\gamma$.
The first term is the no-defect thermal one-point function of $\phi^2$ while the second term is the zero-temperature one-point function $a_{\phi^2}=\lambda^2\kappa^2\pi^2$.
The remaining coefficients of $w^{\hat{\Delta}-2}$ encode the thermal DCFT data of all operators $\hat{\CO}$ in the DOE with scaling dimension $\hat{\Delta}$ and arbitrary transverse spin $s$. 
By considering a general kinematic configuration, we will be able to resolve the contributions of operators with different $s$.

\paragraph{Dispersion relation}
Before moving on to general kinematics, let us make the following remark.
In the absence of a defect ref.~\cite{Barrat:2025nvu} showed that the analytic structure of the thermal two-point functions of local operators with zero space separation generates a dispersion relation.
As a consequence, the correlator can be decomposed as a linear combination of Hurwitz zeta functions, where the coefficients are the zero-temperature three-point functions $\lambda_{\CO_1\CO_2}^{\CO_3}$ and thermal one-point functions $b_{\CO_3}$.
Schematically,
\begin{align}
\vev{\CO_1(\tau,\vec{0})\CO_2(0,\vec{0})}_\beta = &\sum_{\CO_3} \lambda_{\CO_1\CO_2}{}^{\CO_3}b_{\CO_3}\left(\zeta(\Delta_1+\Delta_2-\Delta_3,\tau/\beta) + \zeta(\Delta_1+\Delta_2-\Delta_3,1-\tau/\beta) \right) \nonumber\\ &+ \text{constant}\,.
\end{align}
With a slight abuse of terminology, we will refer to this decomposition as a dispersion relation, although strictly speaking it is a consequence of it.
It is tempting to speculate that a similar decomposition holds for the thermal one-point function in the presence of a non-local operator,
\begin{equation}
\label{eq:dispersion_rel}
\begin{split}
\vev{\CO_{\Delta}(\tau,\vec{0})\mathcal{L}}_\beta = \sum_{\hat{\CO}} \mu_{\CO}{}^{\hat{\CO}}\hat{b}_{\hat{\CO}} \left(\zeta(\Delta-\hat{\Delta},\tau/\beta) + \zeta(\Delta-\hat{\Delta},1-\tau/\beta) \right) + \text{constant}\,.
\end{split}
\end{equation}
For $\CO_\Delta=\phi^2$, only odd-dimensional operators would contribute to this relation since for even dimensional operators one would get $\left(\zeta(2-2n,\tau/\beta) + \zeta(2-2n,1-\tau/\beta) \right) = 0$ for $n\geq2$. 
If eq.~\eqref{eq:dispersion_rel} holds, the one point function $\vev{\phi^2\mathcal{L}}_\beta$ can be written as a sum of Hurwitz zeta functions whose coefficients can be read off from the first line of \eq{phi2_1pt_w_exp}.

To show that eq.~\eqref{eq:dispersion_rel} holds for our one-point functions, we consider $\vev{\phi^2}_\beta$ as given by \eq{phi2_1pt_result1} and write
\begin{equation}
\left(\psi(w)+\psi(1-w)\right)^2
=
\left(\psi(w)-\psi(1-w)\right)^2
+
4\psi(w)\psi(1-w)\,.
\end{equation}
The first term can immediately be expressed as a symmetric combination of Hurwitz zeta functions. 
Since
\begin{equation}
\left(\psi(w)-\psi(1-w)\right)^2
=
\zeta(2,w)+\zeta(2,1-w)-\pi^2\,,
\label{eq:psi_difference_zeta}
\end{equation}
it remains to derive a Hurwitz zeta expansion of the product
$\psi(w)\psi(1-w)$.
This derivation is shown in Appendix~\ref{app:Borel}.
Here, we only report the conclusion that the one-point function admits the asymptotic expansion
\begin{equation}
\begin{split}
\vev{\phi^2\mathcal L}_\beta
\sim{}&
\frac{\lambda^2\kappa^2\pi^2}{\beta^2}
\bigg[
\zeta(2,w)+\zeta(2,1-w)
-4\gamma
\left(
\psi(w)+\psi(1-w)
\right)
\\
&\qquad\qquad
+4\sum_{k=1}^{+\infty}
\zeta(2k+1)
\left(
\zeta(1-2k,w)
+\zeta(1-2k,1-w)
\right)
\bigg]\\
&+\frac{\pi^2\kappa}{3\beta^2}+\frac{\lambda^2\kappa^2\pi^2}{\beta^2}\left(8\gamma_1-\frac{\pi^2}{3}\right),
\label{eq:phi2_exact_Borel_asymptotic}
\end{split}
\end{equation}
where $\gamma_1$ is the Stieltjes constant.
Interpreting $\psi(w)$ as $-$FP$(\zeta(1,w))$, one finds that this expression precisely matches the proposed relation~\eqref{eq:dispersion_rel}.\footnote{
Note that changing the scheme for $\vev{\hat{\phi}}_\beta$ just amounts to changing the coefficient of $\text{FP}\left(\zeta(1,w)+\zeta(1,1-w)\right) = -\psi(w)-\psi(1-w)$, and so the form of the expansion is universal. }

\paragraph{Full space-time dependence}
We now consider the thermal one-point function of $\phi^2$ with general kinematics. 
As before, the one-point function is obtained from the two diagrams in \eq{phi2_diagrams}.
The first diagram is the square of $\vev{\phi(x)\CL}_\beta$, while the second diagram is the no-defect thermal one-point function $\vev{\phi^2}_\beta$.
All dependence on the defect sits in the first part.
Since $\langle\phi(x)\mathcal L\rangle_\beta$ grows logarithmically as $x_\perp\to\infty$, see eq.~\eqref{eq:OnePointPoissonResummed}, $\langle\phi^2\mathcal L\rangle_\beta$ does not reduce to the constant $\langle\phi^2\rangle_\beta$ at large distances. 
Its defect contribution instead grows as $\big(\frac{\lambda T}{2\pi}\big)^2\log^2\frac{Tx_\perp}{2}$. 
This is again an effect of the massless Fourier zero mode discussed above.
At finite distances, $\vev{\phi(x)\CL}_\beta$ is given by (the suitably regularised) \eq{OnePointIntegralFullResult} in terms of an integral.
Therefore it is not immediate to write a closed-form expression for $\vev{\phi^2\CL}_\beta$ for finite $x_\perp$.

Instead, we shall express the thermal one-point function as an expansion in thermal blocks, which will make the short-distance behaviour manifest.
Squaring \eq{OnePointGegenbauerExpansion} and adding the no-defect thermal one-point function of $\phi^2$, we obtain
\begin{equation}
\label{eq:phi2_full_product}
\begin{split}
\vev{\phi^2(\tau,x_\perp)\mathcal L}_\beta
&=
\frac{\lambda^2\kappa^2\pi^2}{r^2}
+\frac{4\lambda^2\kappa^2\pi^2}{\beta^2}\sum_{k=0}^{\infty}
\zeta(2k+1)\varrho^{2k-1}P_{2k}(\eta)
\\
&\phantom{=}+\frac{\pi^2\kappa}{3\beta^2}+
\frac{4\lambda^2\kappa^2\pi^2}{\beta^2}\sum_{p,q=0}^{\infty}
\zeta(2p+1)\zeta(2q+1)\varrho^{2(p+q)}
P_{2p}(\eta)P_{2q}(\eta)\,.
\end{split}
\end{equation}
Both series converge for $\varrho<1$, just like \eq{OnePointGegenbauerExpansion}.
To bring the right-hand side into the form of \eq{thermal_DOE_point_defect}, one needs to
decompose the product of Legendre polynomials in the double series into a sum of Legendre polynomials.
The product can be linearised using
\begin{equation}
P_{2p}(\eta)P_{2q}(\eta)
=
\sum_{j=|p-q|}^{p+q}
(4j+1)
\begin{pmatrix}
2p&2q&2j\\
0&0&0
\end{pmatrix}^{\!2}\,P_{2j}(\eta)\,,
\label{eq:Legendre_linearization}
\end{equation}
where the parenthesis denotes the Wigner $3j$ symbol.
One then finds
\begin{equation}
\label{eq:phi2_lin}
\sum_{p,q=0}^{+\infty}
\zeta(2p+1)\zeta(2q+1)\varrho^{2(p+q)}
P_{2p}(\eta)P_{2q}(\eta)
=
\sum_{N=0}^{\infty}\varrho^{2N}
\sum_{j=0}^{N}
c_{N,j}P_{2j}(\eta)\,,
\end{equation}
where
\begin{equation}
c_{N,j}
=
(4j+1)\sum_{p=0}^{N}
\begin{pmatrix}
2p&2N-2p&2j\\
0&0&0
\end{pmatrix}^{\!2} \zeta(2p+1)\zeta(2N-2p+1)\,.
\label{eq:phi2_convolution_coefficients}
\end{equation}
In the last step we collected terms with the same power of $\varrho$ and extended the sum over $j$ down to $j=0$ since terms with $j<|2p-N|$ vanish by the triangle condition of the $3j$ symbol.
As a quick check of this expression, take the zero spatial separation limit, $\varrho\to w$ and $\eta\to1$.
Using $\sum_{j=0}^N c_{N,j}=\sum_{p=0}^N \zeta(2p+1)\zeta(2q+1)$, \eq{phi2_full_product} then reduces to the small-$w$ expansion~\eqref{eq:phi2_1pt_w_exp} of $\langle\phi^2\CL\rangle_\beta$ in this special kinematics.\footnote{
Similarly to the case of zero spatial separation, we expect that $\langle\phi^2\CL\rangle_\beta$ for general kinematics can be decomposed as in \eq{phi2_exact_Borel_asymptotic}.
However, we will not attempt to show this here.}
Again, $\zeta(1)$ is to be replaced by FP$(\zeta(1))=\gamma$ in our renormalisation scheme. As mentioned in section~\ref{ssec:1ptPhiLine}, we interpret this scheme dependent constant as defining an infinite family of thermal DCFTs.

We now extract transverse spin-resolved thermal DCFT data by comparing \eq{phi2_full_product} with \eq{thermal_DOE_point_defect}.
The first term in \eq{phi2_full_product} is just the zero-temperature one-point function $a_{\phi^2}=\lambda^2\kappa^2\pi^2$.
The series over $k$ in \eq{phi2_full_product} only contains odd powers of $\varrho$ and corresponds schematically to $\p_i^{(\mathrm{even})}\hat{\phi}$ operators with
\begin{subequations}
\begin{equation}
\mu_{\phi^2}{}^{\hat{\mathcal O}}\,
\hat b_{\hat{\mathcal O}}
=
4\pi^2\lambda^2\kappa^2\,\zeta(2k+1)\,,
\end{equation}
where $k$ is a non-negative integer.
Finally, the terms in the second line of \eq{phi2_full_product}, linearised using \eq{phi2_lin}, only have even powers of $\varrho$.
They correspond to degenerate primaries schematically represented by the family $\hat{\phi}\p_\parallel^{2N-2j}\p_{i_1}\ldots\p_{i_{2j}}\hat{\phi}$ with thermal DCFT data
\begin{equation}
\sum_{\substack{\hat{\mathcal O}\\
\hat\Delta=2N+2,\ s=2j}}
\mu_{\phi^2}{}^{\hat{\mathcal O}}\,
\hat b_{\hat{\mathcal O}}
=
4\pi^2\lambda^2\kappa^2\,c_{N,j}
+
\frac{\pi^2\kappa}{3}\delta_{N,0}\delta_{j,0}\,,
\end{equation}
\end{subequations}
where $0\leq j\leq N$.

\paragraph{Bootstrap sum rules. } 
We can now check  that the sum rules~\eqref{eq:DOE_KMS_sum_rules_transverse} are satisfied by the data in the thermal block expansion of $\langle\phi^2\CL\rangle_\beta$.
Since the theory is free, the correlator is composed of the two diagrams in \eq{phi2_diagrams}.
The first contribution is just the square of the thermal one-point function of $\phi$, $(\langle\phi\CL\rangle_\beta)^2$.
As we argued around \eq{sum_rule_phi_nm}, the thermal DCFT data in the thermal block expansion of $\langle\phi\CL\rangle_\beta$ satisfies the sum rules for any $n$ and $m$.
In section~\ref{sec:Bootstr} we derived the sum rules by differentiating $F_\CO=\beta^{\Delta_\CO}\langle\CO\CL\rangle_\beta$ an odd number of times around the point $w=1/2$.
If $F_\CO$ satisfies the sum rule~\eqref{eq:DOE_KMS_sum_rules_full}, i.e.\ $\p_w^{2m+1}F_\CO|_{w=\frac{1}{2}}=0$, then $\p_w^{2m+1}\mathcal{F}[F_\CO]|_{w=\frac{1}{2}}=0$ for any analytic functional $\mathcal{F}$.
In particular, if $\CF[F_\CO]=(F_\CO)^2$, then using the Leibniz rule
\begin{equation}
\p_w^{2m+1}(F_\CO)^2=\sum_{j=0}^{2m+1}\binom{2m+1}{j}\,(\p_w^ jF_\CO)\,(\p_w^{2m+1-j}F_\CO)\,.
\end{equation}
Since each term on the right-hand side contains a factor with an odd number of derivatives acting on $F_\CO$, $\p_w^{2m+1}(F_\CO)^2|_{w=\frac{1}{2}}=0$.
Therefore, the first diagram in \eq{phi2_diagrams} satisfies the sum rules.
The second diagram in \eq{phi2_diagrams} involves the no-defect thermal one-point function of $\phi^2$, which is constant. 
Since the sum rules involve taking an odd number of derivatives in $w$, any constant shift of the thermal one-point function will drop out. 
Consequently, the full thermal one-point function $\langle\phi^2\CL\rangle_\beta$ satisfies the sum rules~\eqref{eq:DOE_KMS_sum_rules_transverse}.

As an explicit check, we will evaluate the $n=0$ family of sum rules at zero spatial separation directly at the level of the thermal DCFT data.
Let $f_\phi=F_\phi/(\lambda\kappa\pi)$.
Then
\begin{equation}
f_\phi(w,0)
=
\sum_{\alpha}d_\alpha w^\alpha,\,
\end{equation}
where the non-vanishing coefficients are
\begin{align}
d_{-1}&=1\,,&
d_0&=2\gamma\,,&
d_{2k}&=2\zeta(2k+1)\,,
\end{align}
and $k$ is a positive integer.
The part of the one-point function of $\phi^2$ connected to the defect is proportional to $(f_\phi(w,0))^2$.
It takes the form
\begin{align}
(f_\phi(w,0))^2&=\sum_A D_A w^A\,,&
D_A
&=
\sum_{\alpha+\beta=A}
d_\alpha d_\beta\,,
\label{eq:phi2_Cauchy_coefficients}
\end{align}
where $D_A\neq0$ when $A\geq-2$ is an integer.
This is nothing but the convolution of thermal blocks derived in \eq{phi2_1pt_w_exp}.

To demonstrate that the $n=0$ sum rule for $\langle\phi^2\CL\rangle_\beta$ holds, consider $(\langle\phi\CL\rangle_\beta)^2$ which corresponds to the first diagram in \eq{phi2_diagrams}.
We wish to show that
\begin{equation}
\sum_{A=-2}^\infty
D_A\,2^{-A}
\frac{\Gamma(A+1)}
{\Gamma(A-2m)}
=0\,,
\label{eq:phi2_n0_sum_rule}
\end{equation}
where $m$ is a non-negative integer.
Let
\begin{equation}
\mathcal{S}_r
\equiv
\sum_\alpha
d_\alpha\,2^{-\alpha}
\frac{\Gamma(\alpha+1)}
{\Gamma(\alpha-r+1)}\,.
\label{eq:phi_Sr_definition}
\end{equation}
For odd non-negative integer $r$, $\mathcal{S}_r$ is the left-hand side of the sum rule~\eqref{eq:DOE_KMS_sum_rules_transverse} with $n=0$ for $\langle\phi\CL\rangle_\beta$.
The ratio of Gamma functions is understood by analytic continuation where necessary. 
As showed around \eq{sum_rule_phi_n0}, $\mathcal{S}_{2m+1}=-2\Gamma(2m+2)+2S_{2m+1}=0$ for non-negative integer $m$.
Here, the first term arises from the contribution of the defect identity to the sum rule, and $S_R$ is the series defined in \eq{identity_sum_rules}.

To show \eq{phi2_n0_sum_rule}, we use \eq{phi2_Cauchy_coefficients} together with the Vandermonde identity
\begin{equation}
\frac{\Gamma(\alpha+\beta+1)}
{\Gamma(\alpha+\beta-r+1)}
=
\sum_{\ell=0}^{r}
\binom{r}{\ell}
\frac{\Gamma(\alpha+1)}
{\Gamma(\alpha-\ell+1)}
\frac{\Gamma(\beta+1)}
{\Gamma(\beta-r+\ell+1)}\,,
\end{equation}
to find
\begin{align}
\sum_A
D_A\,2^{-A}
\frac{\Gamma(A+1)}
{\Gamma(A-2m)}
&=
\sum_{\ell=0}^{2m+1}
\binom{2m+1}{\ell}
\mathcal{S}_\ell \mathcal{S}_{2m+1-\ell}\,.
\label{eq:phi2_sum_rule_factorization}
\end{align}
For integer $m$, one of $\ell$ and $2m+1-\ell$ is necessarily odd.
Every term on the right-hand side of eq.~\eqref{eq:phi2_sum_rule_factorization} therefore vanishes since $\mathcal{S}_r=0$ for odd $r$.

The contribution of the second diagram in \eq{phi2_diagrams} is independent of $w$.
It thus corresponds to an additional term with $A=0$ on the left-hand side of \eq{phi2_n0_sum_rule}. 
However, any such contribution is set to zero in the sum rule due to the factor of $\Gamma(-2m)$ in the denominator.
Therefore the full thermal one-point function $\langle\phi^2\CL\rangle_\beta$ satisfies the $n=0$ family of sum rules at zero space separation.

\subsection{Thermal one-point function of \texorpdfstring{$T_{\mu\nu}$}{T μv}}

Finally, we consider the thermal one-point function of the stress tensor.
For a free massless scalar field in $d$ dimensions, the improved stress tensor in flat space is given by \eq{scalar_T}.
It is symmetric traceless and conserved up to the bulk EOM $\p^2\phi=0$.
Setting $d=4$ and using the EOM, it can be written as
\begin{equation}
T_{\mu\nu}
=
\frac{2}{3}\partial_\mu\phi\,\partial_\nu\phi
-\frac{1}{6}\delta_{\mu\nu}\,
\partial_\rho\phi\,\partial^\rho\phi
-\frac{1}{3}\phi\,\partial_\mu\partial_\nu\phi\,.
\label{eq:line_scalar_stress_tensor}
\end{equation}
To compute the thermal one-point function of $T_{\mu\nu}$, we again proceed diagrammatically.
Similarly to the one-point function of $\phi^2$ in \eq{phi2_diagrams}, there are two types of diagrams. 
The first diagram involves contracting each $\phi$ in $T_{\mu\nu}$ with the defect $\CL$.
Each contraction just gives the thermal one-point function $\langle\phi(x)\CL\rangle_\beta$ as can be seen from \eq{OnePointIntegral}.
Consequently, the contribution of the first diagram is given by replacing each $\phi$ in \eq{line_scalar_stress_tensor} with the  field profile $\Phi(x)\equiv\langle\phi(x)\CL\rangle_\beta$.
We denote the resulting contribution $T_{\mu\nu}[\Phi](x)$.\footnote{
Away from the line defect, the scalar profile in \eq{OnePointIntegralFull} satisfies the free bulk EOM, $\p^2\Phi=0$.
Since the profile is independent of the coordinate along the line, the four-dimensional Laplacian reduces to the three-dimensional one when acting on $\Phi$.
In particular, for each thermal image
\begin{equation*}
\left(
\partial_\tau^2+\partial_{x_\perp^1}^2+\partial_{x_\perp^2}^2
\right)
\frac{1}{\sqrt{(\tau+m\beta)^2+|\vec x_\perp|^2}}
=
-4\pi\,\delta(\tau+m\beta)\delta^{(2)}(\vec x_\perp)\,,
\end{equation*}
such that $\p^2\Phi(x)=0$ away from the defect $\CL$.
Thus, $\delta^{\mu\nu}T_{\mu\nu}[\Phi](x)=0$ and $\partial^\mu T_{\mu\nu}[\Phi](x)=0$ away from $\CL$.}
The second diagram only involves contractions of the two $\phi$'s in the stress tensor without seeing the defect.
This contribution evaluates to the no-defect thermal one-point function $\langle T_{\mu\nu}\rangle_\beta$, which is constant.
We therefore obtain the decomposition
\begin{equation}
\vev{T_{\mu\nu}(x)\mathcal L}_\beta
=
\vev{T_{\mu\nu}}_{\beta}
+
T_{\mu\nu}[\Phi](x)\,,
\label{eq:line_stress_tensor_shift}
\end{equation}
where the entire defect-dependent contribution to the stress-tensor one-point function is determined by the thermal one-point function of the scalar $\phi$ computed in section~\ref{ssec:1ptPhiLine}.

We now compute the thermal one-point function of $T^{\mu\nu}$, expressing the result in terms of the tensor structures in table~\ref{tab:localised_spin_two_structures}.
The non-vanishing ones turn out to be given by the first line of \eq{localised_T_one_point},
\begin{equation}
\begin{split}
\vev{T^{\mu\nu}(x)\mathcal L}_\beta
={}&
\frac{1}{\beta^4}
\left(
e^\mu e^\nu-\frac{g^{\mu\nu}}{4} 
\right)
F_T^{\tau\tau}(w,z)
+
\frac{2e^{(\mu}\hat{x}_\perp^{\nu)}}{\beta^4}
F_T^{\tau\perp}(w,z)
\\
&+
\frac{1}{\beta^4}
\left(
\tilde{N}^{\mu\nu}-\frac{g^{\mu\nu}}{2}
\right)
F_{T,1}^{\perp\perp}(w,z)
+
\frac{1}{\beta^4}
\left(
\hat{x}_\perp^\mu \hat{x}_\perp^\nu-\frac{g^{\mu\nu}}{4}
\right)
F_{T,2}^{\perp\perp}(w,z)\,,
\end{split}
\label{eq:line_stress_tensor_decomposition}
\end{equation}
where $\hat{x}_\perp^\mu=\tilde{N}^\mu{}_\nu x^\nu/|\vec{x}_\perp|$ and $\tilde{N}_{\mu\nu}$ is the projector onto the two spatial transverse directions $\vec{x}_\perp=(x_\perp^1,x_\perp^2)$.
The remaining structures in \eq{localised_T_one_point} involve the unit tangent vector $t^\mu$ along the line or the Levi-Civita tensor $\tilde{n}^{\mu\nu}$ in the two-dimensional spatial plane transverse to the line spanned by $\hat{x}_\perp^\mu$.
Since the pinning-field line defect is invariant under reflection symmetry along the defect, $x_\parallel\to-x_\parallel$ such that $t^\mu\to-t^\mu$, any tensor structure containing a single $t^\mu$ breaks that symmetry.
Similarly, the theory and the defect are invariant under reflections in the two-dimensional spatial plane transverse to the line. 
Since the corresponding Levi-Civita tensor $\tilde{n}^{\mu\nu}$ changes sign under such a reflection, all parity-odd structures containing $\tilde{n}^{\mu\nu}$ are absent. 

It is convenient to introduce the dimensionless scalar profile
\begin{equation}
f(w,z)=\frac{\beta}{\lambda\kappa\pi}\,\Phi(x)=\frac{\beta}{\lambda\kappa\pi}\,\langle\phi(x)\CL\rangle_\beta\,.
\label{eq:line_dimensionless_scalar_profile}
\end{equation}
Using rotational invariance in the plane transverse to the line, we find
\begin{subequations}
\label{eq:line_stress_tensor_functions}
\begin{align}
F_T^{\tau\tau}(w,z)
={}&
b_T
+
(\lambda\kappa\pi)^2
\left[
\frac23 f_w^2
-\frac13 f f_{ww}
\right],
\\
F_T^{\tau\perp}(w,z)
={}&
(\lambda\kappa\pi)^2
\left[
\frac23 f_w f_z
-\frac13 f f_{wz}
\right],
\\
F_{T,1}^{\perp\perp}(w,z)
={}&
-\frac{(\lambda\kappa\pi)^2}{3{z}} f f_z\,,
\\
F_{T,2}^{\perp\perp}(w,z)
={}&
(\lambda\kappa\pi)^2
\left[
\frac23 f_z^2
-\frac13 f f_{zz}
+\frac{1}{3z}f f_z
\right].
\end{align}
\end{subequations}
The constant $b_T$ denotes the thermal one-point coefficient of the stress tensor in the theory without the defect, see \eq{no_defect_1-pt}.
As shown in eq.~\eqref{eq:T_S_4d}, for a single real scalar in 4d, $b_T=-\frac{2\pi^2}{45}$.
We employ the shorthand $f_w=\p_w f$ and $f_{ww}=\p_w\p_w f$, and similarly for $w$ replaced by $z$.
The bulk EOM in this notation becomes $f_{zz}+f_{ww}+f_z/z=0$, which we used to simplify the above expressions.
Written in the form~\eqref{eq:line_stress_tensor_decomposition} with \eq{line_stress_tensor_functions}, the stress tensor thermal one-point is manifestly traceless and conserved away from the defect $\CL$ as a consequence of the EOM.
Note that under a KMS reflection the scalar profile obeys $f(1-w,z)=f(w,z)$. 
Therefore, taking an odd number of $w$-derivatives of $f$ will produce a function that is odd under $w\to1-w$.
It then follows that $F_T^{\tau\perp}(1-w,z)=-F_T^{\tau\perp}$, whereas the remaining structures $F^{\tau\tau}_T$, $F_{T,1}^{\perp\perp}$ and $F_{T,2}^{\perp\perp}$ are even.
This is consistent with the tensorial action of a thermal reflection: 
components carrying an odd number of indices along the thermal direction change sign under $\tau\rightarrow\beta-\tau$ as discussed in section~\ref{sec:localised_1pt_KMS}.

Given \eq{line_stress_tensor_functions}, let us check some kinematic limits of the thermal one-point function of the stress tensor.
First consider the zero transverse space separation limit $z\to0$.
In that limit we can use \eq{OnePointIntegralSmallY} to show that the structures $F^{\tau\perp}_T$ and $F^{\perp\perp}_{T,2}$ vanish.
This is expected as the transverse unit vectors $\hat{x}_\perp^\mu$ appearing as the coefficients of these functions in \eq{line_stress_tensor_decomposition} become ill-defined.
This simplification is a consequence of the enhancement of the stabiliser group in the zero space separation limit discussed in section~\ref{subsec:kinematics_localised}.

If we further take the $w\to0$ limit we should recover the zero-temperature one-point function.
Indeed taking $z=0$ first, the leading behaviour of the thermal one-point function is $(\tilde{N}^{\mu\nu}-g^{\mu\nu}/2)(\lambda\kappa\pi)^2/3\tau^4$.
By comparing with the zero-temperature one-point function~\eqref{eq:T_def_1-pt_fn} in this kinematic limit, we find $a_T=(\lambda\kappa\pi)^2/12$.
It is instructive to see how one recovers the tensor structures appearing in \eq{T_def_1-pt_fn}.
To this end one should take the limit of the thermal stress tensor one-point function along a generic direction.
Consider the limit $\varrho=\sqrt{w^2+z^2}\rightarrow0$ at fixed $\eta=w/\varrho$.
Since the scalar profile $f=1/\varrho+\CO(\varrho^0)$, \eq{line_stress_tensor_functions} becomes
\begin{subequations}
\label{eq:line_stress_near_defect}
\begin{align}
F_T^{\tau\tau}
&=
\frac{(\lambda\kappa\pi)^2}{3\varrho^6}z^2
+\CO(\varrho^{-3})\,,
\\
F_T^{\tau\perp}
&=
-\frac{(\lambda\kappa\pi)^2}{3\varrho^6}w z
+\CO(\varrho^{-3})\,,
\\
F_{T,1}^{\perp\perp}
&=
\frac{(\lambda\kappa\pi)^2}{3\varrho^4}
+\CO(\varrho^{-3})\,,
\\
F_{T,2}^{\perp\perp}
&=
-\frac{(\lambda\kappa\pi)^2}{3\varrho^6}z^{2}
+\CO(\varrho^{-3})\,.
\end{align}
\end{subequations}
Then defining the projector $N^{\mu\nu}$ onto all directions transverse to the line including $\tau$ and the unit position vector
\begin{align}
N^{\mu\nu}&=e^\mu e^\nu+\tilde{N}^{\mu\nu}\,,&
\hat{\bm{x}}^\mu_\perp
&=
\frac{w}{\varrho} e^\mu+\frac{z}{\varrho} \hat{x}_\perp^\mu\,,
\end{align}
respectively, such that $\hat{\bm{x}}_\perp\cdot\hat{\bm{x}}_\perp=1$, we recover \eq{T_def_1-pt_fn} with $a_T=(\lambda\kappa\pi)^2/12$.

In the opposite kinematic regime $z\gg1$, $\langle T^{\mu\nu}\CL\rangle_\beta$ reduces to the no-defect thermal one-point function.
This is most easily seen by using the Poisson re-summed form of the thermal one-point function of $\phi$ given by \eq{OnePointPoissonResummed}.
Since all terms involving the Bessel function are exponentially suppressed at large $z$, the large-distance behaviour is controlled by the term proportional to $\log(2/z)$.
Taking appropriate derivatives to give Poisson re-summed expressions for the functions in \eq{line_stress_tensor_functions}, one finds that $F^{\tau\tau}_T\sim b_T$ and $F^{\tau\perp}_T\sim0$ up to exponentially small corrections.
For the remaining functions one instead finds at leading order $F^{\perp}_{T,1}\,, F^{\perp\perp}_{T,2}\sim \lambda^2 z^{{-}2}\log(2/z)$ up to $\CO(1)$ constant factors.
These contributions vanish in the $z\to \infty$ limit. 
Therefore, the thermal one-point function approaches the no-defect value fixed by $b_T$.
Unlike for $\phi$ and $\phi^2$, the logarithmic growth of the zero mode does not spoil the large-distance behaviour here, since every term in $T_{\mu\nu}[\Phi]$ contains two derivatives.

To illustrate the spatial dependence of the stress tensor, consider e.g.\ its $\tau\tau$-component. 
It is convenient to isolate the defect-dependent contribution by subtracting the constant thermal part.
We then define 
\begin{equation}
\label{eq:line_energy_density}
\begin{split}
\CE&\equiv-\beta^4\left[
    \vev{T_{\tau\tau}(x)\mathcal L}_\beta
    -
    \vev{T_{\tau\tau}}_\beta
    \right]\\
    &=-(\lambda\kappa\pi)^2\left(\frac12 f_w^2
    -\frac16 f_z^2
    -\frac13 f f_{ww}\right),
\end{split}
\end{equation} 
where we used eqs.~\eqref{eq:line_stress_tensor_functions} and \eqref{eq:line_stress_tensor_decomposition} in the second equality.
The quantity $\CE$ is defined identically to the dimensionless energy density studied in sections~\ref{sec:scalar_T}, \ref{sec:F_T}, \ref{sec:HD_T} and \ref{sec:mon_T} up to a constant shift.
Unlike the previous examples, however, the line defect studied here does not wrap the thermal circle.
It is inserted on a single Euclidean-time slice and therefore does not describe a stationary thermal system with a static impurity whose energy density the stress tensor one-point function would measure.
Rather than a \emph{bona fide} energy density, $\mathcal E$ measures the change in the one-point function of the local stress energy tensor's $\tau\tau$-component produced by the insertion of $\mathcal L$.

\begin{figure}
\centering
\includegraphics[width=.98\textwidth]{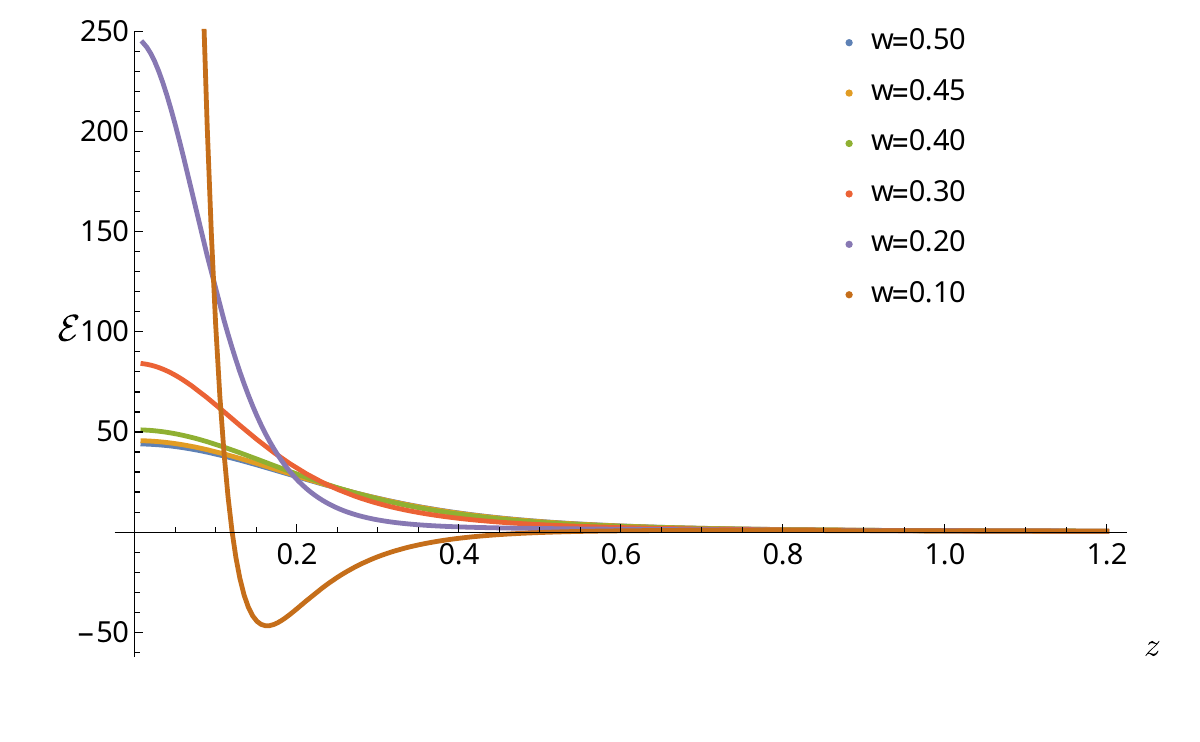}\hfill
\caption{Defect contribution to $\CE$ in units of $(\lambda \kappa \pi )^2$ as a function of the transverse separation $z=x_\perp/\beta$ for several fixed values of the thermal separation $w=\tau/\beta$.
The profiles vanish at large $z$, such that at large distances the full dimensionless energy density is given by its no-defect value. As $w\to0$, the vertical axis intercept and the absolute value of the minimum both increase while the location of the minimum moves towards $z=0$. 
For $w=0$ the minimum coincides with the vertical axis and diverges to $-\infty$.
The function then behaves like $-z^{-4}$ for $z\ll1$ with the coefficient set by its zero-temperature one-point function $a_T$.
}
\label{fig:line_energy_profile}
\end{figure}

We plot $\mathcal E$ in units of $(\lambda \kappa \pi )^2$ in figure~\ref{fig:line_energy_profile} as a function of $z$ for several fixed separations $w$ along the thermal circle. 
To obtain the profiles in figure~\ref{fig:line_energy_profile}, we evaluated the scalar profile $f(w,z)$ using Poisson resummation for the Fourier representation of the Jacobi theta function \eqref{eq:theta3_Fourier_expansion}, and differentiated the resulting Bessel function series term by term to obtain $f_w$, $f_z$ and $f_{ww}$.\footnote{ 
The series converges exponentially for $z>0$.
To produce figure~\ref{fig:line_energy_profile} we truncated the series at $n=1000$, which is sufficient to plot the function numerically to excellent accuracy in the range shown.}
The profiles are computed in the scheme $\mathrm{FP}(\zeta(1))=\gamma$.
As can be seen from the plot, the defect-induced response is largest at zero spatial separation.
It is finite for $w>0$ and becomes increasingly pronounced as the defect is approached as $w\to0$. 
As $w$ is lowered, the correlation function develops a local minimum at finite $z$.
In the limit $w\to0$ and $z\to0$, $\CE$ diverges as $\CO(\varrho^{-4})$ with a coefficient fixed by the zero-temperature one-point function $a_T=(\lambda\kappa\pi)^2/12$.
In the opposite regime, all profiles decay to zero at large $z$, as expected. 

We conclude this section with a remark on eq.~\eqref{eq:line_stress_tensor_decomposition}.
Using the expansion of $\langle\phi\CL\rangle_\beta$ given in \eq{OnePointGegenbauerExpansion}, we can express the stress tensor thermal one-point function in terms of Legendre polynomials.
Since the components of $\langle T^{\mu\nu}\CL\rangle_\beta$ are quadratic functions of the field profile~\eqref{eq:line_dimensionless_scalar_profile}, we can use \eq{Legendre_linearization} to linearise the products of Legendre polynomials.
The functions in \eq{line_stress_tensor_functions} can then be
written in terms of associated Legendre polynomials
\begin{subequations}
\begin{align}
F_T^{\tau\tau}(w,z)
&=
b_T+(\lambda\kappa\pi)^2
\left[
\frac{1}{9\varrho^4}P_2^{(2)}(\eta)
-\frac{4\gamma}{3\varrho^3}P_2^{(0)}(\eta)
+O(\varrho^{-1})
\right],
\\
F_T^{\tau\perp}(w,z)
&=
(\lambda\kappa\pi)^2
\left[
\frac{1}{9\varrho^4}P_2^{(1)}(\eta)
+\frac{2\gamma}{3\varrho^3}P_2^{(1)}(\eta)
+O(\varrho^{-1})
\right],
\\
F_{T,1}^{\perp\perp}(w,z)
&=
(\lambda\kappa\pi)^2
\left[
\frac{1}{3\varrho^4}P_0^{(0)}(\eta)
+\frac{2\gamma}{3\varrho^3}P_0^{(0)}(\eta)
+O(\varrho^{-1})
\right],
\\
F_{T,2}^{\perp\perp}(w,z)
&=
-(\lambda\kappa\pi)^2
\left[
\frac{1}{9\varrho^4}P_2^{(2)}(\eta)
+\frac{2\gamma}{3\varrho^3}P_2^{(2)}(\eta)
+O(\varrho^{-1})
\right],
\end{align}
\label{eq:line_stress_tensor_functions_spin}
\end{subequations}
where we kept all terms in $\langle T^{\mu\nu}\CL\rangle_\beta$ up to order $1/\varrho^2$.
These expressions are the component projections of the expected defect-channel expansion of a spinning bulk operator in tensor spherical harmonics. 
Comparing with \eq{line_stress_near_defect}, we identify the leading term with the contribution of the defect identity. 
Under the branching $SO(4)\to SO(3)$, the bulk spin-2 representation contains transverse scalar, vector and symmetric-traceless rank-two components. 
Reflection symmetry along the line sets the vector component to zero, since $\vev{T^{\parallel\mu}\CL}_\beta$ vanishes for $\mu\neq x_\parallel$, where $\parallel$ is a shorthand for the coordinate along the line $x_\parallel$. 
The contribution of the identity therefore contains only the transverse spin-0 and transverse spin-2 structures.

The next order in $\varrho$, proportional to $\varrho^{-3}$, corresponds to the exchange of $\hat{\phi}$, the unique defect primary with $\hat{\Delta}=1$. 
In the present example, this contribution lies entirely in the transverse $SO(3)$ spin-2 branch and is proportional to
$\hat{\bm{x}}_\perp^\mu\hat{\bm{x}}_\perp^\nu-\frac{1}{3}N^{\mu\nu}$.
The transverse spin-1 branch is forbidden by the combination of translational invariance and reflection symmetry along the line, while the scalar branch vanishes for this exchange due to conservation and translational symmetry.
The coefficient of the remaining structure is proportional to the non-zero thermal one-point coefficient $\hat b_{\hat\phi}$, whose value is fixed by the choice of scheme for $\langle\hat\phi\rangle_\beta$. 

A term of order $1/\varrho^2$ is absent, even though \emph{a priori} there are three operators of dimension $\hat{\Delta}=2$: 
$\hat{\phi}^2$, $\p_\parallel\hat{\phi}$ and $\p_\mu\hat{\phi}$, where $\mu\neq x_\parallel$.
The contributions corresponding to these operators vanish for different reasons. 
The operator $\hat{\phi}^2$ has a vanishing zero-temperature bulk-defect two-point function with $T^{\mu\nu}$, whereas the thermal one-point functions of $\p_\mu\hat{\phi}$ vanish.

At higher orders, the terms $\varrho^{2k}P_{2k}(\eta)$ in the scalar profile generate finite sums of tensor spherical harmonics in the one-point function of $T^{\mu\nu}$. 
The tensor structures within the block associated with a defect primary of transverse spin $s$ follow from the branching of the bulk stress tensor to the transverse $SO(3)$. 
Its scalar component produces one structure with angular momentum $s$, while its symmetric-traceless rank-two component can combine with the exchanged operator to produce structures with angular momenta from $s-2$ to $s+2$, whenever these are non-negative.
The vector component of the stress tensor would instead generate $s-1$ to $s+1$ channels, but for this free DCFT it is forbidden as stated above.
Thus, before imposing conservation or model-specific constraints, a generic block with even $s\geq2$ contains one scalar-type and five rank-two tensor structures, while a scalar exchange contains one structure of each type.
Thermal compactification selects their components that are singlet under $SO(2)$, and the preserved reflection symmetries select which blocks contribute to the one-point function.

\section{Discussion and outlook}
\label{sec:Disc}

In this work, we studied conformal defects of arbitrary dimension and co-dimension at finite temperature.
We considered DCFTs on the thermal manifold $S^1_\beta\times\mathbb{R}^{d-1}$.
Depending on how the defect is placed relative to the thermal circle, we distinguish two configurations: 
defects wrapping $S^1_\beta$ and defects localised on $S^1_\beta$ at a point.
These configurations preserve different symmetries.
For each one, we classified the tensor structures allowed in one-point functions of defect primaries, whose coefficients define thermal DCFT data.
For bulk primaries, we classified the tensor structures up to spin-2.
For bulk scalars, we used the DOE to relate thermal one-point functions to the thermal DCFT data, and derived Euclidean inversion formulas that extract them.
We then discussed how the KMS condition constrains thermal DCFT data. 
While for wrapping defects KMS constrains two-point functions, for localised defects already bulk one-point functions are non-trivially constrained.
In the latter case, this allowed us to derive new thermal bootstrap sum rules.

We illustrated these results in several free field examples.
For unitary and non-unitary BCFTs and for a monodromy defect wrapping the thermal circle, we extracted thermal DCFT data from scalar thermal one-point functions.
The monodromy defect exhibits a rich analytic structure in the cross-ratio $z=|\vec{x}_\perp|/\beta$, with an infinite number of branch points on the imaginary axis.
Additionally, we used the stress tensor one-point function to study thermodynamic quantities at the endpoints of boundary and defect RG flows.
In all cases we found that the free energy is negative and obeys $(-\mathsf{F}^{UV})>(-\mathsf{F}^{IR})$, which is consistent with the expectation that microscopic degrees of freedom are being integrated out.\footnote{
We do not necessarily expect this to hold for general defect RG flows as for bulk QFTs without defects there are known examples where $-\mathsf{F}$ increases along the flow~\cite{Chubukov:1993aau,Sachdev:1993pr}.
}
Finally, we studied a magnetic line defect localised at a point on $S^1_\beta$ in the free scalar CFT when $d=4$.
We showed that the thermal DCFT data in the bulk scalar one-point function satisfies the KMS sum rules and demonstrated that scalar thermal one-point functions can be written as a sum of Hurwitz zeta functions in a kinematic limit.
This is a structure we expect to follow from a dispersion relation similar to that of ref.~\cite{Barrat:2025nvu}.
We also computed the stress-tensor one-point function for this example.
Although its thermodynamic meaning is less apparent, it illustrates how spinning blocks appear in the DOE of spin-two operators.
\\

\noindent
We conclude with several open questions and directions for future research.

\begin{itemize}[label=$\circ$, leftmargin=2em]

\item
An immediate extension of this work is a systematic treatment of spin.
Here we presented the kinematics of thermal one-point functions of bulk operators up to spin-2.
It would be interesting to extend this classification to general representations and to decompose one-point functions of spinning operators into defect-channel thermal blocks.
As a first step, we showed in section~\ref{sec:ex_line} how the stress-tensor one-point function with a localised magnetic line defect can be expanded in associated Legendre polynomials.
For spinning bulk operators, a given defect primary can contribute to several independent tensor structures, each giving additional KMS constraints.
These can help disentangle contributions of degenerate defect operators and constrain defect data that scalar observables cannot see.
For localised defects, the Euclidean inversion formulas derived here for scalar one-point functions may admit a Lorentzian generalisation analytic in transverse spin analogous to refs.~\cite{Caron-Huot:2017vep,Lemos:2017vnx,Iliesiu:2018fao}.

\item
The convergence of the thermal DOE for localised defects is an important open question with direct consequences for the KMS sum rules that we derived.
Radial quantisation guarantees convergence only up to $\beta/2$, beyond which spheres centred on the defect no longer fit on $S^1_\beta$ without self-intersection.
For the magnetic line in free theory, the thermal one-point function coincides with the flat-space one-point function in the presence of a periodic array of line defects at the thermal images.
The DOE then converges up to the nearest image.
It is unclear whether this persists when the bulk is interacting.
In perturbation theory propagators are still sums over images, so the extended radius may survive order by order.
Our derivation of the sum rules in appendix~\ref{app:boots} only requires the radius of convergence to include $\beta/2$, and so the sum rules may hold even if the image picture fails.
Beyond convergence, understanding the large-dimension behaviour of the DOE, along the lines of ref.~\cite{Barrat:2024aoa}, could help control truncations of the KMS sum rules in numerical studies of interacting systems.

\item
Interacting theories would clarify which of the features we observed are universal.
The magnetic line in the $O(N)$ model is a natural first example, since its zero-temperature DCFT data are known in the $\epsilon$-expansion and at large $N$~\cite{Cuomo:2021kfm,Bianchi:2022sbz,Gimenez-Grau:2022ebb}.
Localised on the thermal circle, it would test the KMS sum rules and the DOE convergence properties discussed above.
Boundaries and surface defects in the $O(N)$ model also exhibit boundary or defect RG flows~\cite{McAvity:1995zd,Metlitski:2020cqy,Padayasi:2021sik,Krishnan:2023cff}.
Such flows could test whether the free energy inequality also holds in interacting models.
Additionally, $\mathcal{N}=4$ SYM theory has a wide range of conformal defects.
Since finite temperature breaks supersymmetry, these would have to be studied holographically or perturbatively, see refs.~\cite{Liendo:2016ymz,Bianchi:2020hsz,Ferrero:2021bsb,Artico:2024wut,Artico:2024wnt,Artico:2026pqp} and refs.~therein.
Integrability and localisation nonetheless provide part of the zero-temperature defect data, including defect spectra, one-point functions and DOE coefficients~\cite{Giombi:2018hsx,Grabner:2020nis,Cavaglia:2021bnz,deLeeuw:2015hxa,Buhl-Mortensen:2015gfd,Buhl-Mortensen:2017ind,Komatsu:2020sup,Gombor:2020kgu,Kristjansen:2023ysz,Gombor:2024api,Chalabi:2025nbg}.
Combined with the KMS condition, such input could help constrain the thermal data.

\item
The analytic structure of thermal one-point functions suggests that dispersion relations exist for both defect configurations.
For wrapped defects the monodromy example exhibits branch points on the imaginary axis of the cross-ratio, even though the bulk theory is non-interacting.
For localised defects the one-point functions of the magnetic line decompose into Hurwitz zeta functions, as one would expect from a dispersion relation.
A first goal would be to derive such a relation and to reconstruct one-point functions from their discontinuities.
This would be particularly useful in interacting theories, where diagrammatic computations quickly become difficult.

\item
It would be interesting to extend our analysis to the compact spatial manifold $S^1_\beta\times S^{d-1}$.
Without defects, thermal one-point functions on this geometry decompose into thermal conformal blocks with zero-temperature CFT data as coefficients~\cite{Gobeil:2018fzy,Buric:2024kxo,Buric:2025uqt,Buric:2026pes}.
At high temperature they approach their values on $S^1_\beta\times\mathbb{R}^{d-1}$.
This relates flat space thermal data to averaged zero-temperature CFT data of heavy operators~\cite{El-Showk:2011yvt,Iliesiu:2018fao}.
The same reasoning applies with defects, where the thermal DCFT data studied here should control averaged zero-temperature DCFT data at large scaling dimension.

\item
A complementary approach is the thermal effective action, which, without defects, determines asymptotic properties of the spectrum from simple thermodynamic data~\cite{Benjamin:2023qsc,Allameh:2024qqp}.
It controls the high-temperature behaviour on general backgrounds through a few Wilson coefficients, starting with the thermal free energy density.
The free energy alone then fixes the asymptotic density of operators.
Incorporating defects into the thermal effective action remains largely unexplored beyond effective theories for pairs of conformal defects~\cite{Diatlyk:2024qpr,Diatlyk:2024zkk,Kravchuk:2024qoh}.
One would expect the defect free energies computed here to control the defect contribution to the asymptotic density of defect operators.
Further defect-localised Wilson coefficients would govern sub-leading corrections.

\item
It would be interesting to investigate how the thermal DCFT data studied here constrain real-time dynamics and transport.
For defects localised on the thermal circle, continuing bulk one-point functions to real time describes the evolution of the thermal state after the defect has acted on it.
For boundaries cutting open the thermal cylinder, this becomes a global quantum quench~\cite{Calabrese:2006rx}.
More generally it defines an extended operator version of local quenches~\cite{Nozaki:2014hna,Caputa:2014eta}.\footnote{
See also refs.~\cite{Kawamoto:2022etl,Bianchi:2022ulu,Bianchi:2025fzs} for local quenches on boundaries.}
For lines wrapping the thermal circle, a natural observable is instead the two-point function of the displacement operator, which represents the force on a heavy probe.
The thermal DCFT data in the defect OPE channel control the high-frequency behaviour of the real-time correlator, as for conductivities in thermal CFTs without defects~\cite{Katz:2014rla}. 
For holographic Wilson lines the thermal DCFT data even encode bouncing singularities~\cite{Giombi:2026kdz}.
The low-frequency limit instead determines the heavy-quark momentum diffusion coefficient~\cite{Casalderrey-Solana:2006fio}, which is not captured by the defect OPE at any finite order.
This is part of the broader question of how far short-distance CFT data can constrain long-time dynamics and transport.
\end{itemize}

\acknowledgements
We are grateful to Julien Barrat, Lorenzo Bianchi, Davide Bonomi, Deniz Bozkurt, Bartomeu Fiol, Stefano Galanda, Leonardo Goller, Chris Herzog, Claudio Iuliano, Enrico Marchetto, Alessio Miscioscia, Line Niedeggen, Agostino Patella, Elli Pomoni, Ignacio Salazar Landea, and Philine van Vliet for useful discussions.
The authors would like to thank the Isaac Newton Institute for Mathematical Sciences, Cambridge, for support and hospitality during the programme Quantum field theory with boundaries, impurities, and defects, where work on this paper was undertaken. This work was supported by EPSRC grant EP/Z000580/1.
AC would like to thank the Galileo Galilei Insitute for Theoretical Physics for hospitality during the workshop ``Defects and Extended Excitations in Quantum Field Theory, Quantum Matter and Statistical Models''.
DA is supported by the MUR PRIN contract 2022N9CTAE ``Constraining strongly coupled quantum field theories using symmetry''.
AC is supported by the Italian Ministry of University and Research (MUR) under the FIS grant BootBeyond (CUP: D53C24005470001) and by the INFN ``Iniziativa Specifica'' ST\&\!\! FI.

\appendix

\section{Endpoint derivation of the KMS sum rules}
\label{app:boots}

\begin{figure}[h]
    \centering
    \begin{tikzpicture}[scale=1.15]

        \draw[thick] (0,0) -- (10,0);

        \fill (0,0) circle (2pt);
        \fill (5,0) circle (2pt);
        \fill (10,0) circle (2pt);

        \node[below=5pt] at (0,0) {$w=0$};
        \node[below=5pt] at (5,0) {$w=\frac12$};
        \node[below=5pt] at (10,0) {$w=1$};

        \draw[thick,->] (0.25,0.5) -- (4.75,0.5);
        \node[above=2pt] at (2.5,0.5)
        {DOE centred at $w=0$};

        \draw[thick,->] (9.75,0.5) -- (5.25,0.5);
        \node[above=2pt] at (7.5,0.5)
        {DOE centred at $w=1$};

        \draw[dashed] (5,-0.15) -- (5,1.0);

        \node[above=2pt] at (5,1.0)
        {\footnotesize gluing at fixed order in $z^2$};

    \end{tikzpicture}
    \caption{
    Sketch of the two DOE patches used in the endpoint derivation.
    Each expansion is centred at the nearest thermal image of the defect.
    After expanding in $z^2$, the corresponding transverse Taylor coefficients are smoothly matched at the thermal midpoint $w=1/2$.
    }
    \label{fig:endpoint_DOE_patches}
\end{figure}
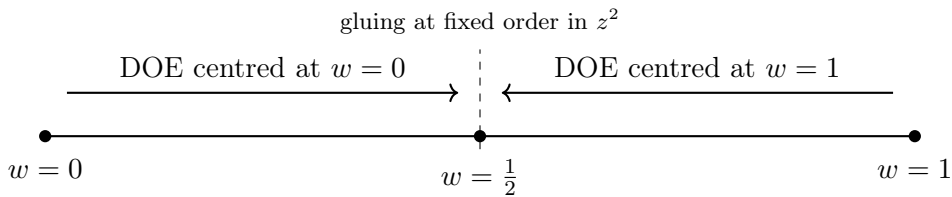

In this appendix we present an alternative derivation of the KMS sum rules introduced in \eq{DOE_KMS_sum_rules_transverse} that apply in time-reversal invariant thermal DCFTs.
Unlike the derivation presented in section~\ref{sec:Bootstr}, this argument does not require the DOE to converge beyond the radius directly suggested by radial quantisation, $\varrho=1/2$.
Instead, we use two DOEs centred around neighbouring thermal images of the defect and match their transverse Taylor coefficients at the midpoint of the thermal circle.
We take the defect to be located at $w=z=0$, while its thermal images are at integer values of $w$ and $z=0$.
This is a manifestation of periodicity of the thermal manifold.
Our procedure is illustrated in figure~\ref{fig:endpoint_DOE_patches}.

In order to match the two DOEs at the midpoint we will require the following assumptions:
\begin{enumerate}
    \item The DOE centred at each thermal image converges in its radial-quantisation domain.
    \item The exact thermal one-point function is sufficiently smooth in $w$ and $z^2$ around $(w,z)=(1/2,0)$.
    \item The $w\to1/2$ limit and the sum in the OPE commute.
    \item The coefficients of the $z^2$ expansions~\eqref{eq:thermal_block_transverse_expansion} for each side, differentiated with respect to $w$, admit finite one-sided limits as $w\to1/2$.
\end{enumerate}
The first two assumptions generally hold in the absence of other operator insertions.
The last condition is a non-trivial assumption about the endpoint convergence of the individual transverse Taylor coefficients.
It does \emph{not} require the full DOE to converge for $z>1/2$ at $w=1/2$ such that $\varrho=\sqrt{1/4+z^2}>1/2$.

Consider a scalar operator inserted at a point $w$ with $z\ll1$.
If $0<w\lesssim1/2$, we can use the DOE of the thermal one-point function $F_\CO$ around $w=0$,
\begin{equation}
    F_{<}(w,z)
    =
    \sum_{\hat{\mathcal O}}
    \mu_{\CO}{}^{\hat{\CO}}\hat{b}_{\hat{\CO}}\,
    \varrho_{<}^{\hat\Delta-\Delta_{\mathcal O}}
    \CP^{(q)}_s(\eta_{<})\,,
    \label{eq:left_DOE_patch}
\end{equation}
where $\varrho_{<}^2=w^2+z^2$ and $\eta_{<}=w/\varrho_{<}$.
If $1/2\lesssim w<1$, we can instead use the DOE centred at the neighbouring thermal image at $w=1$.
Since the displacement from that image has negative Euclidean-time component, we have
\begin{equation}
\begin{split}
    F_{>}(w,z)
    &=
    \sum_{\hat{\mathcal O}}
    \mu_{\CO}{}^{\hat{\CO}}\hat{b}_{\hat{\CO}}\,
    \varrho_{>}^{\hat\Delta-\Delta_{\mathcal O}}
    \CP^{(q)}_s(\eta_{>})
    \\
    &=
    \sum_{\hat{\mathcal O}}
    (-1)^s
    \mu_{\CO}{}^{\hat{\CO}}\hat{b}_{\hat{\CO}}\,
    \varrho_{>}^{\hat\Delta-\Delta_{\mathcal O}}
    \CP^{(q)}_s\left(\frac{1-w}{\varrho_{>}}\right),
    \label{eq:right_DOE_patch}
\end{split}
\end{equation}
where $\varrho_{>}^2=(1-w)^2+z^2$ and $\eta_{>}=-(1-w)/\varrho_{>}$.
In the second equality we used $\CP^{(q)}_s(-\eta)=(-1)^s\CP^{(q)}_s(\eta)$.

We then expand each thermal block around $z=0$.
Terms of order $z^2$ or greater encode information about transverse spin $s$ of the exchanged defect local operator.
Using \eq{thermal_block_transverse_expansion}, the two DOEs become
\begin{align}
    F_{<}(w,z)
    &=
    \sum_{n=0}^{+\infty}z^{2n}
    \sum_{\hat{\mathcal O}}
    \mu_{\CO}{}^{\hat{\CO}}\hat{b}_{\hat{\CO}}\,
    A^{(q)}_n(\hat\Delta-\Delta_{\mathcal O},s)\,
    w^{\hat\Delta-\Delta_{\mathcal O}-2n}\,,
    \label{eq:left_DOE_patch_expanded}
    \\
    F_{>}(w,z)
    &=
    \sum_{n=0}^{+\infty}z^{2n}
    \sum_{\hat{\mathcal O}}
    (-1)^s
    \mu_{\CO}{}^{\hat{\CO}}\hat{b}_{\hat{\CO}}\,
    A^{(q)}_n(\hat\Delta-\Delta_{\mathcal O},s)\,
    (1-w)^{\hat\Delta-\Delta_{\mathcal O}-2n}\,.
    \label{eq:right_DOE_patch_expanded}
\end{align}
As we have discussed below eq.~\eqref{eq:thermal_DOE_point_defect}, the orientation reversing element of $O(q-1)$ forces $\hat{b}_{\hat{\CO}}=0$ when $s$ is odd.
Therefore we can drop the $(-1)^s$ factor in the right DOE~\eqref{eq:right_DOE_patch_expanded}.

We now smoothly glue the two expansions at the midpoint $w=1/2$ by equating the coefficients of the two series in $z^{2n}$.
Taking $r$ derivatives with respect to $w$, the one-sided limit of the left patch gives
\begin{align}
    \lim_{w\to\frac12^-}
    \partial_w^r
    F_{<}^{(n)}
    =
    \sum_{\hat{\mathcal O}}
    \mu_{\CO}{}^{\hat{\CO}}\hat{b}_{\hat{\CO}}\,
    \frac{A^{(q)}_n(\hat\Delta-\Delta_{\mathcal O},s)}{2^{\hat\Delta-\Delta_{\mathcal O}-2n-r}}\,
    \frac{
        \Gamma(\hat\Delta-\Delta_{\mathcal O}-2n+1)
    }{
        \Gamma(\hat\Delta-\Delta_{\mathcal O}-2n-r+1)
    }\,,
    \label{eq:left_endpoint_derivative}
\end{align}
where $F^{(n)}$ denotes the coefficient of $z^{2n}$ in the Taylor expansion of $F$.
For the right patch, every derivative acting on $(1-w)^{\hat\Delta-\Delta_{\mathcal O}-2n}$ produces an additional minus sign. Hence
\begin{align}
    \lim_{w\to\frac12^+}
    \partial_w^r
    F_{>}^{(n)}
    =
    (-1)^r
    \sum_{\hat{\mathcal O}}
    \mu_{\CO}{}^{\hat{\CO}}\hat{b}_{\hat{\CO}}\,
    \frac{A^{(q)}_n(\hat\Delta-\Delta_{\mathcal O},s)}{2^{\hat\Delta-\Delta_{\mathcal O}-2n-r}}\,
    \frac{
        \Gamma(\hat\Delta-\Delta_{\mathcal O}-2n+1)
    }{
        \Gamma(\hat\Delta-\Delta_{\mathcal O}-2n-r+1)
    }\,.
    \label{eq:right_endpoint_derivative}
\end{align}
Smoothness of the exact correlator implies equality of the two one-sided derivatives.
For even $r$, the equality is trivially satisfied.
Setting $r=2m+1$, we obtain
\begin{equation}
\sum_{\hat{\mathcal O}}
\mu_{\CO}{}^{\hat{\CO}}\hat{b}_{\hat{\CO}}\,
2^{2n-\hat\Delta+\Delta_{\mathcal O}}
A^{(q)}_n(\hat\Delta-\Delta_{\mathcal O},s)
\frac{
    \Gamma(\hat\Delta-\Delta_{\mathcal O}-2n+1)
}{
    \Gamma(\hat\Delta-\Delta_{\mathcal O}-2n-2m)
}
=
0\,,
\label{eq:endpoint_spin_sensitive_sum_rules}
\end{equation}
where $m,n$ are non-negative integers.
This is precisely the spin-sensitive family of sum rules~\eqref{eq:DOE_KMS_sum_rules_transverse} obtained in the main text.

\section{Monodromy defect thermal one-point functions of all \texorpdfstring{$T_{\mu\nu}$}{T μv} components}
\label{app:mon}

In this appendix we collect expressions for the thermal one-point function of the stress tensor.

First consider the monodromy defect with $\xi=\tilde{\xi}=0$.
The non-zero components of the stress tensor one-point function can be written as the following integrals
\begin{subequations}
\label{eq:T_mon_s_int}
\begin{align}
\langle T_{\tau\tau}\rangle^{(0)}_\beta &=\sum_{m=-\infty}^\infty\frac{1}{(4\pi)^{\frac{d}{2}}}\int_0^\infty\d s\, e^{-\frac{2\rho^2+(m\beta)^2}{4s}}s^{-\frac{d+2}{2}}\left(1-\frac{(m\beta)^2}{2s}\right)\CI_\alpha\left(\frac{\rho^2}{2s}\right) + \CD^{(0)}_{\tau\tau}\\
\langle T_{xx}\rangle^{(0)}_\beta &=\sum_{m=-\infty}^\infty\frac{1}{(4\pi)^{\frac{d}{2}}}\int_0^\infty\d s\, e^{-\frac{2\rho^2+(m\beta)^2}{4s}}s^{-\frac{d+2}{2}}\CI_\alpha\left(\frac{\rho^2}{2s}\right) + \CD^{(0)}_{xx}\\
\begin{split}
\langle T_{\rho\rho}\rangle^{(0)}_\beta &=\sum_{m=-\infty}^\infty\frac{1}{2(4\pi)^{\frac{d}{2}}}\int_0^\infty\d s\, e^{-\frac{2\rho^2+(m\beta) ^2}{4 s}}s^{-\frac{d+4}{2}} \\
&\quad\times\left(\rho ^2 \CI_\alpha''\left(\frac{\rho ^2}{2 s}\right)-2 \rho ^2 \CI_\alpha'\left(\frac{\rho ^2}{2 s}\right)+2 s \CI_\alpha'\left(\frac{\rho ^2}{2 s}\right)+\rho ^2 \CI_\alpha\left(\frac{\rho ^2}{2 s}\right)\right) + \CD^{(0)}_{\rho\rho}
\end{split}\\
\langle T_{\theta\theta}\rangle^{(0)}_\beta &=\sum_{m=-\infty}^\infty\frac{2}{(4\pi)^{\frac{d}{2}}}\int_0^\infty\d s\, e^{-\frac{2\rho^2+(m\beta)^2}{4s}}s^{-\frac{d}{2}}\CK_\alpha\left(\frac{\rho^2}{2s}\right) + \CD^{(0)}_{\theta\theta}\,,
\end{align}
\end{subequations}
where $\CK_\alpha(\zeta)=\sum_{n}(n-\alpha)^2 I_{|n-\alpha|}(\zeta)$.
Here, $\CD^{(0)}_{\mu\nu}=-\frac{d-2}{2(d-1)}\left(\nabla_\mu\nabla_\nu+\frac{g_{\mu\nu}}{d-2}\nabla^2\right)\langle|\Phi|^2(x)\rangle^{(0)}_\beta$ with $\langle|\Phi|^2(x)\rangle^{(0)}_\beta$ such that
\begin{subequations}
\begin{align}
\CD^{(0)}_{\tau\tau}&=\CD^{(0)}_{xx}=-\frac{1}{2(d-1)}\left(\p_\rho^2+\frac{1}{\rho}\p_\rho\right)\langle|\Phi|^2(x)\rangle^{(0)}_\beta\\
\CD^{(0)}_{\rho\rho}&=-\frac{1}{2(d-1)}\left((d-1)\p_\rho^2+\frac{1}{\rho}\p_\rho\right)\langle|\Phi|^2(x)\rangle^{(0)}_\beta\\
\CD^{(0)}_{\theta\theta}&=-\frac{1}{2(d-1)}\left(\rho^2\p_\rho^2+(d-1)\rho\p_\rho\right)\langle|\Phi|^2(x)\rangle^{(0)}_\beta\,.
\end{align}
\end{subequations}

To perform the $s$-integrals, one can use the integral representation~\eqref{eq:sumI} for $\CI_\alpha$ and its derivatives.
For the sum over Bessel functions $\CK_\alpha(\zeta)=\sum_{n}(n-\alpha)^2 I_{|n-\alpha|}(\zeta)$, we use the fact that $I_\nu(\zeta)$ solves the modified Bessel equation 
\begin{equation}
\zeta^2 I_\nu''(\zeta)+\zeta I_\nu'(\zeta)-(\zeta^2+\nu^2)I_\nu(\zeta)=0
\end{equation}
to re-write
\begin{equation}
\CK_\alpha(\zeta)=\zeta^2\left(\CI_\alpha''(\zeta)-\CI_\alpha(\zeta)\right)+\zeta\CI_\alpha'(\zeta).
\end{equation}

When $d$ is an even integer the sum over Matsubara modes can be performed exactly.
E.g.\ for $d=4$ we find the following rapidly convergent integral representations
\begin{subequations}
\label{eq:T_mon_4d}
\begin{align}
\langle T_{\tau\tau}\rangle^{(0)}_\beta &=-\frac{\pi^2}{15\beta^4}+ \CD^{(0)}_{\tau\tau}-\frac{\sin (\pi  \alpha )}{2\beta^3\rho}\int_0^\infty\d t\,\frac{  \cosh \left(\left(\alpha -\frac{1}{2}\right) t\right) \coth \left(2 \pi  z  \cosh \left(\frac{t}{2}\right)\right) }{ \cosh^2\left(\frac{t}{2}\right)\sinh^2\left(2 \pi  z \cosh \left(\frac{t}{2}\right)\right)}\\
\begin{split}
\langle T_{xx}\rangle^{(0)}_\beta &=\frac{\pi^2}{45\beta^4}+ \CD^{(0)}_{xx}\\
&\quad-\frac{\sin (\pi  \alpha )}{32 \pi ^2 \beta  \rho ^3}\int_0^\infty\d t\,\frac{  \cosh \!\left(\!\left(\alpha -\frac{1}{2}\right) t\right) \left(  \sinh \left(4 \pi  z  \cosh \left(\frac{t}{2}\right)\right)+4 \pi  z  \cosh \left(\frac{t}{2}\right)\right) }{\cosh^4\left(\frac{t}{2}\right)\sinh^2\left(2 \pi  z  \cosh \left(\frac{t}{2}\right)\right)}
\end{split}\\
\begin{split}
\langle T_{\rho\rho}\rangle^{(0)}_\beta &=\frac{\pi^2}{45\beta^4}+ \CD^{(0)}_{\rho\rho}\\
&\quad -\frac{\sin(\pi\alpha)}
{16\pi^{2}\beta^{4}z^{3}}\int_0^\infty\d t\,
\cosh\left(t\left(\alpha-\tfrac{1}{2}\right)\right)
\left\{
\frac{8\pi^{2}z^{2}\,\coth\big(2\pi z\cosh\left(\frac{t}{2}\right)\big)}
{\sinh^{2}\big(2\pi z\cosh\left(\frac{t}{2}\right)\big)}\right.\\
&\quad\left.+\frac{\cosh^{2}\left(\frac{t}{2}\right)+1}{\cosh^{3}\left(\frac{t}{2}\right)}
\left[\frac{2\pi z}{\sinh^{2}\!\big(2\pi z\cosh\left(\frac{t}{2}\right)\big)}
+\frac{\coth\!\big(2\pi z\cosh\left(\frac{t}{2}\right)\big)}{\cosh\left(\frac{t}{2}\right)}\right]
\right\}
\end{split}\\
\begin{split}
\langle T_{\theta\theta}\rangle^{(0)}_\beta &=\frac{\pi^2\rho^2}{45\beta^4}+ \CD^{(0)}_{\theta\theta}\\
&\quad-\frac{\sin(\pi\alpha)}{16\pi^{2}\beta^{2}z}\int_0^\infty\d t\,
\cosh\left(t\left(\alpha-\tfrac{1}{2}\right)\right)
\left\{
\frac{\cosh^{2}\left(\frac{t}{2}\right)-2}{\cosh^{3}\left(\frac{t}{2}\right)}\right.\\
&\quad\times\left[\frac{2\pi z}{\sinh^{2}\left(2\pi z\cosh\left(\frac{t}{2}\right)\right)}
+\frac{\coth\left(2\pi z\cosh\left(\frac{t}{2}\right)\right)}{\cosh\left(\frac{t}{2}\right)}\right]\\
&\left.\quad+\frac{8\pi^{2}z^{2}\tanh^{2}\left(\frac{t}{2}\right)\,\coth\left(2\pi z\cosh\left(\frac{t}{2}\right)\right)}{\sinh^{2}\left(2\pi z\cosh\left(\frac{t}{2}\right)\right)}
\right\},
\end{split}
\end{align}
\end{subequations}
where the $\CD_{\mu\nu}^{(0)}$ are obtained by taking derivatives of \eq{Phi2_4d} with respect to $\rho$.
Note that the first term in each equation is the no-defect thermal one-point function of $T_{\mu\nu}$.
As mentioned in \eq{F_structs_for_T_mon}, we package up $\langle T_{\tau\tau}\rangle_\beta$, $\langle T_{xx}\rangle_\beta$, $\langle T_{\rho\rho}\rangle_\beta$, and $\langle T_{\theta\theta}\rangle_\beta$ into the three functions as in \eq{F_structs_for_T_mon}.
The traceless condition and the conservation equation can be checked numerically.
We find that they hold up to numerical errors of $\CO(10^{-8})$.

When singular modes are included, the contributions proportional to $\xi$ can be computed starting from \eq{T_mon_s_int} by replacing $\CI_\alpha(\zeta) \to I_{-\alpha}(\zeta)-I_\alpha(\zeta)$ and $\CD^{(0)}_{\mu\nu}\to\CD^{\xi}_{\mu\nu}$.
Here, $\CD^{\xi}_{\mu\nu}=-\frac{d-2}{2(d-1)}\left(\nabla_\mu\nabla_\nu+\frac{g_{\mu\nu}}{d-2}\nabla^2\right)\langle|\Phi|^2(x)\rangle^{\xi}_\beta$.
Using the integral representation \eqref{eq:BesselI_int} we first perform the $s$-integration. 
Whenever $d$ is an even integer one can then compute the sum over Matsubara modes in closed form.
We find
\begin{subequations}
\label{eq:T_mon_4d_xi}
\begin{align}
\langle T_{\tau\tau}(x)\rangle^\xi_\beta &= \xi\,\frac{z^3 \sin (\pi  \alpha ) }{\rho ^4}\int_0^\infty\d t\,\frac{ \cosh (\alpha  t) \coth \left(2 \pi  z \cosh \left(\frac{t}{2}\right)\right) }{\cosh\left(\frac{t}{2}\right)\sinh^2\left(2 \pi  z \cosh \left(\frac{t}{2}\right)\right)} + \CD_{\tau\tau}^{\xi}\\
\begin{split}
\langle T_{xx}(x)\rangle^\xi_\beta &=\xi\,\frac{z \sin (\pi  \alpha ) }{16 \pi ^2 \rho ^4}\int_0^\infty\d t\,\frac{ \cosh (\alpha  t) \left(4 \pi  z \cosh \left(\frac{t}{2}\right)+\sinh \left(4 \pi  z \cosh \left(\frac{t}{2}\right)\right)\right) }{\cosh^3\left(\frac{t}{2}\right)\sinh^2\left(2 \pi  z \cosh \left(\frac{t}{2}\right)\right)}\\
&\quad + \CD_{xx}^{\xi}
\end{split}\\
\begin{split}
\langle T_{\rho\rho}(x)\rangle^\xi_\beta &=\xi\,\frac{z\sin(\pi\alpha)}{8\pi^{2}\rho^{4}}\int_0^\infty\d t\,\cosh(t\alpha)
\left\{
\left(1+\operatorname{sech}^{2}\left(\frac{t}{2}\right)\right)\right.\\
&\quad\times
\left[\frac{2\pi z}{\sinh^{2}\left(2\pi z\cosh\left(\frac{t}{2}\right)\right)}
+\frac{\coth\left(2\pi z\cosh\left(\frac{t}{2}\right)\right)}{\cosh\left(\frac{t}{2}\right)}\right]\\
&\quad\left.+\frac{8\pi^{2}z^{2}\cosh\left(\frac{t}{2}\right)\,\coth\left(2\pi z\cosh\left(\frac{t}{2}\right)\right)}{\sinh^{2}\left(2\pi z\cosh\left(\frac{t}{2}\right)\right)}
\right\}+ \CD_{\rho\rho}^{\xi}
\end{split}\\
\langle T_{\theta\theta}(x)\rangle^\xi_\beta &=\xi\,\frac{\alpha ^2 z \sin (\pi  \alpha )}{2 \pi ^2 \rho ^2}\int_0^\infty\d t\,\frac{ \cosh (\alpha  t) \coth \left(2 \pi  z \cosh \left(\frac{t}{2}\right)\right)}{\cosh\left(\frac{t}{2}\right) } + \CD_{\theta\theta}^{\xi}
\end{align}
\end{subequations}
where the $\CD_{\mu\nu}^{\xi}$ are obtained by taking derivatives of \eq{Phi2_4d_xi} with respect to $\rho$.
The terms proportional to $\xi$ also satisfy the traceless condition and conservation.

The contributions proportional to $\tilde{\xi}$ can be computed analogously. 
They can be obtained from the ones proportional to $\xi$ by replacing $\alpha\to1-\alpha$.

\section{Product of digamma functions as a sum of Hurwitz zeta functions}
\label{app:Borel}

Let us show that $\psi(w)\psi(1-w)$ can be written as an infinite sum of Hurwitz-zeta functions. For this purpose, we first introduce the finite triangular sum
\begin{equation}
\mathcal{P}_M(w)
=
\sum_{\substack{m,n\geq 0\\m+n\leq M-1}}
\frac{1}{(m+w)(n+1-w)}\,.
\label{eq:triangular_psi_sum}
\end{equation}
At finite $M$, we use partial fractions to write
\begin{align}
&\mathcal{P}_M(w)
=
\sum_{N=1}^{M}\frac{1}{N}
\sum_{m+n=N-1}
\left(
\frac{1}{m+w}
+
\frac{1}{n+1-w}
\right)
\nonumber\\
&=
\sum_{N=1}^{M}\frac{1}{N}
\big[
\psi(N+w)-\psi(w)
+\psi(N+1-w)-\psi(1-w)
\big]\,,
\label{eq:triangular_psi_sum_diagonal}
\end{align}
where $N=m+n+1$.
However, the triangular cutoff does not coincide with the product $R_M(w)$ of two rectangularly cut-off Dirichlet sums:
their difference is the upper-right corner of the square
$0\leq m,n\leq M-1$,
\begin{align}
R_M(w)-\mathcal P_M(w)
=
\sum_{\substack{0\leq m,n\leq M-1\\m+n\geq M}}
\frac{1}{(m+w)(n+1-w)}
=
\sum_{m=1}^{M-1}
\frac{1}{m+w}
\sum_{n=M-m}^{M-1}
\frac{1}{n+1-w}.
\label{eq:corner_sum}
\end{align}
This difference arise when we want to keep the choice of scheme consistent with the MS choice of the main text.
It is a constant that compensates the different choice of regularizing the infinite sum; to evaluate it we introduce the rescaled variables $x_m=m/M$ and $y_n=n/M$.
For $M\rightarrow\infty$, this is a Riemann sum over the triangular region $0<x<1$ and $1-x<y<1$:
\begin{align}
\lim_{M\rightarrow\infty}
\left[
R_M(w)-\mathcal P_M(w)
\right]
=
\int_0^1\frac{dx}{x}
\int_{1-x}^{1}\frac{dy}{y}
=
-\int_0^1\frac{\log(1-x)}{x}\,dx
=
\frac{\pi^2}{6}.
\end{align}
To manipulate the functions $\psi$ in eq.~\eqref{eq:triangular_psi_sum_diagonal}, we use the Euler-Maclaurin asymptotic expansion
\begin{equation}
\psi(N+x)
\sim
\log N
+
\sum_{k=1}^{+\infty}
\frac{(-1)^{k+1}B_k(x)}{kN^k}\,,
\qquad
N\rightarrow+\infty,
\label{eq:digamma_Euler_Maclaurin}
\end{equation}
where $B_k(x)$ are Bernoulli polynomials.
Using
$
B_k(1-x)=(-1)^kB_k(x)\,,
$
the odd powers cancel in the combination appearing in
eq.~\eqref{eq:triangular_psi_sum_diagonal}, and we obtain
\begin{equation}
\psi(N+w)+\psi(N+1-w)
\sim
2\log N
-
\sum_{k=1}^{+\infty}
\frac{B_{2k}(w)}{kN^{2k}}\,.
\label{eq:symmetric_digamma_asymptotic}
\end{equation}
Substituting this expansion into eq.~\eqref{eq:triangular_psi_sum_diagonal} gives
\begin{align}
\mathcal{P}_M(w)
\sim
2\sum_{N=1}^{M}\frac{\log N}{N}
-
\left[
\psi(w)+\psi(1-w)
\right]
\sum_{N=1}^{M}\frac{1}{N}
-
\sum_{k=1}^{+\infty}
\frac{B_{2k}(w)}{k}
\sum_{N=1}^{M}\frac{1}{N^{2k+1}}\,.
\end{align}
Using
\begin{equation}
\sum_{N=1}^{M}\frac{\log N}{N}
=
\frac{1}{2}\log^2M+\gamma_1+o(1)\,,\qquad
\sum_{N=1}^{M}\frac{1}{N}
=
\log M+\gamma+o(1)\,,
\end{equation}
where $\gamma$ and $\gamma_1$ are respectively the Euler--Mascheroni constant and the first Stieltjes constant, we find
\begin{align}
\underset{M\rightarrow\infty}{\operatorname{FP}}\,
\mathcal{P}_M(w)
\sim
2\gamma_1
-
\gamma
\left[
\psi(w)+\psi(1-w)
\right]
-
\sum_{k=1}^{+\infty}
\zeta(2k+1)\frac{B_{2k}(w)}{k}\,.
\label{eq:triangular_finite_part_asymptotic}
\end{align}
Combining eqs.~\eqref{eq:triangular_psi_sum_diagonal} and
\eqref{eq:triangular_finite_part_asymptotic}, we obtain the formal asymptotic expansion
\begin{align}
\psi(w)\psi(1-w)
\sim
\frac{\pi^2}{6}
+
2\gamma_1
-
\gamma
\left[
\psi(w)+\psi(1-w)
\right]
-
\sum_{k=1}^{+\infty}
\zeta(2k+1)\frac{B_{2k}(w)}{k}\,,
\label{eq:psi_product_Bernoulli_asymptotic}
\end{align}
which, using
$
-\frac{B_{2k}(w)}{k}
=
\zeta(1-2k,w)
+
\zeta(1-2k,1-w)\,,
$
can be written as
\begin{align}
\psi(w)\psi(1-w)
\sim{}&
\frac{\pi^2}{6}
+
2\gamma_1
-
\gamma
\left[
\psi(w)+\psi(1-w)
\right]
\nonumber\\
&+
\sum_{k=1}^{+\infty}
\zeta(2k+1)
\left[
\zeta(1-2k,w)
+
\zeta(1-2k,1-w)
\right].
\label{eq:psi_product_Hurwitz_asymptotic}
\end{align}
The symbol $\sim$ is essential: because of the factorial growth of the Bernoulli polynomials, the series in eq.~\eqref{eq:psi_product_Hurwitz_asymptotic} is not convergent at fixed $w$.
It is instead an Euler--Maclaurin asymptotic expansion.
\paragraph{Borel resummation}
In the following, we show that the asymptotic sum nevertheless admits a natural Borel resummation that reconstructs the exact product
$\psi(w)\psi(1-w)$.
For convenience, we define
\begin{equation}
\mathcal{T}(w)
\equiv
-\sum_{k=1}^{+\infty}
\zeta(2k+1)\frac{B_{2k}(w)}{k}
=
-\sum_{n=1}^{+\infty}\frac{1}{n}
\sum_{k=1}^{+\infty}
\frac{B_{2k}(w)}{k n^{2k}}.
\label{eq:T_double_series}
\end{equation}
For each fixed $n$, we define the Borel--Laplace resummation of the inner series by
\begin{equation}
\mathcal{B}_n(w)
=
2\int_0^{+\infty}\frac{dt}{t}\,e^{-nt}
\left[
\sum_{k=0}^{+\infty}
\frac{B_{2k}(w)t^{2k}}{(2k)!}
-1
\right].
\label{eq:Bn_Borel_definition}
\end{equation}
The even part of the generating function of the Bernoulli polynomials is\footnote{This identity follows by taking the even part under $t\rightarrow-t$ of the generating function for Bernoulli polynomials, DLMF \href{https://dlmf.nist.gov/24.2}{eq.~(24.2.3).}}
\begin{equation}
\sum_{k=0}^{+\infty}
\frac{B_{2k}(w)t^{2k}}{(2k)!}
=
\frac{t}{2}
\frac{
e^{-w t}
+
e^{-(1-w)t}
}{
1-e^{-t}
}.
\label{eq:Bernoulli_even_generating}
\end{equation}
Substituting eq.~\eqref{eq:Bernoulli_even_generating} into
eq.~\eqref{eq:Bn_Borel_definition}, we obtain
\begin{equation}
\mathcal{B}_n(w)
=
\int_0^{+\infty}dt\,e^{-nt}
\left[
\frac{
e^{-w t}
+
e^{-(1-w)t}
}{
1-e^{-t}
}
-\frac{2}{t}
\right].
\label{eq:Bn_integral}
\end{equation}
For $0<w<1$, the integrand is regular at $t=0$ and exponentially suppressed for large $t$.
The integral is therefore unambiguous along the positive real axis.
Using the integral representation of the digamma function together with
\begin{equation}
\log n
=
\int_0^{+\infty}\frac{dt}{t}
\left(
e^{-t}-e^{-nt}
\right),
\end{equation}
eq.~\eqref{eq:Bn_integral} evaluates to
\begin{equation}
\mathcal{B}_n(w)
=
2\log n
-
\psi(n+w)
-
\psi(n+1-w).
\label{eq:Bn_digamma}
\end{equation}
Taking into account the minus sign in eq.~\eqref{eq:T_double_series}, the
Borel resummation of $\mathcal{T}(w)$ is thus
\begin{equation}
\mathcal{S}_{\mathrm B}\mathcal{T}(w)
=
\sum_{n=1}^{+\infty}
\frac{
\psi(n+w)
+
\psi(n+1-w)
-
2\log n
}{n}.
\label{eq:T_Borel_convergent_sum}
\end{equation}
In contrast to the original series over $k$, the sum in
eq.~\eqref{eq:T_Borel_convergent_sum} is convergent.
Indeed, the symmetric large-$n$ expansion gives
\begin{equation}
\psi(n+w)+\psi(n+1-w)-2\log n
=
-\frac{B_2(w)}{n^2}
+
O\left(n^{-4}\right),
\end{equation}
so that the summand in eq.~\eqref{eq:T_Borel_convergent_sum} behaves as $\CO(n^{-3})$.
We conclude that the Hurwitz-zeta expansion identifies similar structures and coefficients to ref.~\cite{Barrat:2025nvu}, although it is not convergent in the ordinary sense.

\bibliography{auxi/thermal_bib.bib}
\bibliographystyle{auxi/JHEP}

\end{document}

%% file: auxi/commands.tex
\pgfdeclareverticalshading{rainbow}{100bp}
{color(0bp)=(red); color(25bp)=(red); color(35bp)=(yellow);
color(45bp)=(green); color(55bp)=(cyan); color(65bp)=(blue);
color(75bp)=(violet); color(100bp)=(violet)}
\definecolor{DarkBlueGrey}{RGB}{76,94,107}
\definecolor{MediumBlueGrey}{RGB}{110,135,153}
\definecolor{LightBlueGrey}{RGB}{134,163,184}
\definecolor{BalancedOrange}{RGB}{242,146,29}
\definecolor{DarkRed}{RGB}{179,48,48}
\definecolor{SalmonPink}{RGB}{255,172,172}
\definecolor{SEColor}{RGB}{134,163,184}

\newcommand{\Tr}{\operatorname{Tr}}

\DeclareMathOperator{\Res}{Res}

\newcommand{\vev}[1]{\langle\, #1 \, \rangle}

\def\Mm{{\mathcal{M}}}

\def\Rds{{\mathbb{R}}}

\newif\ifstartcompletesineup
\newif\ifendcompletesineup
\pgfkeys{
    /pgf/decoration/.cd,
    start up/.is if=startcompletesineup,
    start up=true,
    start up/.default=true,
    start down/.style={/pgf/decoration/start up=false},
    end up/.is if=endcompletesineup,
    end up=true,
    end up/.default=true,
    end down/.style={/pgf/decoration/end up=false}
}
\pgfdeclaredecoration{complete sines}{initial}
{
    \state{initial}[
        width=+0pt,
        next state=upsine,
        persistent precomputation={
            \ifstartcompletesineup
                \pgfkeys{/pgf/decoration automaton/next state=upsine}
                \ifendcompletesineup
                    \pgfmathsetmacro\matchinglength{
                        0.5*\pgfdecoratedinputsegmentlength / (ceil(0.5* \pgfdecoratedinputsegmentlength / \pgfdecorationsegmentlength) )
                    }
                \else
                    \pgfmathsetmacro\matchinglength{
                        0.5 * \pgfdecoratedinputsegmentlength / (ceil(0.5 * \pgfdecoratedinputsegmentlength / \pgfdecorationsegmentlength ) - 0.499)
                    }
                \fi
            \else
                \pgfkeys{/pgf/decoration automaton/next state=downsine}
                \ifendcompletesineup
                    \pgfmathsetmacro\matchinglength{
                        0.5* \pgfdecoratedinputsegmentlength / (ceil(0.5 * \pgfdecoratedinputsegmentlength / \pgfdecorationsegmentlength ) - 0.4999)
                    }
                \else
                    \pgfmathsetmacro\matchinglength{
                        0.5 * \pgfdecoratedinputsegmentlength / (ceil(0.5 * \pgfdecoratedinputsegmentlength / \pgfdecorationsegmentlength ) )
                    }
                \fi
            \fi
            \setlength{\pgfdecorationsegmentlength}{\matchinglength pt}
        }] {}
    \state{downsine}[width=\pgfdecorationsegmentlength,next state=upsine]{
        \pgfpathsine{\pgfpoint{0.5\pgfdecorationsegmentlength}{0.5\pgfdecorationsegmentamplitude}}
        \pgfpathcosine{\pgfpoint{0.5\pgfdecorationsegmentlength}{-0.5\pgfdecorationsegmentamplitude}}
    }
    \state{upsine}[width=\pgfdecorationsegmentlength,next state=downsine]{
        \pgfpathsine{\pgfpoint{0.5\pgfdecorationsegmentlength}{-0.5\pgfdecorationsegmentamplitude}}
        \pgfpathcosine{\pgfpoint{0.5\pgfdecorationsegmentlength}{0.5\pgfdecorationsegmentamplitude}}
}
    \state{final}{}
}

\tikzset{
corner/.style={line width=1pt,dashed,draw=black,dash pattern=on 6pt off 4pt},
scalar/.style={line width=1pt,draw=black},
gluon/.style={line width=1pt,decorate, draw=GluonColor,
    decoration={complete sines,aspect=0,amplitude=1.25mm,segment length=1.5mm,start up,end up}},
gluontwo/.style={line width=1pt,decorate, draw=GluonColor,
    decoration={complete sines,aspect=0,amplitude=.7mm,segment length=1mm,start up,end up}},
ghost/.style={line width=1pt,loosely dotted,draw=black},
wilson/.style={line width=2pt,draw=black},
 }
\NewDocumentCommand\semiloop{O{black}mmmO{}O{above}}
{%
\draw[#1] let \p1 = ($(#3)-(#2)$) in (#3) arc (#4:({#4+180}):({0.5*veclen(\x1,\y1)})node[midway, #6] {#5};)
}

\newcommand{\ThermalLineOnePointScalar}{%
\begin{tikzpicture}[scale=1.3, transform shape,
 baseline={([yshift=-1.3ex]current bounding box.center)}]

\draw[thin] (0,.45) -- (1.8,.45);
\draw[thin] (0,-.45) -- (1.8,-.45);

\draw[thin]
  (0,.45) arc (90:270:.18 and .45);
\draw[thin]
  (0,-.45) arc (270:450:.18 and .45);

\draw[thin,dotted]
  (1.8,.45) arc (90:270:.18 and .45);
\draw[thin]
  (1.8,-.45) arc (270:450:.18 and .45);

\coordinate (D) at (.67,-.18);
\coordinate (O) at (1.15,.31);

\draw[wilson] (0.16,-.18) -- (1.96,-.18);

\draw[scalar]
  (D)
  .. controls (.76,.02) and (.96,.27) ..
  (O);

\draw[fill=orange,orange] (D) circle (.075);
\draw[fill=white] (O) circle (.075);
\node[anchor=west] at (1.20,.20) {\footnotesize $\phi$};
\end{tikzpicture}%
}

\newcommand{\ThermalLineOnePointPhiTwoConnected}{%
\begin{tikzpicture}[scale=1.3, transform shape,
 baseline={([yshift=-1.3ex]current bounding box.center)}]

\draw[thin] (0,.45) -- (1.8,.45);
\draw[thin] (0,-.45) -- (1.8,-.45);

\draw[thin]
  (0,.45) arc (90:270:.18 and .45);
\draw[thin]
  (0,-.45) arc (270:450:.18 and .45);

\draw[thin,dotted]
  (1.8,.45) arc (90:270:.18 and .45);
\draw[thin]
  (1.8,-.45) arc (270:450:.18 and .45);

\draw[wilson] (0.16,-.18) -- (1.96,-.18);

\coordinate (D1) at (.52,-.18);
\coordinate (D2) at (1.48,-.18);
\coordinate (O)  at (1.02,.29);

\draw[scalar,blue]
  (D1) .. controls (.64,.04) and (.82,.25) .. (O);

\draw[scalar,blue]
  (D2) .. controls (1.37,.04) and (1.22,.25) .. (O);

\draw[fill=orange,orange] (D1) circle (.075);
\draw[fill=orange,orange] (D2) circle (.075);

\draw[fill=white] (O) circle (.085);

\end{tikzpicture}%
}

\newcommand{\ThermalLineOnePointPhiTwoDisconnected}{%
\begin{tikzpicture}[scale=1.3, transform shape,
 baseline={([yshift=-1.3ex]current bounding box.center)}]

\draw[thin] (0,.45) -- (1.8,.45);
\draw[thin] (0,-.45) -- (1.8,-.45);

\draw[thin]
  (0,.45) arc (90:270:.18 and .45);
\draw[thin]
  (0,-.45) arc (270:450:.18 and .45);

\draw[thin,dotted]
  (1.8,.45) arc (90:270:.18 and .45);
\draw[thin]
  (1.8,-.45) arc (270:450:.18 and .45);

\draw[wilson] (0.16,-.18) -- (1.96,-.18);

\coordinate (O) at (1.00,.29);

\draw[scalar,blue]
  (O)
  .. controls (.30,-.08) and (1.70,-.08) ..
  (O);

\draw[fill=white] (O) circle (.085);

\end{tikzpicture}%
}

%% file: Tatreez/tatreez.tex
\pdfoutput=1
\usepackage{jheppub} 
\usepackage[T1]{fontenc} 
\usepackage{empheq}
\usepackage{enumitem}  
\usepackage{amssymb}
\usepackage{amsmath}
\usepackage{bbm}
\usepackage{bm}
\usepackage{slashed}
\usepackage{tabstackengine}
\usepackage{natbib}
\usepackage{url}
\usepackage{tikz-cd}
\usepackage{tikz}
\usetikzlibrary{decorations.pathmorphing}
\usetikzlibrary{decorations.markings}
\usepgflibrary{shapes.geometric}

\usepackage[bb=libus]{mathalpha}

\stackMath

\renewcommand{\d}{\text{d}}

\usepackage{tikz}
\usepackage{xcolor}

\newcommand{\motifpixelsize}{0.05em}
\newcommand{\motifcrosswidth}{0.3pt}

\definecolor{motifred}{HTML}{98002E}
\definecolor{motifblue}{HTML}{2864B4}
\definecolor{motifgreen}{HTML}{287A3D}

\ExplSyntaxOn

\seq_new:N \l__pixel_rows_seq

\int_new:N \l__pixel_x_int
\int_new:N \l__pixel_y_int
\int_new:N \l__pixel_draw_y_int
\int_new:N \l__pixel_cols_int
\int_new:N \l__pixel_nrows_int

\cs_new_protected:Npn \pixel_cross:nnn #1#2#3
{
  \draw[
    draw=#1,
    line~width=\motifcrosswidth,
    line~cap=round
  ]
    (#2+0.15,#3-0.15)
    --
    (#2+0.85,#3-0.85)

    (#2+0.15,#3-0.85)
    --
    (#2+0.85,#3-0.15);
}

\cs_new_protected:Npn \pixel_glyph:nn #1#2
{
  \group_begin:

  \seq_set_split:Nnn \l__pixel_rows_seq {,} {#2}
  \seq_get_left:NN \l__pixel_rows_seq \l_tmpa_tl

  \int_set:Nn \l__pixel_cols_int
    {\tl_count:N \l_tmpa_tl}

  \int_set:Nn \l__pixel_nrows_int
    {\seq_count:N \l__pixel_rows_seq}

  \begin{tikzpicture}[
    x=#1,
    y=#1,
    baseline=-0.2ex
  ]

    \path[use~as~bounding~box]
      (0,0)
      rectangle
      (
        \int_use:N \l__pixel_cols_int,
        \int_use:N \l__pixel_nrows_int
      );

    \int_zero:N \l__pixel_y_int

    \seq_map_inline:Nn \l__pixel_rows_seq
    {
      \int_zero:N \l__pixel_x_int

      \str_map_inline:nn {##1}
      {
        \int_set:Nn \l__pixel_draw_y_int
        {
          \l__pixel_nrows_int-\l__pixel_y_int
        }

        \str_case:nnF {####1}
        {
          {1}{
            \pixel_cross:nnn
              {.}
              {\int_use:N \l__pixel_x_int}
              {\int_use:N \l__pixel_draw_y_int}
          }

          {R}{
            \pixel_cross:nnn
              {motifred}
              {\int_use:N \l__pixel_x_int}
              {\int_use:N \l__pixel_draw_y_int}
          }

          {B}{
            \pixel_cross:nnn
              {motifblue}
              {\int_use:N \l__pixel_x_int}
              {\int_use:N \l__pixel_draw_y_int}
          }

          {G}{
            \pixel_cross:nnn
              {motifgreen}
              {\int_use:N \l__pixel_x_int}
              {\int_use:N \l__pixel_draw_y_int}
          }
        }
        {
        }

        \int_incr:N \l__pixel_x_int
      }

      \int_incr:N \l__pixel_y_int
    }

  \end{tikzpicture}

  \group_end:
}

\NewDocumentCommand{\DeclarePixelMotif}{mm}
{
  \cs_set_protected:cpn {#1}
  {
    \pixel_glyph:nn {\motifpixelsize} {#2}
  }
}

\cs_new_protected:Npn \motifTree
{
\pixel_glyph:nn{\motifpixelsize}{
00000011011000000,%
00000001010000000,%
00000000100000000,%
00000000100000000,%
00000001010000000,%
00000100100100000,%
00000110101100000,%
00000010101000000,%
00001001010010000,%
00001101010110000,%
00000101010100000,%
00010011011001000,%
00011011011011000,%
00001011011010000,%
00000111011100000,%
00000111011100000,%
00000011011000000,%
00000000100000000,%
00000001010000000,%
00000000100000000,%
00000000100000000}
}

\cs_new_protected:Npn \motifTreeRed
{
\pixel_glyph:nn{\motifpixelsize}{
000000RR0RR000000,%
0000000R0R0000000,%
00000000R00000000,%
00000000R00000000,%
0000000R0R0000000,%
00000R00R00R00000,%
00000RR0R0RR00000,%
000000R0R0R000000,%
0000R00R0R00R0000,%
0000RR0R0R0RR0000,%
00000R0R0R0R00000,%
000R00RR0RR00R000,%
000RR0RR0RR0RR000,%
0000R0RR0RR0R0000,%
00000RRR0RRR00000,%
00000RRR0RRR00000,%
000000RR0RR000000,%
00000000R00000000,%
0000000R0R0000000,%
00000000R00000000,%
00000000R00000000}
}

\cs_new_protected:Npn \motifOlive
{
\pixel_glyph:nn{\motifpixelsize}{
00000110000000,
00000101111000,
01001011111100,
01101011111110,
01101011111110,
00101011111010,
00010001111010,
11101000111100,
01110100000000,
00000110000000,
01111001100100,
10111101101100,
10111110101100,
11111110101000,
11111110010111,
01111110101110,
00111101100000,
00000011000000}
}

\cs_new_protected:Npn \motifOliveGreen
{
\pixel_glyph:nn{\motifpixelsize}{
00000GG0000000,
00000G0GGGG000,
0G00G0GGGGGG00,
0GG0G0GGGGGGG0,
0GG0G0GGGGGGG0,
00G0G0GGGGG0G0,
000G000GGGG0G0,
GGG0G000GGGG00,
0GGG0G00000000,
00000GG0000000,
0GGGG00GG00G00,
G0GGGG0GG0GG00,
G0GGGGG0G0GG00,
GGGGGGG0G0G000,
GGGGGGG00G0GGG,
0GGGGGG0G0GGG0,
00GGGG0GG00000,
000000GG000000}
}

\cs_new_protected:Npn \motifStar
{
\pixel_glyph:nn{\motifpixelsize}{
0001000001000,%
0001100011000,%
0001110111000,%
1110110110111,%
0111010101110,%
0011101011100,%
0000010100000,%
0011101011100,%
0111010101110,%
1110110110111,%
0001110111000,%
0001100011000,%
0001000001000}
}

\cs_new_protected:Npn \motifStarRed
{
\pixel_glyph:nn{\motifpixelsize}{
000R00000R000,%
000RR000RR000,%
000RRR0RRR000,%
RRR0RR0RR0RRR,%
0RRR0R0R0RRR0,%
00RRR0R0RRR00,%
00000R0R00000,%
00RRR0R0RRR00,%
0RRR0R0R0RRR0,%
RRR0RR0RR0RRR,%
000RRR0RRR000,%
000RR000RR000,%
000R00000R000}
}

\cs_new_protected:Npn \motifStarr
{
\pixel_glyph:nn{\motifpixelsize}{
00001000000010000,
00001100000110000,%
00001110001110000,%
00001111011110000,%
11110111011101111,%
01111011011011110,%
00111101010111100,%
00011110101111000,%
00000001010000000,%
00011110101111000,%
00111101010111100,%
01111011011011110,%
11110111011101111,%
00001111011110000,%
00001110001110000,%
00001100000110000,
00001000000010000}
}

\cs_new_protected:Npn \motifStarrRed
{
\pixel_glyph:nn{\motifpixelsize}{
0000R0000000R0000,
0000RR00000RR0000,%
0000RRR000RRR0000,%
0000RRRR0RRRR0000,%
RRRR0RRR0RRR0RRRR,%
0RRRR0RR0RR0RRRR0,%
00RRRR0R0R0RRRR00,%
000RRRR0R0RRRR000,%
0000000R0R0000000,%
000RRRR0R0RRRR000,%
00RRRR0R0R0RRRR00,%
0RRRR0RR0RR0RRRR0,%
RRRR0RRR0RRR0RRRR,%
0000RRRR0RRRR0000,%
0000RRR000RRR0000,%
0000RR00000RR0000,
0000R0000000R0000}
}

\cs_new_protected:Npn \motifStarrr
{
\pixel_glyph:nn{\motifpixelsize}{
10101001010010101,
01101100100110110,%
11101110101110111,%
00001111011110000,%
11110111011101111,%
01111011011011110,%
00111101010111100,%
10011110101111001,%
01100001010000110,%
10011110101111001,%
00111101010111100,%
01111011011011110,%
11110111011101111,%
00001111011110000,%
11101110101110111,%
01101100100110110,
10101001010010101}
}

\cs_new_protected:Npn \motifStarrrColour
{
\pixel_glyph:nn{\motifpixelsize}{
1010R0010100R0101,
0110RR00100RR0110,%
1110RRR010RRR0111,%
0000RRRR0RRRR0000,%
RRRR0RRR0RRR0RRRR,%
0RRRR0RR0RR0RRRR0,%
00RRRR0R0R0RRRR00,%
100RRRR0R0RRRR001,%
0110000R0R0000110,%
100RRRR0R0RRRR001,%
00RRRR0R0R0RRRR00,%
0RRRR0RR0RR0RRRR0,%
RRRR0RRR0RRR0RRRR,%
0000RRRR0RRRR0000,%
1110RRR010RRR0111,%
0110RR00100RR0110,
1010R0010100R0101}
}

\cs_new_protected:Npn \motifHawthorn
{
\pixel_glyph:nn{\motifpixelsize}{
100000000000,
110000000000,
111000000000,
111100011100,
111110111110,
011110111110,
001110111010,
000010111100,%
001111100000,%
000001000001,%
000011000011,%
000010000111,%
011100001111,%
111110011110,%
111110111110,%
101110111100,%
011100111000,%
000010100000,
000011111100,
000001000000}
}

\cs_new_protected:Npn \motifTile
{
\pixel_glyph:nn{\motifpixelsize}{
10100110001100101,%
01101011011010110,%
11100001010000111,%
00011011011011000,%
01010111011101010,%
10001011011010001,%
11011101010111011,%
01111110101111110,%
00000001010000000,%
01111110101111110,%
11011101010111011,%
10001011011010001,%
01010111011101010,%
00011011011011000,%
11100001010000111,%
01101011011010110,%
10100110001100101}
}

\cs_new_protected:Npn \motifBird
{
\pixel_glyph:nn{\motifpixelsize}{
0011100000000000000,%
0110100000000000000,%
1111100001000000000,%
0011100011100000000,%
0011100111000000000,%
0011101010100000000,%
0111111101010000000,%
0111111010101000000,%
0111111101010100000,%
0011111110101010000,%
0001111011010101000,%
0001010011100010100,%
0001010001110001010,%
0001010000111000001,%
0010101000011100000}
}

\cs_new_protected:Npn \motifMill
{
\pixel_glyph:nn{\motifpixelsize}{
1010110101010110101,
0110101010101010110,
1110100101010010111,
0001100010100011000,
1111100001000011111,
1000010001000100001,
0100001001001000010,
1010000101010000101,
0101000010100001010,
1010111100011110101,
0101000010100001010,
1010000101010000101,
0100001001001000010,
1000010001000100001,
1111100001000011111,
0001100010100011000,
1110100101010010111,
0110101010101010110,
1010110101010110101}
}

\DeclarePixelMotif{motifColourTest}{
00R00,%
0RBR0,%
RB1BR,%
0RGR0,%
00G00}

\ExplSyntaxOff

\abstract{...}



%% file: auxi/thermal_bib.bib
@article{Appelquist:1999hr,
	title        = {{A New constraint on strongly coupled gauge theories}},
	author       = {Appelquist, Thomas and Cohen, Andrew G. and Schmaltz, Martin},
	year         = 1999,
	journal      = {Phys. Rev. D},
	volume       = 60,
	pages        = {045003},
	doi          = {10.1103/PhysRevD.60.045003},
	eprint       = {hep-th/9901109},
	archiveprefix = {arXiv},
	reportnumber = {SLAC-PUB-8045, BUHEP-99-2, YCTP-P2-99}
}

@article{Barrat:2024aoa,
	title        = {{Conformal line defects at finite temperature}},
	author       = {Barrat, Julien and Fiol, Bartomeu and Marchetto, Enrico and Miscioscia, Alessio and Pomoni, Elli},
	year         = 2024,
	month        = 7,
	journal      = {SciPost Phys.},
	volume       = 18,
	number       = 1,
	pages        = {018},
	doi          = {10.21468/SciPostPhys.18.1.018},
	eprint       = {2407.14600},
	archiveprefix = {arXiv},
	primaryclass = {hep-th},
	reportnumber = {DESY-24-099}
}

@article{Barrat:2025nvu,
	title        = {{The analytic bootstrap at finite temperature}},
	author       = {Barrat, Julien and Bozkurt, Deniz N. and Marchetto, Enrico and Miscioscia, Alessio and Pomoni, Elli},
	year         = 2025,
	month        = 6,
	eprint       = {2506.06422},
	archiveprefix = {arXiv},
	primaryclass = {hep-th},
	reportnumber = {DESY-25-078}
}

@article{Barrat:2025wbi,
	title        = {{Thermal Bootstrap for the Critical O(N) Model}},
	author       = {Barrat, Julien and Marchetto, Enrico and Miscioscia, Alessio and Pomoni, Elli},
	year         = 2025,
	journal      = {Phys. Rev. Lett.},
	volume       = 134,
	number       = 21,
	pages        = 211604,
	doi          = {10.1103/PhysRevLett.134.211604},
	eprint       = {2411.00978},
	archiveprefix = {arXiv},
	primaryclass = {hep-th},
	reportnumber = {DESY-24-167}
}

@article{Barrat:2025twb,
	title        = {{Analytic thermal bootstrap meets holography}},
	author       = {Barrat, Julien and Bozkurt, Deniz N. and Marchetto, Enrico and Miscioscia, Alessio and Pomoni, Elli},
	year         = 2025,
	month        = 10,
	eprint       = {2510.20894},
	archiveprefix = {arXiv},
	primaryclass = {hep-th},
	reportnumber = {DESY-25-139 , YITP-SB-2025-16}
}

@article{Bianchi:2015liz,
	title        = {{R{\'e}nyi entropy and conformal defects}},
	author       = {Bianchi, Lorenzo and Meineri, Marco and Myers, Robert C. and Smolkin, Michael},
	year         = 2016,
	journal      = {JHEP},
	volume       = {07},
	pages        = {076},
	doi          = {10.1007/JHEP07(2016)076},
	eprint       = {1511.06713},
	archiveprefix = {arXiv},
	primaryclass = {hep-th},
	reportnumber = {DESY-15-229}
}

@article{Bianchi:2019sxz,
	title        = {{Superconformal surfaces in four dimensions}},
	author       = {Bianchi, Lorenzo and Lemos, Madalena},
	year         = 2020,
	journal      = {JHEP},
	volume       = {06},
	pages        = {056},
	doi          = {10.1007/JHEP06(2020)056},
	eprint       = {1911.05082},
	archiveprefix = {arXiv},
	primaryclass = {hep-th},
	reportnumber = {CERN-TH-2019-190}
}

@article{Bianchi:2021snj,
	title        = {{Monodromy defects in free field theories}},
	author       = {Bianchi, Lorenzo and Chalabi, Adam and Proch{\'a}zka, Vladim{\'\i}r and Robinson, Brandon and Sisti, Jacopo},
	year         = 2021,
	journal      = {JHEP},
	volume       = {08},
	pages        = {013},
	doi          = {10.1007/JHEP08(2021)013},
	eprint       = {2104.01220},
	archiveprefix = {arXiv},
	primaryclass = {hep-th},
	reportnumber = {UUITP- 16/21}
}

@article{Bianchi:2022ppi,
	title        = {{Conformal dispersion relations for defects and boundaries}},
	author       = {Bianchi, Lorenzo and Bonomi, Davide},
	year         = 2023,
	journal      = {SciPost Phys.},
	volume       = 15,
	number       = 2,
	pages        = {055},
	doi          = {10.21468/SciPostPhys.15.2.055},
	eprint       = {2205.09775},
	archiveprefix = {arXiv},
	primaryclass = {hep-th}
}

@article{Billo:2016cpy,
	title        = {{Defects in conformal field theory}},
	author       = {Bill{\`o}, Marco and Gon{\c{c}}alves, Vasco and Lauria, Edoardo and Meineri, Marco},
	year         = 2016,
	journal      = {JHEP},
	volume       = {04},
	pages        = {091},
	doi          = {10.1007/JHEP04(2016)091},
	eprint       = {1601.02883},
	archiveprefix = {arXiv},
	primaryclass = {hep-th}
}

@article{Casalderrey-Solana:2006fio,
	title        = {{Heavy quark diffusion in strongly coupled N=4 Yang-Mills}},
	author       = {Casalderrey-Solana, Jorge and Teaney, Derek},
	year         = 2006,
	journal      = {Phys. Rev. D},
	volume       = 74,
	pages        = {085012},
	doi          = {10.1103/PhysRevD.74.085012},
	eprint       = {hep-ph/0605199},
	archiveprefix = {arXiv}
}

@article{Chalabi:2021jud,
	title        = {{Weyl anomalies of four dimensional conformal boundaries and defects}},
	author       = {Chalabi, Adam and Herzog, Christopher P. and O'Bannon, Andy and Robinson, Brandon and Sisti, Jacopo},
	year         = 2022,
	journal      = {JHEP},
	volume       = {02},
	pages        = 166,
	doi          = {10.1007/JHEP02(2022)166},
	eprint       = {2111.14713},
	archiveprefix = {arXiv},
	primaryclass = {hep-th},
	reportnumber = {UUITP-58/21}
}

@article{Chalabi:2022qit,
	title        = {{Boundaries in free higher derivative conformal field theories}},
	author       = {Chalabi, Adam and Herzog, Christopher P. and Ray, Krishnendu and Robinson, Brandon and Sisti, Jacopo and Stergiou, Andreas},
	year         = 2023,
	journal      = {JHEP},
	volume       = {04},
	pages        = {098},
	doi          = {10.1007/JHEP04(2023)098},
	eprint       = {2211.14335},
	archiveprefix = {arXiv},
	primaryclass = {hep-th},
	reportnumber = {UUITP-53/22}
}

@article{Cheung:2026dng,
	title        = {{Thermal Positivity}},
	author       = {Cheung, Clifford and Rosen, Rachel A.},
	year         = 2026,
	month        = 6,
	eprint       = {2606.05136},
	archiveprefix = {arXiv},
	primaryclass = {hep-th}
}

@article{Chubukov:1993aau,
	title        = {{Theory of two-dimensional quantum Heisenberg antiferromagnets with a nearly critical ground state}},
	author       = {Chubukov, Andrey V. and Sachdev, Subir and Ye, Jinwu},
	year         = 1994,
	journal      = {Phys. Rev. B},
	volume       = 49,
	pages        = {11919--11961},
	doi          = {10.1103/PhysRevB.49.11919},
	eprint       = {cond-mat/9304046},
	archiveprefix = {arXiv}
}

@article{Copetti:2026ncv,
    author = "Copetti, Christian",
    title = "{When Symmetries Twist: Anomaly Inflow on Monodromy Defects}",
    eprint = "2605.16482",
    archivePrefix = "arXiv",
    primaryClass = "hep-th",
    month = "5",
    year = "2026"
}

@article{Cuomo:2021kfm,
	title        = {{Localized magnetic field in the O(N) model}},
	author       = {Cuomo, Gabriel and Komargodski, Zohar and Mezei, M{\'a}rk},
	year         = 2022,
	journal      = {JHEP},
	volume       = {02},
	pages        = 134,
	doi          = {10.1007/JHEP02(2022)134},
	eprint       = {2112.10634},
	archiveprefix = {arXiv},
	primaryclass = {hep-th}
}

@article{Diatlyk:2026oxm,
	title        = {{$\mathcal{PT}$-symmetric Field Theories at Finite Temperature}},
	author       = {Diatlyk, Oleksandr and Katsevich, Andrei and Popov, Fedor K.},
	year         = 2026,
	month        = 4,
	eprint       = {2604.08459},
	archiveprefix = {arXiv},
	primaryclass = {hep-th}
}

@article{Dowker:1978md,
	title        = {{Finite Temperature and Boundary Effects in Static Space-Times}},
	author       = {Dowker, J. S. and Kennedy, Gerard},
	year         = 1978,
	journal      = {J. Phys. A},
	volume       = 11,
	pages        = 895,
	doi          = {10.1088/0305-4470/11/5/020},
	reportnumber = {PRINT-78-0112 (MANCHESTER)}
}

@article{Gaikwad:2023gef,
	title        = {{Boundary Liouville conformal field theory in four dimensions}},
	author       = {Gaikwad, Adwait and Kislev, Amitay C. and Levy, Tom and Oz, Yaron},
	year         = 2024,
	journal      = {JHEP},
	volume       = {07},
	pages        = 271,
	doi          = {10.1007/JHEP07(2024)271},
	eprint       = {2312.14744},
	archiveprefix = {arXiv},
	primaryclass = {hep-th}
}

@article{Gimenez-Grau:2022ebb,
	title        = {{Probing magnetic line defects with two-point functions}},
	author       = {Gimenez-Grau, Aleix},
	year         = 2022,
	month        = 12,
	eprint       = {2212.02520},
	archiveprefix = {arXiv},
	primaryclass = {hep-th}
}

@article{Giombi:2021uae,
	title        = {{Monodromy defects from hyperbolic space}},
	author       = {Giombi, Simone and Helfenberger, Elizabeth and Ji, Ziming and Khanchandani, Himanshu},
	year         = 2022,
	journal      = {JHEP},
	volume       = {02},
	pages        = {041},
	doi          = {10.1007/JHEP02(2022)041},
	eprint       = {2102.11815},
	archiveprefix = {arXiv},
	primaryclass = {hep-th}
}

@article{Giombi:2026kdz,
	title        = {{Bouncing singularities and thermal correlators on line defects}},
	author       = {Giombi, Simone and Li, Yue-Zhou and Shan, Jieru},
	year         = 2026,
	month        = 3,
	eprint       = {2603.11012},
	archiveprefix = {arXiv},
	primaryclass = {hep-th}
}

@article{Gobeil:2018fzy,
	title        = {{Thermal Conformal Blocks}},
	author       = {Gobeil, Yan and Maloney, Alexander and Ng, Gim Seng and Wu, Jie-qiang},
	year         = 2019,
	journal      = {SciPost Phys.},
	volume       = 7,
	number       = 2,
	pages        = {015},
	doi          = {10.21468/SciPostPhys.7.2.015},
	eprint       = {1802.10537},
	archiveprefix = {arXiv},
	primaryclass = {hep-th}
}

@article{Graham:1992,
	title        = {Conformally invariant powers of the {Laplacian}, {I}: Existence},
	author       = {Graham, C Robin and Jenne, Ralph and Mason, Lionel J and Sparling, George AJ},
	year         = 1992,
	journal      = {Journal of the London Mathematical Society},
	publisher    = {Oxford Academic},
	volume       = 2,
	number       = 3,
	pages        = {557--565},
	doi          = {10.1112/jlms/s2-46.3.557}
}

@article{Iliesiu:2018fao,
	title        = {{The Conformal Bootstrap at Finite Temperature}},
	author       = {Iliesiu, Luca and Kolo\u{g}lu, Murat and Mahajan, Raghu and Perlmutter, Eric and Simmons-Duffin, David},
	year         = 2018,
	journal      = {JHEP},
	volume       = 10,
	pages        = {070},
	doi          = {10.1007/JHEP10(2018)070},
	eprint       = {1802.10266},
	archiveprefix = {arXiv},
	primaryclass = {hep-th},
	reportnumber = {CALT-TH-2018-013, PUPT-2550}
}

@article{Jensen:2018rxu,
	title        = {{From the Weyl Anomaly to Entropy of Two-Dimensional Boundaries and Defects}},
	author       = {Jensen, Kristan and O'Bannon, Andy and Robinson, Brandon and Rodgers, Ronnie},
	year         = 2019,
	journal      = {Phys. Rev. Lett.},
	volume       = 122,
	number       = 24,
	pages        = 241602,
	doi          = {10.1103/PhysRevLett.122.241602},
	eprint       = {1812.08745},
	archiveprefix = {arXiv},
	primaryclass = {hep-th}
}

@article{Kapustin:2005py,
	title        = {{Wilson-'t Hooft operators in four-dimensional gauge theories and S-duality}},
	author       = {Kapustin, Anton},
	year         = 2006,
	journal      = {Phys. Rev. D},
	volume       = 74,
	pages        = {025005},
	doi          = {10.1103/PhysRevD.74.025005},
	eprint       = {hep-th/0501015},
	archiveprefix = {arXiv},
	reportnumber = {CALT-68-2536}
}

@article{Kennedy:1979ar,
	title        = {{Finite Temperature Field Theory with Boundaries: Stress Tensor and Surface Action Renormalization}},
	author       = {Kennedy, Gerard and Critchley, Raymond and Dowker, J. S.},
	year         = 1980,
	journal      = {Annals Phys.},
	volume       = 125,
	pages        = 346,
	doi          = {10.1016/0003-4916(80)90138-4},
	reportnumber = {Print-79-0519 (MANCHESTER)}
}

@article{Lauria:2018klo,
	title        = {{Spinning operators and defects in conformal field theory}},
	author       = {Lauria, Edoardo and Meineri, Marco and Trevisani, Emilio},
	year         = 2019,
	journal      = {JHEP},
	volume       = {08},
	pages        = {066},
	doi          = {10.1007/JHEP08(2019)066},
	eprint       = {1807.02522},
	archiveprefix = {arXiv},
	primaryclass = {hep-th}
}

@article{Lewkowycz:2014jia,
	title        = {{Universality in the geometric dependence of Renyi entropy}},
	author       = {Lewkowycz, Aitor and Perlmutter, Eric},
	year         = 2015,
	journal      = {JHEP},
	volume       = {01},
	pages        = {080},
	doi          = {10.1007/JHEP01(2015)080},
	eprint       = {1407.8171},
	archiveprefix = {arXiv},
	primaryclass = {hep-th}
}

@article{Marchetto:2023fcw,
	title        = {{Broken (super) conformal Ward identities at finite temperature}},
	author       = {Marchetto, Enrico and Miscioscia, Alessio and Pomoni, Elli},
	year         = 2023,
	month        = 6,
	journal      = {JHEP},
	volume       = 12,
	pages        = 186,
	doi          = {10.1007/JHEP12(2023)186},
	eprint       = {2306.12417},
	archiveprefix = {arXiv},
	primaryclass = {hep-th},
	reportnumber = {DESY-23-083}
}

@article{Marchetto:2023xap,
	title        = {{Sum rules {\&} Tauberian theorems at finite temperature}},
	author       = {Marchetto, Enrico and Miscioscia, Alessio and Pomoni, Elli},
	year         = 2024,
	journal      = {JHEP},
	volume       = {09},
	pages        = {044},
	doi          = {10.1007/JHEP09(2024)044},
	eprint       = {2312.13030},
	archiveprefix = {arXiv},
	primaryclass = {hep-th},
	reportnumber = {DESY-23-224}
}

@article{McAvity:1993ue,
	title        = {{Energy momentum tensor in conformal field theories near a boundary}},
	author       = {McAvity, D. M. and Osborn, H.},
	year         = 1993,
	journal      = {Nucl. Phys. B},
	volume       = 406,
	pages        = {655--680},
	doi          = {10.1016/0550-3213(93)90005-A},
	eprint       = {hep-th/9302068},
	archiveprefix = {arXiv},
	reportnumber = {DAMTP-93-01}
}

@article{McAvity:1995zd,
	title        = {{Conformal field theories near a boundary in general dimensions}},
	author       = {McAvity, D. M. and Osborn, H.},
	year         = 1995,
	journal      = {Nucl. Phys. B},
	volume       = 455,
	pages        = {522--576},
	doi          = {10.1016/0550-3213(95)00476-9},
	eprint       = {cond-mat/9505127},
	archiveprefix = {arXiv},
	reportnumber = {DAMTP-95-1, UBC-TP-95-002}
}

@article{Niarchos:2025cdg,
	title        = {{Deep finite temperature bootstrap}},
	author       = {Niarchos, V. and Papageorgakis, C. and Stratoudakis, A. and Woolley, M.},
	year         = 2025,
	journal      = {Phys. Rev. D},
	volume       = 112,
	number       = 12,
	pages        = 126012,
	doi          = {10.1103/qrjx-w4md},
	eprint       = {2508.08560},
	archiveprefix = {arXiv},
	primaryclass = {hep-th},
	reportnumber = {CCTP-2025-8, QMUL-PH-25-16, ITCP-2025/8}
}

@article{Osborn:2016bev,
	title        = {{C$_{T}$ for non-unitary CFTs in higher dimensions}},
	author       = {Osborn, Hugh and Stergiou, Andreas},
	year         = 2016,
	journal      = {JHEP},
	volume       = {06},
	pages        = {079},
	doi          = {10.1007/JHEP06(2016)079},
	eprint       = {1603.07307},
	archiveprefix = {arXiv},
	primaryclass = {hep-th},
	reportnumber = {DAMTP-2016-25}
}

@article{Paci:2025pxo,
	title        = {{Auxiliary-field formalism for higher-derivative boundary CFTs}},
	author       = {Paci, Gregorio and Solodukhin, Sergey N.},
	year         = 2026,
	journal      = {Nucl. Phys. B},
	volume       = 1026,
	pages        = 117452,
	doi          = {10.1016/j.nuclphysb.2026.117452},
	eprint       = {2512.18017},
	archiveprefix = {arXiv},
	primaryclass = {hep-th}
}

@article{Sachdev:1993pr,
	title        = {{Polylogarithm identities in a conformal field theory in three-dimensions}},
	author       = {Sachdev, Subir},
	year         = 1993,
	journal      = {Phys. Lett. B},
	volume       = 309,
	pages        = {285--288},
	doi          = {10.1016/0370-2693(93)90935-B},
	eprint       = {hep-th/9305131},
	archiveprefix = {arXiv},
	reportnumber = {PRINT-93-0431 (YALE)}
}

@article{Stergiou:2022qqj,
	title        = {{Weyl covariance and the energy momentum tensors of higher-derivative free conformal field theories}},
	author       = {Stergiou, Andreas and Vacca, Gian Paolo and Zanusso, Omar},
	year         = 2022,
	journal      = {JHEP},
	volume       = {06},
	pages        = 104,
	doi          = {10.1007/JHEP06(2022)104},
	eprint       = {2202.04701},
	archiveprefix = {arXiv},
	primaryclass = {hep-th}
}

@article{Sun:2026mib,
	title        = {{Analytic Bootstrap for $O(N)$ Boundary Conformal Field Theories with Interacting Boundaries}},
	author       = {Sun, Xinyu and Jian, Shao-Kai and Yao, Hong},
	year         = 2026,
	month        = 5,
	eprint       = {2605.28933},
	archiveprefix = {arXiv},
	primaryclass = {hep-th}
}

@article{Kubo:1957mj,
	title        = {{Statistical mechanical theory of irreversible processes. 1. General theory and simple applications in magnetic and conduction problems}},
	author       = {Kubo, Ryogo},
	year         = 1957,
	journal      = {J. Phys. Soc. Jap.},
	volume       = 12,
	pages        = {570--586},
	doi          = {10.1143/JPSJ.12.570}
}

@article{Martin:1959jp,
	title        = {{Theory of many particle systems. 1.}},
	author       = {Martin, Paul C. and Schwinger, Julian S.},
	year         = 1959,
	journal      = {Phys. Rev.},
	volume       = 115,
	pages        = {1342--1373},
	doi          = {10.1103/PhysRev.115.1342},
	editor       = {Milton, K. A.}
}

@article{Artico:2026pqp,
	title        = {{New Exotic Operators in the Spectrum of Wilson Lines in General Representations}},
	author       = {Artico, Daniele and Meneghelli, Carlo and Savi, Michele and Treilis, Rudolfs},
	year         = 2026,
	month        = 6,
	eprint       = {2606.07738},
	archiveprefix = {arXiv},
	primaryclass = {hep-th}
}

@article{Artico:2024wut,
	title        = {{Perturbative bootstrap of the Wilson-line defect CFT: multipoint correlators}},
	author       = {Artico, Daniele and Barrat, Julien and Peveri, Giulia},
	year         = 2025,
	journal      = {JHEP},
	volume       = {02},
	pages        = 190,
	doi          = {10.1007/JHEP02(2025)190},
	eprint       = {2410.08271},
	archiveprefix = {arXiv},
	primaryclass = {hep-th},
	reportnumber = {HU-EP-24/28-RTG | DESY-24-165, HU-EP-24/28-RTG}
}

@article{Ferrero:2021bsb,
	title        = {{Bootstrapping the half-BPS line defect CFT in N=4 supersymmetric Yang-Mills theory at strong coupling}},
	author       = {Ferrero, Pietro and Meneghelli, Carlo},
	year         = 2021,
	journal      = {Phys. Rev. D},
	volume       = 104,
	number       = 8,
	pages        = {L081703},
	doi          = {10.1103/PhysRevD.104.L081703},
	eprint       = {2103.10440},
	archiveprefix = {arXiv},
	primaryclass = {hep-th}
}

@article{Cardy:1984bb,
	title        = {{Conformal Invariance and Surface Critical Behavior}},
	author       = {Cardy, John L.},
	year         = 1984,
	journal      = {Nucl. Phys. B},
	volume       = 240,
	pages        = {514--532},
	doi          = {10.1016/0550-3213(84)90241-4}
}

@article{Chalabi:2025nbg,
	title        = {{Integrable corners in the space of Gukov-Witten surface defects}},
	author       = {Chalabi, Adam and Kristjansen, Charlotte and Su, Chenliang},
	year         = 2025,
	journal      = {Phys. Lett. B},
	volume       = 866,
	pages        = 139512,
	doi          = {10.1016/j.physletb.2025.139512},
	eprint       = {2503.22598},
	archiveprefix = {arXiv},
	primaryclass = {hep-th}
}

@article{Gukov:2006jk,
	title        = {{Gauge Theory, Ramification, And The Geometric Langlands Program}},
	author       = {Gukov, Sergei and Witten, Edward},
	year         = 2006,
	month        = 12,
	eprint       = {hep-th/0612073},
	archiveprefix = {arXiv}
}

@article{Barrat:2022psm,
	title        = {{A dispersion relation for defect CFT}},
	author       = {Barrat, Julien and Gimenez-Grau, Aleix and Liendo, Pedro},
	year         = 2023,
	journal      = {JHEP},
	volume       = {02},
	pages        = 255,
	doi          = {10.1007/JHEP02(2023)255},
	eprint       = {2205.09765},
	archiveprefix = {arXiv},
	primaryclass = {hep-th},
	reportnumber = {HU-EP-22/18-RTG}
}

@article{Liendo:2016ymz,
	title        = {{Bootstrap equations for $ \mathcal{N} $ = 4 SYM with defects}},
	author       = {Liendo, Pedro and Meneghelli, Carlo},
	year         = 2017,
	journal      = {JHEP},
	volume       = {01},
	pages        = 122,
	doi          = {10.1007/JHEP01(2017)122},
	eprint       = {1608.05126},
	archiveprefix = {arXiv},
	primaryclass = {hep-th}
}

@article{Rattazzi:2008pe,
    author = "Rattazzi, Riccardo and Rychkov, Vyacheslav S. and Tonni, Erik and Vichi, Alessandro",
    title = "{Bounding scalar operator dimensions in 4D CFT}",
    eprint = "0807.0004",
    archivePrefix = "arXiv",
    primaryClass = "hep-th",
    doi = "10.1088/1126-6708/2008/12/031",
    journal = "JHEP",
    volume = "12",
    pages = "031",
    year = "2008"
}

@article{Pappadopulo:2012jk,
    author = "Pappadopulo, Duccio and Rychkov, Slava and Espin, Johnny and Rattazzi, Riccardo",
    title = "{OPE Convergence in Conformal Field Theory}",
    eprint = "1208.6449",
    archivePrefix = "arXiv",
    primaryClass = "hep-th",
    reportNumber = "LPTENS-12-31",
    doi = "10.1103/PhysRevD.86.105043",
    journal = "Phys. Rev. D",
    volume = "86",
    pages = "105043",
    year = "2012"
}

@article{Zamolodchikov:1986gt,
    author = "Zamolodchikov, A. B.",
    title = "{Irreversibility of the Flux of the Renormalization Group in a 2D Field Theory}",
    journal = "JETP Lett.",
    volume = "43",
    pages = "730--732",
    year = "1986"
}

@article{Cardy:1988cwa,
    author = "Cardy, John L.",
    title = "{Is There a c Theorem in Four-Dimensions?}",
    doi = "10.1016/0370-2693(88)90054-8",
    journal = "Phys. Lett. B",
    volume = "215",
    pages = "749--752",
    year = "1988"
}

@article{Komargodski:2011vj,
    author = "Komargodski, Zohar and Schwimmer, Adam",
    title = "{On Renormalization Group Flows in Four Dimensions}",
    eprint = "1107.3987",
    archivePrefix = "arXiv",
    primaryClass = "hep-th",
    doi = "10.1007/JHEP12(2011)099",
    journal = "JHEP",
    volume = "12",
    pages = "099",
    year = "2011"
}

@article{Jafferis:2011zi,
    author = "Jafferis, Daniel L. and Klebanov, Igor R. and Pufu, Silviu S. and Safdi, Benjamin R.",
    title = "{Towards the F-Theorem: N=2 Field Theories on the Three-Sphere}",
    eprint = "1103.1181",
    archivePrefix = "arXiv",
    primaryClass = "hep-th",
    reportNumber = "PUPT-2366",
    doi = "10.1007/JHEP06(2011)102",
    journal = "JHEP",
    volume = "06",
    pages = "102",
    year = "2011"
}

@article{Klebanov:2011gs,
    author = "Klebanov, Igor R. and Pufu, Silviu S. and Safdi, Benjamin R.",
    title = "{F-Theorem without Supersymmetry}",
    eprint = "1105.4598",
    archivePrefix = "arXiv",
    primaryClass = "hep-th",
    reportNumber = "PUPT-2377",
    doi = "10.1007/JHEP10(2011)038",
    journal = "JHEP",
    volume = "10",
    pages = "038",
    year = "2011"
}

@article{Casini:2004bw,
    author = "Casini, H. and Huerta, M.",
    title = "{A Finite entanglement entropy and the c-theorem}",
    eprint = "hep-th/0405111",
    archivePrefix = "arXiv",
    doi = "10.1016/j.physletb.2004.08.072",
    journal = "Phys. Lett. B",
    volume = "600",
    pages = "142--150",
    year = "2004"
}

@article{Casini:2012ei,
    author = "Casini, H. and Huerta, Marina",
    title = "{On the RG running of the entanglement entropy of a circle}",
    eprint = "1202.5650",
    archivePrefix = "arXiv",
    primaryClass = "hep-th",
    doi = "10.1103/PhysRevD.85.125016",
    journal = "Phys. Rev. D",
    volume = "85",
    pages = "125016",
    year = "2012"
}

@article{Casini:2017vbe,
    author = "Casini, Horacio and Test{\'e}, Eduardo and Torroba, Gonzalo",
    title = "{Markov Property of the Conformal Field Theory Vacuum and the a Theorem}",
    eprint = "1704.01870",
    archivePrefix = "arXiv",
    primaryClass = "hep-th",
    doi = "10.1103/PhysRevLett.118.261602",
    journal = "Phys. Rev. Lett.",
    volume = "118",
    number = "26",
    pages = "261602",
    year = "2017"
}

@article{Giombi:2014xxa,
    author = "Giombi, Simone and Klebanov, Igor R.",
    title = "{Interpolating between $a$ and $F$}",
    eprint = "1409.1937",
    archivePrefix = "arXiv",
    primaryClass = "hep-th",
    reportNumber = "PUPT-2472",
    doi = "10.1007/JHEP03(2015)117",
    journal = "JHEP",
    volume = "03",
    pages = "117",
    year = "2015"
}

@article{Heemskerk:2009pn,
    author = "Heemskerk, Idse and Penedones, Joao and Polchinski, Joseph and Sully, James",
    title = "{Holography from Conformal Field Theory}",
    eprint = "0907.0151",
    archivePrefix = "arXiv",
    primaryClass = "hep-th",
    reportNumber = "NSF-KITP-09-110",
    doi = "10.1088/1126-6708/2009/10/079",
    journal = "JHEP",
    volume = "10",
    pages = "079",
    year = "2009"
}

@article{Guo:2025edk,
    author = "Guo, Yongwei and Li, Wenliang",
    title = "{Boundary anomalous dimensions from BCFT: O(N)-symmetric {\ensuremath{\phi}}2n theories with a boundary and higher-derivative generalizations}",
    eprint = "2504.16844",
    archivePrefix = "arXiv",
    primaryClass = "hep-th",
    doi = "10.1103/pqs4-hs43",
    journal = "Phys. Rev. D",
    volume = "112",
    number = "6",
    pages = "065001",
    year = "2025"
}

@article{Guo:2026vmq,
    author = "Guo, Yongwei and Li, Wenliang",
    title = "{Boundary anomalous dimensions from BCFT: $\phi^{3}$ theories with a boundary and higher-derivative generalizations}",
    eprint = "2605.16119",
    archivePrefix = "arXiv",
    primaryClass = "hep-th",
    month = "5",
    year = "2026"
}

@article{Fitzpatrick:2011dm,
    author = "Fitzpatrick, A. Liam and Kaplan, Jared",
    title = "{Unitarity and the Holographic S-Matrix}",
    eprint = "1112.4845",
    archivePrefix = "arXiv",
    primaryClass = "hep-th",
    reportNumber = "SLAC-PUB-14979",
    doi = "10.1007/JHEP10(2012)032",
    journal = "JHEP",
    volume = "10",
    pages = "032",
    year = "2012"
}

@article{David:2023uya,
    author = "David, Justin R. and Kumar, Srijan",
    title = "{Thermal one-point functions: CFT{\textquoteright}s with fermions, large d and large spin}",
    eprint = "2307.14847",
    archivePrefix = "arXiv",
    primaryClass = "hep-th",
    doi = "10.1007/JHEP10(2023)143",
    journal = "JHEP",
    volume = "10",
    pages = "143",
    year = "2023"
}

@article{Herzog:2022jlx,
    author = "Herzog, Christopher P. and Schaub, Vladimir",
    title = "{Fermions in boundary conformal field theory: crossing symmetry and E-expansion}",
    eprint = "2209.05511",
    archivePrefix = "arXiv",
    primaryClass = "hep-th",
    doi = "10.1007/JHEP02(2023)129",
    journal = "JHEP",
    volume = "02",
    pages = "129",
    year = "2023"
}

@article{Penedones:2010ue,
    author = "Penedones, Joao",
    title = "{Writing CFT correlation functions as AdS scattering amplitudes}",
    eprint = "1011.1485",
    archivePrefix = "arXiv",
    primaryClass = "hep-th",
    doi = "10.1007/JHEP03(2011)025",
    journal = "JHEP",
    volume = "03",
    pages = "025",
    year = "2011"
}

@article{Liendo:2012hy,
    author = "Liendo, Pedro and Rastelli, Leonardo and van Rees, Balt C.",
    title = "{The Bootstrap Program for Boundary CFT$_d$}",
    eprint = "1210.4258",
    archivePrefix = "arXiv",
    primaryClass = "hep-th",
    reportNumber = "YITP-SB-12-37",
    doi = "10.1007/JHEP07(2013)113",
    journal = "JHEP",
    volume = "07",
    pages = "113",
    year = "2013"
}

@article{Herzog:2024zxm,
    author = "Herzog, Christopher P. and Zhou, Yanjun",
    title = "{An interacting, higher derivative, boundary conformal field theory}",
    eprint = "2409.11072",
    archivePrefix = "arXiv",
    primaryClass = "hep-th",
    doi = "10.1007/JHEP12(2024)133",
    journal = "JHEP",
    volume = "12",
    pages = "133",
    year = "2024"
}

@article{Affleck:1991tk,
    author = "Affleck, Ian and Ludwig, Andreas W. W.",
    title = "{Universal noninteger 'ground state degeneracy' in critical quantum systems}",
    reportNumber = "UBCTP-91-007",
    doi = "10.1103/PhysRevLett.67.161",
    journal = "Phys. Rev. Lett.",
    volume = "67",
    pages = "161--164",
    year = "1991"
}

@article{Friedan:2003yc,
    author = "Friedan, Daniel and Konechny, Anatoly",
    title = "{On the boundary entropy of one-dimensional quantum systems at low temperature}",
    eprint = "hep-th/0312197",
    archivePrefix = "arXiv",
    reportNumber = "RU-NHETC-2003-39",
    doi = "10.1103/PhysRevLett.93.030402",
    journal = "Phys. Rev. Lett.",
    volume = "93",
    pages = "030402",
    year = "2004"
}

@article{Cuomo:2021rkm,
    author = "Cuomo, Gabriel and Komargodski, Zohar and Raviv-Moshe, Avia",
    title = "{Renormalization Group Flows on Line Defects}",
    eprint = "2108.01117",
    archivePrefix = "arXiv",
    primaryClass = "hep-th",
    doi = "10.1103/PhysRevLett.128.021603",
    journal = "Phys. Rev. Lett.",
    volume = "128",
    number = "2",
    pages = "021603",
    year = "2022"
}

@article{Jensen:2015swa,
    author = "Jensen, Kristan and O'Bannon, Andy",
    title = "{Constraint on Defect and Boundary Renormalization Group Flows}",
    eprint = "1509.02160",
    archivePrefix = "arXiv",
    primaryClass = "hep-th",
    reportNumber = "OUTP-15-19P, YITP-SB-15-33",
    doi = "10.1103/PhysRevLett.116.091601",
    journal = "Phys. Rev. Lett.",
    volume = "116",
    number = "9",
    pages = "091601",
    year = "2016"
}

@article{Shachar:2022fqk,
    author = "Shachar, Tom and Sinha, Ritam and Smolkin, Michael",
    title = "{RG flows on two-dimensional spherical defects}",
    eprint = "2212.08081",
    archivePrefix = "arXiv",
    primaryClass = "hep-th",
    doi = "10.21468/SciPostPhys.15.6.240",
    journal = "SciPost Phys.",
    volume = "15",
    number = "6",
    pages = "240",
    year = "2023"
}

@article{Gaiotto:2014gha,
    author = "Gaiotto, Davide",
    title = "{Boundary F-maximization}",
    eprint = "1403.8052",
    archivePrefix = "arXiv",
    primaryClass = "hep-th",
    month = "3",
    year = "2014"
}

@article{Wang:2021mdq,
    author = "Wang, Yifan",
    title = "{Defect a-theorem and a-maximization}",
    eprint = "2101.12648",
    archivePrefix = "arXiv",
    primaryClass = "hep-th",
    doi = "10.1007/JHEP02(2022)061",
    journal = "JHEP",
    volume = "02",
    pages = "061",
    year = "2022"
}

@article{Kobayashi:2018lil,
    author = "Kobayashi, Nozomu and Nishioka, Tatsuma and Sato, Yoshiki and Watanabe, Kento",
    title = "{Towards a $C$-theorem in defect CFT}",
    eprint = "1810.06995",
    archivePrefix = "arXiv",
    primaryClass = "hep-th",
    reportNumber = "UT-18-22, IPMU18-0166",
    doi = "10.1007/JHEP01(2019)039",
    journal = "JHEP",
    volume = "01",
    pages = "039",
    year = "2019"
}

@article{Casini:2023kyj,
    author = "Casini, Horacio and Salazar Landea, Ignacio and Torroba, Gonzalo",
    title = "{Irreversibility, QNEC, and defects}",
    eprint = "2303.16935",
    archivePrefix = "arXiv",
    primaryClass = "hep-th",
    doi = "10.1007/JHEP07(2023)004",
    journal = "JHEP",
    volume = "07",
    pages = "004",
    year = "2023"
}

@article{Casini:2018nym,
    author = "Casini, Horacio and Salazar Landea, Ignacio and Torroba, Gonzalo",
    title = "{Irreversibility in quantum field theories with boundaries}",
    eprint = "1812.08183",
    archivePrefix = "arXiv",
    primaryClass = "hep-th",
    doi = "10.1007/JHEP04(2019)166",
    journal = "JHEP",
    volume = "04",
    pages = "166",
    year = "2019"
}

@article{Casini:2016udt,
    author = "Casini, Horacio and Teste, Eduardo and Torroba, Gonzalo",
    title = "{Relative entropy and the RG flow}",
    eprint = "1611.00016",
    archivePrefix = "arXiv",
    primaryClass = "hep-th",
    doi = "10.1007/JHEP03(2017)089",
    journal = "JHEP",
    volume = "03",
    pages = "089",
    year = "2017"
}

@article{Casini:2016fgb,
    author = "Casini, Horacio and Salazar Landea, Ignacio and Torroba, Gonzalo",
    title = "{The g-theorem and quantum information theory}",
    eprint = "1607.00390",
    archivePrefix = "arXiv",
    primaryClass = "hep-th",
    doi = "10.1007/JHEP10(2016)140",
    journal = "JHEP",
    volume = "10",
    pages = "140",
    year = "2016"
}

@article{Casini:2022bsu,
    author = "Casini, Horacio and Salazar Landea, Ignacio and Torroba, Gonzalo",
    title = "{Entropic g Theorem in General Spacetime Dimensions}",
    eprint = "2212.10575",
    archivePrefix = "arXiv",
    primaryClass = "hep-th",
    doi = "10.1103/PhysRevLett.130.111603",
    journal = "Phys. Rev. Lett.",
    volume = "130",
    number = "11",
    pages = "111603",
    year = "2023"
}

@article{Nozaki:2012qd,
    author = "Nozaki, Masahiro and Takayanagi, Tadashi and Ugajin, Tomonori",
    title = "{Central Charges for BCFTs and Holography}",
    eprint = "1205.1573",
    archivePrefix = "arXiv",
    primaryClass = "hep-th",
    reportNumber = "YITP-12-42, IPMU12-0087",
    doi = "10.1007/JHEP06(2012)066",
    journal = "JHEP",
    volume = "06",
    pages = "066",
    year = "2012"
}

@article{Li:2026udh,
    author = "Li, Yucheng and Nakayama, Haruki and Nishioka, Tatsuma",
    title = "{Conformal defects of general dimensions at finite temperature}",
    eprint = "2609.12720",
    archivePrefix = "arXiv",
    primaryClass = "hep-th",
    reportNumber = "OU-HET-1327",
    month = "9",
    year = "2026"
}

@article{Parwani:1991gq,
    author = "Parwani, Rajesh R.",
    title = "{Resummation in a hot scalar field theory}",
    eprint = "hep-ph/9204216",
    archivePrefix = "arXiv",
    reportNumber = "ITP-SB-91-64",
    doi = "10.1103/PhysRevD.45.4695",
    journal = "Phys. Rev. D",
    volume = "45",
    pages = "4695",
    year = "1992",
    note = "[Erratum: Phys.Rev.D 48, 5965 (1993)]"
}

@book{Laine:2016hma,
    author = "Laine, Mikko and Vuorinen, Aleksi",
    title = "{Basics of Thermal Field Theory}",
    eprint = "1701.01554",
    archivePrefix = "arXiv",
    primaryClass = "hep-ph",
    doi = "10.1007/978-3-319-31933-9",
    publisher = "Springer",
    volume = "925",
    year = "2016"
}

@article{Dolan:1973qd,
    author = "Dolan, L. and Jackiw, R.",
    title = "{Symmetry Behavior at Finite Temperature}",
    reportNumber = "MIT-CTP-406",
    doi = "10.1103/PhysRevD.9.3320",
    journal = "Phys. Rev. D",
    volume = "9",
    pages = "3320--3341",
    year = "1974"
}

@article{Gaiotto:2013nva,
    author = "Gaiotto, Davide and Mazac, Dalimil and Paulos, Miguel F.",
    title = "{Bootstrapping the 3d Ising twist defect}",
    eprint = "1310.5078",
    archivePrefix = "arXiv",
    primaryClass = "hep-th",
    doi = "10.1007/JHEP03(2014)100",
    journal = "JHEP",
    volume = "03",
    pages = "100",
    year = "2014"
}

@article{Gliozzi:2015qsa,
    author = "Gliozzi, Ferdinando and Liendo, Pedro and Meineri, Marco and Rago, Antonio",
    title = "{Boundary and Interface CFTs from the Conformal Bootstrap}",
    eprint = "1502.07217",
    archivePrefix = "arXiv",
    primaryClass = "hep-th",
    reportNumber = "HU-EP-15-08, HU-EP-15/08",
    doi = "10.1007/JHEP05(2015)036",
    journal = "JHEP",
    volume = "05",
    pages = "036",
    year = "2015",
    note = "[Erratum: JHEP 12, 093 (2021)]"
}

@article{Bissi:2018mcq,
    author = {Bissi, Agnese and Hansen, Tobias and S{\"o}derberg, Alexander},
    title = "{Analytic Bootstrap for Boundary CFT}",
    eprint = "1808.08155",
    archivePrefix = "arXiv",
    primaryClass = "hep-th",
    doi = "10.1007/JHEP01(2019)010",
    journal = "JHEP",
    volume = "01",
    pages = "010",
    year = "2019"
}

@article{Barrat:2026jfg,
    author = "Barrat, Julien and Bozkurt, Deniz N. and Marchetto, Enrico and Miscioscia, Alessio and Pomoni, Elli",
    title = "{Analytic thermal bootstrap in momentum space: From thermal OPE to QNMs}",
    eprint = "2607.24919",
    archivePrefix = "arXiv",
    primaryClass = "hep-th",
    reportNumber = "DESY-26-094, YITP-SB-2026-13",
    month = "7",
    year = "2026"
}

@article{Gaiotto:2014kfa,
    author = "Gaiotto, Davide and Kapustin, Anton and Seiberg, Nathan and Willett, Brian",
    title = "{Generalized Global Symmetries}",
    eprint = "1412.5148",
    archivePrefix = "arXiv",
    primaryClass = "hep-th",
    doi = "10.1007/JHEP02(2015)172",
    journal = "JHEP",
    volume = "02",
    pages = "172",
    year = "2015"
}

@article{El-Showk:2011yvt,
    author = "El-Showk, Sheer and Papadodimas, Kyriakos",
    title = "{Emergent Spacetime and Holographic CFTs}",
    eprint = "1101.4163",
    archivePrefix = "arXiv",
    primaryClass = "hep-th",
    doi = "10.1007/JHEP10(2012)106",
    journal = "JHEP",
    volume = "10",
    pages = "106",
    year = "2012"
}

@article{Katz:2014rla,
    author = "Katz, Emanuel and Sachdev, Subir and S{\o}rensen, Erik S. and Witczak-Krempa, William",
    title = "{Conformal field theories at nonzero temperature: Operator product expansions, Monte Carlo, and holography}",
    eprint = "1409.3841",
    archivePrefix = "arXiv",
    primaryClass = "cond-mat.str-el",
    doi = "10.1103/PhysRevB.90.245109",
    journal = "Phys. Rev. B",
    volume = "90",
    number = "24",
    pages = "245109",
    year = "2014"
}

@article{Iliesiu:2018zlz,
    author = "Iliesiu, Luca and Kolo{\u{g}}lu, Murat and Simmons-Duffin, David",
    title = "{Bootstrapping the 3d Ising model at finite temperature}",
    eprint = "1811.05451",
    archivePrefix = "arXiv",
    primaryClass = "hep-th",
    reportNumber = "CALT-TH-2018-049, PUPT-2573",
    doi = "10.1007/JHEP12(2019)072",
    journal = "JHEP",
    volume = "12",
    pages = "072",
    year = "2019"
}

@book{HardyRiesz1915,
    author    = {Hardy, G. H. and Riesz, Marcel},
    title     = {The General Theory of Dirichlet's Series},
    series    = {Cambridge Tracts in Mathematics and Mathematical Physics},
    number    = {18},
    publisher = {Cambridge University Press},
    address   = {Cambridge},
    year      = {1915}
}

@article{Mack:1975je,
    author = "Mack, G.",
    title = "{All unitary ray representations of the conformal group SU(2,2) with positive energy}",
    reportNumber = "DESY-75-50",
    doi = "10.1007/BF01613145",
    journal = "Commun. Math. Phys.",
    volume = "55",
    pages = "1",
    year = "1977"
}

@article{Mack:1976pa,
    author = "Mack, G.",
    title = "{Convergence of Operator Product Expansions on the Vacuum in Conformal Invariant Quantum Field Theory}",
    reportNumber = "DESY-76-30",
    doi = "10.1007/BF01609130",
    journal = "Commun. Math. Phys.",
    volume = "53",
    pages = "155",
    year = "1977"
}

@article{Polyakov:1974gs,
    author = "Polyakov, A. M.",
    title = "{Non-Hamiltonian approach to conformal quantum field theory}",
    journal = "Zh. Eksp. Teor. Fiz.",
    volume = "66",
    number = "1",
    pages = "23--42",
    year = "1974"
}

@article{Ferrara:1973yt,
    author = "Ferrara, S. and Grillo, A. F. and Gatto, R.",
    title = "{Tensor representations of conformal algebra and conformally covariant operator product expansion}",
    doi = "10.1016/0003-4916(73)90446-6",
    journal = "Annals Phys.",
    volume = "76",
    pages = "161--188",
    year = "1973"
}

@article{Mack:1975jr,
    author = "Mack, G.",
    title = "{Duality in quantum field theory}",
    reportNumber = "DESY-75-44",
    doi = "10.1016/0550-3213(77)90238-3",
    journal = "Nucl. Phys. B",
    volume = "118",
    pages = "445--457",
    year = "1977"
}

@article{Poland:2011ey,
    author = "Poland, David and Simmons-Duffin, David and Vichi, Alessandro",
    title = "{Carving Out the Space of 4D CFTs}",
    eprint = "1109.5176",
    archivePrefix = "arXiv",
    primaryClass = "hep-th",
    doi = "10.1007/JHEP05(2012)110",
    journal = "JHEP",
    volume = "05",
    pages = "110",
    year = "2012"
}

@article{Simmons-Duffin:2015qma,
    author = "Simmons-Duffin, David",
    title = "{A Semidefinite Program Solver for the Conformal Bootstrap}",
    eprint = "1502.02033",
    archivePrefix = "arXiv",
    primaryClass = "hep-th",
    doi = "10.1007/JHEP06(2015)174",
    journal = "JHEP",
    volume = "06",
    pages = "174",
    year = "2015"
}

@article{El-Showk:2012cjh,
    author = "El-Showk, Sheer and Paulos, Miguel F. and Poland, David and Rychkov, Slava and Simmons-Duffin, David and Vichi, Alessandro",
    title = "{Solving the 3D Ising Model with the Conformal Bootstrap}",
    eprint = "1203.6064",
    archivePrefix = "arXiv",
    primaryClass = "hep-th",
    reportNumber = "LPTENS-12-07",
    doi = "10.1103/PhysRevD.86.025022",
    journal = "Phys. Rev. D",
    volume = "86",
    pages = "025022",
    year = "2012"
}

@article{Fitzpatrick:2012yx,
    author = "Fitzpatrick, A. Liam and Kaplan, Jared and Poland, David and Simmons-Duffin, David",
    title = "{The Analytic Bootstrap and AdS Superhorizon Locality}",
    eprint = "1212.3616",
    archivePrefix = "arXiv",
    primaryClass = "hep-th",
    doi = "10.1007/JHEP12(2013)004",
    journal = "JHEP",
    volume = "12",
    pages = "004",
    year = "2013"
}

@article{Komargodski:2012ek,
    author = "Komargodski, Zohar and Zhiboedov, Alexander",
    title = "{Convexity and Liberation at Large Spin}",
    eprint = "1212.4103",
    archivePrefix = "arXiv",
    primaryClass = "hep-th",
    doi = "10.1007/JHEP11(2013)140",
    journal = "JHEP",
    volume = "11",
    pages = "140",
    year = "2013"
}

@article{Caron-Huot:2017vep,
    author = "Caron-Huot, Simon",
    title = "{Analyticity in Spin in Conformal Theories}",
    eprint = "1703.00278",
    archivePrefix = "arXiv",
    primaryClass = "hep-th",
    doi = "10.1007/JHEP09(2017)078",
    journal = "JHEP",
    volume = "09",
    pages = "078",
    year = "2017"
}

@article{Caron-Huot:2020adz,
    author = "Caron-Huot, Simon and Mazac, Dalimil and Rastelli, Leonardo and Simmons-Duffin, David",
    title = "{Dispersive CFT Sum Rules}",
    eprint = "2008.04931",
    archivePrefix = "arXiv",
    primaryClass = "hep-th",
    doi = "10.1007/JHEP05(2021)243",
    journal = "JHEP",
    volume = "05",
    pages = "243",
    year = "2021"
}

@article{Mazac:2019shk,
    author = "Maz{\'a}{\v{c}}, Dalimil and Rastelli, Leonardo and Zhou, Xinan",
    title = "{A basis of analytic functionals for CFTs in general dimension}",
    eprint = "1910.12855",
    archivePrefix = "arXiv",
    primaryClass = "hep-th",
    doi = "10.1007/JHEP08(2021)140",
    journal = "JHEP",
    volume = "08",
    pages = "140",
    year = "2021"
}

@article{Carmi:2019cub,
    author = "Carmi, Dean and Caron-Huot, Simon",
    title = "{A Conformal Dispersion Relation: Correlations from Absorption}",
    eprint = "1910.12123",
    archivePrefix = "arXiv",
    primaryClass = "hep-th",
    doi = "10.1007/JHEP09(2020)009",
    journal = "JHEP",
    volume = "09",
    pages = "009",
    year = "2020"
}

@article{Kos:2016ysd,
    author = "Kos, Filip and Poland, David and Simmons-Duffin, David and Vichi, Alessandro",
    title = "{Precision Islands in the Ising and $O(N)$ Models}",
    eprint = "1603.04436",
    archivePrefix = "arXiv",
    primaryClass = "hep-th",
    reportNumber = "CERN-TH-2016-050",
    doi = "10.1007/JHEP08(2016)036",
    journal = "JHEP",
    volume = "08",
    pages = "036",
    year = "2016"
}

@article{Diehl:1996kd,
    author = "Diehl, H. W.",
    title = "{The Theory of boundary critical phenomena}",
    eprint = "cond-mat/9610143",
    archivePrefix = "arXiv",
    doi = "10.1142/S0217979297001751",
    journal = "Int. J. Mod. Phys. B",
    volume = "11",
    pages = "3503--3523",
    year = "1997"
}

@article{Affleck:1995ge,
    author = "Affleck, Ian",
    editor = "Nowak, Maciej A. and Wegrzyn, Pawel",
    title = "{Conformal field theory approach to the Kondo effect}",
    eprint = "cond-mat/9512099",
    archivePrefix = "arXiv",
    journal = "Acta Phys. Polon. B",
    volume = "26",
    pages = "1869--1932",
    year = "1995"
}

@article{Cardy:1991tv,
    author = "Cardy, John L. and Lewellen, David C.",
    title = "{Bulk and boundary operators in conformal field theory}",
    reportNumber = "UCSB-TH-90-70, NSF-ITP-91-15",
    doi = "10.1016/0370-2693(91)90828-E",
    journal = "Phys. Lett. B",
    volume = "259",
    pages = "274--278",
    year = "1991"
}

@article{Lemos:2017vnx,
    author = "Lemos, Madalena and Liendo, Pedro and Meineri, Marco and Sarkar, Sourav",
    title = "{Universality at large transverse spin in defect CFT}",
    eprint = "1712.08185",
    archivePrefix = "arXiv",
    primaryClass = "hep-th",
    reportNumber = "DESY 17-239, HU-EP-17/31, DESY-17-239, HU-EP-17-31",
    doi = "10.1007/JHEP09(2018)091",
    journal = "JHEP",
    volume = "09",
    pages = "091",
    year = "2018"
}

@article{Mazac:2018biw,
    author = "Maz{\'a}{\v{c}}, Dalimil and Rastelli, Leonardo and Zhou, Xinan",
    title = "{An analytic approach to BCFT$_{d}$}",
    eprint = "1812.09314",
    archivePrefix = "arXiv",
    primaryClass = "hep-th",
    reportNumber = "PUPT-2580; YITP-44",
    doi = "10.1007/JHEP12(2019)004",
    journal = "JHEP",
    volume = "12",
    pages = "004",
    year = "2019"
}

@article{Kaviraj:2018tfd,
    author = "Kaviraj, Apratim and Paulos, Miguel F.",
    title = "{The Functional Bootstrap for Boundary CFT}",
    eprint = "1812.04034",
    archivePrefix = "arXiv",
    primaryClass = "hep-th",
    doi = "10.1007/JHEP04(2020)135",
    journal = "JHEP",
    volume = "04",
    pages = "135",
    year = "2020"
}

@article{Liendo:2019jpu,
    author = "Liendo, Pedro and Linke, Yannick and Schomerus, Volker",
    title = "{A Lorentzian inversion formula for defect CFT}",
    eprint = "1903.05222",
    archivePrefix = "arXiv",
    primaryClass = "hep-th",
    reportNumber = "DESY-19-039",
    doi = "10.1007/JHEP08(2020)163",
    journal = "JHEP",
    volume = "08",
    pages = "163",
    year = "2020"
}

@article{Petkou:2018ynm,
    author = "Petkou, Anastasios C. and Stergiou, Andreas",
    title = "{Dynamics of Finite-Temperature Conformal Field Theories from Operator Product Expansion Inversion Formulas}",
    eprint = "1806.02340",
    archivePrefix = "arXiv",
    primaryClass = "hep-th",
    reportNumber = "CERN-TH-2018-132",
    doi = "10.1103/PhysRevLett.121.071602",
    journal = "Phys. Rev. Lett.",
    volume = "121",
    number = "7",
    pages = "071602",
    year = "2018"
}

@article{Manenti:2019wxs,
    author = "Manenti, Andrea",
    title = "{Thermal CFTs in momentum space}",
    eprint = "1905.01355",
    archivePrefix = "arXiv",
    primaryClass = "hep-th",
    doi = "10.1007/JHEP01(2020)009",
    journal = "JHEP",
    volume = "01",
    pages = "009",
    year = "2020"
}

@article{Alday:2020eua,
    author = "Alday, Luis F. and Kologlu, Murat and Zhiboedov, Alexander",
    title = "{Holographic correlators at finite temperature}",
    eprint = "2009.10062",
    archivePrefix = "arXiv",
    primaryClass = "hep-th",
    reportNumber = "CERN-TH-2020-155",
    doi = "10.1007/JHEP06(2021)082",
    journal = "JHEP",
    volume = "06",
    pages = "082",
    year = "2021"
}

@article{Allameh:2024qqp,
    author = "Allameh, Kuroush and Shaghoulian, Edgar",
    title = "{Modular invariance and thermal effective field theory in CFT}",
    eprint = "2402.13337",
    archivePrefix = "arXiv",
    primaryClass = "hep-th",
    doi = "10.1007/JHEP01(2025)200",
    journal = "JHEP",
    volume = "01",
    pages = "200",
    year = "2025"
}

@article{Benjamin:2023qsc,
    author = "Benjamin, Nathan and Lee, Jaeha and Ooguri, Hirosi and Simmons-Duffin, David",
    title = "{Universal asymptotics for high energy CFT data}",
    eprint = "2306.08031",
    archivePrefix = "arXiv",
    primaryClass = "hep-th",
    reportNumber = "CALT-TH 2023-014, IPMU 23-0020",
    doi = "10.1007/JHEP03(2024)115",
    journal = "JHEP",
    volume = "03",
    pages = "115",
    year = "2024"
}

@article{Sachdev:1999,
author = {Subir Sachdev  and Chiranjeeb Buragohain  and Matthias Vojta },
title = {Quantum Impurity in a Nearly Critical Two-Dimensional Antiferromagnet},
journal = {Science},
volume = {286},
number = {5449},
pages = {2479-2482},
year = {1999},
doi = {10.1126/science.286.5449.2479},
URL = {https://www.science.org/doi/abs/10.1126/science.286.5449.2479}
}

@article{Brandhuber:1998bs,
    author = "Brandhuber, A. and Itzhaki, N. and Sonnenschein, J. and Yankielowicz, S.",
    title = "{Wilson loops in the large N limit at finite temperature}",
    eprint = "hep-th/9803137",
    archivePrefix = "arXiv",
    reportNumber = "TAUP-2499-98",
    doi = "10.1016/S0370-2693(98)00730-8",
    journal = "Phys. Lett. B",
    volume = "434",
    pages = "36--40",
    year = "1998"
}

@article{Rey:1998bq,
    author = "Rey, Soo-Jong and Theisen, Stefan and Yee, Jung-Tay",
    title = "{Wilson-Polyakov loop at finite temperature in large N gauge theory and anti-de Sitter supergravity}",
    eprint = "hep-th/9803135",
    archivePrefix = "arXiv",
    reportNumber = "SNUTP-98-018",
    doi = "10.1016/S0550-3213(98)00471-4",
    journal = "Nucl. Phys. B",
    volume = "527",
    pages = "171--186",
    year = "1998"
}

@article{Witten:1998zw,
    author = "Witten, Edward",
    editor = "Bergstrom, L. and Lindstrom, U.",
    title = "{Anti-de Sitter space, thermal phase transition, and confinement in gauge theories}",
    eprint = "hep-th/9803131",
    archivePrefix = "arXiv",
    reportNumber = "IASSNS-HEP-98-21",
    doi = "10.4310/ATMP.1998.v2.n3.a3",
    journal = "Adv. Theor. Math. Phys.",
    volume = "2",
    pages = "505--532",
    year = "1998"
}

@article{Vojta:2000tld,
    author = "Vojta, Matthias and Buragohain, Chiranjeeb and Sachdev, Subir",
    title = "{Quantum impurity dynamics in two-dimensional antiferromagnets and superconductors}",
    doi = "10.1103/PhysRevB.61.15152",
    journal = "Phys. Rev. B",
    volume = "61",
    number = "22",
    pages = "15152",
    year = "2000"
}

@article{Guo:2026xyl,
    author = "Guo, Yongwei and Li, Zhijin and Shen, Tinghong",
    title = "{Thermal Polyakov bootstrap}",
    eprint = "2609.28627",
    archivePrefix = "arXiv",
    primaryClass = "hep-th",
    month = "9",
    year = "2026"
}

@article{Metlitski:2020cqy,
    author = "Metlitski, Max A.",
    title = "{Boundary criticality of the O(N) model in d = 3 critically revisited}",
    eprint = "2009.05119",
    archivePrefix = "arXiv",
    primaryClass = "cond-mat.str-el",
    doi = "10.21468/SciPostPhys.12.4.131",
    journal = "SciPost Phys.",
    volume = "12",
    number = "4",
    pages = "131",
    year = "2022"
}

@article{Padayasi:2021sik,
    author = "Padayasi, Jaychandran and Krishnan, Abijith and Metlitski, Max A. and Gruzberg, Ilya A. and Meineri, Marco",
    title = "{The extraordinary boundary transition in the 3d O(N) model via conformal bootstrap}",
    eprint = "2111.03071",
    archivePrefix = "arXiv",
    primaryClass = "cond-mat.stat-mech",
    doi = "10.21468/SciPostPhys.12.6.190",
    journal = "SciPost Phys.",
    volume = "12",
    number = "6",
    pages = "190",
    year = "2022"
}

@article{Krishnan:2023cff,
    author = "Krishnan, Abijith and Metlitski, Max A.",
    title = "{A plane defect in the 3d O(N) model}",
    eprint = "2301.05728",
    archivePrefix = "arXiv",
    primaryClass = "cond-mat.str-el",
    doi = "10.21468/SciPostPhys.15.3.090",
    journal = "SciPost Phys.",
    volume = "15",
    number = "3",
    pages = "090",
    year = "2023"
}

@article{Giombi:2018hsx,
    author = "Giombi, Simone and Komatsu, Shota",
    title = "{More Exact Results in the Wilson Loop Defect CFT: Bulk-Defect OPE, Nonplanar Corrections and Quantum Spectral Curve}",
    eprint = "1811.02369",
    archivePrefix = "arXiv",
    primaryClass = "hep-th",
    reportNumber = "PUTP-2572",
    doi = "10.1088/1751-8121/ab046c",
    journal = "J. Phys. A",
    volume = "52",
    number = "12",
    pages = "125401",
    year = "2019"
}

@article{Cavaglia:2021bnz,
    author = "Cavagli{\`a}, Andrea and Gromov, Nikolay and Julius, Julius and Preti, Michelangelo",
    title = "{Integrability and conformal bootstrap: One dimensional defect conformal field theory}",
    eprint = "2107.08510",
    archivePrefix = "arXiv",
    primaryClass = "hep-th",
    doi = "10.1103/PhysRevD.105.L021902",
    journal = "Phys. Rev. D",
    volume = "105",
    number = "2",
    pages = "L021902",
    year = "2022"
}

@article{Grabner:2020nis,
    author = "Grabner, David and Gromov, Nikolay and Julius, Julius",
    title = "{Excited States of One-Dimensional Defect CFTs from the Quantum Spectral Curve}",
    eprint = "2001.11039",
    archivePrefix = "arXiv",
    primaryClass = "hep-th",
    doi = "10.1007/JHEP07(2020)042",
    journal = "JHEP",
    volume = "07",
    pages = "042",
    year = "2020"
}

@article{deLeeuw:2015hxa,
    author = "de Leeuw, Marius and Kristjansen, Charlotte and Zarembo, Konstantin",
    title = "{One-point Functions in Defect CFT and Integrability}",
    eprint = "1506.06958",
    archivePrefix = "arXiv",
    primaryClass = "hep-th",
    reportNumber = "NORDITA-2015-72, UUITP-12-15",
    doi = "10.1007/JHEP08(2015)098",
    journal = "JHEP",
    volume = "08",
    pages = "098",
    year = "2015"
}

@article{Buhl-Mortensen:2015gfd,
    author = "Buhl-Mortensen, Isak and de Leeuw, Marius and Kristjansen, Charlotte and Zarembo, Konstantin",
    title = "{One-point Functions in AdS/dCFT from Matrix Product States}",
    eprint = "1512.02532",
    archivePrefix = "arXiv",
    primaryClass = "hep-th",
    reportNumber = "NORDITA-2015-132, UUITP-26-15",
    doi = "10.1007/JHEP02(2016)052",
    journal = "JHEP",
    volume = "02",
    pages = "052",
    year = "2016"
}

@article{Komatsu:2020sup,
    author = "Komatsu, Shota and Wang, Yifan",
    title = "{Non-perturbative defect one-point functions in planar $\mathcal{N}=4$  super-Yang-Mills}",
    eprint = "2004.09514",
    archivePrefix = "arXiv",
    primaryClass = "hep-th",
    doi = "10.1016/j.nuclphysb.2020.115120",
    journal = "Nucl. Phys. B",
    volume = "958",
    pages = "115120",
    year = "2020"
}

@article{Kristjansen:2023ysz,
    author = "Kristjansen, Charlotte and Zarembo, Konstantin",
    title = "{{\textquoteright}t Hooft loops and integrability}",
    eprint = "2305.03649",
    archivePrefix = "arXiv",
    primaryClass = "hep-th",
    doi = "10.1007/JHEP08(2023)184",
    journal = "JHEP",
    volume = "08",
    pages = "184",
    year = "2023"
}

@article{Buhl-Mortensen:2017ind,
    author = "Buhl-Mortensen, Isak and de Leeuw, Marius and Ipsen, Asger C. and Kristjansen, Charlotte and Wilhelm, Matthias",
    title = "{Asymptotic One-Point Functions in Gauge-String Duality with Defects}",
    eprint = "1704.07386",
    archivePrefix = "arXiv",
    primaryClass = "hep-th",
    doi = "10.1103/PhysRevLett.119.261604",
    journal = "Phys. Rev. Lett.",
    volume = "119",
    number = "26",
    pages = "261604",
    year = "2017"
}

@article{Gombor:2020kgu,
    author = "Gombor, Tamas and Bajnok, Zoltan",
    title = "{Boundary states, overlaps, nesting and bootstrapping AdS/dCFT}",
    eprint = "2004.11329",
    archivePrefix = "arXiv",
    primaryClass = "hep-th",
    doi = "10.1007/JHEP10(2020)123",
    journal = "JHEP",
    volume = "10",
    pages = "123",
    year = "2020"
}

@article{Gombor:2024api,
    author = "Gombor, Tamas and Bajnok, Zolt{\'a}n",
    title = "{Dual overlaps and finite coupling {\textquoteright}t Hooft loops}",
    eprint = "2408.14901",
    archivePrefix = "arXiv",
    primaryClass = "hep-th",
    doi = "10.1007/JHEP12(2024)034",
    journal = "JHEP",
    volume = "12",
    pages = "034",
    year = "2024"
}

@article{Buric:2024kxo,
    author = "Buric, Ilija and Russo, Francesco and Schomerus, Volker and Vichi, Alessandro",
    title = "{Thermal one-point functions and their partial wave decomposition}",
    eprint = "2408.02747",
    archivePrefix = "arXiv",
    primaryClass = "hep-th",
    doi = "10.1007/JHEP12(2024)021",
    journal = "JHEP",
    volume = "12",
    pages = "021",
    year = "2024"
}

@article{Buric:2025uqt,
    author = "Buri{\'c}, Ilija and Mangialardi, Francesco and Russo, Francesco and Schomerus, Volker and Vichi, Alessandro",
    title = "{Heavy-heavy-light asymptotics from thermal correlators}",
    eprint = "2506.21671",
    archivePrefix = "arXiv",
    primaryClass = "hep-th",
    doi = "10.1007/JHEP04(2026)027",
    journal = "JHEP",
    volume = "04",
    pages = "027",
    year = "2026"
}

@article{Buric:2026pes,
    author = "Buri{\'c}, Ilija and Mangialardi, Francesco and Russo, Francesco and Schomerus, Volker and Vichi, Alessandro",
    title = "{Thermal One-point Functions and Asymptotic CFT Data: QFT in AdS}",
    eprint = "2606.17167",
    archivePrefix = "arXiv",
    primaryClass = "hep-th",
    month = "6",
    year = "2026"
}

@article{Diatlyk:2024qpr,
    author = "Diatlyk, Oleksandr and Khanchandani, Himanshu and Popov, Fedor K. and Wang, Yifan",
    title = "{Effective Field Theory of Conformal Boundaries}",
    eprint = "2406.01550",
    archivePrefix = "arXiv",
    primaryClass = "hep-th",
    doi = "10.1103/PhysRevLett.133.261601",
    journal = "Phys. Rev. Lett.",
    volume = "133",
    number = "26",
    pages = "261601",
    year = "2024"
}

@article{Diatlyk:2024zkk,
    author = "Diatlyk, Oleksandr and Khanchandani, Himanshu and Popov, Fedor K. and Wang, Yifan",
    title = "{Defect fusion and Casimir energy in higher dimensions}",
    eprint = "2404.05815",
    archivePrefix = "arXiv",
    primaryClass = "hep-th",
    doi = "10.1007/JHEP09(2024)006",
    journal = "JHEP",
    volume = "09",
    pages = "006",
    year = "2024"
}

@article{Kravchuk:2024qoh,
    author = "Kravchuk, Petr and Radcliffe, Alex and Sinha, Ritam",
    title = "{Effective theory for fusion of conformal defects}",
    eprint = "2406.04561",
    archivePrefix = "arXiv",
    primaryClass = "hep-th",
    doi = "10.1088/1751-8121/ae14c5",
    journal = "J. Phys. A",
    volume = "58",
    number = "46",
    pages = "465402",
    year = "2025"
}

@article{Calabrese:2006rx,
    author = "Calabrese, Pasquale and Cardy, John L.",
    title = "{Time-dependence of correlation functions following a quantum quench}",
    eprint = "cond-mat/0601225",
    archivePrefix = "arXiv",
    doi = "10.1103/PhysRevLett.96.136801",
    journal = "Phys. Rev. Lett.",
    volume = "96",
    pages = "136801",
    year = "2006"
}

@article{Kawamoto:2022etl,
    author = "Kawamoto, Taishi and Mori, Takato and Suzuki, Yu-ki and Takayanagi, Tadashi and Ugajin, Tomonori",
    title = "{Holographic local operator quenches in BCFTs}",
    eprint = "2203.03851",
    archivePrefix = "arXiv",
    primaryClass = "hep-th",
    reportNumber = "KEK-TH-2399, YITP-22-21, IPMU22-0005",
    doi = "10.1007/JHEP05(2022)060",
    journal = "JHEP",
    volume = "05",
    pages = "060",
    year = "2022"
}

@article{Bianchi:2022ulu,
    author = "Bianchi, Lorenzo and De Angelis, Stefano and Meineri, Marco",
    title = "{Radiation, entanglement and islands from a boundary local quench}",
    eprint = "2203.10103",
    archivePrefix = "arXiv",
    primaryClass = "hep-th",
    reportNumber = "QMUL-PH-22-10",
    doi = "10.21468/SciPostPhys.14.6.148",
    journal = "SciPost Phys.",
    volume = "14",
    number = "6",
    pages = "148",
    year = "2023"
}

@article{Bianchi:2025fzs,
    author = "Bianchi, Lorenzo and Mattiello, Andrea and Sisti, Jacopo",
    title = "{The entropy of radiation for local quenches in higher dimensions}",
    eprint = "2502.00105",
    archivePrefix = "arXiv",
    primaryClass = "hep-th",
    doi = "10.1007/JHEP06(2025)038",
    journal = "JHEP",
    volume = "06",
    pages = "038",
    year = "2025"
}

@article{Nozaki:2014hna,
    author = "Nozaki, Masahiro and Numasawa, Tokiro and Takayanagi, Tadashi",
    title = "{Quantum Entanglement of Local Operators in Conformal Field Theories}",
    eprint = "1401.0539",
    archivePrefix = "arXiv",
    primaryClass = "hep-th",
    reportNumber = "IPMU13-0244; YITP-13-132, IPMU13-0244, YITP-13-132",
    doi = "10.1103/PhysRevLett.112.111602",
    journal = "Phys. Rev. Lett.",
    volume = "112",
    pages = "111602",
    year = "2014"
}

@article{Caputa:2014eta,
    author = "Caputa, Pawel and Sim{\'o}n, Joan and {\v{S}}tikonas, Andrius and Takayanagi, Tadashi",
    title = "{Quantum Entanglement of Localized Excited States at Finite Temperature}",
    eprint = "1410.2287",
    archivePrefix = "arXiv",
    primaryClass = "hep-th",
    reportNumber = "YITP-14-76, IPMU14-0311, YITP-14-76, IPMU14-0311",
    doi = "10.1007/JHEP01(2015)102",
    journal = "JHEP",
    volume = "01",
    pages = "102",
    year = "2015"
}

@article{Drukker:2022pxk,
    author = "Drukker, Nadav and Kong, Ziwen and Sakkas, Georgios",
    title = "{Broken Global Symmetries and Defect Conformal Manifolds}",
    eprint = "2203.17157",
    archivePrefix = "arXiv",
    primaryClass = "hep-th",
    doi = "10.1103/PhysRevLett.129.201603",
    journal = "Phys. Rev. Lett.",
    volume = "129",
    number = "20",
    pages = "201603",
    year = "2022"
}

@article{Lauria:2017wav,
    author = "Lauria, Edoardo and Meineri, Marco and Trevisani, Emilio",
    title = "{Radial coordinates for defect CFTs}",
    eprint = "1712.07668",
    archivePrefix = "arXiv",
    primaryClass = "hep-th",
    doi = "10.1007/JHEP11(2018)148",
    journal = "JHEP",
    volume = "11",
    pages = "148",
    year = "2018"
}

@article{Billo:2013jda,
    author = "Bill{\'o}, M. and Caselle, M. and Gaiotto, D. and Gliozzi, F. and Meineri, M. and Pellegrini, R.",
    title = "{Line defects in the 3d Ising model}",
    eprint = "1304.4110",
    archivePrefix = "arXiv",
    primaryClass = "hep-th",
    doi = "10.1007/JHEP07(2013)055",
    journal = "JHEP",
    volume = "07",
    pages = "055",
    year = "2013"
}

@article{Artico:2024wnt,
    author = "Artico, Daniele and Barrat, Julien and Xu, Yingxuan",
    title = "{Perturbative bootstrap of the Wilson-line defect CFT: Bulk-defect-defect correlators}",
    eprint = "2410.08273",
    archivePrefix = "arXiv",
    primaryClass = "hep-th",
    reportNumber = "HU-EP-24/27-RTG | DESY-24-166, HU-EP-24/27-RTG",
    doi = "10.1007/JHEP03(2025)191",
    journal = "JHEP",
    volume = "03",
    pages = "191",
    year = "2025"
}

@article{Brust:2016gjy,
    author = "Brust, Christopher and Hinterbichler, Kurt",
    title = "{Free {\ensuremath{\square}}$^{k}$ scalar conformal field theory}",
    eprint = "1607.07439",
    archivePrefix = "arXiv",
    primaryClass = "hep-th",
    doi = "10.1007/JHEP02(2017)066",
    journal = "JHEP",
    volume = "02",
    pages = "066",
    year = "2017"
}

@article{Bianchi:2020hsz,
    author = "Bianchi, Lorenzo and Bliard, Gabriel and Forini, Valentina and Griguolo, Luca and Seminara, Domenico",
    title = "{Analytic bootstrap and Witten diagrams for the ABJM Wilson line as defect CFT$_{1}$}",
    eprint = "2004.07849",
    archivePrefix = "arXiv",
    primaryClass = "hep-th",
    reportNumber = "HU-EP-20/07",
    doi = "10.1007/JHEP08(2020)143",
    journal = "JHEP",
    volume = "08",
    pages = "143",
    year = "2020"
}

@article{Bianchi:2022sbz,
    author = "Bianchi, Lorenzo and Bonomi, Davide and de Sabbata, Elia",
    title = "{Analytic bootstrap for the localized magnetic field}",
    eprint = "2212.02524",
    archivePrefix = "arXiv",
    primaryClass = "hep-th",
    doi = "10.1007/JHEP04(2023)069",
    journal = "JHEP",
    volume = "04",
    pages = "069",
    year = "2023"
}

@article{vanTonder:2006ye,
    author = "van Tonder, Andre and Dorca, Miquel",
    title = "{Non-perturbative quantization of phantom and ghost theories: Relating definite and indefinite representations}",
    eprint = "hep-th/0610185",
    archivePrefix = "arXiv",
    doi = "10.1142/S0217751X07036580",
    journal = "Int. J. Mod. Phys. A",
    volume = "22",
    pages = "2563--2608",
    year = "2007"
}
